\documentstyle[11pt,epsf]{article}
\def\la{\mathrel{\mathpalette\fun <}}
\def\ga{\mathrel{\mathpalette\fun >}}
\def\fun#1#2{\lower3.6pt\vbox{\baselineskip0pt\lineskip.9pt
  \ialign{$\mathsurround=0pt#1\hfil##\hfil$\crcr#2\crcr\sim\crcr}}}

\def\cm{\,{\rm cm}}
\def\km{\,{\rm km}}

\def\yr{\,{\rm yr}}

\def\pc{\,{\rm pc}}
\def\kpc{\,{\rm kpc}}
\def\mpc{\,{\rm Mpc}}

\def\ev{\,{\rm eV}}

\def\gev{\,{\rm GeV}}
\def\tev{\,{\rm TeV}}
\def\G{\,{\rm G}}
\def\sr{\,{\rm sr}}
\def\km{\,{\rm km}}
\def\sec{\,{\rm sec}}

\newcommand{\beq}{\begin{equation}}
\newcommand{\eeq}{\end{equation}}
\newcommand{\beqarray}{\begin{eqnarray}}
\newcommand{\eeqarray}{\end{eqnarray}}

\begin{document}

\begin{titlepage}
\hbox to\hsize{To appear in Physics Reports\hfil E-Print
astro-ph/9811011}

\begin{center}

{\Large \bf Origin and Propagation of Extremely High Energy Cosmic Rays\\}

\vspace{.3in}

{\bf Pijushpani Bhattacharjee}\footnote{e-mail: 
pijush@iiap.ernet.in}\\

\vspace{.2in}

Indian Institute of Astrophysics, Bangalore-560 034, India.\\

\vspace{.3in}

{\bf G\"unter Sigl}\footnote{e-mail: sigl@humboldt.uchicago.edu}\\

\vspace{.2in}

Astronomy \& Astrophysics Center,
Enrico Fermi Institute, University of Chicago,\\
5640 South Ellis Avenue, Chicago, IL 60637, USA\\

\end{center}

\vspace{.2in}

\begin{abstract}
Cosmic ray particles with energies in excess of $10^{20}\,$eV have 
been detected. The sources as well as the physical mechanism(s) responsible 
for endowing cosmic ray particles with such enormous energies are unknown. 
This report gives a review of the physics and astrophysics associated with
the questions of origin and propagation of these Extremely High Energy
(EHE) cosmic rays in the
Universe. After a brief review of the
observed cosmic rays in general and their possible sources and
acceleration mechanisms, a detailed discussion is given of possible 
``top-down'' ({\it non-acceleration}) scenarios of origin of EHE cosmic
rays through {\it decay} of sufficiently massive particles originating from
processes in the early Universe. The massive particles can
come from collapse and/or annihilation of cosmic topological defects
(such as monopoles, cosmic strings, etc.) associated with Grand Unified
Theories or they could be some long-lived metastable
supermassive relic particles that were created in the early Universe and
are decaying in the current epoch. The highest 
energy end of the cosmic ray spectrum can thus be used as a 
probe of new fundamental physics beyond Standard Model. 
We discuss the role of existing and proposed 
cosmic ray, gamma-ray and neutrino experiments in this context. 
We also discuss how observations with next generation experiments
of images and spectra of EHE cosmic ray sources 
can be used to obtain new information on Galactic and
extragalactic magnetic fields and possibly their origin.
\end{abstract}

\end{titlepage}

\tableofcontents

\newpage

\section{Introduction and Scope of This Review} 
The cosmic rays (CR) of Extremely High Energy (EHE) --- those with energy
$\ga10^{20}\ev$~\cite{volcano,sugar,haverah,yakutsk,fe,agasa}  --- 
pose a serious
challenge for conventional theories of origin of CR based on acceleration
of charged particles in powerful astrophysical objects. The question of
origin of these extremely high energy cosmic rays (EHECR)\footnote{We
shall use the abbreviation EHE to specifically denote energies
$E\ga10^{20}\ev$, while the abbreviation UHE for ``Ultra-High Energy''
will sometimes be used to denote $E\ga$ 1 EeV, where 1 EeV =
$10^{18}\ev$. Clearly UHE includes EHE but not vice versa.} is, therefore,
currently a subject of much intense debate and
discussions; see Refs.~\cite{icrr90,icrr96,owl-proc}, and
Ref.~\cite{cronin} for a recent brief review.

The current theories of origin of EHECR can be broadly categorized
into two distinct ``scenarios'': the ``bottom-up'' acceleration
scenario, and the ``top-down'' decay scenario, with various different models
within each scenario. As the names suggest, the
two scenarios are in a sense exact opposite of each other. In the
bottom-up scenario, charged particles are accelerated from lower energies
to the requisite high energies in certain special astrophysical
environments. Examples are acceleration in shocks associated with supernova
remnants, active galactic nuclei (AGNs), powerful radio galaxies, and so
on, or acceleration in the strong electric fields generated 
by rotating neutron stars with high surface magnetic fields, for
example. In the top-down scenario, on the other hand, the energetic
particles arise simply from decay of certain sufficiently massive
particles originating from physical processes in the early
Universe, and no acceleration mechanism is needed. 

The problems encountered in trying to explain the EHECR in terms of
acceleration mechanisms have been well-documented in a number of studies;
see, e.g., Refs.~\cite{hillas-araa,ssb,norman,blandford}. Even if it is
possible, in principle, to accelerate particles to EHECR energies of order
100 EeV in some astrophysical sources, it is generally extremely difficult
in most cases to get the particles come out of the dense regions in and/or 
around the sources without losing much energy. Currently, the most
favorable sources in this
regard are perhaps a class of powerful radio galaxies (see, e.g., 
Refs.~\cite{takahara-rev,bier-rev1,bier-rev2,bier-rev3,proth-rev,kirk-duffy}
for recent reviews and 
references to literature), although the values of the relevant parameters 
required for acceleration to energies $\ga$ 100 EeV are somewhat on the
side of extreme~\cite{norman}. However, even
if the requirements of energetics are met, the main problem with radio
galaxies as sources of EHECR is that most of them seem to 
lie at large cosmological distances, $\gg$ 100 Mpc, from Earth. 
This is a major problem if EHECR
particles are conventional particles such as nucleons or heavy nuclei. The
reason is that nucleons above $\simeq$ 70 EeV lose energy drastically
during their propagation from the source to Earth due to
Greisen-Zatsepin-Kuzmin (GZK)
effect~\cite{greisen,zat-kuz}, namely, 
photo-production of pions when the nucleons collide with photons of the
cosmic microwave background (CMB), the mean-free path for which is $\sim$ 
few Mpc~\cite{stecker-gzk}. This process limits the possible
distance of any source of EHE nucleons to $\la$ 100 Mpc.
If the particles were heavy nuclei, 
they would be photo-disintegrated~\cite{stecker-nuc1,psb} in
the CMB and infrared (IR) background within similar 
distances (see Sect.~4 for details). Thus, nucleons or heavy
nuclei originating in distant radio galaxies are unlikely to survive with
EHECR energies at Earth with any significant flux, even if they were 
accelerated to energies of order 100 EeV at source. In addition, 
since EHECR are hardly deflected by the intergalactic and/or Galactic
magnetic fields, their arrival directions should point back to their
sources in the sky (see Sect.~4 for details). Thus, EHECR offer
us the unique opportunity of doing
charged particle astronomy. Yet, for the observed EHECR events so far, no
powerful sources along the arrival directions of individual events
are found within about 100 Mpc~\cite{elb-som,ssb}.\footnote{Very recently,
it has been suggested by Boldt and Ghosh~\cite{boldt-ghosh} that particles 
may be accelerated to energies $\sim10^{21}\ev$ near the event horizons of
spinning supermassive black holes associated with presently {\it inactive}
quasar remnants whose numbers within the local cosmological universe
(i.e., within a GZK distance of order 50 Mpc) may be sufficient to explain
the observed EHECR flux. This would solve the problem of absence of
suitable currently {\it active} sources associated with EHECR. A detailed
model incorporating this suggestion, however, remains to be worked out.} 

There are, of course,
ways to avoid the distance restriction imposed by the GZK effect,
provided
the problem of energetics is somehow solved separately and provided one
allows new physics beyond the Standard Model of particle physics; we shall
discuss those suggestions later in this review. 

In the top-down scenario, on the other hand, the problem of energetics is
trivially solved from the beginning. Here, the EHECR particles owe their
origin to decay of some supermassive ``X'' particles of mass
$m_X\gg10^{20}\ev$, so that their decay products, envisaged as
the EHECR particles, can have energies all the way up to $\sim m_X$. Thus,
no acceleration mechanism is needed. The sources of the massive X
particles could be topological defects such as cosmic strings or magnetic
monopoles that could be produced in the early Universe during 
symmetry-breaking phase transitions envisaged in Grand Unified
Theories (GUTs). In an inflationary early Universe, the relevant
topological defects could be formed at a phase transition at the end of
inflation. Alternatively, the X particles could be certain
supermassive metastable relic particles of lifetime comparable to
or larger than the age of the Universe, which could be
produced in the early Universe through, for example, particle production
processes associated with inflation. Absence of nearby powerful
astrophysical objects such as AGNs or radio galaxies is not a problem in
the top-down scenario because the X particles or their sources need not
necessarily be associated with any specific active astrophysical objects.
In certain models, the X particles
themselves or their sources may be clustered in galactic halos, in
which case the dominant contribution to the EHECR observed at Earth would
come from the X particles clustered within our Galactic Halo, for which 
the GZK restriction on source distance would be of no concern. 

In this report we review our current understanding of some of the major
theoretical issues concerning the origin
and propagation of EHECR with special emphasis on the top-down scenario of
EHECR origin. The principal reason for focusing primarily on the
top-down scenario is that there already exists a large number of
excellent reviews which discuss the question of origin of ultra-high
energy cosmic rays (UHECR) in general and of EHECR in particular within
the general bottom-up acceleration scenario in details; see, e.g.,
Refs.~\cite{hillas-araa,wolf-wdow,bier-rev1,bier-rev2,bier-rev3,proth-rev,
kirk-duffy}. However, for completeness, we shall briefly discuss the
salient features
of the standard acceleration mechanisms and the predicted maximum
energy achievable for various proposed sources of UHECR. 

We would like to emphasize here that this is
primarily a theoretical review; we do not discuss the experimental issues
(for the obvious reason of lack of expertise), although, again, for
completeness, we shall mention the major experimental techniques
and briefly review the experimental {\it results}
concerning the EHECR. For an excellent
historical account of the early experimental developments and the first
claim of detection of an EHECR event, see Ref.~\cite{linsley-owl}. For
reviews on UHECR experiments in general
and various kinds of experimental techniques used in detecting UHECR, see 
Refs.~\cite{sokolsky-book,sokolsky-physrep,rao-sreek-book}.  
For a review of the current experimental
situation concerning EHECR, see the recent review by Yoshida and
Dai~\cite{yoshida-dai}. Overviews of the various currently operating,
up-coming, as well as proposed future EHECR experiments can be found,
e.g., in Refs.~\cite{icrr96,owl-proc}.  

By focusing primarily on the top-down scenario, we do not wish to give
the wrong impression that the top-down scenario explains all
aspects of EHECR. In fact, as we shall see below, essentially each of the 
specific top-down models that have been studied so
far has its own peculiar set of problems. Indeed, the main problem of
the top-down scenario in general is that it is highly model dependent and
invariably involves as-yet untested physics beyond the Standard Model of
particle physics. On the other hand, it is precisely because of this
reason that the scenario is also attractive --- it brings
in ideas of new physics beyond the Standard Model of particle physics
(such as Grand Unification) as well as ideas of early-Universe cosmology
(such as topological defects and/or massive particle production in
inflation) into the realms of EHECR where these ideas have the
potential to be tested by future EHECR experiments. 

The physics and astrophysics of UHECR are intimately linked with
the emerging field of neutrino astronomy as well as with the already
established field of $\gamma-$ray astronomy which in turn are important
subdisciplines of particle astrophysics (for a review see, e.g.,
Ref.~\cite{mannheim4}). Indeed, as we shall see, all
scenarios of UHECR origin, including the top-down models, are severely
constrained by neutrino and $\gamma-$ray observations and limits. We shall
also discuss how EHECR observations have the potential to yield important
information on Galactic and extragalactic magnetic fields. 

The plan of this review is summarized in the Table of Contents.

Unless otherwise stated or obvious from the context, we use natural units
with $\hbar=c=k_{\rm B}=1$ throughout.  

\section{The Observed Cosmic Rays}
In this section we give a brief overview of CR
observations in general. Since this is a very rich topic with a
tradition of almost 90 years, only the most important facts can
be summarized. For more details the reader is referred to recent
monographs on CR~\cite{bbdgp,gaisser} and to rapporteur
papers presented at the biennial International Cosmic Ray
Conference (ICRC) (see, e.g., Refs.~\cite{icrc24,icrc25,icrc26})
for updates on the data situation. The relatively young field of
$\gamma-$ray astrophysics which has now become an important subfield
of CR astrophysics, can only be skimmed even more superficially and for
more information the reader is referred to the proceedings of
the Compton $\gamma-$ray symposia and the ICRC and to
Refs.~\cite{gammarev,sinnis,cw}, for example. We will only
mention those $\gamma-$ray 
issues that are relevant for UHECR physics. 
Similarly, the emerging field of neutrino astrophysics~\cite{ghs} 
will be discussed only in the context of 
ultra-high energies for which a possible neutrino component and its
potential detection will be discussed later in this review.

\subsection{Detection Methods at Different Energies}
The CR primaries are shielded by the Earth's
atmosphere and near the ground reveal their existence only by
indirect effects such as ionization. Indeed, it was the height
dependence of this latter effect which lead to the discovery of
CR by Hess in 1912. Direct observation of CR primaries is only
possible from space by flying detectors with balloons or
spacecraft. Naturally, such detectors are very limited in size
and because the differential CR spectrum is a steeply falling function of 
energy, roughly in accord with a power-law with index $-2.7$ up to an
energy of $\simeq2\times10^{16}\,$eV (see Fig.~\ref{F2.1}),
direct observations run out of statistics typically around
a few $100\,$TeV ($=10^{14}\,$eV)~\cite{shibata}. For the neutral
component, i.e. $\gamma-$rays, whose flux at a given energy is
lower than the charged CR flux by several orders of magnitude, this
statistical limit occurs at even lower energies, for example
around $100\,$GeV for the instruments on board the Compton Gamma
Ray Observatory (CGRO)~\cite{cgro}. The
space based detectors of charged
CR traditionally use nuclear emulsion stacks such as in the JACEE
experiment~\cite{jacee}; now-a-days, spectrometric techniques are also
used which are 
advantageous for measuring the chemical composition. For
$\gamma-$rays, for example, the Energetic
Gamma Ray Experiment Telescope (EGRET) on board the CGRO uses
spark chambers combined with a NaI calorimeter.

Above roughly $100\,$TeV, the showers of secondary particles
created in the interactions of the primary CR
with the atmosphere are extensive enough to be detectable from
the ground. In the most traditional technique, charged hadronic
particles, as well as electrons and muons in these Extensive Air
Showers (EAS) are recorded on the
ground~\cite{petrera} with standard instruments 
such as water Cherenkov detectors used in the old
Volcano Ranch~\cite{volcano} and Haverah Park~\cite{haverah}
experiments, and scintillation detectors which are
used now-a-days. Currently operating ground
arrays for UHECR EAS are the Yakutsk experiment
in Russia~\cite{yakutsk} and the Akeno
Giant Air Shower Array (AGASA) near Tokyo, Japan, which is
the largest one, covering an
area of roughly $100\,{\rm km}^2$ with about 100
detectors mutual separated by about $1\,$km~\cite{agasa}.
The Sydney University Giant Air Shower Recorder (SUGAR)~\cite{sugar}
operated until 1979 and was the largest array in the Southern
hemisphere. The ground array technique allows one to
measure a lateral cross section of the shower profile.
The energy of the shower-initiating primary particle is estimated by
appropriately parametrizing it in terms a measurable parameter;
traditionally this parameter is taken to be the particle density at 600 m
from the shower core, which
is found to be quite insensitive to the primary composition
and the interaction model used to simulate air
showers~\cite{hillas2}.

The detection of secondary photons from EAS represents a
complementary technique. The experimentally most important light
sources are the fluorescence of air nitrogen excited by the charged
particles in the EAS
and the Cherenkov radiation from the charged particles that travel
faster than the speed of light in the atmosphere. The first
source is practically isotropic whereas the second one produces
light strongly concentrated on the surface of a cone around the
propagation direction of the charged source. The fluorescence
technique can be used equally well for both charged and neutral
primaries and was first used by the Fly's Eye detector~\cite{fe}
and will be part of several future projects on UHECR
(see Sect.~2.6). The primary energy can be estimated from
the total fluorescence yield. Information on the primary
composition is contained in the column depth $X_{\rm max}$ 
(measured in g$\,{\rm cm}^{-2}$) at which the shower reaches maximal
particle density. The average of $X_{\rm max}$ is related to the primary
energy $E$ by
\begin{equation}
  \left\langle X_{\rm max}\right\rangle = X_0^\prime\,\ln
  \left(\frac{E}{E_0}\right)\,.\label{elongation}
\end{equation}
Here, $X_0^\prime$ is called the elongation rate and $E_0$
is a characteristic energy that depends on the primary composition.
Therefore, if $X_{\rm max}$ and $X_0^\prime$ are  
determined from the longitudinal shower profile measured
by the fluorescence detector, then $E_0$ and thus
the composition, can be extracted after determining the
$E$ from the total fluorescence yield. 
Comparison of CR spectra measured with the ground array
and the fluorescence technique indicate systematic errors in
energy calibration that are generally smaller than $\sim$ 40\%.
For a more detailed discussion of experimental EAS analysis with the
ground array and the fluorescence technique see, e.g., the recent review
by Yoshida and Dai~\cite{yoshida-dai} and
Refs.~\cite{sokolsky-book,sokolsky-physrep,rao-sreek-book}. 

In contrast to the fluorescence light, for a given primary energy,
the output in Cherenkov light is
much larger for $\gamma-$ray primaries than for charged CR
primaries. In combination with the
so called imaging technique --- in which the Cherenkov light image
of an electromagnetic cascade in the upper atmosphere (and thus
also the primary arrival direction) is reconstructed~\cite{iact} --- the
Cherenkov technique is one of the best tools available to
discriminate
$\gamma-$rays from point sources against the strong background
of charged CR. This technique is used, for example, by the
High Energy Gamma Ray Astronomy (HEGRA)
experiment (now 5 telescopes of 8.5$\,m^2$ mirror area)~\cite{hegra1}
and by the 10 meter Whipple
telescope~\cite{whipple} with threshold energies of
$\simeq500\,$GeV and $200\,$GeV, respectively. In the
Southern hemisphere, the Collaboration of Australia and Nippon
(Japan) for a GAmma Ray Observatory in the Outback (CANGAROO)
experiment~\cite{cangaroo} currently consists of two 7 meter imaging
atmospheric Cherenkov telescopes at Woomera, Australia, with
an energy threshold of 200 GeV.

Another new experiment which is at the completion stages of construction
and testing is the Multi-Institution Los Alamos Gamma Ray Observatory
(MILAGRO)~\cite{milagro} which is a water (rather than
atmospheric)-Cherenkov detector that detects electrons, photons, hadrons
and muons in EAS, has a 24-hour duty cycle, ``all-sky'' coverage, and good
angular resolution ($\leq 0.4^\circ$ at 10 TeV), and is sensitive to
$\gamma-$rays in the energy range from $\sim 200\gev$ to $\sim100\tev$. 
For $\gamma-$rays, therefore, an as yet unexplored window between a
few tens of GeV and $\simeq200\,$GeV remains which may soon be
closed by large-area atmospheric Cherenkov detectors~\cite{ong}.
For a detailed review of this field of very high energy
$\gamma-$ray astronomy see, e.g., Refs.~\cite{gammarev,sinnis,cw}.

Finally, muons of a few hundred GeV and above have penetration
depths of the order of a kilometer even in rock and can thus be
detected underground. The Monopole Astrophysics and Cosmic Ray
Observatory (MACRO) experiment~\cite{macro}, for example, located
in the Gran Sasso laboratory near Rome, Italy, has a rock
overburden of about $1.5\,$km,
consists of $\simeq600\,$tons of liquid scintillator and acts as
a giant time-of-flight counter. Operated in coincidence with the
Cherenkov telescope array EAS-TOP~\cite{eastop} located above it,
it can for instance
be used to study the primary CR composition around the ``knee''
region~\cite{macro1} (see Fig.~\ref{F2.1}). A similar combination is
represented
by the
Antarctic Muon And Neutrino Detector Array (AMANDA) detector and
the South Pole Air ShowEr array (SPASE) of scintillation detectors.
AMANDA consists of strings of photomultiplier tubes of a
few hundred meters in length deployed in the antarctic ice at
depths of up to $2\,$km, and reconstructs tracks of muons of
energies in the TeV range~\cite{amanda}.

\subsection{The Measured Energy Spectrum}

\begin{figure}[ht]
\centering\leavevmode
\epsfxsize=6.0in
\epsfbox{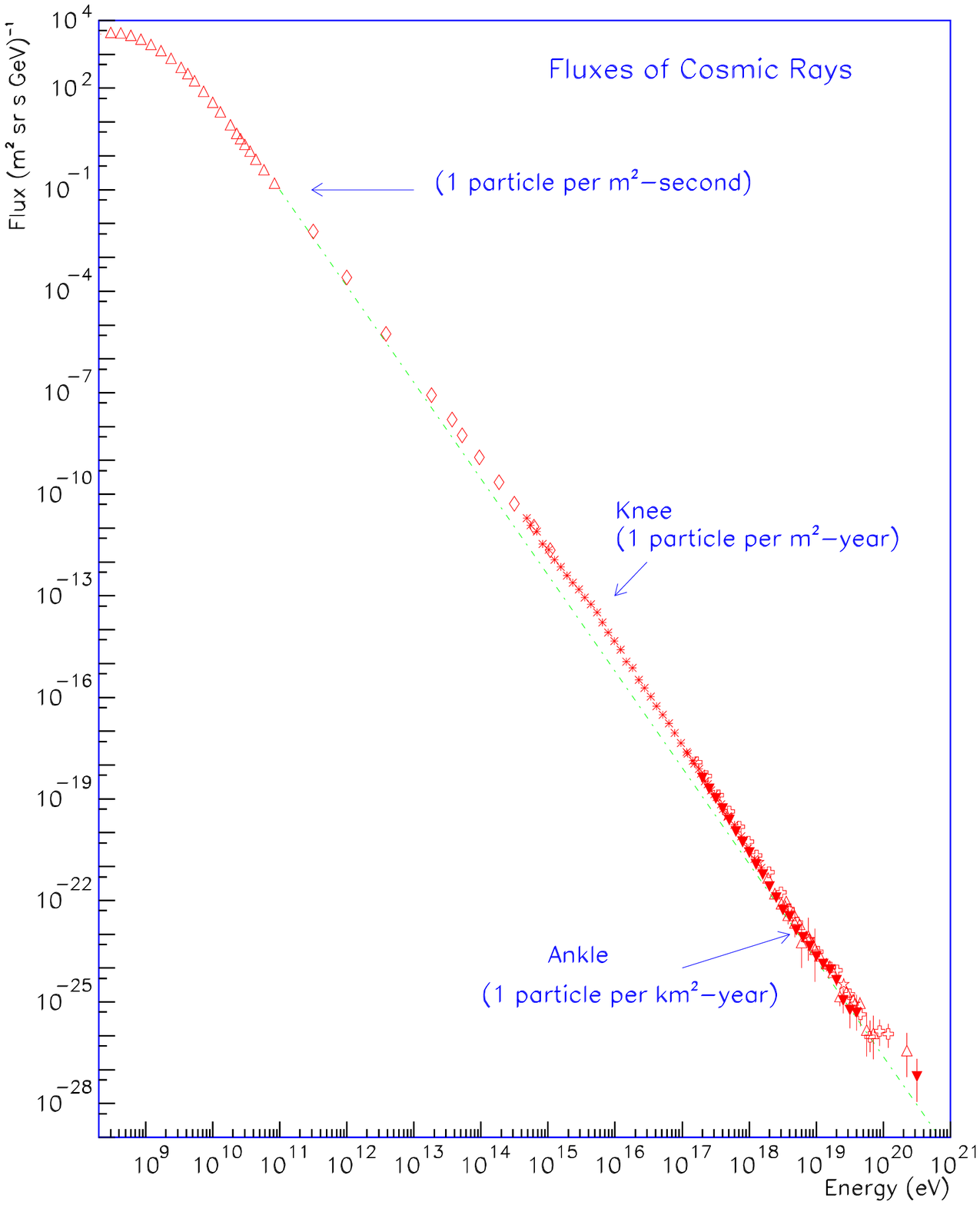}
\bigskip
\caption[...]{The CR all particle spectrum~\cite{swordy2}.
Approximate integral fluxes are also shown.}
\label{F2.1}
\end{figure}

Fig.~\ref{F2.1} shows a compilation of the CR all-particle
spectrum over the whole range of energies observed through
different experimental strategies as discussed in
Sect.~2.1. The spectrum exhibits power law behavior over a wide
range of energies, but comparison with a fit to a single power
law (dashed line in Fig.~\ref{F2.1}) shows significant breaks at
the ``knee'' at $\simeq4\times10^{15}\ev\,$ and, to a somewhat
lesser extent, at the ``ankle'' at
$\simeq5\times10^{18}\,$eV. The sharpness of the knee feature is
a not yet resolved experimental issue, particularly because it
occurs in the transition region between the energy range where direct
measurements are available and the energy range where the data come from
indirect detection by the ground array techniques whose energy
resolution is typically 20\% or worse~\cite{shibata}.
For example, the EAS-TOP array observed a sharp spectral
break at the knee within their experimental resolution~\cite{eastop3},
whereas the AGASA~\cite{agasa1} and CASA-MIA~\cite{casa3} data
support a softer transition.

\begin{figure}[ht]
\centering\leavevmode
\epsfxsize=6.0in
\epsfbox{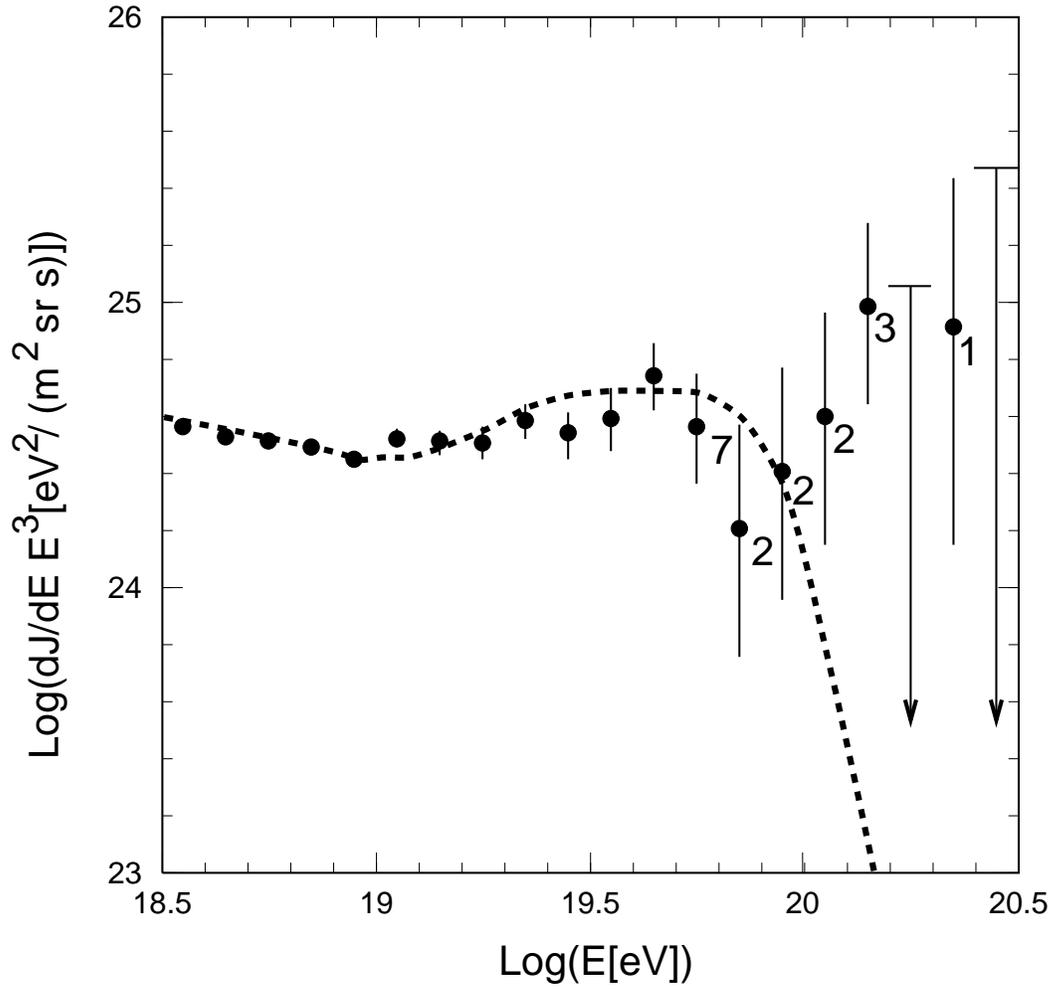}
\bigskip
\caption[...]{Energy spectrum of UHECR measured by the AGASA
experiment. The dashed curve represents the
spectrum expected for extragalactic sources distributed uniformly in
the Universe.
The numbers attached to the data points are the number of events observed
in the corresponding energy bins. (From M.~Takeda et al~\cite{agasa}.)}
\label{F2.2}
\end{figure}

\begin{figure}[ht]
\centering\leavevmode
\epsfxsize=6.0in
\epsfbox{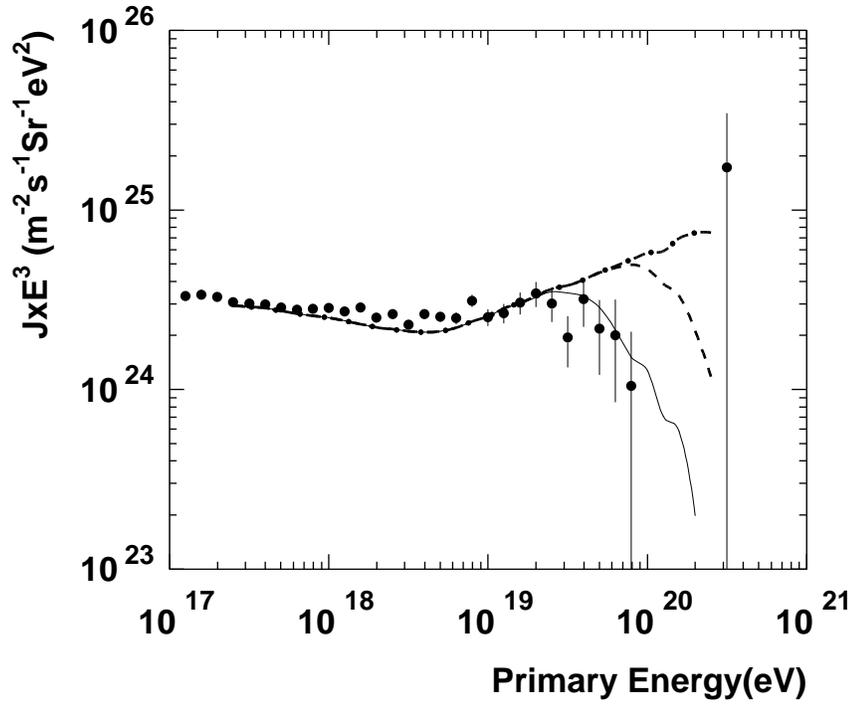}
\bigskip
\caption[...]{Fly's Eye monocular energy spectrum. Dots: data. Lines:
predicted spectra for source energy cutoff at different
energies. Solid line: cutoff at $10^{19.6}\ev$. Dashed line: cutoff at
$10^{20}\ev$. Chain line: cutoff at $10^{21}\ev$. 
(From Yoshida and Dai~\cite{yoshida-dai}).}
\label{F2.3}
\end{figure}

\begin{figure}[ht]
\centering\leavevmode
\epsfxsize=6.0in
\epsfbox{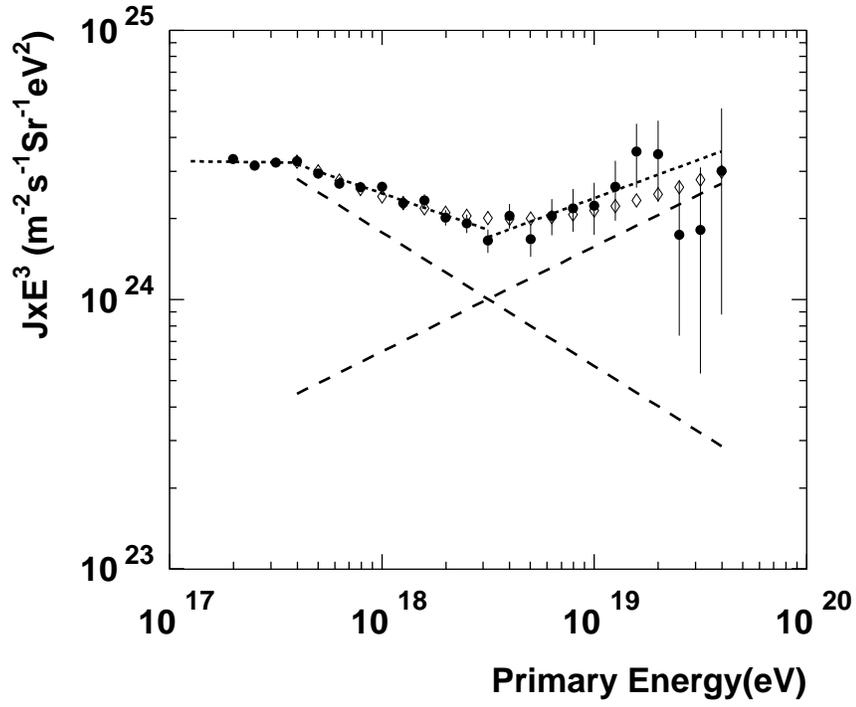}
\bigskip
\caption[...]{Fly's Eye stereo energy spectrum. Dots: data. Dotted
line: best fit in each region. Dashed lines: a two-component
fit. (From Yoshida and Dai~\cite{yoshida-dai}). }
\label{F2.4}
\end{figure}

\begin{figure}[ht]
\centering\leavevmode
\epsfxsize=6.0in
\epsfbox{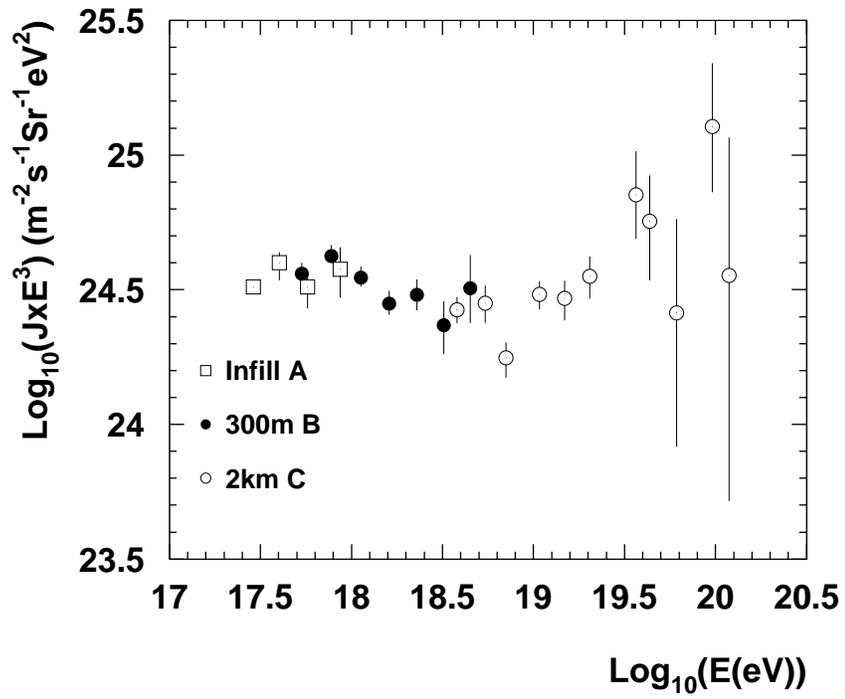}
\bigskip
\caption[...]{The Haverah Park energy spectrum. 
(From Yoshida and Dai~\cite{yoshida-dai}). }
\label{F2.5}
\end{figure}

\begin{figure}[ht]
\centering\leavevmode
\epsfxsize=6.0in
\epsfbox{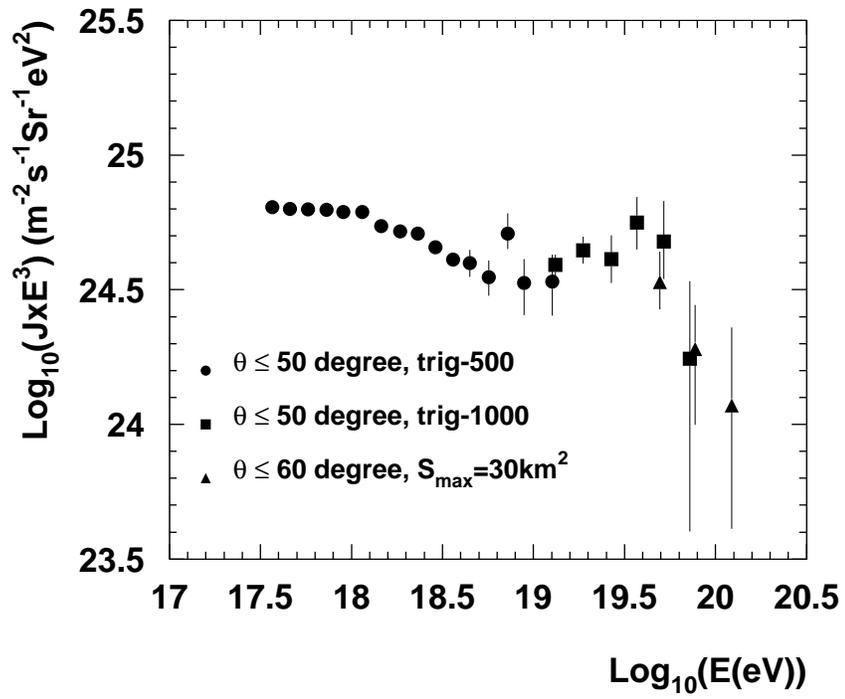}
\bigskip
\caption[...]{The Yakutsk energy spectrum. 
(From Yoshida and Dai~\cite{yoshida-dai}).}
\label{F2.6}
\end{figure}

Figs.~\ref{F2.2}--\ref{F2.6} show the CR data above
$10^{17}\,$eV measured by different experiments. 
The ankle feature was first discussed in detail by the Fly's Eye
experiment~\cite{fe}. The slope between the knee and up to
$\simeq4\times10^{17}\,$eV is very close to 3.0 (Fig.~\ref{F2.1}); 
then it seems
to steepen to about 3.2 up to the dip at $\simeq3\times10^{18}\,$eV,
after which it flattens to about 2.7 above the dip. As will be
discussed in Sect.~2.4, the Fly's Eye also found evidence for
a change in composition to a lighter component above the ankle,
that is correlated with the change in spectral slope.

The situation at the high end of the CR spectrum is
as yet inconclusive and represents the main subject of the recent
strong increase of theoretical and experimental activities in UHECR
physics
which also motivated the present review. The present data (see
Figs.~\ref{F2.2}--\ref{F2.6}) 
seem to reveal a steepening just below $10^{20}\,$eV, but above
that energy significantly more events have been seen than
expected from an extrapolation of the GZK ``cutoff'' at
$\simeq10^{20}\,$eV. This is perhaps the most puzzling and hence
interesting aspect of UHECR because a
cutoff is expected at least for extragalactic nucleon primaries
irrespective of the production mechanism (see Sect.~4.1). Even
for conventional local sources, the maximal energy to which charged 
primaries can be accelerated is expected to be limited (see Sect.~5)
and it is generally hard to achieve energies beyond the cutoff energy. 

\subsection{Events above $10^{20}\,$eV}

The first published event above $10^{20}\,$eV was observed by the
Volcano Ranch experiment~\cite{volcano}. The Haverah Park experiment
reported 8 events around $10^{20}\,$eV~\cite{haverah}, and
the Yakutsk array saw one event above this energy~\cite{yakutsk}.
The SUGAR array in Australia reported 8 events
above $10^{20}\,$eV~\cite{sugar}, the highest one at
$2\times10^{20}\,$eV. The world record holder is
still a $3.2\times10^{20}\,$eV event which was the only event
above $10^{20}\,$eV observed by the Fly's Eye experiment~\cite{fe},
on 15 October 1991. Probably the second highest event at
$\simeq2.1\times10^{20}\,$eV in the world data set was seen by
the AGASA experiment~\cite{agasa} which meanwhile detected
a total of 6 events above $10^{20}\,$eV (see Fig.~\ref{F2.2}). The Fly's
Eye and the
AGASA events have been documented in detail in the literature
and it seems unlikely that their energy has been overestimated
by more than 30\%. For more detailed experimental information
see, e.g., the review~\cite{yoshida-dai}. Theoretical and astrophysical
implications of these events are a particular focus of the
present review. For an overview of specific source searches for these
events see Sect.~4.6.

\subsection{Composition}
We will discuss the question of composition here only for CR detected by
ground based EAS detectors, i.e., for CR above $\sim100\,$TeV, only.
Information on the
chemical composition is mainly provided by the muon content in
case of ground arrays and by the depth of shower maximum for
optical observation of the EAS. Just to indicate the qualitative
trend we mention that, for a given primary energy, a heavier
nucleus produces EAS with a higher muon content and a shower
maximum higher up in the atmosphere on average compared to those for a
proton shower. The latter property can
be understood by viewing a nucleus as a collection of
independent nucleons whose interaction probabilities add,
leading to a faster development of the shower on average. 
The higher muon content in a heavy nucleus shower is due to the fact that,  
because the shower develops relatively higher up in the atmosphere where
the atmosphere is less dense, it is relatively easier for the charged
pions in a heavy nucleus shower to decay to muons before interacting
with the medium. 
 
The spectral and compositional behavior around the knee at
$\simeq4\times10^{15}\,$eV may play a crucial role in attempts to
understand 
the origin and nature of CR in this energy range, as will be
discussed in little more detail in Sect.~3. Indeed, there are
indications that the chemical composition becomes heavier with
increasing energies below the knee~\cite{shibata}. Around
the knee the situation becomes less clear and most of the experimental
results, such as from the SOUDAN-2~\cite{soudan2}, the
HEGRA~\cite{hegra3}, and the KArlsruhe Shower Core and Array
DEtector (KASCADE)~\cite{kascade} experiments,
seem to indicate a substantial proton component and no significant
increase in primary mass.
Recent results, for example, from the Dual Imaging Cherenkov Experiment
(DICE) seem to indicate a lighter composition above the knee which may
hint to a transition to a different component~\cite{dice},
but evidence for an inreasingly heavy composition above
the knee has also been reported by the KASCADE
collaboration~\cite{kascade1} and by HEGRA~\cite{hegra4}.

Based on the analysis discussed above, Eq.~(\ref{elongation}),
the Fly's Eye collaboration reported a composition change
from a heavy component below the ankle to a light component
above, that is correlated with the spectral changes around
the ankle~\cite{fe}. However, this was not confirmed by
the AGASA experiment~\cite{agasa,yoshida-dai}. In addition, there
have been suggestions that the observed energy dependence
of $\left\langle X_{\rm max}\right\rangle$ could be caused
by air shower physics rather than an actual composition
change~\cite{psv}.

One signature of a heavy nucleus primary would be the almost simultaneous
arrival of a pair of EASs at the Earth. 
Such pairs would be produced by
photodisintegration of nuclei by solar photons and could
be used to measure their mass, as was pointed out quite
early on~\cite{gerasimova}. This effect has been reconsidered
recently in light of existing and proposed UHECR
detectors~\cite{mtv,epele}.

At the highest energies, observed EAS seem to be consistent
with nucleon primaries, but due to poor statistics and
large fluctuations from shower to shower, the issue is not
settled yet. Some scenarios of EHECR origin, such as the top-down
scenario discussed in Sects.~6 and 7, predict the EHECR primaries to be
dominated by photons and neutrinos rather than nucleons. 
Distinguishing between photon and nucleon induced showers is, however,
extremely difficult at UHE and EHE regions --- the standard muon-poorness
criterion of photon induced showers relative to nucleon induced showers,
applicable at lower ($10^{14}$ -- $10^{16}\,$eV) energies, does not apply
to the UHE region. It has been claimed that the highest energy Fly's
Eye event is inconsistent with a $\gamma-$ray primary~\cite{hvsv}.
It should be noted, however, that
at least for electromagnetic showers, EAS simulation at EHE
is complicated by the Landau-Pomeranchuk-Migdal (LPM) effect
and by the influence of the geomagnetic field~\cite{kasahara}.
Furthermore, in the simulations, EAS development depends
to some extent on the hadronic interaction event
generator which complicates data interpretation~\cite{eas-simu}.
Definite conclusions on the composition of the EHECR, therefore, have to
await data from next generation experiments. 
Together with certain characteristic features of the photon induced
EHECR showers due to geomagnetic effects~\cite{kasahara}, 
the large event statistics expected from the next
generation experiments will hopefully allow to distinguish between photon
and nucleon EHECR primaries. In turn, accelerator data together
with EAS data can be used to constrain, for example, the
cross section of protons with air nuclei at center of mass
energies of 30 TeV~\cite{proton-air}.

The hypothesis of neutrinos or new neutral
particles as EHECR primaries will be discussed in Sect.~4.3.1 and
4.3.2, respectively.

\subsection{Anisotropy}
\begin{figure}[ht]
\centering\leavevmode
\epsfxsize=6.0in
\epsfbox{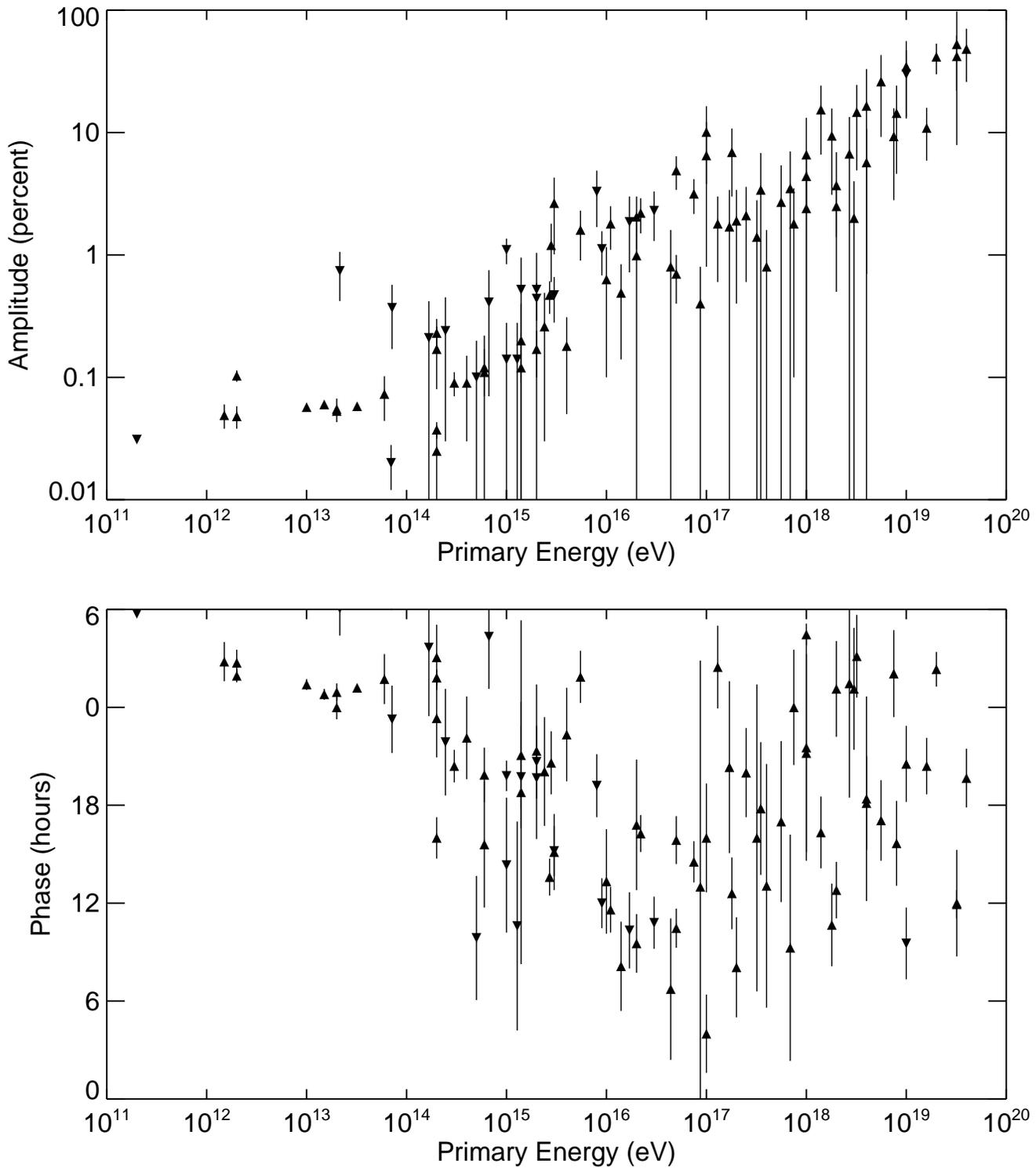}
\bigskip
\caption[...]{A compilation of anisotropy measurements (first harmonic
Fourier amplitude and phase). Northern and southern hemisphere results
are denoted by upward-pointing and downward-pointing triangles
respectively. (From Clay and Smith~\cite{smith-clay})} 
\label{F2.7} 
\end{figure}

For a recent compilation and discussion of anisotropy
measurements see Ref.~\cite{smith-clay}. Fig.~\ref{F2.7} shows the
summary figure from that reference. Implications of these anisotropy
measurements will be discussed briefly in the next section where we
discuss the origin of CR in general. For discussions of subtleties 
involved in the measurements and interpretation of anisotropy data, 
choice of coordinate systems used in presenting anisotropy results
etc., see e.g., Ref.~\cite{sokolsky-book}.  

The anisotropy amplitude is defined as
\begin{equation}
  |\delta|=\frac{I_{\rm max}-I_{\rm min}}
  {I_{\rm max}+I_{\rm min}}\,,\label{deltaan}
\end{equation}
where $I_{\rm min}$ and $I_{\rm max}$ are the minimum and
maximum CR intensity as a function of arrival direction. 
Very recently, results have been presented on the anisotropy of
the CR flux above $\simeq10^{17}\,$eV
from the Fly's Eye~\cite{fe-aniso} and the AGASA~\cite{agasa-aniso}
experiments. Both experiments report a small but statistically
significant anisotropy of the order of 4\% 
in terms of Eq.~(\ref{deltaan}) toward the Galactic plane 
at energies around $10^{18}\,$eV. These analyses did
not reveal a significant correlation with the Supergalactic Plane,
whereas earlier work seemed to indicate some enhancement of the
flux from this plane~\cite{sblrw,kcd,haya2}.

In addition, the newest data seem to indicate that also the
events above $10^{20}\,$eV are consistent with an isotropic
distribution on large scales~\cite{hillas-nature}, as far as
that is possible to tell from about 15 events in the world data set.
At the same time, there seems to be significant small scale
clustering~\cite{agasa2}.

\subsection{Next-generation Experiments on Ultrahigh Energy Cosmic
Ray, $\gamma-$Ray, and Neutrino Astrophysics}

As an upscaled version of the old Fly's Eye Cosmic Ray experiment, the
High Resolution Fly's Eye detector is currently under construction
at Utah, USA~\cite{hires}. Taking into account a duty cycle of about
10\% (a fluorescence detector requires clear, moonless nights),
the effective aperture of this instrument will be
$\simeq600\,{\rm km}^2\,{\rm sr}$, about 10 times the AGASA
aperture, with a threshold around $10^{17}\,$eV. Another
project utilizing the fluorescence technique
is the Japanese Telescope Array~\cite{tel_array} which is currently
in the proposal stage. Its effective aperture will be about
15-20 times that of AGASA above $10^{17}\,$eV, and
it can also be used as a Cherenkov detector for TeV $\gamma-$ray
astrophysics. Probably the largest up-coming project is 
the international Pierre Auger Giant Array
Observatories~\cite{auger} which will be a combination of a ground array
of about 1700 particle detectors mutually separated from each
other by about 1.5 km and covering about $3000\,{\rm km}^2$, and one or
more fluorescence Fly's Eye type detectors. The ground array component
will have a duty cycle of nearly 100\%, leading to an effective
aperture about 200 times as large as the AGASA array, and an
event rate of 50--100 events per year above $10^{20}\,$eV. About 10\% 
of the events will be detected by both the ground array
and the fluorescence component and can be used for cross
calibration and detailed EAS studies. The energy threshold will
be around $10^{19}\,$eV. For maximal sky coverage
it is furthermore planned to construct one site in each hemisphere.
The southern site will be in Argentina, and the northern site
probably in Utah, USA.

Recently NASA initiated a concept study for detecting EAS
from space~\cite{owl} by observing their fluorescence light
from an Orbiting Wide-angle Light-collector (OWL). This would
provide an increase by another factor $\sim50$ in aperture
compared to the Pierre Auger Project, corresponding to an
event rate of up to a few thousand events per year above
$10^{20}\,$eV. Similar concepts such as the AIRWATCH~\cite{linsley}
and Maximum-energy air-Shower Satellite (MASS)~\cite{mass} missions
are also being discussed. The energy threshold of such instruments
would be between $10^{19}$ and $10^{20}\,$eV. This technique
would be especially suitable for detection of very small
event rates such as those caused by UHE neutrinos which
would produce horizontal air showers (see Sect.~7.4). For
more details on these recent experimental considerations
see Ref.~\cite{owl-proc}.

New experiments are also planned in $\gamma-$ray astrophysics.
The Gamma ray Large Area Space Telescope (GLAST)~\cite{glast}
detector is planned by NASA
as an advanced version of the EGRET experiment, with an about
100 fold increase in sensitivity at energies between 10 MeV
and 200 GeV. For new ground based $\gamma-$ray experiments
we mention the Very Energetic Radiation Imaging Telescope Array
System (VERITAS) project~\cite{veritas} which consists of eight
10 meter optical reflectors which will be about two orders 
of magnitude more sensitive between 50 GeV and 50 TeV
than WHIPPLE. A similar next generation atmospheric imaging
Cherenkov system with up to 16 planned telescopes is the
High Energy Stereoscopic System (HESS) project~\cite{hess}.
Furthermore, the Mayor Atmospheric Gamma-ray Imaging Cherenkov Telescope
(MAGIC) project~\cite{magic} aims to build a very large atmospheric
imaging Cherenkov telescope with $220\,{\rm m}^2$ mirror area for
detection of $\gamma-$rays between 10 GeV and 300 GeV, i.e.
within the as yet unexplored window of $\gamma-$ray astrophysics.
The CANGAROO experiment in Australia plans to upgrade to
four 10 meter telescopes and lower the threshold to 100 GeV.
Finally, another strategy to explore this window utilizes
existing solar heliostat arrays, and is represented by the Solar
Tower Atmospheric
Cherenkov Effect Experiment (STACEE)~\cite{stacee} in the USA,
the ChErenkov Low Energy Sampling \& Timing Experiment
(CELESTE) in France~\cite{celeste}, and the German-Spanish
Gamma Ray Astrophysics at ALmeria (GRAAL) experiment~\cite{graal}.

High energy neutrino astronomy is aiming towards a kilometer
scale neutrino observatory. The major technique is the optical
detection of Cherenkov light emitted by muons created in charged current
reactions of neutrinos with nucleons either in water
or in ice. The largest pilot experiments representing
these two detector media are the now defunct Deep Undersea Muon
and Neutrino Detection (DUMAND) experiment~\cite{dumand} in the
deep sea near Hawai and the AMANDA experiment~\cite{amanda} in the South
Pole ice. Another water based experiment is situated at
Lake Baikal~\cite{baikal}. Next generation deep sea projects
include the French Astronomy with a Neutrino Telescope and
Abyss environmental RESearch (ANTARES)~\cite{antares}
and the underwater Neutrino Experiment SouthwesT Of GReece
(NESTOR) project in the Mediterranean~\cite{nestor},
whereas ICECUBE~\cite{icecube} represents the planned kilometer scale
version of the AMANDA detector. Also under consideration are neutrino
detectors utilizing techniques to detect the radio pulse from the
electromagnetic showers created by neutrino
interactions in ice~\cite{radio-technique}. This technique
could possibly be scaled up to an effective area of
$10^4\,{\rm km}^2$ and a prototype is represented by the
Radio Ice Cherenkov Experiment (RICE) experiment at the
South Pole~\cite{rice}. The radio technique might also have
some sensitivity to the flavor of the primary neutrino~\cite{amvz}.
Neutrinos can also initiate
horizontal EAS which can be detected by giant ground arrays
such as the Pierre Auger Project~\cite{bpvz,pz}. Furthermore,
as mentioned above, horizontal EAS could be detected from
space by instruments such as the the proposed OWL detector~\cite{owl}.
Finally, the search for pulsed radio emission
from cascades induced by neutrinos or cosmic rays above
$\sim10^{19}\,$eV in the lunar regolith could also lead
to interesting limits~\cite{gln}.
More details on neutrino astronomy detectors are contained
in Refs.~\cite{learned,ghs,klein-mann}, and some recent overviews on
neutrino astronomy can be found in Ref.~\cite{halzen,protheroe1}.

\section{Origin of Bulk of the Cosmic Rays: General Considerations} 
The question of origin of cosmic rays continues to be regarded as an
``unsolved problem'' even after almost ninety years of research since
the announcement of their discovery in 1912. Although 
the general aspects of the question of CR origin are regarded as fairly
well-understood now, major gaps and uncertainties remain,  
the level of uncertainty being in general a function that increases with
energy of the cosmic rays. 

The total CR energy density measured above the
atmosphere is dominated by particles with energies between about 1 and
10 GeV. At energies below $\simeq1\,$GeV the intensities are temporally
correlated with the solar activity which is a direct evidence for an
origin
at the Sun. At higher energies, however, the flux observed at Earth
exhibits a temporal anticorrelation with solar activity and a screening
whose efficiency increases with the strength of the solar wind,
indicating an origin outside the solar system. Several arguments
involving energetics, composition, and secondary $\gamma-$ray production
suggest that the bulk of the CR between
1 GeV and at least up to the knee region (see Fig.~\ref{F2.1}) is confined
to the
Galaxy and is probably produced in supernova remnants (SNRs). Between
the knee and the ankle the situation becomes less clear,
although the ankle is sometimes interpreted as a cross over from
a Galactic to an extragalactic component. Finally, beyond
$\simeq10\,$EeV, CR are generally expected to have an extragalactic
origin due to their apparent isotropy, but ways around this reasoning have
also been suggested.

In the following, we give a somewhat more detailed account of these
general considerations, separating the discussion into issues related to
energetics, Galactic versus extragalactic origin, and
acceleration mechanisms and the possible sources of CR. 
We reserve a more comprehensive discussion of the origin
of UHECR above $\sim10^{17}\,$eV (which make only a small part
of the total CR energy density, but are the main focus of this review)  
for Sects.~5 and 6.

\subsection{Energetics}
As mentioned above, the bulk of the CR 
observed at the Earth is of extrasolar origin. The average energy density 
of CR is thus expected to be
uniform at least throughout most of the Galaxy. If CR are 
universal, their density should be constant throughout
the whole Universe. As a curiosity we note in this context that
the mean energy density of CR, 
$u_{\rm CR}$, is comparable to the energy density of the CMB. 
It is not clear, however, what physical
process could lead to such an ``equilibration'', which is thus most
likely just a coincidence. We will see in the following that
indeed a universal origin of bulk of the CR is now-a-days
not regarded as a likely possibility. 

If the CR accelerators are Galactic, they must replenish for the escape of
CR from the Galaxy in order to sustain the observed Galactic CR differential
intensity $j(E)$. Their total luminosity in CR must therefore satisfy
$L_{\rm CR}=(4\pi/c)\int dEdV t_{\rm CR}(E)^{-1}Ej(E)$, 
where $t_{\rm CR}(E)$ is the mean residence time of CR
with energy $E$ in the Galaxy and $V$ is the volume. $t_{\rm
CR}(E)$ can be estimated from the mean column
density, $X(E)$, of gas in the interstellar medium that
Galactic CR with energy $E$ have traversed. Interaction of the primary CR
particles with the gas in the interstellar medium leads to production of
various secondary species. From the secondary to
primary abundance ratios of Galactic CR it was infered
that~\cite{swordy}
\begin{equation}
X(E)=\rho_g t_{\rm CR}(E)\simeq 6.9
\left(\frac{E}{20Z\,{\rm GeV}}\right)^{-0.6}\,{\rm g}\,{\rm cm}^{-2}\,,
\end{equation}
where $\rho_g$ is the mean density of interstellar gas and $Z$ is the
mean charge number of the CR particles.
The mean energy density of CR and the total mass of gas in the Milky Way
that have been inferred from the diffuse Galactic $\gamma-$ray,
X-ray and radio emissions are $u_{\rm CR}=(4\pi/c)\int dE Ej(E)\simeq
1\,{\rm eV}\,{\rm cm}^{-3}$
and  $M_{g}\sim\rho_g V \sim 4.8\times 10^9M_\odot$,
respectively. Hence, simple integration yields 
\begin{equation}
L_{\rm CR}\sim M_{g}\int dE\frac{Ej(E)}{X(E)}
\sim 1.5\times10^{41}\,{\rm erg}\,{\rm sec}^{-1}\,.\label{lumcr}
\end{equation}
This is about 10\% of the estimated total power output in the form
of kinetic energy of the ejected material in    
Galactic supernovae which, from the energetics point of view,
could therefore account for most of the CR. We note
that the energy release from other Galactic sources,
e.g. ordinary stars~\cite{bbdgp} or isolated neutron
stars~\cite{vmo} is expected to be too small, even for
UHECR\footnote{This conclusion may, however, change somewhat
with the recent detection of certain soft-gamma repeaters~\cite{kouveliotou}
which seem to indicate the existence of a subclass
of pulsars with dipole magnetic fields as large as a few times
$10^{14}\,$G. This may increase the available magnetic energy budget
from pulsars by two to three orders of magnitude.}. Together
with other considerations (see Sect.~3.2) this leads to the
widely held notion that CR at least up to the knee
predominantly originate from first-order Fermi acceleration (see below) 
in SNRs.

Another interesting observation is that the energy density in the form of 
CR is comparable both to the energy density in the Galactic
magnetic field ($\sim10^{-6}\,$G) as well as that in the
turbulent motion of the gas,
\begin{equation}
  u_{\rm CR}\sim\frac{B^2}{8\pi}\sim\frac{1}{2}\,\rho_g v_t^2
  \,,\label{crb}
\end{equation}
where $\rho_g$ and $v_t$ are the density and turbulent velocity
of the gas, respectively.
This can be expected from a pressure equilibrium between the
(relativistic) CR, the magnetic field, and the gas flow. If
Eq.~(\ref{crb}) roughly holds not only in the Galaxy but also
throughout extragalactic space, then we would expect the
extragalactic CR energy density to be considerably
smaller than the Galactic one which is another argument in favor
of a mostly Galactic origin of the CR observed near
Earth (see Sect.3.2). We note, however, that, in order for
Eq.~(\ref{crb}) to hold, typical CR diffusion time-scale over
the size of the system under consideration must be smaller than
its age. This is not the case, for example, in clusters of
galaxies if the bulk of CR are produced in the member galaxies or in 
cluster accretion shocks~\cite{bbp}.

\subsection{Galactic versus Extragalactic Origin of the Bulk of the CR}
The energetical considerations mentioned above already
provide some arguments in favor of a Galactic origin of the
bulk of the CR. Another argument involves the
production of secondary $\gamma-$rays from the decays of neutral
pions produced in interactions of CR 
with the baryonic gas throughout the Universe: For given
densities of the CR and the gas, 
the resulting $\gamma-$ray flux can be calculated quite
reliably~\cite{ginzburg} and the predictions can be compared with
observations. This has been done, for example, for the
Small Magellanic Cloud (SMC). The observed upper limits~\cite{sreekumar}
turn out to be a factor of a few below the predictions assuming a
universal CR density. The CR density at the SMC should, therefore, be at
least a factor of a few smaller than the local Galactic density.

As a second test we mention the search for a CR
gradient (e.g.,~\cite{dsw,stecker75,ddv}: For a
Galactic CR origin one expects a decrease of CR intensity with
increasing distances from the Galactic center which should be
encoded in the secondary $\gamma-$ray emission that can be
measured by space based instruments such as EGRET. The
observational situation is, however, not completely settled
yet~\cite{egret_grad}. Whereas the spatial variation of the
$\gamma-$ray flux fits rather well, the observed spectrum
appears to be too flat compared to the one expected from the
average CR spectrum. Since the average CR spectrum
throughout the Galaxy is generally steepened (compared to the 
spectrum at the source) by diffusion in Galactic
magnetic fields, the observed relatively flat secondary $\gamma-$ray
spectrum may be interpreted as if the secondary $\gamma-$rays
are produced by CR interactions mainly at the (Galactic) sources
rather than in the interstellar medium. This interpretation, however,
requires that the escape time of CR from their 
sources be energy-independent. 

Some information on CR origin is in principle also contained in
the distribution of their arrival directions which has been
discussed in Sect.~2.5 (see Fig.~\ref{F2.7}). Below
$\sim10^{14}\,$eV, the amplitude of the observed anisotropy,
$\sim10^{-3}$, is statistically significant and roughly energy
independent. Above $10^{14}\,$eV, observed
anisotropy amplitudes are generally statistically insignificant
with possible exceptions between $\simeq10^{15}\,$eV and
$10^{16}\,$eV~\cite{smith-clay} and again close to $10^{18}\,$eV, the latter
correlated with the Galactic plane~\cite{fe-aniso,agasa-aniso}. A possible
clustering towards the Supergalactic Plane for energies above a few tens of
EeV was claimed~\cite{sblrw,kcd,haya2}, but has not been confirmed by more
recent studies~\cite{fe-aniso,agasa-aniso}. Since charged CR
at these energies are hardly deflected by the Galactic magnetic
field, the apparent lack of any significant anisotropy associated with the
Galactic plane implies that the high energy end of the CR spectrum is
most likely to have an extragalactic origin (see Sect.~4.6).

For Galactic sources, detailed models of CR diffusion and $\gamma-$ray
production in the Galaxy have been developed (see, e.g., Ref.~\cite{sm}).
These models are generally based on the energy loss - diffusion equation
that will be discussed below in the context of UHECR propagation
[see Eq.~(\ref{cel_d})], with the diffusion constant generalized to a
diffusion tensor.
This tensor and other parameters in these models can be obtained
from fits to the observed abundances of nuclear isotopes. It is often
sufficient to consider a simplified model, the so called ``leaky box'' 
model (see Refs.~\cite{bbdgp,gaisser} for detailed discussions) in which
the diffusion term is approximated by a loss term involving a CR
containment time $t_{\rm CR}$.
Fits to the data lead to $t_{\rm CR}\sim10^7\,$yr below
$\sim10^{16}\,$eV with only a weak energy dependence. This is in
turn consistent with observed anisotropies which, below
$\sim10^{14}\,$eV, can be interpreted by the
Compton-Getting effect~\cite{cg} which describes the effect of
the motion of the observer relative to an isotropic
distribution of CR. In this case the relative motion is a
combination of the motion of the
solar system within the Galaxy and the drift motion of the
charged CR diffusing and/or convecting in the interstellar
medium. The magnitude and the weak energy dependence of the anisotropy in
this energy range can be interpreted as arising out of diffusion of CR
predominantly along the tangled
incoherent component of the Galactic magnetic field. In summary, CR
composition and anisotropy data provide further evidence for a
Galactic origin for energies at least up to the knee region of the
spectrum.

In this context, the knee itself is often interpreted as a magnetic
deconfinement effect such that CR above the knee leave the
Galaxy relatively faster, leading to steepening of the spectrum
above the knee. In addition, the maximum 
energy achieved in shock acceleration is proportional to the
primary charge and could also lead to a spectral steepening
(see Sect.~3.3). Alternatively, the knee has also been interpreted
as being caused by the flux contribution from a strong single
source~\cite{elw}.

Finally, since the range of electrons above $\sim10^{11}\,$eV
becomes smaller than $t_{\rm CR}$ due to synchrotron and inverse
Compton losses, the electronic CR component at such energies, 
which is about 1\% of the hadronic flux, is undoubtedly of
Galactic origin. This can also be explained by acceleration in
SNRs.

\subsection{Acceleration Mechanisms and Possible Sources}
There are basically two kinds of acceleration mechanisms considered in
connection with CR acceleration: (1) direct acceleration of charged
particles by an electric field, and (2) statistical acceleration
(Fermi acceleration) in a magnetized plasma.  

In the direct acceleration mechanism, the electric field in
question can be due, for example, to a rotating magnetic neutron star
(pulsar) or, a (rotating) accretion disk threaded by magnetic fields,
etc. The details of the actual acceleration process and the maximum 
energy to which a particle can be accelerated depend on the
particular physical situation under consideration. For a variety of
reasons, the direct acceleration mechanisms are, however, not widely
favored these days as the CR acceleration mechanism.
Apart from disagreements among authors about the crucial details of the
various models, a major disadvantage of the mechanism in general is that
it is difficult to obtain the characteristic power-law spectrum of the
observed CR in any natural way. However, as pointed out by
Colgate~\cite{colgate}, a power law spectrum does not necessarily
point to Fermi acceleration, but results whenever a fractional
gain in energy of a few particles is accompanied by a significantly
larger fractional loss in the number of remaining particles.
We will not discuss the direct acceleration mechanism in any more
details, and refer the reader to reviews, e.g.,
in Ref.~\cite{bbdgp,hillas-araa,takahara-rev,colgate}.

The basic idea of the statistical acceleration mechanism originates from
a paper by Fermi~\cite{fermi} in 1949: Even though the average
electric field may vanish, there can still be a net
transfer of macroscopic kinetic energy of moving magnetized plasma to
individual charged particles (``test particles'') in the medium 
due to repeated collisionless scatterings (``encounters'') of the
particles either with randomly moving inhomogeneities of the
turbulent magnetic field or with shocks in the medium. 
Fermi's original paper~\cite{fermi} considered the former case, i.e.,
scattering with randomly moving magnetized ``clouds'' in the interstellar
medium. In this case, although in each individual encounter the particle
may either gain or lose energy, there is on average a net gain of
energy after many encounters. The original Fermi mechanism is
now-a-days called ``second-order'' Fermi mechanism, because the 
average fractional energy gain in this case is proportional
to $(u/c)^2$, where $u$ is the relative velocity of the cloud with respect
to the frame in which the CR ensemble is isotropic, and $c$ is the
velocity of light. Because of the dependence on the square of the cloud
velocity ($u/c < 1$), the second-order Fermi mechanism is not a very
efficient acceleration process. Indeed, for typical interstellar clouds
in the Galaxy, the acceleration time scale turns out to be much larger
than the typical escape time ($\sim10^7$ years) of CR in the Galaxy
deduced from observed isotopic ratios of CR. In addition, although the
resulting spectrum of particles happens to be a power-law in energy, the
power-law index depends on the cloud velocity, and so the superposed
spectrum due to many different sources with widely different cloud
velocities would not in general have a power-law form.

A more efficient version of Fermi mechanism is realized when one considers
encounters of particles with plane shock fronts. 
In this case, the average fractional energy gain of a particle per
encounter (defined as a cycle of one crossing and then a
re-crossing of the shock after the particle is turned back by the magnetic
field) is of first order in the relative velocity between the
shock front and the isotropic-CR frame. Currently, the ``standard'' theory
of CR acceleration --- the so-called 
``Diffusive Shock Acceleration Mechanism'' (DSAM) is, therefore, based on 
this first-order Fermi acceleration mechanism at shocks. For reviews and
references to original literature on DSAM, see, e.g.,
Refs.~\cite{drury,bland-eich,jones-ellison,takahara-rev,proth-rev,kirk-duffy}.
An important feature of DSAM is that particles emerge out of the
acceleration site with a characteristic power-law spectrum with a 
power-law index that depends only on the shock compression ratio, and not
on the shock velocity. Shocks are ubiquitous in astrophysical
situations: in the interplanetary space, in supernovae in interstellar
medium, and even in cosmological situations as in radio-galaxies.
The basic ideas of the DSAM have received impressive 
confirmation from in-situ observations in the solar system, in particular,
from observations of high energy particles accelerated at the Earth's bow
shock generated by collision of the solar wind with the Earth's
magnetosphere; see, again the reviews in
Refs.~\cite{drury,bland-eich,jones-ellison,takahara-rev} for
references. We will discuss the DSAM again in connection with UHECR in
Sect.~5. Here we only note that
for a given acceleration site, there is a maximum energy achievable,
$E_{\rm max}$, which is limited either by the
size of the shock (which has to be larger than the gyroradius of
the particles being accelerated) or by the time scale of acceleration up
to this energy (which has to be smaller than the lifetime of the shock and
also smaller than the shortest time-scale of energy losses). 

From a theoretical point of view, SNRs are not only attractive
(and maybe the only serious) candidate of Galactic CR origin in
terms of power (see Sect.~3.1) but also in terms of the maximum
achievable CR energy, which is estimated to lie somewhere between
$10^{12}\,$eV and $10^{17}\,$eV. In addition, the observed constant
beryllium-to-iron abundance ratio in the atmospheres of stars
of different metallicity is another indicator that at least the
carbon, nitrogen and oxygen CR, that produce beryllium by spallation
with interstellar hydrogen and helium (this being the main production
channel for beryllium), have to be accelerated in SNRs~\cite{rkl}.
For recent discussions of the relevance of composition for the origin
of Galactic CR, see Refs.~\cite{libeb} for lithium, beryllium, and boron
in particular, and Refs.~\cite{galactic-cr} in general.

The DSAM theory of CR
acceleration in SNRs has been worked out in considerable 
details; see, e.g., Refs.~\cite{drury,bland-eich,lagage-ces,bier-strom}. 
Support to the shock-acceleration scenario for hadronic CR is
given by experimental indications that while the composition below the
knee region becomes heavier with energy (see Sect.~2.4), the composition
is relatively less dependent on 
rigidity ($\equiv pc/Ze$, where $p$ is the momentum and 
$Ze$ is the charge, and $c$ is the speed of light). This is expected for
shock acceleration for which the maximum rigidity should be equal for all
nuclei. Furthermore, the observed X-ray emission from SNRs seems
to be caused by synchrotron radiation of electrons with energies
up to $\sim100\,$TeV. Assuming that nuclei are accelerated
as well, this implies fluxes consistent with Galactic CR
acceleration in SNR shocks~\cite{agp}.

As another effect, the interactions with the surrounding matter of
protons accelerated in SNRs produce neutral
pions, and the resulting flux of secondary $\gamma-$rays from SNRs has
been predicted as well; see,
e.g., Ref~\cite{berez-ptus,dav,naito-takahara,gps,berezhko-volk,bergg}. 
Nowadays, given the existence
of space and ground based $\gamma-$ray detecting systems (see Sect.~2.1),
the SNR acceleration paradigm for Galactic CR origin can also be tested
by searching for these secondary $\gamma-$rays. As of now, the
situation is still somewhat inconclusive since no firm detection
of such $\gamma-$rays has been reported (see, e.g.,
Ref.~\cite{buckley,hillas3}). Furthermore, the SNR scenario almost
certainly does not explain UHECR which consequently would
constitute a separate component.
Pulsars and neutron stars in close binary systems
have also been discussed as alternative Galactic CR sources for which the
maximum energy in principle may even reach the UHECR energy range. 
However, an origin of the bulk of the cosmic rays in X-ray
binary systems is contradicted by the complete absence of
detectable TeV radiation from Cygnus X-3 and Hercules
X-1, as reported by the Chicago Air Shower Array-MIchigan
Anti (CASA-MIA) experiment~\cite{casa1}. 

A comprehensive scenario for the origin of CR based exclusively
on first-order Fermi acceleration has been proposed by
Biermann~\cite{bier-rev1}. In this scenario, the sources are (a) supernovae
exploding into the interstellar medium, for energies up to
$\sim10^{15}\,$eV, (b) supernovae exploding into a predecessor stellar
wind, for energies up to $\sim10^{17}\,$eV, and (c) the hot spots of
powerful radio-galaxies for the highest energies. It is claimed that
this scenario meets every observational test to date.

A criticism of shock acceleration as the origin of CR has been
given by Colgate~\cite{colgate}. Instead, acceleration in the
electric fields produced by reconnection of twisted magnetic
fields has been suggested as a mechanism that could operate
in a much larger fraction in space than shock acceleration
and up to the highest observed CR energies.
This is due to the wide-spread presence of helical magnetic
fields carrying excess angular momentum from mass condensations
in the Universe. Apart from proposed laboratory
experiments~\cite{colgate}, it
is, however, presently not clear how to observationally
discriminate this scenario of CR origin from the shock
acceleration scenario. Also, the power law index
of the predicted spectra does not fall out of this scenario
naturally and may strongly depend on the specific environment.

Plaga~\cite{plaga} has presented a scenario where all
extrasolar hadronic CR are extragalactic in origin and
accumulate in the Galaxy due to ``magnetic flux trapping''.
It was claimed that the $\gamma-$ray flux levels from the
Magellanic clouds is not a suitable test of this scenario
and that the ankle in the energy spectrum appears as
a natural consequence of this scenario.

The opposite possibility that all CR nuclei above a few GeV
and up to the highest energies observed,
and all electrons and $\gamma-$rays above a few MeV
are of Galactic origin has also been put forward
by Dar and collaborators~\cite{dar-galactic}. In this
scenario the acceleration sources have been suggested to be
the hot spots in the highly
relativistic jets from merger and accretion induced collapse
of compact stellar objects, the so called microblazars, within
our own Galaxy and its halo. The same objects 
in external galaxies could also give rise
to cosmological $\gamma-$ray bursts.

Finally, to close this short summary with a very speculative
possibility of CR origin, we note that it is known
that charged and/or polarizable particles interacting with the
electromagnetic zero-point fluctuations are accelerated
stochastically~\cite{rueda,cole}. The discussion of this effect
goes back to Einstein and Hopf~\cite{eh} who investigated
classical atoms interacting with classical thermal
radiation. The acceleration rate $\Omega\equiv dE/dt$ for a
proton is given by~\cite{rueda}
\begin{equation}
  \Omega\simeq\frac{3}{5\pi}\frac{\Gamma^2\omega_0^5}{m_N}
  \la10^{13}\,{\rm eV}\,{\rm sec}^{-1}\,,\label{ruedarate}
\end{equation}
where $\Gamma=2e^2/3m_N$ is the radiation damping constant,
$m_N$ the nucleon mass, and $\omega_0$ is a frequency that is
smaller than the Compton frequencies of the quarks. In an energy
range where energy losses are negligible, the resulting
acceleration spectrum must have the form $j(E)\propto E^{-1}$
due to the Lorentz invariance of the spectrum of the vacuum
fluctuations. The latter is also the reason that a net
acceleration results because it implies the absence of a drag
force. The spectrum typically cuts off exponentially at energies
where the acceleration time $T_{\rm acc}\simeq E/\Omega$ becomes
larger than the proton attenuation time at that same energy due
to loss processes. It seems, however, unlikely that this
acceleration process plays a significant role in CR production
because, for given typical baryon densities, the predicted hard
spectrum tends to overproduce CR fluxes at high energies. 

\section{Propagation and Interactions of Ultra-High Energy Radiation}
Since implications and predictions of the spectrum of UHECR 
depend on their composition which is uncertain, we
will in this chapter review the propagation of all types of
particles that could play the role of UHECR. We start with the hadronic
component, continue with discussion on electromagnetic
cascades initiated by UHE photons in extragalactic space, 
and then comment on more exotic options such as UHE neutrinos
and new neutral particles predicted in certain
supersymmetric models of particle physics. 
We then discuss how propagation can be
influenced by cosmic magnetic fields and what constraints on the
location of UHECR sources are implied. The
role played by these constraints in the search for sources of
EHECR beyond $10^{20}\,$eV is discussed. Finally, the
formal description of CR propagation by transport equations
is briefly reviewed, with an account of the literature on
analytical and numerical approaches to their solution.

Before proceeding, we set up some general notation. The
interaction length $l(E)$ of a CR of energy $E$ and mass $m$ propagating
through 
a background of particles of mass $m_b$ 
is given by
\begin{equation}
  l(E)^{-1}=\int d\varepsilon n_b(\varepsilon)\int_{-1}^{+1}d\mu
  \frac{1-\mu\beta\beta_b}{2}\,\sigma(s)\,,\label{intlength}
\end{equation}
where $n_b(\varepsilon)$ is the number density of the background particles
per unit energy at energy $\varepsilon$,  
$\beta_b=(1-m_b^2/\varepsilon^2)^{1/2}$ and
$\beta=(1-m^2/E^2)^{1/2}$ are the velocities of the background
particle and the CR, respectively, $\mu$ is the cosine
of the angle between the incoming momenta, and 
$\sigma(s)$ is the total cross section of the relevant process
for the squared center of mass (CM) energy
\begin{equation}
  s=m_b^2+m^2+2\varepsilon E\left(1-\mu\beta\beta_b\right)
  \,.\label{s}
\end{equation}
The most important background particles turn out to be photons
with energies in the infrared and optical (IR/O) range or below,
so that we will usually have $m_b=0$, $\beta_b=1$. A review of
the universal photon background has been given in Ref.~\cite{rt}.

It proves convenient to also introduce an energy attenuation
length $l_E(E)$ that is obtained from Eq.~(\ref{intlength}) by
multiplying the integrand with the inelasticity, i.e. the
fraction of the energy transferred from the incoming
CR to the recoiling final state particle of interest. The
inelasticity $\eta(s)$ is given by
\begin{equation}
  \eta(s)\equiv1-\frac{1}{\sigma(s)}\int dE^\prime
  E^\prime\frac{d\sigma}{dE^\prime}\,
  (E^\prime,s)\,,\label{etas}
\end{equation}
where $E^\prime$ is the energy of the recoiling
particle considered in units of the incoming CR energy
$E$. Here by recoiling particle we usually mean the ``leading''
particle, i.e. the one which carries most of the energy.

If one is mostly interested in this leading particle, the
detailed transport equations (see Sect.~4.7) for the local density of
particles per unit energy, $n(E)$, are often
approximated by the simple ``diffusion equation''  
\begin{equation}
  \partial_t n(E)=-\partial_E\left[b(E)n(E)\right]+
  \Phi(E)\label{cel}
\end{equation}
in terms of the energy loss rate $b(E)=E/l_E(E)$ and the local
injection spectrum $\Phi(E)$. Eq.~(\ref{cel}) applies to a particle
which loses energy at a rate $dE/dt=b(E)$, and is often referred to as the
continuous energy loss (CEL) approximation. 
The CEL approximation is in general good if the non-leading
particle is of a different nature than the leading particle, and
if the inelasticity is small, $\eta(s)\ll1$. 
For an isotropic source distribution $\Phi(E,z)$ in the
matter-dominated regime for a flat Universe ($\Omega_0=1$),
Eq.~(\ref{cel}) yields a differential flux today at energy $E$, $j(E)$,
as 
\begin{equation}
  j(E)=\frac{3}{8\pi}t_0\int_0^{z_{i,\rm max}}dz_i(1+z_i)^{-11/2}
  \frac{dE_i(E,z_i)}{dE}\,\Phi(E,z_i)\,,\label{cel_sol}
\end{equation}
where $t_0$ is the age of the Universe, $E_i(E,z_i)$ is the
energy at injection redshift $z_i$ in the CEL approximation, i.e. the
solution of $dE/dt=b(E)$ (with $b(E)$ including loss due to
redshifting), $E_i(E,0)=E$ with $t=t_0/(1+z)^{3/2}$. The maximum redshift
$z_{i,\rm max}$ corresponds either to an absolute cutoff of the source
spectrum at $E_{\rm max}=E_i(E,z_{i,\rm max})$ or to the earliest epoch 
when the source became active, whichever is smaller. 
For a homogeneous production spectrum $\Phi(E)$, this simplifies
to
\begin{equation}
  j(E)\simeq\frac{1}{4\pi}l_E(E)\Phi(E)\,,\label{cel_diff}
\end{equation}
if $l_E(E)$ is much smaller than the horizon size such that
redshift and evolution effects can be ignored. Eqs.~(\ref{cel_sol})
and~(\ref{cel_diff}) are often used in the literature for
approximate flux calculations.

\subsection{Nucleons, Nuclei, and the Greisen-Zatsepin-Kuzmin Cutoff}

\begin{figure}[ht]
\centering\leavevmode
\epsfxsize=6.0in
\epsfbox{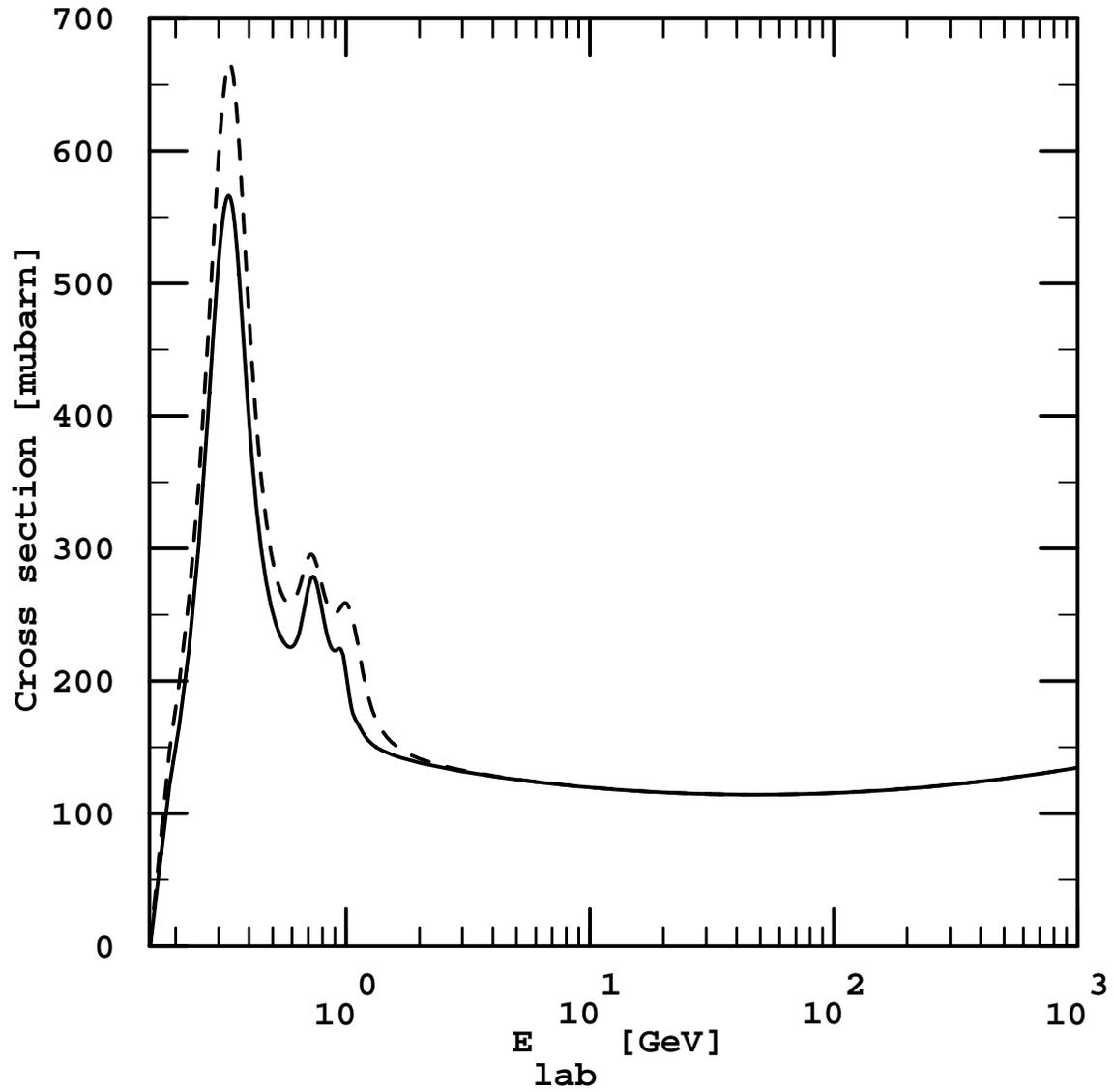}
\bigskip
\caption[...]{The total photo-pion production cross section for protons
(solid line) and neutrons (dashed line) 
as a function of the photon energy in the nucleon rest frame, 
$E_{\rm lab}$.}
\label{F4.1} 
\end{figure}

\begin{figure}[ht]
\centering\leavevmode
\epsfxsize=6.0in
\epsfbox{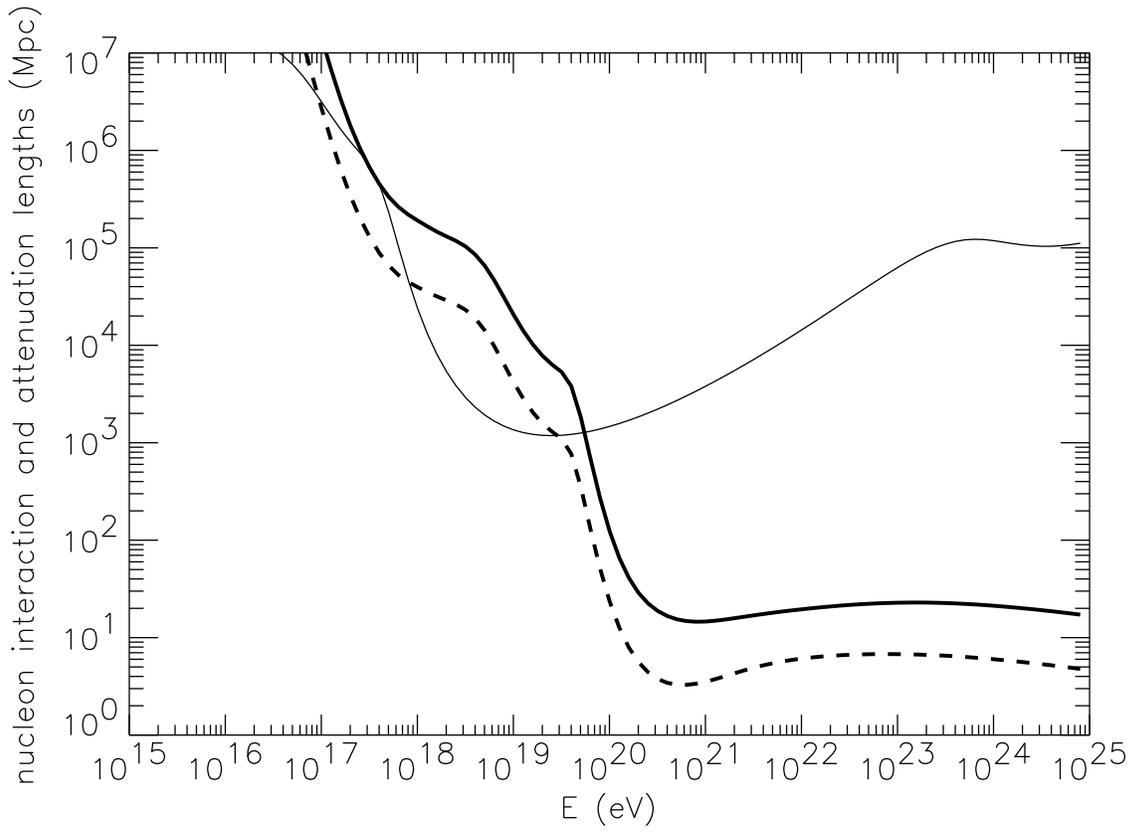}
\bigskip
\caption[...]{The nucleon interaction length (dashed line)
and attenuation length (solid line) for photo-pion production
and the proton attenuation length for pair production (thin solid
line) in the combined CMB and the estimated total
extragalactic radio background intensity shown in Fig.~\ref{F4.3}
below.}
\label{F4.2} 
\end{figure}

Shortly after its discovery, it was pointed out by
Greisen~\cite{greisen} and by Zatsepin \& Kuzmin~\cite{zat-kuz}
that the cosmic microwave background (CMB) radiation field has
profound consequences for UHECR: With respect to the
rest frame of a nucleon that has a sufficiently high energy in
the cosmic rest frame (CRF, defined as the frame in which the CMB is
isotropic), a substantial fraction of the CMB photons will
appear as $\gamma-$rays above the threshold energy for photo-pion
production, $E_\gamma^{\rm lab,thr}\equiv
m_\pi+m_\pi^2/(2m_N)\simeq160\,$MeV. The total cross section for this
process as a function of the $\gamma-$ray energy in the nucleon
rest frame, $E_\gamma^{\rm lab}$, is shown in Fig.~\ref{F4.1}.
Near the threshold the cross section exhibits a pronounced resonance
associated with single pion production, whereas in the
limit of high energies it increases 
logarithmically with $s=m_N^2+2m_N E_\gamma^{\rm
lab}$~\cite{databook}. The long tail beyond the first resonance
is essentially dominated by multiple pion production,
$N\gamma_b\to N(n\pi)$, $n>1$ ($\gamma_b$ stands for the
background photon). For a background photon of energy
$\varepsilon$ in the CRF, the threshold energy $E_\gamma^{\rm lab,thr}$
translates into a corresponding threshold for the nucleon
energy,
\begin{equation}
  E_{th}=\frac{m_\pi(m_N+m_\pi/2)}{\varepsilon}\simeq
  6.8\times10^{16}\left(\frac{\varepsilon}{{\rm eV}}\right)^{-1}
  \,{\rm eV}\,.\label{pionprod}
\end{equation}
Typical CMB photon energies are
$\varepsilon\sim10^{-3}\,$eV, leading to the so called
Greisen-Zatsepin-Kuzmin (GZK) ``cutoff'' at a few tens of EeV
where the nucleon interaction length drops to about
$6\,$Mpc as can be seen in Fig.~\ref{F4.2}. Detailed investigations
of differential cross sections, extending into the multiple
pion production regime, have been performed in the literature, mainly for
the purpose of calculating secondary $\gamma-$ray and neutrino production;  
for recent discussions and references to earlier literature 
see, e.g., Refs.~\cite{gazizov,lee,mreps}. 

Below this energy range, the dominant loss mechanism for protons
is production of electron-positron pairs on the CMB, $p\gamma_b\to
pe^+e^-$, down to the corresponding threshold
\begin{equation}
  E_{th}=\frac{m_e(m_N+m_e)}{\varepsilon}\simeq4.8\times10^{14}
  \left(\frac{\varepsilon}{{\rm eV}}\right)^{-1}\,{\rm eV}
  \,.\label{ppp}
\end{equation}
Therefore, pair production by protons (PPP) in the CMB ensues at
a proton energy $E\sim5\times10^{17}\,$eV. The first detailed
discussion of PPP in astrophysics was given by
Blumenthal~\cite{blumenthal}. PPP is very similar to triplet
pair production by electrons, $e\gamma_b\to e e^+e^-$
(see Sect.~4.2.), where ``electron'', $e$, means either an
electron or a positron in the following. Away from the threshold
the total cross
section for a nucleus of charge $Z$ is well approximated by the
one for triplet pair production, multiplied by $Z^2$. Parametric
fits to the total cross section and the
inelasticity for PPP over the whole energy range were given in
Ref.~\cite{czs}. The resulting proton attenuation length is
shown in Fig.~\ref{F4.2}. The next important loss mechanism
which starts to dominate near and below the PPP threshold is
redshifting due to the cosmic expansion. Indeed, all other
loss processes are negligible, except possibly in very dense
central regions of galaxies: The interaction length due to hadronic
processes which have total cross sections of the order of 0.1
barn in the energy range of interest, for example, is
$l\simeq3\times10^5(\Omega_b h^2)^{-1}\,{\rm Mpc}\ga10^7\,$Mpc,
where $0.009\la\Omega_b h^2\la0.02$~\cite{cst}, with $\Omega_b$
the average cosmic baryon density in units of the critical
density, and $h$ the Hubble constant $H_0$ in units of $100\,{\rm
km}\,{\rm sec}^{-1}{\rm Mpc}^{-1}$.

For neutrons, $\beta-$decay ($n\to pe^-\bar{\nu}_e$) is the
dominant loss process for $E\la10^{20}\,$eV. The neutron decay
rate $\Gamma_n=m_N/(\tau_n E)$, with
$\tau_n\simeq888.6\pm3.5\,$sec the laboratory lifetime, implies
a neutron range of propagation
\begin{equation}
  R_n=\tau_n\frac{E}{m_N}\simeq0.9\left(\frac{E}{10^{20}\,{\rm
  eV}}\right)\,{\rm Mpc}\,.\label{Rn}
\end{equation}

The dominant loss process for nuclei of energy $E\ga10^{19}\,$eV
is photodisintegration~\cite{stecker-nuc1,psb,tww,kara-tkaczyk} in the CMB
and the IR background (IRB) due to
the giant dipole resonance. Early calculations~\cite{psb} suggested a loss
length of a few Mpc. Recent observations of multi-TeV $\gamma-$rays from
the BL Lac objects Mrk 421 and Mrk 501 suggest~\cite{irb,mannheim2},  
however, an IRB roughly a factor 10 lower than previously
assumed, which is also consistent with recent independent
calculation~\cite{malkan-stecker} of the intensity and spectral energy
distribution of the IRB based on empirical data primarily from IRAS
galaxies. This tends to increase the loss length for
nuclei~\cite{stecker-nuc2}. Recent detailed Monte Carlo
simulations~\cite{epele-roulet,stecker-nuc3,stecker-sala-nuc} 
indicate that, with the reduced IR background, the CMB becomes the
dominant photon background responsible for photodisintegration and, for
example, leads to
a loss length of $\simeq10\,$Mpc at $2\times10^{20}\,$eV.
This loss length plays an important role
for scenarios in which the highest energy events observed are
heavy nuclei that have been accelerated to UHE (see, e.g.,
Ref.~\cite{elb-som}): The accelerators can not be much further
away than a few tens of Mpc. Specific flux calculations for the
source NGC 253 have been performed in Ref.~\cite{arc}.
Apart from photodisintegration, nuclei are subject to the
same loss processes as nucleons, where the respective thresholds
are given by substituting $m_N$ by the mass of the nucleus in
Eqs.~(\ref{pionprod}) and (\ref{ppp}).

\subsection{UHE Photons and Electromagnetic Cascades}

As in the case of UHE nucleons and nuclei, the propagation of UHE photons
(and electrons/positrons) is also governed by their interaction with the
cosmic photon background. The dominant interaction processes are the
attenuation (absorption) of UHE photons due to pair-production (PP) on the
background photons $\gamma_b$: $\gamma\gamma_b\to
e^+e^-$~\cite{ppics_old}, and inverse Compton scattering (ICS) of the
electrons (positrons) on the background photons. Early studies of the
effect of PP attenuation on the cosmological
UHE $\gamma$-ray flux can be found, e.g., in Refs.~\cite{stecker-pp}. 

The $\gamma-$ray threshold energy for PP on a
background photon of energy $\varepsilon$ is
\begin{equation}
  E_{th}=\frac{m_e^2}{\varepsilon}\simeq2.6\times10^{11}
  \left(\frac{\varepsilon}{{\rm eV}}\right)^{-1}\,{\rm eV}
  \,,\label{pp}
\end{equation}
whereas ICS has no threshold. In the high energy limit,
the total cross sections for PP and ICS are
\begin{equation}
  \sigma_{\rm PP}\simeq2\sigma_{\rm ICS}\simeq
  \frac{3}{2}\,\sigma_T\frac{m_e^2}{s}\,
  \ln\frac{s}{2m_e^2}\quad(s\gg m_e^2)\,.\label{sigmappics}
\end{equation}
For $s\ll m_e^2$, $\sigma_{\rm ICS}$ approaches the Thomson
cross section $\sigma_T\equiv8\pi\alpha^2/3m_e^2$ ($\alpha$ is
the fine structure constant), whereas $\sigma_{\rm PP}$
peaks near the threshold Eq.~(\ref{pp}).
Therefore, the most efficient targets for electrons and
$\gamma-$rays of energy $E$ are background photons of
energy $\varepsilon\simeq m_e^2/E$. For UHE this corresponds
to $\varepsilon\la10^{-6}\,{\rm eV}\simeq100\,$MHz. Thus,
radio background photons play an important role in UHE $\gamma$-ray
propagation through extragalactic space. 

\begin{figure}[ht]
\centering\leavevmode
\epsfxsize=6.0in
\epsfbox{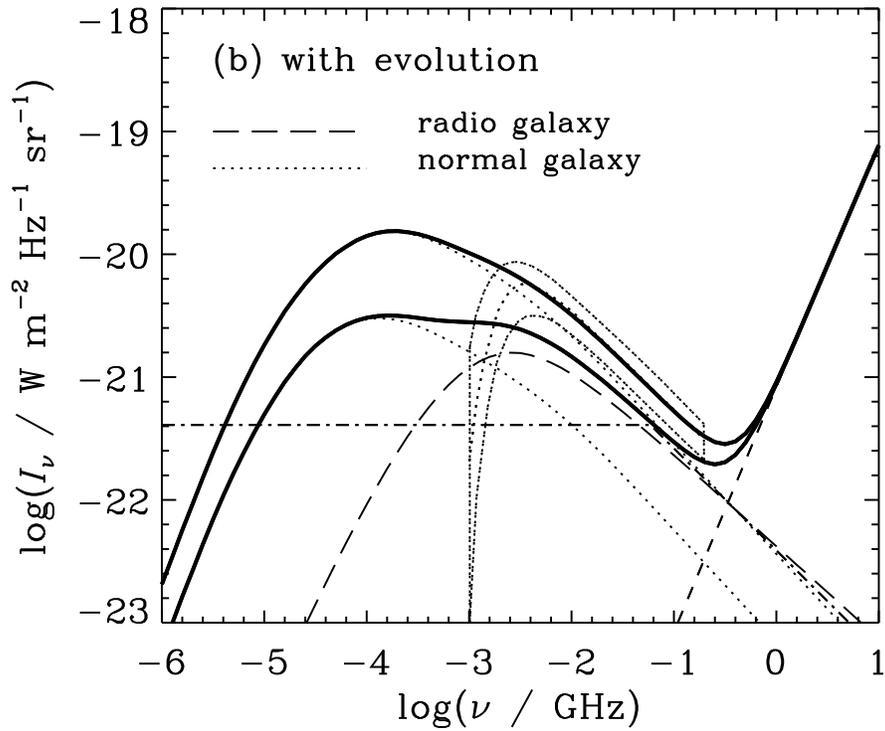}
\bigskip
\caption[...]{Contributions of normal galaxies (dotted curves),
radio galaxies (long dashed curve),
and the cosmic microwave background (short dashed curve)
to the extragalactic radio background intensity (thick solid curves)
with pure luminosity evolution for all sources (upper curves),
and for radio galaxies only (lower curves), from Ref.~\cite{pb}.
Dotted band gives an observational estimate of the total
extragalactic radio background intensity~\cite{cba}
and the dot-dash curve gives an earlier theoretical
estimate~\cite{bere1} (From Protheroe and Biermann~\cite{pb}).}
\label{F4.3}
\end{figure}

Unfortunately, the universal radio background (URB) is not very well
known mostly because it is difficult to disentangle the Galactic
and extragalactic components. Observational estimates have been given in
Ref.~\cite{cba}, and an early theoretical estimate was given in
Ref.~\cite{bere1}. Recently, an attempt has been made to calculate
the contribution to the URB from radio-galaxies and
AGNs~\cite{pb}, and also from clusters of galaxies~\cite{bbs}
which tends to give higher estimates. The issue
does not seem to be settled, however. At frequencies somewhere
below $1\,$MHz the URB is expected to cut off exponentially due to
free-free absorption. The exact location of the cut-off
depends on the abundance and clustering of electrons in the
intergalactic medium and/or the radio source and is uncertain
between about $0.1-2\,$MHz. Fig.~\ref{F4.3} compares results from
Ref.~\cite{pb} with Ref.~\cite{bere1} and the observational estimate
from Ref.~\cite{cba}.

\begin{figure}[ht]
\centering\leavevmode
\epsfxsize=6.0in
\epsfbox{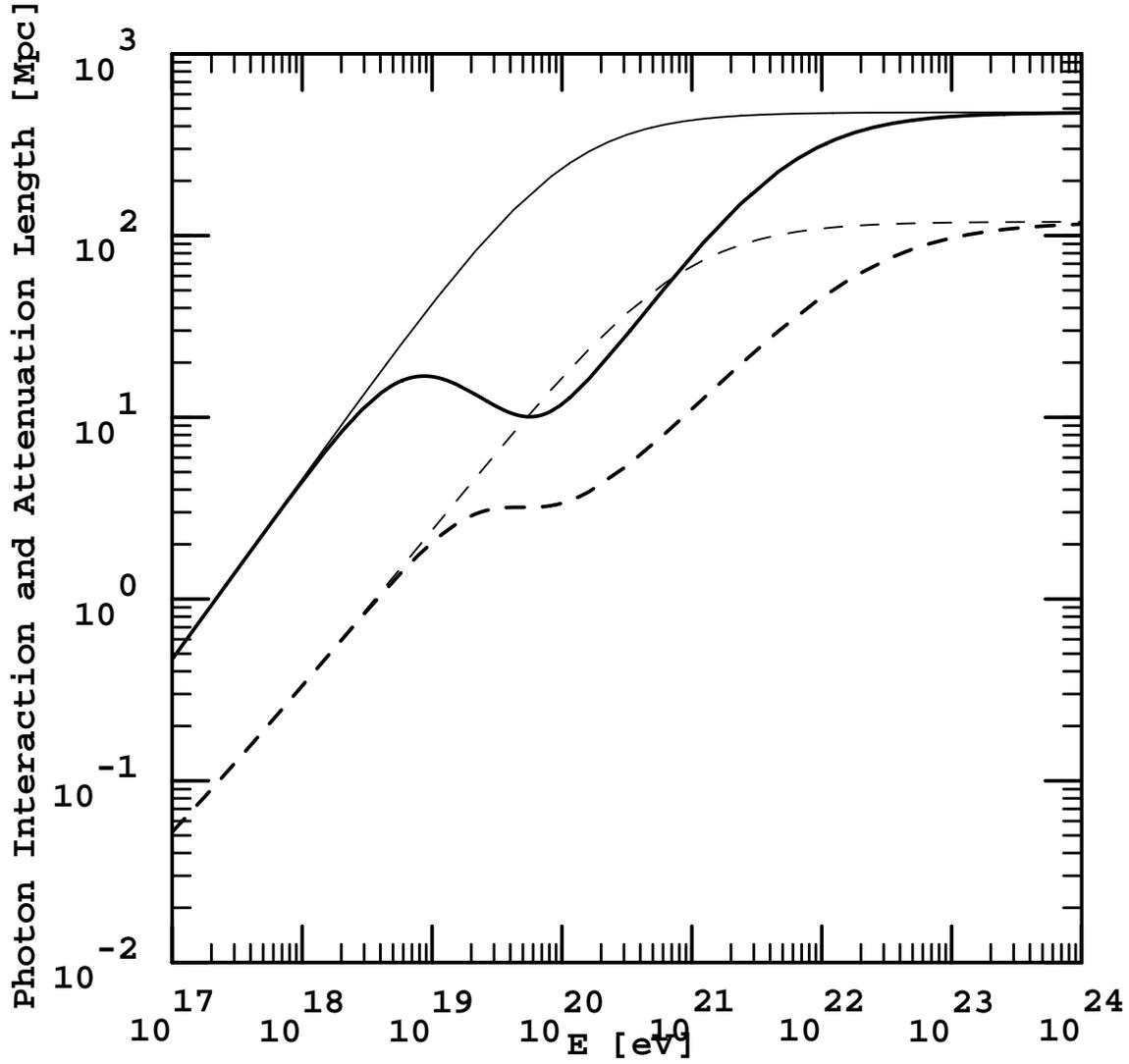}
\bigskip
\caption[...]{Interaction lengths (dashed lines) and energy
attenuation lengths (solid lines) of $\gamma-$rays in the CMB
(thin lines) and in the total low energy photon background
spectrum shown in Fig.~5.3 with the observational URB estimate
from Ref.~\cite{cba} (thick lines),
respectively. The interactions taken into account are single and
double pair production.}
\label{F4.4}
\end{figure}

\begin{figure}[ht]
\centering\leavevmode
\epsfxsize=6.0in
\epsfbox{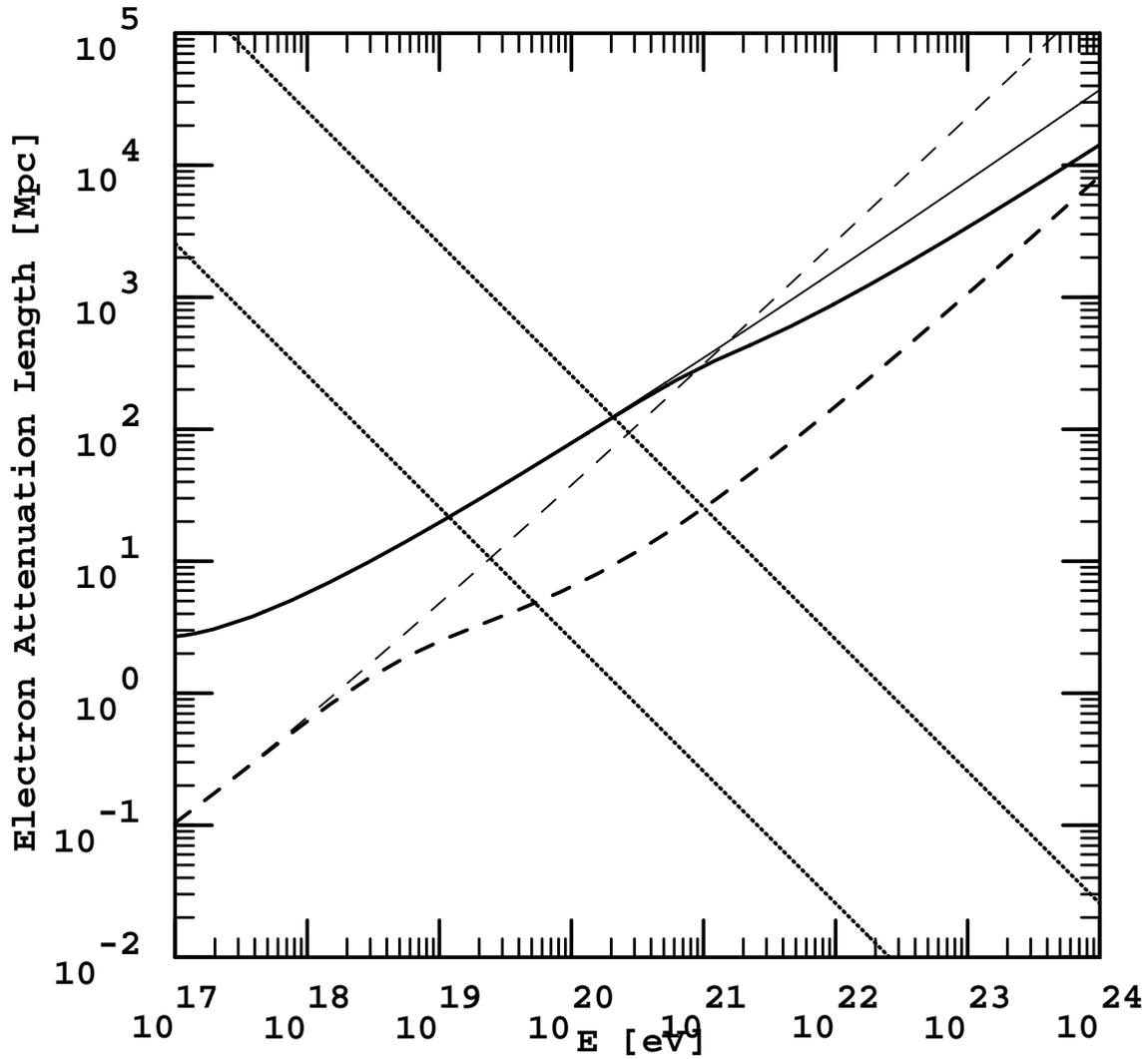}
\bigskip
\caption[...]{Energy attenuation lengths of electrons for various
processes: Solid lines are for triplet pair production, and dashed
lines for inverse Compton scattering in the CMB (thin lines) and
in the total low energy photon background spectrum shown in
Fig.~5.3 with the observational URB estimate
from Ref.~\cite{cba} (thick lines). The dotted lines are for
synchrotron emission losses in a large-scale extragalactic
magnetic field of r.m.s. strength of $10^{-11}\,$G (upper curve)
and $10^{-10}\,$G (lower curve), respectively.}
\label{F4.5}
\end{figure}

In the extreme Klein-Nishina limit, $s\gg
m_e^2$, either the electron or the positron produced in the
process $\gamma\gamma_b\to e^+e^-$ carries most of the energy of
the initial UHE photon. This leading electron can
then undergo ICS whose inelasticity (relative to the
electron) is close to 1 in the Klein-Nishina limit. As
a consequence, the upscattered photon which is now the leading
particle after this two-step cycle still carries most of the
energy of the original $\gamma-$ray, and can initiate a fresh cycle of 
PP and ICS interactions. This leads to the development of an {\it
electromagnetic (EM) cascade} which plays an important
role in the resulting observable $\gamma-$ray spectra. An important
consequence of the EM cascade development is that 
the effective penetration depth of
the EM cascade, which can be characterized by the energy
attenuation length of the leading particle (photon or electron/positron),
is considerably greater than just the interaction
lengths~\cite{bonometto}; see
Figs.~\ref{F4.4} and~\ref{F4.5}). As a result, the predicted flux of UHE
photons can be considerably larger than that calculated by considering
only the absorption of UHE photons due to PP. 

EM cascades play an important role particularly in some 
exotic models of UHECR origin such as collapse or annihilation
of topological defects (see Sect.~6) in which the EHECR
injection spectrum is predicted to be dominated by
$\gamma$-rays~\cite{abs}. Even if
only UHE nucleons and nuclei are produced in the first place, for
example, via conventional shock acceleration (see Sect.~5),
EM cascades can be produced by the secondaries coming from the
decay of pions which are created in interactions of UHE nucleons with the
low energy photon background.

The EM cascading process and the resulting diffuse $\gamma$-ray fluxes
in the conventional acceleration scenarios of UHECR origin were 
calculated in the '70s; see, e.g., Refs.~\cite{em-cascade,stecker-cel,gr}.
The EM cascades
initiated by ``primary'' $\gamma$-rays and their effects on the diffuse
UHE $\gamma$-ray flux in the topological defect scenario of UHECR were 
first considered in Ref.~\cite{abs}. All these calculations were
performed within the CEL approximation which, as described above, deals
with only the leading particle. However, the contribution of
non-leading particles to the flux can be substantial for cascades that are
not fully developed. A reliable calculation of the flux at energies much
smaller than the maximal injection energy should
therefore go beyond the CEL approximation, i.e., one should solve the
relevant Boltzmann equations for propagation; this is discussed 
in Sect.~4.7.

Cascade development
accelerates at lower energies due to the decreasing interaction
lengths (see Figs.~\ref{F4.4} and~\ref{F4.5}) until most of the
$\gamma-$rays fall below the PP threshold on the low energy
photon background at which point they pile up with a
characteristic $E^{-1.5}$ spectrum below this
threshold~\cite{bbdgp,zdziarski,hpsv,ww}. The source of these $\gamma-$rays
are predominantly the ICS photons of average energy
$\left\langle E_\gamma\right\rangle=E_e(1-4\left\langle
s\right\rangle/3m_e^2)$ arising from interactions of electrons
of energy $E_e$ with the background at
average squared CM energy $\left\langle s\right\rangle$
in the Thomson regime. The
relevant background for cosmological propagation is constituted
by the universal IR/O background, corresponding to
$\varepsilon\la1\,$eV in Eq.~(\ref{pp}), or
$E_{th}\simeq10^{11}\,$eV. Therefore, most of the energy of
fully developed EM cascades ends up below $\simeq100\,$GeV where it
is constrained by measurements of the diffuse
$\gamma-$ray flux by EGRET on board the CGRO~\cite{cdkf} and
other effects (see
Sect.~7). Flux predictions involving EM cascades are therefore
an important source of constraints of UHE energy injection on
cosmological scales. This is further discussed in Sects.~6 and~7.

It should be mentioned here that the development of EM cascades depends 
sensitively on the strength of the extragalactic magnetic fields (EGMFs)
which is rather uncertain. The EGMF typically inhibits cascade
development because of the synchrotron cooling of the $e^+e^-$ pairs
produced in the PP process. For a sufficiently strong EGMF the 
synchrotron cooling time scale of the leading electron (positron) may be
small compared to the time
scale of ICS interaction, in which case, the electron (positron)
synchrotron cools before it can undergo ICS, and thus cascade
development stops. In this case, the UHE $\gamma$-ray flux is determined
mainly by the ``direct'' $\gamma$-rays, i.e., the ones that originate at
distances less than the absorption length due to PP process. The energy
lost through synchrotron cooling does not, however, disappear; rather, 
it reappears at lower energies and can even initiate fresh EM cascades
there depending on the remaining path length and the strength of the
relevant background photons. Thus, the overall effect of a relatively
strong EGMF is to deplete the UHE $\gamma$-ray flux above some energy and
increase the flux below a corresponding energy in the ``low'' (typically
few tens to hundreds of GeV) energy region. These issues are further
discussed in Sect. 4.4.1.  

The lowest order cross sections, Eq.~(\ref{sigmappics}), fall off
as $\ln s/s$ for $s\gg m_e^2$. Therefore, at EHE, higher order
processes with more than two final state particles start to
become important because the mass scales of these particles can
enter into the corresponding cross section which typically is
asymptotically constant or proportional to powers of $\ln s$.

Double pair production (DPP), $\gamma\gamma_b\to e^+e^-e^+e^-$,
is a higher order QED process that affects UHE photons. The DPP
total cross section is a sharply rising function of $s$ near the
threshold that is given by Eq.~(\ref{pp}) with $m_e\to2m_e$, and
quickly approaches its asymptotic value~\cite{bhmm}
\begin{equation}
  \sigma_{\rm DPP}\simeq\frac{172\alpha^4}{36\pi m_e^2}
  \simeq6.45\,\mu{\rm barn}\quad(s\gg m_e^2)\,.\label{sigmadpp}
\end{equation}
DPP begins to dominate over PP above $\sim10^{21}-10^{23}\,$eV,
where the higher values apply for stronger URB (see Fig.~\ref{F4.4}).

For electrons, the relevant higher order process is triplet pair
production (TPP), $e\gamma_b\to ee^+e^-$. This process has been discussed
in some detail in Refs.~\cite{tpp} and its asymptotic high
energy cross section is
\begin{equation}
  \sigma_{\rm TPP}\simeq\frac{3\alpha}{8\pi}\,\sigma_T
  \left(\frac{28}{9}\ln\frac{s}{m_e^2}-\frac{218}{27}\right)
  \quad(s\gg m_e^2)\,,\label{sigmatpp}
\end{equation}
with an inelasticity of
\begin{equation}
  \eta\simeq1.768\left(\frac{s}{m_e^2}\right)^{-3/4}
  \quad(s\gg m_e^2)\,.\label{etatpp}
\end{equation}
Thus, although the total cross section for TPP on CMB photons
becomes comparable to the ICS cross section already around
$10^{17}\,$eV, the energy attenuation is not important up to
$\sim10^{22}\,$eV because $\eta\la10^{-3}$ (see
Fig.~\ref{F4.5}). The main effect of TPP between these energies
is to create a considerable number of electrons and channel them
to energies below the UHE range. However, TPP is dominated over by
synchrotron cooling (see Sect.~4.4), and therefore negligible,
if the electrons propagate in a magnetic field of r.m.s. strength
$\ga10^{-12}\,$G, as can be seen from Fig.~\ref{F4.5}.

Various possible processes other than those discussed above --- e.g.,
those involving the production of one or more
muon, tau, or pion pairs, double Compton
scattering ($e\gamma_b\to e\gamma\gamma$), $\gamma-\gamma$
scattering ($\gamma\gamma_b\to\gamma\gamma$), Bethe-Heitler pair
production ($\gamma X\to Xe^+e^-$, where $X$ stands for an atom, an
ion, or a free electron), the process $\gamma\gamma_b\to
e^+e^-\gamma$, and photon interactions with magnetic fields such
as pair production ($\gamma
B\to e^+e^-$) --- are in general negligible in EM cascade
development. The total cross section for the production of a
single muon pair ($\gamma\gamma_b\to\mu^+\mu^-$), for example,
is smaller than that for electron pair production by about a factor
10. Energy loss rate contributions for TPP involving pairs of
heavier particles of mass $m$ are suppressed by a factor
$\simeq(m_e/m)^{1/2}$ for $s\gg m^2$. Similarly, DPP involving
heavier pairs is also negligible~\cite{bhmm}. The cross section
for double Compton scattering is of order $\alpha^3$ and must be
treated together with the radiative corrections to ordinary
Compton scattering of the same order. Corrections to the lowest
order ICS cross section from processes involving $m_\gamma$
additional photons in the final state, $e\gamma_b\to
e+(m_\gamma+1)\gamma$, $m_\gamma\geq1$, turn out to be smaller
than 10\% in the UHE range~\cite{gould}. A similar remark
applies to corrections to the lowest order PP cross section from
the processes $\gamma\gamma_b\to e^+e^-+m_\gamma\gamma$,
$m_\gamma\geq1$. Photon$-$photon scattering can only play a role
at redshifts beyond $\simeq100$ and at energies below the
redshift-dependent pair production threshold given by 
Eq.~(\ref{pp})~\cite{sz,zs,kribs}. A similar remark applies to
Bethe-Heitler pair production~\cite{zs}. Photon interactions
with magnetic fields of typical galactic strength, $\sim10^{-6}\,$G,
are only relevant for $E\ga10^{24}\,$eV~\cite{erber}. For
extragalactic magnetic fields (EGMFs) the critical energy for such
interactions is even higher.

\subsection{Propagation and Interactions of Neutrinos and
``Exotic'' Particles}

\subsubsection{Neutrinos}

{\bf Neutrino propagation}

The propagation of UHE neutrinos is governed mainly by their interaction
with the relic neutrino background (RNB). 
The average squared CM energy for interaction of an UHE
neutrino of energy $E$ with a relic neutrino of energy
$\varepsilon$ is given by
\begin{equation}
  \left\langle s\right\rangle\simeq(45\,{\rm GeV})^2
  \left(\frac{\varepsilon}{10^{-3}\,{\rm eV}}\right)
  \left(\frac{E}{10^{15}\,{\rm GeV}}\right)\,.\label{snu}
\end{equation}
If the relic neutrino is relativistic, then
$\varepsilon\simeq3T_\nu(1+\eta_b/4)$ in Eq.~(\ref{snu}), where
$T_\nu\simeq1.9(1+z)\,{\rm K}=1.6\times10^{-4}(1+z)\,$eV is the
temperature at redshift $z$ and $\eta_b\la50$ is the dimensionless
chemical potential of relativistic relic neutrinos. For
nonrelativistic relic neutrinos of mass $m_\nu\la20\,$eV,
$\varepsilon\simeq\max\left[3T_\nu,m_\nu\right]$. Note that
Eq.~(\ref{snu}) implies interaction energies
that are typically smaller than electroweak energies even for UHE
neutrinos, except for energies near the Grand Unification scale,
$E\ga10^{15}\,$GeV, or if $m_\nu\ga1\,$eV. In this energy range,
the cross sections are given by the Standard Model of
electroweak interactions which are well confirmed
experimentally. Physics beyond the Standard Model is, therefore,
not expected to play a significant role in UHE neutrino interactions
with the low energy relic backgrounds.

The dominant interaction mode of UHE neutrinos with the RNB is the
exchange of a
$W^\pm$ boson in the t-channel ($\nu_i+\bar\nu_j\to l_i+\bar{l}_j$), 
or of a $Z^0$ boson in either the s-channel ($\nu_i+\bar\nu_i\to
f\bar{f}$) or the t-channel
($\nu_i+\bar\nu_j\to\nu_i+\bar\nu_j$)~\cite{weiler1,roulet,yoshida,ydjs}.
Here, $i,j$ stands for either the electron, muon, or tau flavor, where
$i\neq j$ for the first reaction, $l$ denotes a
charged lepton, and $f$ any charged fermion. If the latter is
a quark, it will, of course, subsequently fragment into hadrons.
As an example, the differential cross section for s-channel
production of $Z^0$ is given by
\begin{equation}
  \frac{d\sigma_{\nu_i+\bar\nu_j\to Z^0\to f\bar{f}}}{d\mu}
  =\frac{G_{\rm F}^2s}{4\pi}\,\frac{M_Z^2}{(s-M_Z^2)^2+M_Z^2
  \Gamma_Z^2}\left[g_L^2(1+\mu^*)^2+g_R^2(1-\mu^*)^2\right]
  \,,\label{Zres}
\end{equation}
where $G_{\rm F}$ is the Fermi constant, $M_Z$ and $\Gamma_Z$ are
mass and lifetime of the $Z^0$, $g_L$ and $g_R$ are the usual
dimensionless left- and right-handed coupling constants for $f$,
and $\mu^*$ is the cosine of the scattering angle in the CM system.

The t-channel processes have cross sections that rise linearly
with $s$ up to $s\simeq M_W^2$, with $M_W$ the $W^\pm$ mass,
above which they are roughly constant with a value
$\sigma_t(s\ga M_W)\sim G_{\rm F}^2M_W^2\sim10^{-34}\,{\rm cm}^2$.
Using Eq.~(\ref{snu}) this yields the rough estimate
\begin{eqnarray}
  \sigma_t(E,\varepsilon)&\sim&\min\left[10^{-34},10^{-44}
  \left(\frac{s}{{\rm MeV}^2}\right)\right]\,{\rm cm}^2
  \label{ewcross}\\
  &\sim&\min\left[10^{-34},3\times10^{-39}
  \left(\frac{\varepsilon}{10^{-3}\,{\rm eV}}\right)
  \left(\frac{E}{10^{20}\,{\rm eV}}\right)\right]\,{\rm cm}^2
  \,.\nonumber
\end{eqnarray}
In contrast, within the Standard Model the neutrino-nucleon cross
section roughly behaves as
\begin{equation}
  \sigma_{\nu N}(E)\sim10^{-31}(E/10^{20}
  \,{\rm eV})^{0.4}\,{\rm cm}^2\label{cccross1}
\end{equation}
for $E\ga10^{15}\,$eV (see discussion below at end of Sect.~4.3.1).
Interactions of UHE neutrinos with nucleons are, however, still
negligible compared to interactions with the RNB because the
RNB particle density is about ten orders of magnitude larger 
than the baryon density. The only exception could occur near
Grand Unification scale energies and at high redshifts and/or
if contributions to the neutrino-nucleon cross section from
physics beyond the Standard Model dominate at these energies
(see below at end of Sect.~4.3.1).

\begin{figure}[ht]
\centering\leavevmode
\epsfxsize=6.0in
\epsfbox{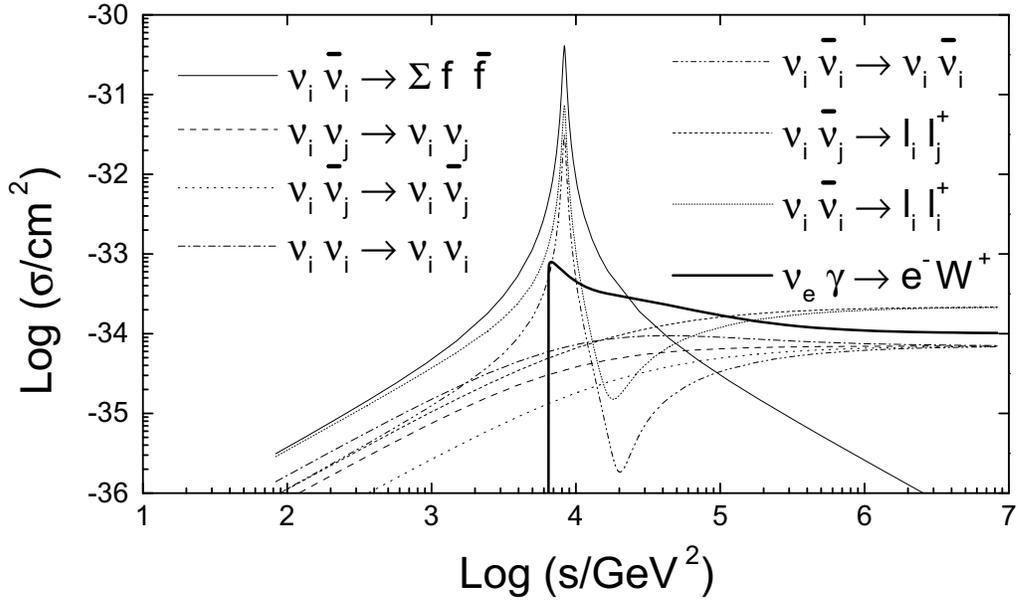}
\bigskip
\caption[...]{Various cross sections relevant for neutrino
propagation as a function of $s$~\cite{roulet,seckel}.  
The sum $\sum_jf_j\bar f_j$ does not include $f_j = \nu_i\,,l_i\,,t\,,W$,
or $Z$. (From D.~Seckel~\cite{seckel}).}
\label{F4.6}
\end{figure}

It has recently been pointed out~\cite{seckel} that above
the threshold for $W^\pm$ production the
process $\nu+\gamma\to lW^+$ becomes comparable to the
$\nu\nu$ processes discussed above. Fig.~\ref{F4.6} compares
the cross sections relevant for neutrino propagation
at CM energies around the electroweak scale. Again,
for UHE neutrino interactions with the RNB
the relevant CM energies can only be reached if 
(a) the UHE neutrino energy is close to the Grand Unification scale,
or (b) the RNB neutrinos have masses in the eV regime, or (c) at
redshifts $z\ga10^3$. Even then the $\nu\gamma$ process
never dominates over the $\nu\nu$ process.

At lower energies
there is an additional $\nu\gamma$ interaction that was
recently discussed as potentially important besides the
$\nu\nu$ processes: Using an effective Lagrangian derived
from the Standard Model, Ref.~\cite{dr} obtained the
result $\sigma_{\gamma+\nu\to\gamma+\gamma+\nu}(s)\simeq
9\times10^{-56}\,(s/{\rm MeV}^2)^5\,{\rm cm}^2$, supposed
to be valid at least up to $s\la10\,{\rm MeV}^2$. Above
the electron pair production threshold the cross section
has not been calculated because of its complexity but is
likely to level off and eventually decrease. Nevertheless,
if the $s^5$ behavior holds up to $s\simeq$ a few hundred
MeV$^2$, comparison with Eq.~(\ref{ewcross}) shows that
the process $\gamma+\nu\to\gamma+\gamma+\nu$ would start
to dominate and influence neutrino propagation around
$E\sim3\times10^{17}\left(\varepsilon/10^{-3}\,{\rm eV}\right)
\,$eV, as was pointed out in Ref.~\cite{hwt}.

For a given source distribution, the contribution of the ``direct''
neutrinos to the flux can be computed by integrating Eq.~(\ref{cel_sol})
up to the interaction redshift
$z(E)$, i.e. the average redshift from which a neutrino of
present day energy $E$ could have propagated without
interacting. This approximation neglects the secondary neutrinos
and the decay products of the leptons created
in the neutral current and charged current reactions of UHE
neutrinos with the RNB discussed above.
Similarly to the EM case, these secondary particles can lead to
neutrino cascades developing over cosmological
redshifts~\cite{yoshida}.

Approximate expressions for the interaction redshift for the
processes discussed above
have been given in Refs.~\cite{bbdgp,bhs} for CM energies
below the electroweak scale, assuming relativistic,
nondegenerate relic neutrinos, $m_\nu\la T_\nu$, and
$\eta_b\ll1$. Approaching the electroweak scale,
a resonance occurs in the interaction cross section for
s-channel $Z^0$ exchange at the $Z^0$ mass,
$s=M_Z^2\simeq(91\,{\rm GeV})^2$, see Eq.~(\ref{Zres}).
The absorption redshift for
the corresponding neutrino energy, $E\simeq10^{15}\,{\rm
GeV}(\varepsilon/10^{-3}\,{\rm eV})^{-1}$ drops to a few (or
less for a degenerate, relativistic RNB) and asymptotically
approaches constant values of a few tens at higher energies.

An interesting situation arises if the RNB consists of massive
neutrinos with $m_\nu\sim1\,$eV: Such neutrinos would constitute
hot dark matter which is expected to cluster~\cite{cowsik-mccl}, 
for example, in galaxy clusters. This would potentially increase 
the interaction probability for any neutrino of energy within the
width of the $Z^0$ resonance at $E=M_Z^2/2m_\nu=4\times10^{21}({\rm
eV}/m_\nu)\,$eV. Recently it has
been suggested that the stable end products of the ``Z-bursts''
that would thus be induced at close-by distances ($\la50\mpc$) from
Earth may explain the
highest energy cosmic rays~\cite{fms,weiler2} and may also
provide indirect evidence for neutrino hot dark matter. These end
products would be mostly nucleons and $\gamma-$rays with
average energies a factor of $\simeq5$ and $\simeq40$ lower, respectively,  
than the original UHE neutrino. As a consequence, if the UHE neutrino
was produced as a secondary of an accelerated proton, the energy
of the latter would have to be at least a few
$10^{22}\,$eV~\cite{fms}, making Z-bursts above GZK energies
more likely to play a role in the context of non-acceleration
scenarios (see Sects.~6,7). Moreover, it has
subsequently been pointed out~\cite{wax3} that Z production is
dominated by annihilation on the non-clustered massive RNB
compared to annihilation with neutrinos clustering in
the Galactic halo or in nearby galaxy clusters.
As a consequence, for a significant contribution of
neutrino annihilation to the observed EHECR flux, a new
class of neutrino sources, unrelated to UHECR sources,
seems necessary. This has been confirmed by more detailed
numerical simulations~\cite{ysl} where it has, however, also been
demonstrated that the most significant contribution could come
from annihilation on neutrino dark matter clustering in the
Local Supercluster by amounts consistent with expectations.
In the absence of any assumptions on the neutrino sources,
the minimal constraint comes from the unavoidable
production of secondary $\gamma-$rays contributing to the
diffuse flux around 10 GeV measured by EGRET: If the Z-burst
decay products are to explain EHECR, the massive neutrino
overdensity $f_\nu$ over a length scale $l_\nu$ has to
satisfy $f_\nu\ga20\,(l_\nu/5\,{\rm Mpc})^{-1}$, provided
that only neutrinos leave the source, a situation that may
arise in top-down models if the X particles decay exclusively
into neutrinos (see Ref.~\cite{slby} for a model involving
topological defects and Ref.~\cite{gk2} for a scenario involving
decaying superheavy relic particles). If, instead, the
total photon source luminosity is comparable to the total
neutrino luminosity, as in most models, the EGRET constraint
translates into the more stringent requirement
$f_\nu\ga10^3(l_\nu/5\,{\rm Mpc})^{-1}$. This bound can
only be relaxed if most of the EM energy is radiated in the TeV
range where the Universe is more transparent~\cite{ysl}.
Furthermore, the Z-burst scenario requires sources that
are optically thick for accelerated protons with respect to
photo-pion production because otherwise the observable
proton flux below the GZK cutoff would be comparable
to the neutrino flux~\cite{wax3}. A systematic parameter study of
the required overdensity, based on analytical flux estimates,
has been performed in Ref.~\cite{bpvz1}.
Recently it has been noted that a degenerate relic neutrino
background would increase the interaction probability and
thereby make the Z-burst scenario more
promising~\cite{gelmini-kusenko}. A neutrino asymmetry of
order unity is not excluded phenomenologically~\cite{pk}
and can be created in the early Universe, for example,
through the Affleck-Dine baryogenesis mechanism~\cite{affleck-dine} or due
to neutrino oscillations. The authors of Ref.~\cite{gelmini-kusenko}
pointed out that for a neutrino mass $m_\nu\simeq0.07\,$eV,
a value suggested by the Super-Kamiokande experiment~\cite{superk},
and for sources at redshifts of a few,
the flux of secondary Z-decay products is maximal for a
RNB density parameter $\Omega_\nu\simeq0.01$. Such neutrino masses,
however, require the sources to produce neutrinos at least up
to $10^{22}\,$eV.

UHE neutrinos from the decay of pions, that are produced
by interactions of accelerated protons in astrophysical sources, must have
originated within redshifts of a few. Moreover, in most
conventional models their flux is
expected to fall off rapidly above $10^{20}\,$eV. Examples are
production in active galactic nuclei within
hadronic models~\cite{agn-nu,mannheim,hz,protheroe2,protheroe,wb2}, and
``cosmogenic'' neutrinos from interactions of UHECR nucleons (near or 
above the GZK cutoff) with the CMB (see,
e.g.,Refs.~\cite{stecker-neut,hs}). 
The latter source is the only one that is
guaranteed to exist due to existence of UHECR near the GZK
cutoff, but the fluxes are generally quite small. 
Therefore, interaction of these UHE neutrinos with the RNB, that could
reveal the latter's existence, can, 
if at all, be important only if the relic neutrinos have a mass
$m_\nu\ga1\,$eV~\cite{weiler1}. Due to the continuous release of
UHE neutrinos up to much higher redshifts, most top-down scenarios would imply
substantially higher fluxes that also extend to much
higher energies~\cite{bhs}. Certain features in the UHE neutrino spectrum
predicted within such top-down scenarios, such as a change of slope for
massless neutrinos~\cite{yoshida} or a dip structure for
relic neutrino masses of order $1\,$eV~\cite{ydjs,weiler2}, have
therefore been proposed as possibly the only way to detect the RNB.
However, some of the scenarios at the high end of
neutrino flux predictions have recently been ruled out
based on constraints on the accompanying energy release into
the EM channel (see Sect.~7).

Since in virtually all models UHE neutrinos are
created as secondaries from pion decay, i.e. as electron or muon
neutrinos, $\tau-$neutrinos can only be produced by a flavor
changing $W^\pm$ t-channel interaction with the RNB.
The flux of UHE $\tau-$neutrinos is
therefore usually expected to be substantially smaller than the
one of electron and muon neutrinos, if no neutrino oscillations
take place at these energies. However, the recent evidence
from the Superkamiokande experiment for nearly maximal
mixing between muon and $\tau-$neutrinos with $|\Delta m^2|=
|m^2_{\nu_\mu}-m^2_{\nu_\tau}|\simeq
5\times10^{-3}\,{\rm eV}^2$~\cite{superk}
would imply an oscillation length of $L_{osc}=2E/|\Delta m^2|=
2.6\times10^{-6}(E/{\rm PeV})(|\Delta m^2|/5\times10^{-3}\,{\rm eV}^2)^{-1}
\,$pc and, therefore, a rough equilibration between muon
and $\tau-$neutrino fluxes from any source at a distance
larger than $L_{osc}$~\cite{mannheim3}. Turning this around, one
sees that a source at distance $d$ emitting neutrinos of energy
$E$ is sensitive to neutrino mixing with $|\Delta m^2|=2E/d\simeq
1.3\times10^{-16}\,(E/{\rm PeV})(d/100\,{\rm Mpc})^{-1}\,
{\rm eV}^2$~\cite{pakvasa,halzen-saltzberg}. Under certain
circumstances, resonant conversion in the
potential provided by the RNB clustering in galactic halos
may also influence the flavor composition of UHE neutrinos
from extraterrestrial sources~\cite{horvat}. In addition,
such huge cosmological baselines can be sensitive probes
of neutrino decay~\cite{kmp}.

{\bf Neutrino detection}

We now turn to a discussion of UHE neutrino interactions
with matter relevant for neutrino detection.
UHE neutrinos can be detected by detecting the muons produced 
in ordinary matter via
charged-current reactions with nucleons; see,
e.g., Refs.~\cite{fmr,gqrs,gkr} for recent discussions.
Corresponding cross sections are
calculated by folding the fundamental
standard model quark-neutrino cross section with the
distribution function of the partons in the nucleon.
These cross sections are most sensitive to the abundance of
partons of fractional momentum $x\simeq M_W^2/2m_N E$, where
$E$ is the neutrino energy. For
the relevant squared momentum transfer, $Q^2\sim M_W^2$, these
parton distribution functions have been measured down to
$x\simeq0.02$~\cite{hera}. (It has been suggested that observation
of the atmospheric neutrino flux with future neutrino telescopes
may probe parton distribution functions at much smaller $x$
currently inaccessible to colliders~\cite{ggv}).
Currently, therefore, neutrino-nucleon cross
sections for $E\ga10^{14}\,$eV can be obtained only by
extrapolating the 
parton distribution functions to lower $x$. Above
$10^{19}\,$eV, the resulting uncertainty has been estimated
to be a factor 2~\cite{gqrs}, whereas within the dynamical
radiative parton model it has been claimed to be at most
20 \%~\cite{gkr}. An intermediate estimate using the CTEQ4-DIS
distributions can roughly be parameterized by~\cite{gqrs}
\begin{equation}
  \sigma_{\nu N}(E)\simeq2.36\times10^{-32}(E/10^{19}
  \,{\rm eV})^{0.363}\,{\rm cm}^2\quad(10^{16}\,{\rm eV}\la
  E\la10^{21}\,{\rm eV})\,.\label{cccross2}
\end{equation}
Improved calculations including non-leading logarithmic
contributions in $1/x$ have recently been performed in
Ref.~\cite{kms}. The results for the neutrino-nucleon
cross section differ by less than a factor 1.5 with
Refs.~\cite{gqrs,gkr} even at $10^{21}\,$eV.
Neutral-current neutrino-nucleon cross sections are
expected to be a factor 2-3 smaller than charged-current cross
sections at UHE and interactions with electrons only play a
role at the Glashow resonance, $\bar\nu_e e\to W$, at
$E=6.3\times10^{15}\,$eV. Furthermore, cross sections of
neutrinos and anti-neutrinos are basically identical at UHE.
Radiative corrections influence the total cross section
negligibly compared to the parton distribution uncertainties,
but may lead to an increase of the average inelasticity in
the outgoing lepton from $\simeq0.19$ to $\simeq0.24$ at
$E\sim10^{20}\,$eV~\cite{sigl}, although this would
probably hardly influence the shower character.

Neutrinos propagating through the Earth start to be attenuated
above $\simeq100\,$TeV due to the increasing Standard Model
cross section as indicated by Eq.~(\ref{cccross2}).
Detailed integrations of the relevant transport equations for
muon neutrinos above a TeV have been presented in Ref.~\cite{kms},
and, for a general cold medium, in Ref.~\cite{np}.
In contrast, $\tau-$neutrinos with energy up to $\simeq100\,$PeV 
can penetrate the Earth due to their
regeneration from $\tau$ decays~\cite{halzen-saltzberg}.
As a result, a primary UHE $\tau-$neutrino beam propagating
through the Earth would cascade down below $\simeq100\,$TeV
and in a neutrino telescope could give rise to a higher total
rate of upgoing events as compared to downgoing events for
the same beam arriving from above the horizon. As mentioned
above, a primary $\tau-$neutrino beam could arise even in
scenarios based on pion decay, if $\nu_\mu-\nu_\tau$ mixing
occurs with the parameters suggested by the Super-Kamiokande
results~\cite{mannheim3}. In the PeV range, $\tau-$neutrinos
can produce characteristec "double-bang" events where the first
bang would be due to the charged-current production by the
$\tau-$neutrino of a $\tau$ whose decay at a typical distance
$\simeq$ 100\ m would produce the second bang~\cite{pakvasa}. These
effects have also been suggested as an independent astrophysical
test of the neutrino oscillation hypothesis. In addition,
isotropic neutrino fluxes in the energy range between 10 TeV and
10 PeV have been suggested as probes of the Earth's density
profile, whereby neutrino telescopes could be used for
neutrino absorption tomography~\cite{jrf}.

{\bf New Interactions}

It has been suggested that the neutrino-nucleon
cross section could be enhanced by new physics beyond the
electroweak scale in the CM, or above about a PeV in the
nucleon rest frame; see Eq.~(\ref{snu}). For the lowest partial wave
contribution to the cross section of a point-like particle this would
violate unitarity~\cite{bhg}. However, two major possibilities
have been discussed in the literature for which unitarity
bounds seem not to be violated. In the first,
a broken SU(3) gauge symmetry dual to the unbroken SU(3) color gauge group
of strong interaction is introduced as the ``generation symmetry'' such
that the three generations of leptons and quarks represent the quantum
numbers of this generation symmetry. In this scheme, neutrinos can have
effectively strong interaction with quarks and, in addition, neutrinos can
interact coherently with all partons in the nucleon, resulting in
an effective cross section
comparable to the geometrical nucleon cross section~\cite{bhfpt}.  
However, the massive neutral gauge bosons of the broken
generation symmetry would also mediate flavor changing neutral current
(FCNC) processes, and experimental bounds on these processes indicate
that the scale of any such new interaction must be above $\sim100$ TeV. 
The second possibility is that there may be a large increase in the
number of degrees of freedom above the electroweak 
scale~\cite{kovesi-domokos}. A specific implementation
of this idea is given in theories with $n$ additional large
compact dimensions and a quantum gravity scale $M_{4+n}\sim\,$TeV
that has recently received much attention in the literature~\cite{tev-qg}
because it provides an alternative solution (i.e., without 
supersymmetry) to the hierarchy problem
in Grand Unifications of gauge interactions. 
In such scenarios, the exchange of bulk gravitons (Kaluza-Klein
modes) leads to an extra contribution to any two-particle cross section
given by~\cite{ns}
\begin{equation}
  \sigma_{g}\simeq\frac{4\pi s}{M^4_{4+n}}\simeq
  10^{-27}\left(\frac{M_{4+n}}{{\rm TeV}}\right)^{-4}
  \left(\frac{E}{10^{20}\,{\rm eV}}\right)\,{\rm cm}^2\,,
  \label{sigma_graviton}
\end{equation}
where the last expression applies to a neutrino
of energy $E$ hitting a nucleon at rest. Note that a neutrino
would typically start to interact in the atmosphere
for $\sigma_{\nu N}\ga10^{-27}\,{\rm cm}^2$, i.e. for
$E\ga10^{20}\,$eV, assuming $M_{4+n}\simeq1\,$TeV.
The neutrino therefore becomes a primary candidate for the
observed EHECR events. A specific signature of this scenario
in neutrino telescopes based on ice or water as detector
medium would be the absence of events above the energy $E_c$ where
$\sigma_g$ grows beyond $\simeq10^{-27}\,{\rm cm}^2$. The corresponding
signature in atmospheric detectors such as the Pierre Auger detectors
would be a hardening of the spectrum above the energy $E_c$. 

There are, however, astrophysical constraints on $M_{4+n}$ which result
from limiting the emission of bulk gravitons into the extra dimensions. 
The strongest constraints in this regard come from nucleon-nucleon
bremsstrahlung
in type II supernovae~\cite{astro-extra-dim}. These contraints read
$M_6\ga50\,$TeV, $M_7\ga4\,$TeV, and
$M_8\ga1\,$TeV, for $n=2,3,4$, respectively, and,
therefore, $n\geq4$ is required if neutrino primaries
are to serve as a primary candidate for the EHECR events observed
above $10^{20}\,$eV. This assumes that all extra
dimensions have the same size given by
\begin{equation}
  r_n\simeq M^{-1}_{4+n}\left(\frac{M_{\rm Pl}}{M_{4+n}}\right)^{2/n}
  \simeq2\times10^{-17}\left(\frac{{\rm TeV}}{M_{4+n}}\right)
  \left(\frac{M_{\rm Pl}}{M_{4+n}}\right)^{2/n}\,{\rm cm}
  \,,\label{rextra}
\end{equation}
where $M_{\rm Pl}$ denotes the Planck mass. 
The above lower bounds on $M_{4+n}$ thus translate into the corresponding
upper bounds $r_n\la3\times10^{-4}\,$mm, $r_n\la4\times10^{-7}\,$mm,
and $r_n\la 2\times10^{-8}\,$mm, respectively.
The neutrino primary hypothesis of EHECR together with other astrophysical
and cosmological constraints thus provides an interesting testing
ground for theories involving large compact extra dimensions representing 
one possible kind of physics beyond the Standard Model. 
In this context, we mention that in theories with large compact extra
dimensions mentioned above, Newton's law of gravity is expected to be
modified at distances smaller than the length scale given by
Eq.~(\ref{rextra}). Indeed, there are laboratory 
experiments measuring gravitational interaction at small
distances (for a recent review of such experiments see
Ref.~\cite{lcp}), which also probe these theories. Thus, future EHECR
experiments and gravitational experiments in the laboratory together 
have the potential of providing rather strong tests of these theories. 
These tests would be complementary to constraints
from collider experiments~\cite{coll-extra-dim}.

In the context of conventional astrophysical sources, the relevant UHE
neutrino primaries
could, of course, only be produced as secondaries in interactions of
accelerated protons of energies at least $10^{21}\,$eV
with matter or with low energy photons. This implies strong requirements
on the possible sources (see Sect.~5).

\subsubsection{Supersymmetric Particles}

Certain supersymmetric particles have been suggested as
candidates for the EHECR events. For example, if the gluino
is light and has a lifetime long compared to the strong
interaction time scale, because it carries color charge, it will bind
with quarks, anti-quarks and/or gluons to form color-singlet hadrons,
so-called R-hadrons. This can occur in supersymmetric theories 
involving gauge-mediated supersymmetry (SUSY) breaking~\cite{raby} where
the resulting gluino mass arises dominantly from radiative corrections
and can vary between $\sim1\,$GeV and $\sim100\,$GeV.
In these scenarios, the gluino can be the lightest supersymmetric
particle (LSP). There are also arguments against a light
quasi-stable gluino~\cite{vo}, mainly based on constraints on
the abundance of anomalous heavy isotopes of hydrogen and oxygen
which could be formed as bound states of these nuclei and the
gluino. However, the case of a light quasi-stable gluino does
not seem to be settled.

In the context of such scenarios a specific case has been suggested
in which the gluino mass lies between 0.1 and $1\,$GeV~\cite{farrar}.
The lightest gluino-containing baryon, $uds\tilde{g}$, denoted $S^0$,
could then be long-lived or stable, and the kinematical threshold for
$\gamma_b$ -- $S^0$ ``GZK'' interaction would be higher than for nucleons,
at an energy given
by substituting the $S^0$ mass $M_{S^0}$ for the nucleon mass in
Eq.~(\ref{pionprod})~\cite{cfk}. Furthermore, the cross section
for $\gamma_b$ -- $S^0$ interaction peaks at an energy higher by a facor
$(m_{S^0}/m_N)(m_*-m_{S^0})/(m_\Delta-m_N)$ where the ratio
of the mass splittings between
the primary and the lowest lying resonance of the $S^0$ (of mass $m_*$) 
and the nucleon satisfies $(m_*-m_{S^0})/(m_\Delta-m_N)\ga2$.
As a result of this and a somewhat smaller interaction cross
section of $S^0$ with photons, the effective GZK threshold is
higher by factors of a few and sources of events above $10^{19.5}\,$eV
could be 15-30 times further away than for the case of
nucleons. The existence of such events was, therefore,
proposed as a signal of supersymmetry~\cite{cfk}.
In fact, Farrar and Biermann reported a possible correlation between
the arrival direction of the five highest energy CR events and compact radio
quasars at redshifts between 0.3 and 2.2~\cite{fb}, as might be expected
if these quasars were sources of massive neutral particles.
Undoubtedly, with the present amount of data the interpretation
of such evidence for a correlation remains somewhat subjective,
as is demonstrated by the criticism of the statistical analysis
in Ref.~\cite{fb} by Hoffman~\cite{hoffman} and the reply by Farrar
and Biermann~\cite{fb-reply}). Still, it can be expected that
only a few more events could confirm or rule out the quasar 
hypothesis.

Meanwhile, however, accelerator constraints have become
more stringent~\cite{e761,ktev} and seem to be inconsistent with the
scenario from Ref.~\cite{farrar}. However, the scenario with
a ``tunable'' gluino mass~\cite{raby} still seems possible
and suggests either the gluino--gluon bound state $g\tilde g$,
called glueballino $R_0$, or the isotriplet $\tilde g-(u\bar u-
d\bar d)_8$, called $\tilde\rho$, as the lightest quasi-stable
R-hadron. For a summary of scenarios with light gluinos
consistent with accelerator constraints see Ref.~\cite{clavelli}.

Similar to the neutrino primary hypothesis in the context
of acceleration sources (see Sect.~4.3.1), a specific difficulty
of this scenario is the fact that, of course,
the neutral R-hadron can not be accelerated, but rather has to be
produced as a secondary of an accelerated proton interacting
with the ambient matter. As a consequence, protons must
be accelerated to at least $10^{21}\,$eV at the source in
order for the secondary $S^0$ particles to explain the EHECR
events. Furthermore, secondary production would also include
neutrinos and especially $\gamma-$rays, leading to fluxes from
powerful discrete acceleration sources that may be
detectable in the GeV range by space-borne $\gamma-$ray instruments
such as EGRET and GLAST, and in the TeV range by ground
based $\gamma-$ray detectors such as HEGRA and WHIPPLE
and the planned VERITAS, HESS, and MAGIC projects. At least
the latter three
ground based instruments should have energy thresholds low
enough to detect $\gamma-$rays from the postulated sources
at redshift $z\sim1$.
Such observations in turn imply constraints on the required
branching ratio of proton interactions into the R-hadron which,
very roughly, should be larger than $\sim0.01$. These
constraints, however, will have to be investigated in more
detail for specific sources. It was also suggested to search
for heavy neutral baryons in the data from Cherenkov
instruments in the TeV range in this context~\cite{plaga2}.

A further constraint on new, massive strongly interacting
particles in general comes from the character of the air
showers created by them: The observed EHECR air showers
are consistent with nucleon primaries and limits the
possible primary rest mass to less than
$\simeq50\,$GeV~\cite{afk}. With the statistics expected
from upcoming experiments such as the Pierre Auger Project,
this upper limit is likely to be lowered down to
$\simeq10\,$GeV.

It is interesting to note in this context that in case of
a confirmation of the existence of new neutral particles
in UHECR, a combination of accelerator, air shower, and
astrophysics data would be highly restrictive in terms of
the underlying physics: In the above scenario, for example,
the gluino would have to be in a narrow mass range, 1--10
GeV, and the newest accelerator constraints on the Higgs
mass, $m_h\ga90\,$GeV, would require the presence of
a D term of an anomalous $U(1)_X$ gauge symmetry, in
addition to a gauge-mediated contribution to SUSY breaking
at the messenger scale~\cite{raby}.

Finally, SUSY could also play a role in top-down scenarios
where it would modify the spectra of particles resulting from the
decay of the X particles (see Sect.~6.2.1).

\subsubsection{Other Particles}
Recently it was suggested that QCD instanton induced interactions
between quarks can lead to a stable, strong bound state of two
$\Lambda=uds$ particles, a so called $uuddss$ H-dibaryon state
with a mass $M_H\simeq1700\,$MeV~\cite{kochelev}. This particle
would have properties similar to the sypersymmetric $S^0$ particle
discussed in the previous section, i.e. it is neutral and
its spin is zero. Its effective GZK cutoff would, therefore,
also be considerably higher than for nucleons, at approximately
$7.3\times10^{20}\,$eV, according to Ref.~\cite{kochelev}.
It would thus also be a primary candidate for the observed
EHECR events that could be produced at high redshift sources.

\subsection{Signatures of Galactic and Extragalactic Magnetic
Fields in UHECR Spectra and Images}
Cosmic magnetic fields can have several implications for UHECR
propagation that may leave signatures in the observable spectra
which could in turn be used to constrain or even measure the
magnetic fields in the halo of our Galaxy and/or the extragalactic
magnetic field (EGMF).

\subsubsection{Synchrotron Radiation and Electromagnetic
Cascades}
As already mentioned in Sect.~4.2, the development of EM cascades
strongly depends on
presence and strength of magnetic fields via the synchrotron
loss of its electronic component: For a particle of mass $m$ and
charge $qe$ ($e$ is the electron charge) the energy loss rate in a
field of squared r.m.s. strength $B^2$ is
\begin{equation}
  \frac{dE}{dt}=-\frac{4}{3}\,\sigma_T\,\frac{B^2}{8\pi}
  \left(\frac{qm_e}{m}\right)^4\left(\frac{E}{m_e}\right)^2
  \,.\label{synch}
\end{equation}
For UHE protons this is negligible, whereas for
UHE electrons the synchrotron losses eventually dominate over
their attenuation (due to interaction with the background photons) above
some critical energy
$E_{\rm tr}\sim10^{20}(B/10^{-10}\,{\rm G})^{-1}\,$eV that depends
somewhat on the URB (see Fig.~\ref{F4.5}). Cascade development
above that energy is essentially blocked because the electrons
lose their energy through synchrotron radiation almost instantaneously
once they are
produced. In this energy range, $\gamma-$ray propagation is
therefore governed basically by absorption due to PP or DPP,
and the observable flux is dominated by the ``direct'' or
``first generation'' $\gamma-$rays, and their flux can be
calculated by integrating Eq.~(\ref{cel_sol}) up to the
absorption length (or redshift). Since this length is much
smaller than the Hubble radius, for a homogeneous source
distribution this reduces to Eq.~(\ref{cel_diff}), with
$l_E(E)$ replaced by the interaction length
$l(E)$.

Thus, for a given injection spectrum of $\gamma-$rays and
electrons for a source beyond a few Mpc, the observable cascade spectrum
depends on the EGMF. As mentioned in
Sect.~4.2, the hadronic part of UHECR is a continuous source of
secondary photons whose spectrum may therefore contain
information on the large scale magnetic fields~\cite{los}. This
spectrum should be measurable down to $\simeq10^{19}\,$eV if
$\gamma-$rays can be discriminated from nucleons at the
$\sim1\%$ level. In more speculative models of UHECR origin
such as the topological defect scenario that predict domination
of $\gamma-$rays above $\sim10^{20}\,$eV, EGMFs can have even
more direct consequences for UHECR fluxes and constraints on
such scenarios (see Sect.~7.1).

The photons coming from the synchrotron radiation of electrons of
energy $E$ have a typical energy given by
\begin{equation}
  E_{\rm syn}\simeq6.8\times10^{13}
  \left(\frac{E}{10^{21}\,{\rm eV}}\right)^2
  \left(\frac{B}{10^{-9}\,{\rm G}}\right)\,{\rm eV}
  \,,\label{esynch}
\end{equation}
which is valid in the classical limit, $E_{\rm syn}\ll
E$. Constraints can arise when this energy falls in a range
where there exist measurements of the diffuse $\gamma-$ray flux,
such as from EGRET around $1\,$GeV~\cite{cdkf}, or upper
limits on it, such as at $50-100\,$TeV from HEGRA~\cite{hegra2},
and between $\simeq6\times10^{14}\,$eV and
$\simeq6\times10^{16}\,$eV from CASA-MIA~\cite{casa2}.
For example, certain strong discrete sources of UHE $\gamma-$rays
such as massive topological defects with an almost monoenergetic
injection spectrum in a $10^{-9}\,$G EGMF would predict
$\gamma-$ray fluxes that are larger than the charged cosmic ray
flux for some energies above $\simeq10^{16}\,$eV and can
therefore be ruled out~\cite{pj}.

\subsubsection{Deflection and Delay of Charged Hadrons}
Whereas for electrons synchrotron loss is more important than
deflection in the EGMF, for charged hadrons the opposite is the
case. A relativistic particle of charge $qe$ and energy $E$
has a gyroradius $r_g\simeq E/(qeB_\perp)$ where $B_\perp$ is the
field component perpendicular to the particle momentum. If this
field is constant over a distance $d$, this leads to a
deflection angle
\begin{equation}
  \theta(E,d)\simeq\frac{d}{r_g}\simeq0.52^\circ q
  \left(\frac{E}{10^{20}\,{\rm eV}}\right)^{-1}
  \left(\frac{d}{1\,{\rm Mpc}}\right)
  \left(\frac{B_\perp}{10^{-9}\,{\rm G}}\right)
  \,.\label{gyro}
\end{equation}

Magnetic fields beyond the Galactic disk are poorly known and
include a possible extended field in the halo of our Galaxy and
a large scale EGMF. In both cases, the magnetic
field is often characterized by an r.m.s. strength $B$ and a
correlation length $l_c$, i.e. it is assumed that
its power spectrum has a cut-off in wavenumber space at
$k=2\pi/l_c$ and in real space it is smooth on scales below
$l_c$. If we neglect energy loss processes for the moment, then
the r.m.s. deflection angle over a distance $d$ in such a field
is
\begin{equation}
  \theta(E,d)\simeq\frac{(2dl_c/9)^{1/2}}{r_g}\simeq0.8^\circ\,
  q\left(\frac{E}{10^{20}\,{\rm eV}}\right)^{-1}
  \left(\frac{d}{10\,{\rm Mpc}}\right)^{1/2}
  \left(\frac{l_c}{1\,{\rm Mpc}}\right)^{1/2}
  \left(\frac{B}{10^{-9}\,{\rm G}}\right)\,,\label{deflec}
\end{equation}
for $d\ga l_c$, where the numerical prefactors were calculated
from the analytical treatment in Ref.~\cite{wm}. There it was
also pointed
out that there are two different limits to distinguish: For
$d\theta(E,d)\ll l_c$, particles of all energies 
``see'' the same magnetic field realization during their
propagation from a discrete source to the observer. In this
case, Eq.~(\ref{deflec}) gives the typical coherent deflection
from the line-of-sight source direction, and the spread in arrival
directions of particles of different energies 
is much smaller. In contrast, for $d\theta(E,d)\gg l_c$, the
image of the source is washed out over a typical angular extent
again given by Eq.~(\ref{deflec}), but in this case it is 
centered on the true source direction. If $d\theta(E,d)\simeq
l_c$, the source may even have several images, similar to the
case of gravitational lensing. Therefore, observing
images of UHECR sources and identifying counterparts in other
wavelengths would allow one to distinguish these limits and thus
obtain information on cosmic magnetic fields. If $d$ is
comparable to or larger than the interaction length for stochastic
energy loss due to photo-pion production or photodisintegration,
the spread in deflection angles is always comparable to the
average deflection angle.

Deflection also implies an average time delay of
\begin{equation}
  \tau(E,d)\simeq d\theta(E,d)^2/4\simeq1.5\times10^3\,q^2
  \left(\frac{E}{10^{20}\,{\rm eV}}\right)^{-2}
  \left(\frac{d}{10\,{\rm Mpc}}\right)^{2}
  \left(\frac{l_c}{1\,{\rm Mpc}}\right)
  \left(\frac{B}{10^{-9}\,{\rm G}}\right)^2\,{\rm yr}
  \label{delay}
\end{equation}
relative to rectilinear propagation with the speed of light. It was 
pointed
out in Ref.~\cite{mw} that, as a consequence, the observed UHECR spectrum
of a bursting source at a given time can be
different from its long-time average and would typically peak
around an energy $E_0$, given by equating $\tau(E,d)$ with the time of
observation relative to the time of arrival for vanishing time
delay. Higher energy particles would have passed the observer
already, whereas lower energy particles would not have arrived
yet. Similarly to the behavior of deflection angles, the width
of the spectrum around $E_0$ would be much smaller than $E_0$ if
both $d$ is smaller than the interaction length for stochastic
energy loss and $d\theta(E,d)\ll l_c$. In all other cases the
width would be comparable to $E_0$.

Constraints on magnetic fields from deflection and time delay
cannot be studied separately from the characteristics of the
``probes'', namely the UHECR sources, at least as long as their
nature is unknown. An approach to the general case is discussed
in Sect.~4.7.

\subsection{Constraints on EHECR Source Locations}
Nucleons, nuclei, and $\gamma-$rays above a few $10^{19}\,$eV
cannot have originated much further away than
$\simeq50\,$Mpc. For nucleons this follows from the GZK effect
(see Fig.~\ref{F4.2}, the range of nuclei is limited mainly by
photodisintegration on the CMB (see Sect.~4.1), whereas photons are
restricted by PP and DPP on the CMB and URB (see
Fig.~\ref{F4.4}). Together with Eq.~(\ref{deflec}) this implies
that above a few $10^{19}\,$eV the arrival direction of such
particles should in general point back to their source within a few
degrees~\cite{ssb}. This argument is often made in the literature
and follows from the Faraday rotation bound on the EGMF and a
possible extended field in the halo of our Galaxy, which in its
original form reads
$Bl_c^{1/2}\la10^{-9}\,{\rm G}\,{\rm Mpc}^{1/2}$~\cite{kronberg,vallee},
as well as from the known strength and
scale height of the field in the disk of our Galaxy,
$B_g\simeq3\times10^{-6}\,$G, $l_g\la1\,$kpc. Furthermore, the
deflection in the disk of our Galaxy can be corrected for in
order to reconstruct the extragalactic arrival direction: Maps
of such corrections as a function of arrival direction have been
calculated in Refs.~\cite{stanev,tph2} for plausible models of
the Galactic magnetic field. The deflection of UHECR trajectories in
the Galactic magnetic field may, however, also give rise to several other 
important effects~\cite{hmr} such as (de)magnification of the UHECR
fluxes due to the magnetic lensing effect mentioned in the previous section 
(which can modify the UHECR spectrum from individual sources), 
formation of multiple images of a source, and apparent ``blindness'' of
the Earth towards certain regions of the sky with regard to UHECR. These
effects may in turn have important implications for UHECR source
locations.

However, important modifications of the Faraday rotation bound
on the EGMF have recently been discussed in the literature:
The average electron density which enters estimates of the EGMF
from rotation measures, can now be more reliably estimated from
the baryon density $\Omega_bh^2\simeq0.02$, whereas in the original
bound the closure density was used. Assuming an unstructured
Universe and $\Omega_0=1$ results in the much weaker bound~\cite{bbo}
\begin{equation}
  B\la3\times10^{-7}\left(\frac{\Omega_bh^2}{0.02}\right)^{-1}
  \left(\frac{h}{0.65}\right)
  \left(\frac{l_c}{{\rm Mpc}}\right)^{-1/2}
  \,{\rm G}\,,\label{newFhom}
\end{equation}
which suggests much stronger deflection. However, taking into
account the large scale structure of the Universe in the form
of voids, sheets, filaments etc., and assuming flux freezing
of the magnetic fields whose strength then approximately scales
with the 2/3 power of the local density, leads to more stringent bounds:
Using the Lyman $\alpha$ forest to model the density distribution
yields~\cite{bbo}
\begin{equation}
  B\la10^{-9}-10^{-8}\,{\rm G}\label{newF}
\end{equation}
for the large scale EGMF for coherence scales between the Hubble
scale and 1 Mpc. This estimate is closer to the original Faraday
rotation limit. However, in this scenario the maximal fields
in the sheets and voids can be as high as a
$\mu\,$G~\cite{rkb,bbo,fp}.

Therefore, according to Eq.~(\ref{deflec}) and~(\ref{newF}),
deflection of UHECR nucleons is still expected to be on
the degree scale if the local large scale structure around the
Earth is not strongly magnetized. However,
rather strong deflection can occur if the Supergalactic Plane
is strongly magnetized, for particles originating in
nearby galaxy clusters where magnetic
fields can be as high as $10^{-6}\,$G~\cite{kronberg,vallee,eilek-dbl}
(see Sect.~4.6) and/or for
heavy nuclei such as iron~\cite{elb-som}. In this case, magnetic
lensing in the EGMF can also play an important role in
determining UHECR source locations~\cite{slb,lsb}.

\subsection{Source Search for EHECR Events}

The identification of sources of EHECR has been attempted in it
least two different ways:

First, it has been tried to associate some of the EHE events
with discrete sources. For the $300\,$EeV Fly's Eye event,
potential extragalactic sources have been discussed in
Ref.~\cite{elb-som}. Prominent objects that are within the range
of nuclei and nucleons typically require strong magnetic
bending, such as Cen A at $\simeq3\,$Mpc and $\simeq136^\circ$
from the arrival direction, Virgo A ($13-26\,$Mpc,
$\simeq87^\circ$), and M82 ($3.5\,$Mpc, $\simeq37^\circ$). The
Seyfert galaxy MCG 8-11-11 at $62-124\,$Mpc and the
radio galaxy 3C134 of Fanaroff-Riley (FR) class II are within
about $10^\circ$ of the arrival direction. Due to
Galactic obscuration, the redshift (and thus the distance) of the
latter is, however, not known with certainty, and estimates range between
30 and $500\,$Mpc~\cite{rachen1}. A powerful quasar, 3C147,
within the Fly's Eye event error box at redshift $z\simeq0.5$
has been suggested as a neutrino source~\cite{hvsv}. A potential
problem of this option is that standard neutrino-nucleon cross
sections predict an interaction probability of neutrinos near
$10^{20}\,$eV of $\sim10^{-5}$ in the atmosphere.
As long as deflection in the EGMF is
not too strong [see Eq.~(\ref{deflec})], the required
large neutrino flux would most likely imply a comparable nucleon
flux below the GZK cutoff that is not observed~\cite{slproc}. We
note, however, that, in contrast to interactions with the RNB,
the CM energy of a neutrino-nucleon collision at that energy is a
few hundred TeV where new physics beyond the electroweak scale
could enhance the neutrino-nucleon cross section (see
discussion in Sect.~4.3.1). For the highest energy AGASA event, a
potential source for the neutrino option is the FR-II galaxy
3C33 at $\simeq300\,$Mpc distance, whereas the FR-I galaxy NGC
315 at $\simeq100\,$Mpc is a candidate in case of a nucleon
primary. A Galactic origin for both the highest energy Fly's Eye
and AGASA event seems only possible in case of iron primaries
and an extended Galactic halo magnetic field~\cite{gz}.

Second, identification of UHECR sources with classes of
astrophysical objects has been attempted by testing statistical
correlations between arrival directions and the locations of
such objects. The Haverah park data set and some data from the
AGASA, the Volcano Ranch, and the Yakutsk experiments were
tested for correlation with the Galactic and Supergalactic
plane, and positive result at a level of almost $3\sigma$ was
found for the latter case for events above
$4\times10^{19}\,$eV~\cite{sblrw}. An analysis of the SUGAR data
from the southern hemisphere, however, did not give 
significant correlations~\cite{kcd}. More recently, a possible
correlation of a subset of about $20\%$ of the events above
$4\times10^{19}\,$eV among each other and with the Supergalactic
Plane was reported by the AGASA experiment, whereas
the rest of the events seemed consistent with an isotropic
distribution~\cite{haya2,agasa2}. Results
from a similar analysis combining data from the Volcano
Ranch, the Haverah Park, the Yakutsk, and the Akeno surface arrays
in the northern hemisphere~\cite{uchihori1}, as well as from
these and the Fly's Eye experiment~\cite{uchihori2} were found
consistent with that, although no final conclusions can be drawn
presently yet. These findings give support to the
hypothesis that at least part of the EHECR are accelerated in
objects associated
with the Supergalactic Plane. However, it was subsequently
pointed out~\cite{wfp} that the Supergalactic Plane correlation
at least of the Haverah Park data seems to be too strong for an
origin of these particles in objects 
associated with the large-scale galaxy structure because, within
the range of the corresponding nucleon primaries, galaxies
beyond the Local Supercluster become relevant as well. As a
possible resolution it was suggested~\cite{bkrproc,rachen1} that
the possible existence of strong magnetic fields with strengths
up to $\mu\,$G and coherence lengths in the Mpc range, aligned
along the large-scale structure~\cite{rkb}, could produce a focusing
effect of UHECR along the sheets and filaments of galaxies. A recent
study claims, however, consistency of the arrival directions of
UHECR with the distribution of galaxies within $50\,$Mpc from
the Cfa Redshift Catalog~\cite{medinatanco}. The case of UHECR
correlations with the large scale structure of galaxies,
therefore, does not seem to be settled yet.

Correlations between arrival directions of UHECR above
$4\times10^{19}\,$eV and $\gamma-$ray burst (GRB) locations have
also been investigated. Although the arrival directions of the two
highest energy events are within the error boxes of two strong
GRBs detected by BATSE~\cite{mu}, no significant positive result was
found for the larger UHECR sample~\cite{ssw}. This may be
evidence against an association of UHECR with GRBs if their
distance scale is Galactic, but not if they have an
extragalactic origin because of the implied large time delays of
UHECR relative to GRB photons (see Sect.~5.3). Furthermore, whereas no
enhancement of the TeV
$\gamma-$ray flux has been found in the direction of the Fly's
Eye event in Ref.~\cite{Akerlof}, a weak excess was recently
reported in Ref.~\cite{lindner}.

Finally, a statistically significant correlation between the arrival
directions of UHECR events in the energy range 
(0.8---4)$\times10^{19}\,$eV and directions of pulsars along
the Galactic magnetic field lines has been claimed for the Yakutsk EAS
data in Ref.~\cite{pulsar-corr}. It would be interesting to look for
similar correlations for the data sets from other UHECR experiments. 

\subsection{Detailed Calculations of Ultra-High Energy Cosmic
Ray Propagation}

In order to obtain accurate predictions of observable CR spectra
for given production scenarios, one has to solve the equations
of motion for the total and differential cross sections for the
loss processes discussed in Sects.~4.1$-$4.4. If deviations
from rectilinear propagation are unimportant, for example, if
one is only interested in time averaged fluxes, one typically
solves the coupled Boltzmann equations for CR transport in one
spatial dimension either directly or by Monte Carlo simulation.
In contrast, if it is important to follow
3-dimensional trajectories, for example, to compute images of discrete
UHECR sources in terms of 
energy and time and direction of arrival 
in the presence of magnetic fields, the only feasible approach
for most purposes is a Monte Carlo simulation. We describe both
cases briefly in the following. 

\subsubsection{Average Fluxes and Transport Equations in One Dimension}

Computation of time averaged fluxes from transport equations or
one-dimensional Monte Carlo simulation is most relevant for
diffuse fluxes from many sources
and for spectra from discrete sources that emit constantly over long
time periods. This is applicable at sufficiently high energies
such that deflection angles in potential magnetic fields are
much smaller than unity. Formally, the Boltzmann equations for
the evolution of a set of species with local densities per
energy $n_i(E)$ are given by
\begin{eqnarray}
  \partial_t n_i(E)&=&-n_i(E)\int d\varepsilon n_b(\varepsilon)
  \int_{-1}^{+1}d\mu\frac{1-\beta_b\beta_i}{2}\sum_j
  \sigma_{i\to j}\left[s=\varepsilon E(1-\beta_b\beta_i)\right]
  \label{transport}\\
  &&+\int dE^\prime\int d\varepsilon n_b(\varepsilon)\int_{-1}^{+1}
  d\mu\sum_j\frac{1-\beta_b\beta^\prime_j}{2}n_j(E^\prime)
  \frac{d\sigma_{j\to i}}{dE}\left[s=\varepsilon E^\prime
  (1-\beta_b\beta_j),E\right]+\Phi_i\nonumber
\end{eqnarray}
for an isotropic background distribution (here assumed to be only
one species) with our notation (see Eqs.~7, 8) extended to several
species. We briefly summarize work on solving these equations 
for the propagation of nucleons, nuclei, $\gamma-$rays, electrons,
and neutrinos in turn.

{\bf Nucleons and Nuclei}

\noindent
Motivated by conventional acceleration models (see Sect.~5),
many studies on propagation of nucleons and nuclei have been
published in the literature.
Approximate analytical solutions of the transport equations can
only be found for very specific
situations, for example, for the propagation of nucleons
near the GZK cutoff (e.g., ~\cite{bg,akv,rb,gqw}) and/or under
certain
simplifying assumptions such as the CEL
approximation Eqs.~(\ref{cel})$-$(\ref{cel_diff}) for
nucleons (e.g., ~\cite{wax1,ades}) and $\gamma-$rays (e.g., ~\cite{abs}).
The CEL approximation is excellent for PPP because of its small
inelasticity. For pion production, due to its
stochastic nature implied by its large inelasticity, the CEL
approximation tends to produce
a sharper pile-up right below the GZK cutoff compared to
exact solutions~\cite{yt}. It still works reasonably well
as long as many pion production events take place on
average, i.e. for continuous source distributions and
for distant discrete sources. Numerical solutions
for nucleons solve the transport equations either
directly~\cite{hs1,yt,lee} or through Monte Carlo
simulation~\cite{ac,elb-som,slo,sl}. Monte Carlo studies of the
photodisintegration histories of nuclei have first been
performed in Ref.~\cite{psb} and subsequently in
Refs.~\cite{elb-som,epele-roulet,stecker-sala-nuc}.

{\bf Electromagnetic Cascades}

\noindent
Numerical calculations of average $\gamma-$ray fluxes
from EM cascades beyond the analytical CEL approximation are
more demanding due to the exponential growth of the number of
electrons and photons and are
usually not feasible within a pure Monte Carlo approach. Such
simulations have been performed mainly in the context of
topological defect models of UHECR origin (see Sect.~6).
Calculations of the photon flux between $\simeq100\,$MeV and
$\simeq10^{16}\,$GeV (the Grand Unification Scale) have been
presented in Ref.~\cite{pj,ps1} where a hybrid
Monte Carlo matrix doubling method~\cite{ps2} was used, and in
Ref.~\cite{lee,ysl,slby} where the transport equations are solved by an
implicit numerical method. Such calculations play an important
role in deriving constraints on top-down models from a comparison of the
predicted and observed photon flux down to energies of
$\simeq100\,$MeV (see Sect.~7). EM cascade simulations are also
relevant for the secondary $\gamma-$ray flux produced from
interactions of primary hadrons~\cite{yt} and its dependence on
cosmic magnetic fields~\cite{los}. Under certain
circumstances, this secondary flux can become comparable to the
primary flux~\cite{ww}.

Analytical calculations have been performed for saturated EM
cascades~\cite{zdziarski}. These calculations show that 
the cascade spectrum below the pair production threshold 
has a generic shape. 
This has also been used to derive constraints on energy
injection based on direct observation of this cascade flux
or on a comparison of its side effects, for example, on light element
abundances, with observations. We will discuss these issues
in Sects.~7.1 and~7.2.

\begin{figure}[ht]
\centering\leavevmode
\epsfxsize=6.0in
\epsfbox{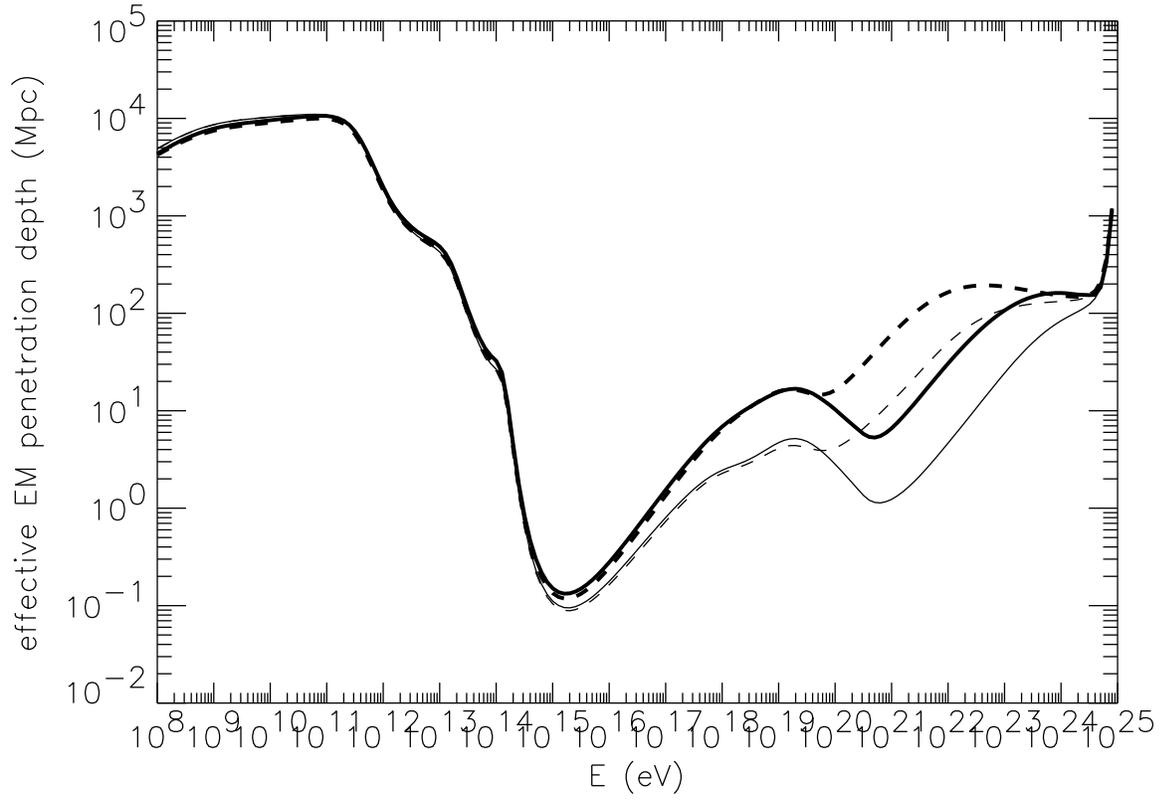}
\bigskip
\caption[...]{Effective penetration depth of EM cascades, as
defined in the text, for the strongest theoretical URB estimate
(solid lines), and the observational URB estimate
from Ref.~\cite{cba} (dashed lines), as shown in Fig.~\ref{F4.3},
and for an EGMF $\ll10^{-11}\,$G (thick lines), and $10^{-9}\,$G
(thin lines), respectively.}
\label{F4.7}
\end{figure}

As a first application of numerical transport calculations we
present the effective penetration depth of EM
cascades, which we define as the coefficient $l_E(E)$ in
Eq.~(\ref{cel_diff}), where $j(E)$ is the $\gamma-$ray flux
resulting after propagating a homogeneous injection flux
$\Phi(E)$. Fig.~\ref{F4.7} shows results computed for the
new estimates of the IR background from Ref.~\cite{irb}, and
for some combinations of the URB and the EGMF.

{\bf Neutrino Fluxes}

\noindent
Accurate predictions for the UHE neutrino flux have become
more relevant recently due to several proposals for a km$^3$
scale neutrino observatory~\cite{ghs}.
Fluxes of secondary neutrinos from photo-pion production
by UHECR have been calculated numerically, e.g., in
Refs.~\cite{yt,ydjs,lee}, by solving the full transport equations
for nucleons. Because of the small redshifts involved, the
neutrinos can be treated as interaction-free, and the main
uncertainties come from the poorly known injection history of
the primary nucleons (see Fig.~\ref{F7.5}). In top-down scenarios,
neutrinos are continuously produced up to very high redshifts
and secondaries produced in neutrino interactions can enhance
the UHE neutrino fluxes compared to the simple absorption
approximation used in Refs.~\cite{bhs,slsc}. By solving the full
Boltzmann equations for the neutrino cascade, unnormalized spectral
shapes of neutrino fluxes from topological defects have been
calculated in Ref.~\cite{yoshida}, and absolute fluxes in
Ref.~\cite{ydjs}. Semianalytical calculations of $\gamma-$ray,
nucleon, and neutrino fluxes for a specific class of cosmic
string models predicting an absolute normalization of the
UHECR injection rate (see Sect.~6.4.6) have been performed
in Ref.~\cite{wmgb}.

Recently an integrated code has been developed
which solves the coupled full transport equations for all
species, i.e, nucleons, $\gamma-$rays, electrons, and
neutrinos concurrently~\cite{ysl,slby}. This allows, for example,
to make detailed predictions for the spectra of the nucleons
and $\gamma-$rays produced by resonant $Z^0$ production of
UHE neutrinos on a massive RNB which could serve as a signature
of hot dark matter~\cite{fms,weiler2} (see Sect.~4.3.1).

\subsubsection{Angle-Time-Energy Images of Ultra-High-Energy
Cosmic Ray Sources}

In Sect.~4.4 we gave simple analytical estimates for the average
deflection and time delay of nucleons propagating in a cosmic
magnetic field. Here we review approaches that have been taken
in the literature to compute effects of magnetic fields on both
spectra and angular images (and their time dependence) of
sources of UHE nucleons.

{\bf Strong Deflection}

\noindent
An exact analytical expression for the
distribution of time delays that applies in the limit
$d\theta(E,d)\gg l_c$ for $E\la4\times10^{19}\,$eV where
photo-pion production is negligible has been given in
Ref.~\cite{wm}. The consequences for the spectra in this
energy range and their temporal behavior, especially for
the possibility of bursting sources such as cosmological
GRBs (see Sect.~5.3), have been discussed there.

Indications for rather strong magnetic fields in the range
between $10^{-8}\,$G up to $10^{-6}\,$G have
been observed near large mass agglomerations such as
clusters of galaxies or even the filaments and sheets
connecting them~\cite{kronberg,vallee}. UHECR deflection in
such regions could be strong enough for the diffusion
approximation to become applicable. The uncertainties in
strength and spectrum of the magnetic fields translate directly
into a corresponding uncertainty in the energy dependent diffusion
coefficient $D(E)$ which is often obtained by simply fitting
calculated fluxes to the data. At $\sim10^{19}\,$eV estimated
values of $D(E)$ range between $\simeq5\times10^{33}\,{\rm cm}^2\,
{\rm sec}^{-1}$ and $\simeq3\times10^{35}\,{\rm cm}^2\,
{\rm sec}^{-1}$, and energy dependence like $D(E)\propto E^{\alpha}$ with
$\alpha$ in the range $1/3$ and 1 have been 
suggested~\cite{bbdgp,diff_refs}.

Several approximate treatments for calculations of fluxes
in the diffusion approximation have been pursued in the
literature: If pion production is treated
in the CEL approximation, the problem reduces to solving
Eq.~(\ref{cel}) with an additional, in general location and
energy dependent diffusion coefficient $D({\bf r},E)$:
\begin{equation}
  \partial_t n({\bf r},E)=-\partial_E
  \left[b(E)n({\bf r},E)\right]+
  \hbox{\boldmath$\nabla$}\left[D({\bf r},E)
  \hbox{\boldmath$\nabla$}n({\bf r},E)\right]
  +\Phi(E)\,.\label{cel_d}
\end{equation}
If $D({\bf r},E)$ is independent of ${\bf r}$, an analytical
solution of this ``energy loss-diffusion equation'' given by
Syrovatskii~\cite{syrovatskii} can be
employed. This solution or approximations to it have been used
in Refs.~\cite{diff_refs,bbdgp,bo}
to compute the expected spectra from discrete sources in an EGMF
of a few $10^{-8}\,$G for energies up to $\simeq10^{20}\,$eV 
(the typical range of validity of the diffusion approximation).
In some sense a complementary approach has been taken in
Ref.~\cite{ac} where the effects of diffusion were taken
into account by using an average propagation time roughly
given by $d^2/D(E)$ and treating pion production exactly
by Monte Carlo. Ref.~\cite{lcd} improved on that by representing
the EGMF (assumed to be homogeneous) by a finite number of modes
and following trajectories explicitly. This paper did, however,
not present any spectra directly.

Once Eq.~(\ref{cel_d})
has been solved, the anisotropy defined in Eq.~(\ref{deltaan})
can be calculated from the relation~\cite{bbdgp}
\begin{equation}
  \delta(E)=3\,\frac{D({\bf r},E)}{n({\bf r},E)}\,
  |\hbox{\boldmath$\nabla$}n({\bf r},E)|\,.\label{delta_cel_d}
\end{equation}
As mentioned in Sect.~3.2, Eq.~(\ref{cel_d}) and its generalization
to an anisotropic diffusion tensor plays a prominent
role also in models of Galactic CR propagation. We stress here
that while this equation provides a good description of the
propagation of Galactic CR for energies up to the knee, it has rather 
limited applicability in studying   
UHECR propagation which often takes place in the transition
regime between diffusion and rectilinear propagation (see below).

{\bf Small Deflection}

\noindent
For small deflection angles and if photo-pion
production is important, one has to resort to numerical Monte
Carlo simulations in 3 dimensions. Such simulations have been
performed in Ref.~\cite{tph} for the case $d\theta(E,d)\gg l_c$
and in Refs.~\cite{lsos,slo,sl} for the general case.

\begin{figure}[ht]
\centering\leavevmode
\epsfxsize=6.0in
\epsfbox{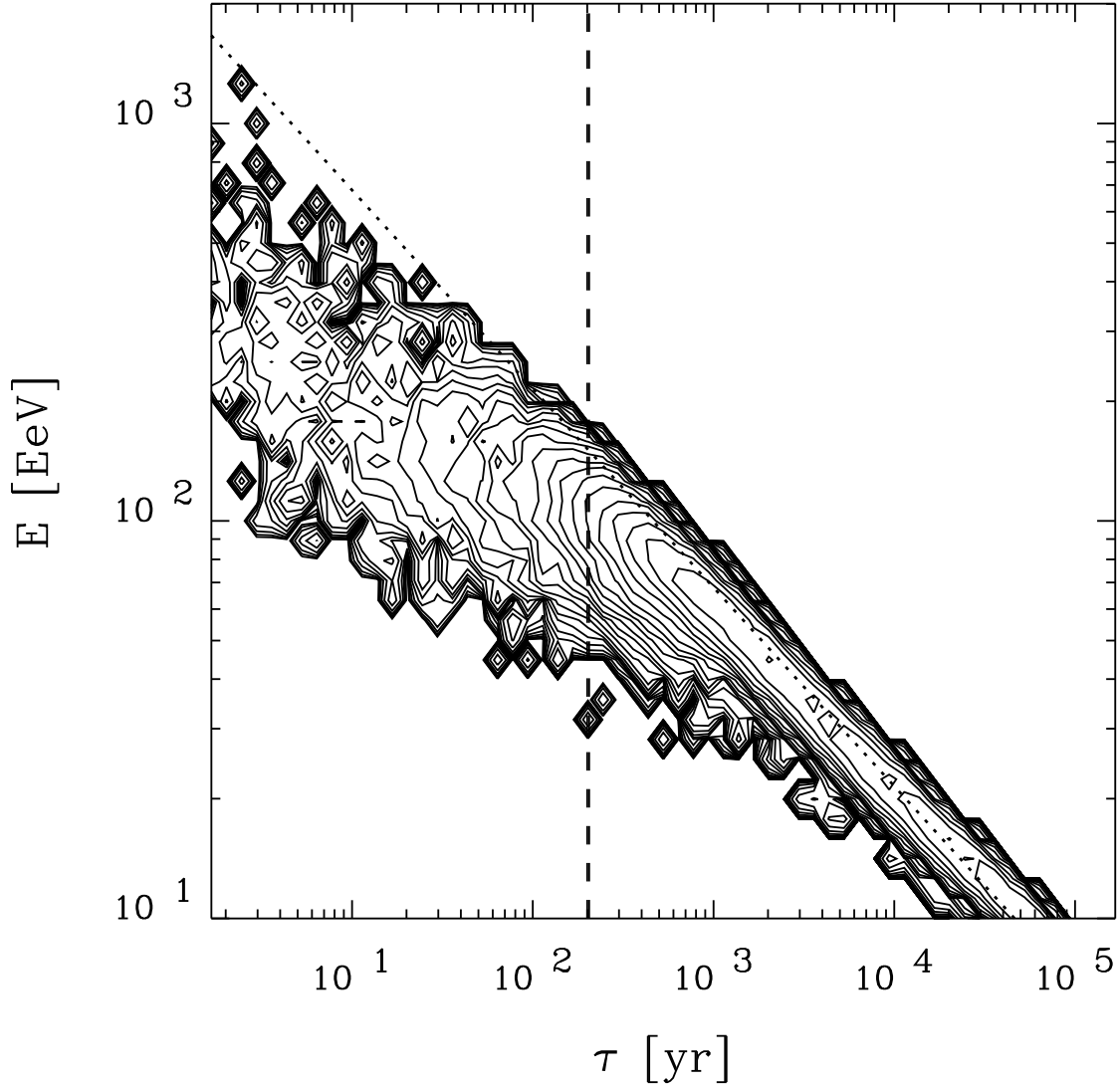}
\bigskip
\caption[...]{Contour plot of the UHECR image of a bursting source
at $d=30\,$Mpc, projected onto the time-energy plane, with
$B=2\times10^{-10}\,$G, $l_c=1\,$Mpc, from Ref.~\cite{lsos}.
The contours decrease in steps of 0.2 in the logarithm to base 10.
The dotted line indicates the energy-time delay correlation 
$\tau(E,d)\propto E^{-2}$ as would be obtained in the absence of 
pion production losses. Clearly, $d\theta(E,d)\ll l_c$ in this
example, since for $E<4\times10^{19}\,$eV, the width of the
energy distribution at any given time is much smaller than the
average (see Sect.~4.4). The dashed lines, which are not resolved 
here, indicate the location (arbitrarily chosen) of the
observational window, of length $T_{obs}=5\,$yr.}
\label{F4.8}
\end{figure}

In Refs.~\cite{lsos,slo,sl} the Monte Carlo simulations were
performed in the following way:
The magnetic field was represented as Gaussian random field
with zero mean and a power spectrum
with $\left\langle B^2(k)\right\rangle\propto
k^{n_H}$ for $k<k_c$ and $\left\langle B^2(k)\right\rangle=0$
otherwise, where $k_c=2\pi/l_c$ characterizes the numerical
cut-off scale and the r.m.s. strength is
$B^2=\int_0^\infty\,dk\,k^2\left\langle B^2(k)
\right\rangle$. The field is then calculated on a
grid in real space via Fourier transformation.
For a given magnetic field realization and source, nucleons with
a uniform logarithmic distribution of injection energies are
propagated between two given points (source and observer) on the
grid. This is done by solving the equations of motion in the
magnetic field interpolated between the grid points, and
subjecting nucleons to stochastic production of pions and (in
case of protons) continuous loss of energy due to PP.
Upon arrival, injection and detection energy, and time
and direction of arrival are recorded. From many (typically
40000) propagated particles, a histogram of average number of
particles detected as a function of time and energy of
arrival is constructed for any given injection spectrum by
weighting the injection energies correspondingly. This histogram
can be scaled to any desired total fluence at the detector and,
by convolution in time, can be constructed for arbitrary
emission time scales of the source. An example for the
distribution of arrival times and energies of UHECR from a
bursting source is given in Fig.~\ref{F4.8}.

We adopt the following notation for the parameters: $\tau_{100}$ 
denotes the time delay due to magnetic deflection at $E=100\,$EeV
and is given by Eq.~(\ref{delay}) in terms of the magnetic field
parameters;
$T_S$ denotes the emission time scale of the source; $T_S\ll1$yr 
correspond to a burst, and $T_S\gg1$yr (roughly speaking) to a 
continuous source; $\gamma$ is the differential index of the 
injection energy spectrum; $N_0$ denotes the fluence of the 
source with respect to the detector, {\it i.e.}, the total number 
of particles that the detector would detect from the source on 
an infinite time scale; finally, ${\cal L}$ is the likelihood, 
function of the above parameters.

\begin{figure}[ht]
\centering\leavevmode
\epsfxsize=6.0in
\epsfbox{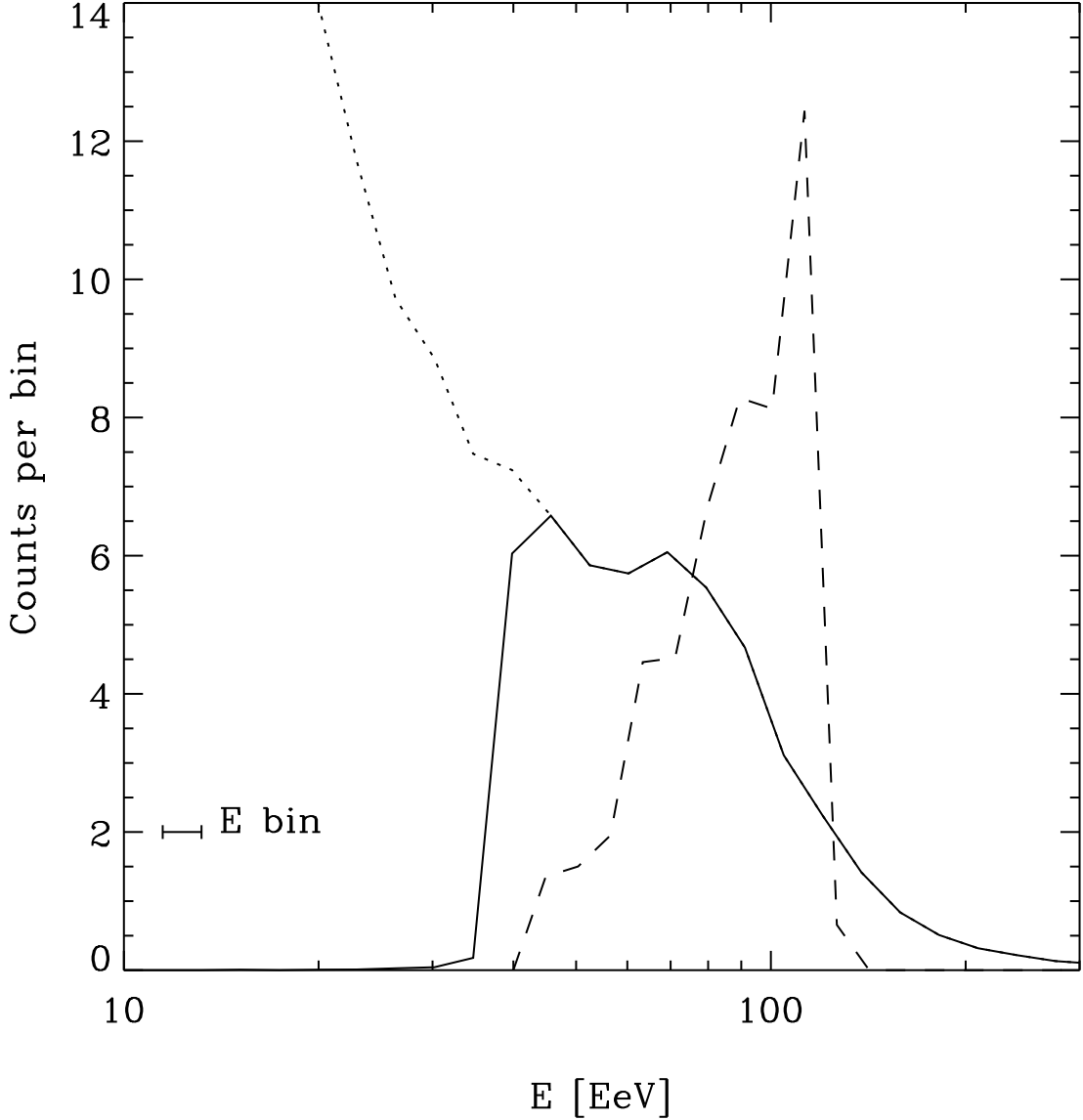}
\bigskip
\caption[...]{Energy spectra for a continuous source (solid 
line), and for a burst (dashed line), from
Ref.~\cite{lsos}. Both spectra are
normalized to a total of 50 particles detected. The parameters 
corresponding to the continuous source case are: $T_S=10^4\,$yr, 
$\tau_{100}=1.3\times10^3\,$yr, and the time of observation is 
$t=9\times10^3\,$yr, relative to rectilinear propagation with the speed
of light. A low energy cutoff results at the energy
$E_S=4\times10^{19}\,$eV where $\tau_{E_S}=t$. The dotted line
shows how the spectrum would continue if $T_S\ll10^4\,$yr. The
case of a bursting source corresponds to a slice of the image in 
the $\tau_E-E$ plane, as indicated in Fig.~\ref{F4.8} by dashed lines. 
For both spectra, $d=30\,$Mpc, and $\gamma=2.$.}
\label{F4.9}
\end{figure}

By putting windows of width equal to the time
scale of observation over these histograms one obtains expected
distributions of
events in energy and time and direction of arrival
for a given magnetic field realization, source distance and
position, emission time scale, total fluence, and injection
spectrum. Examples of the resulting energy spectrum are shown in
Fig.~\ref{F4.9}. By dialing Poisson statistics on such
distributions, one can simulate corresponding observable event
clusters.

Conversely, for any given real or simulated event cluster, one
can construct a likelihood of the observation as a function of 
the time delay, the emission time scale, the differential 
injection spectrum index, the fluence, and the distance. In
order to do so, and to obtain the maximum of the likelihood, 
one constructs histograms for many different parameter
combinations as described above, randomly puts observing time
windows over the histograms, calculates the likelihood function
from the part of the histogram within the window and the cluster
events, and averages over different window locations and magnetic
field realizations.

\begin{figure}[ht]
\centering\leavevmode
\epsfxsize=6.0in
\epsfbox{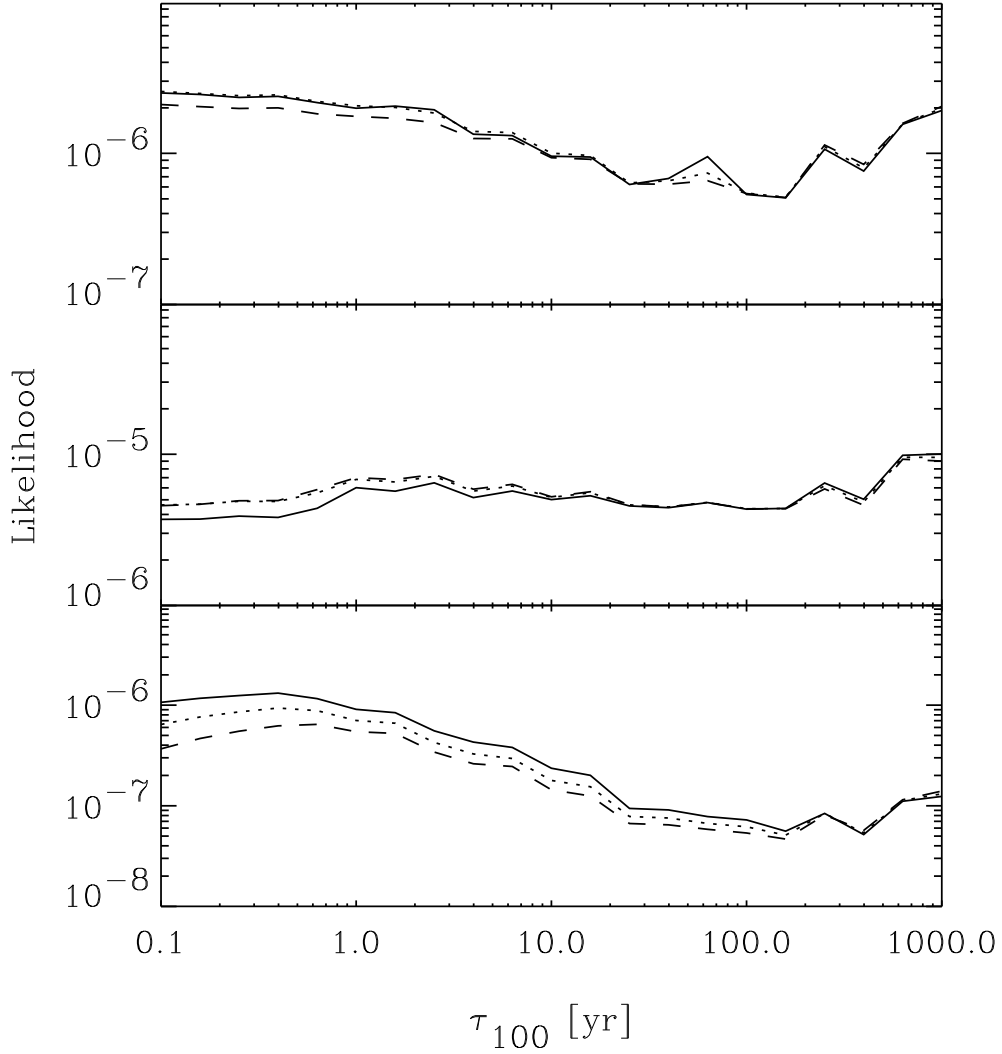}
\bigskip
\caption[...]{The logarithm of the likelihood, $\log_{10}{\cal L}$,
marginalized over $T_S$ and $N_0$ as a function of the average
time delay at $10^{20}\,$eV, $\tau_{100}$, assuming a source
distance $d=30\,$Mpc. The panels are for pair \# 3 through \# 1,
from top to bottom, 
of the AGASA pairs~\cite{haya2} (see Sect.~4.7.2).  
Solid lines are for $\gamma=1.5$, dotted
lines for $\gamma=2.0$, and dashed lines for $\gamma=2.5$.}
\label{F4.10}
\end{figure}

In Ref.~\cite{slo} this approach has been applied to and discussed in
detail for the three pairs observed by the AGASA
experiment~\cite{haya2}, under the assumption that all events
within a pair were produced by the same discrete source.
Although the inferred angle between the momenta of the paired
events acquired in the EGMF is several degrees~\cite{medinatanco2},
this is not necessarily evidence against a common source,
given the uncertainties in the Galactic field and the angular
resolution of AGASA which is $\simeq2.5^\circ$. As a result
of the likelihood analysis, these pairs do not seem to follow
a common characteristic; one of them seems to favor a burst, another
one seems to be more consistent with a continuously emitting source.
The current data, therefore, does not allow one to rule out any of
the models of UHECR sources. Furthermore, two of the three pairs are
insensitive to the time delay. However, the pair which contains the
$200\,$EeV event seems to significantly favor a comparatively
small average time delay, $\tau_{100}\la10\,$yr, as can be
seen from the likelihood function marginalized over $T_S$ and $N_0$
(see Fig.~\ref{F4.10}).
According to Eq.~(\ref{delay}) this
translates into a tentative bound for the r.m.s. magnetic field, namely, 
\begin{equation}
  B\la2\times10^{-11}
  \left(\frac{l_c}{1\,{\rm Mpc}}\right)^{-1/2}
  \left(\frac{d}{30\,{\rm Mpc}}\right)^{-1}\,{\rm G}
  \,,\label{limit}
\end{equation}
which also applies to magnetic fields in the halo of our Galaxy
if $d$ is replaced by the lesser of the source distance and the linear
halo extent. If confirmed by future data, this bound would be at least
two orders of magnitude more restrictive than the best existing
bounds which come from Faraday rotation measurements [see
Eq.~(\ref{newF})] and, for a homogeneous EGMF, from CMB
anisotropies~\cite{bfs}. UHECR are therefore at least as sensitive
a probe of cosmic magnetic fields as other measures in the range
near existing limits such as the polarization~\cite{lk} and the
small scale anisotropy~\cite{sb} of the CMB.

More generally, confirmation of a clustering of
EHECR would provide significant information on both the nature
of the sources and on large-scale magnetic
fields~\cite{sslh}. This has been shown quantitatively~\cite{sl} by
applying the hybrid Monte Carlo likelihood analysis discussed
above to simulated clusters of a few tens of events as they
would be expected from next generation experiments~\cite{icrr96} such
as the High Resolution Fly's Eye~\cite{hires}, the Telescope
Array~\cite{tel_array}, and most notably, the Pierre Auger
Project~\cite{auger} (see Sect.~2.6), provided the clustering recently
suggested by the AGASA experiment~\cite{haya2,agasa2} is real.
The proposed satellite observatory
concept for an Orbiting Wide-angle Light collector (OWL)~\cite{owl}
might even allow one to detect clusters of hundreds of such events. 

Five generic situations of the time-energy images of 
UHECR were discussed in Ref.~\cite{sl}, classified according to
the values of the time delay 
$\tau_E$ induced by the magnetic field, the emission timescale of the 
source $T_{\rm S}$, as compared to the lifetime of the experiment.
The likelihood calculated for the simulated clusters in these
cases presents different degeneracies between
different parameters, which complicates the analysis. As an example,
the likelihood is degenerate in the ratios $N_0/T_{\rm S}$, or
$N_0/\Delta\tau_{100}$, where $N_0$ is the total fluence, and
$\Delta\tau_{100}$ is the spread in arrival time; these ratios
represent rates of detection. Another example is given by the
degeneracy between the distance $d$ and the injection energy 
spectrum index $\gamma$. Yet another is the ratio
$(d\tau_E)^{1/2}/l_c$,
that controls the size of the scatter around the mean of the 
$\tau_E-E$ correlation. Therefore, in most general cases, values for 
the different parameters cannot be pinned down, and generally, only 
domains of validity are found. In the following the reconstruction
quality of the main parameters considered is summarized.

The distance to the source can be obtained from the pion 
production signature, above the GZK cut-off, when
the emission timescale of the source dominates over the time delay. 
Since the time delay decreases with increasing energy, the lower the 
energy $E_{\rm C}$, defined by $\tau_{E_{\rm C}}\simeq T_{\rm S}$,
the higher the accuracy on the distance $d$. The error on $d$ is,
in the best case, typically a factor 2, for one cluster of
$\simeq40$ events.
In this case, where the emission timescale dominates over the time 
delay at all observable energies,
information on the magnetic field is only contained in the
angular image, which was not systematically included in the
likelihood analysis of Ref.~\cite{sl} due to computational limits.
Qualitatively, the size of the angular image is 
proportional to $B(dl_c)^{1/2}/E$, whereas the structure
of the image, {\it i.e.}, the number of separate images, is
controlled by the ratio $d^{3/2}B/l_c^{1/2}/E$. Finally, in the
case when the time delay 
dominates over the emission timescale, with a time delay shorter than 
the lifetime of the experiment, one can also estimate the distance 
with reasonable accuracy.

Some sensitivity to the injection spectrum index $\gamma$ exists
whenever events are recorded over a sufficiently broad energy
range. At least if the distance $d$ is known, it is in general
comparatively easy to rule out a hard injection spectrum if the
actual $\gamma\ga2.0$, but much harder to distinguish between
$\gamma=2.0$ and 2.5.

If the lifetime of the experiment is the largest time scale
involved, the strength of the magnetic field can only be obtained
from the time-energy image because the angular image
will not be resolvable. When the time delay 
dominates over the emission timescale, and is, at the same time,
larger than the lifetime of the experiment, only a lower limit 
corresponding to this latter timescale, can be placed on the time 
delay and hence on the strength of the magnetic field. When combined
with the Faraday rotation upper limit Eq.~(\ref{newF}),
this would nonetheless allow one to bracket the r.m.s. magnetic
field strength within a
few orders of magnitude. In this case also, significant information is
contained in the angular image. If the emission time scale is
larger then the delay time, the angular image is obviously the
only source of information on the magnetic field strength.

The coherence length $l_c$ enters in the ratio
$(d\tau_E)^{1/2}/l_c$
that controls the scatter around the mean of the $\tau_E-E$
correlation in the time-energy image. It can therefore be estimated
from the width of this image, provided the emission timescale is
much less than $\tau_E$ (otherwise the correlation would not be seen),
and some prior information on $d$ and $\tau_E$ is available.

\begin{figure}[ht]
\centering\leavevmode
\epsfxsize=6.0in
\epsfbox{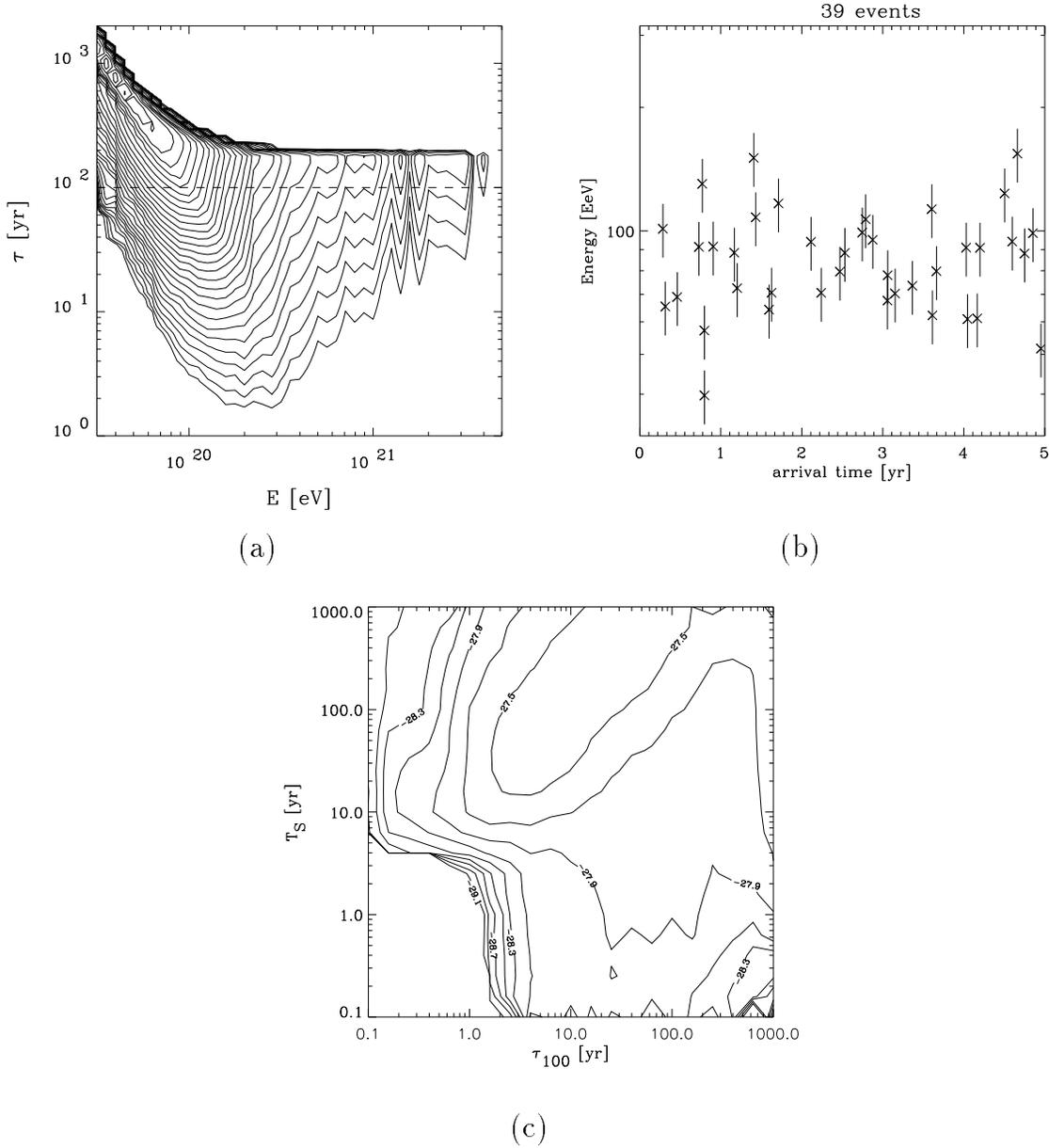}
\bigskip
\caption[...]{(a) Arrival time-energy histogram for
$\gamma=2.0$, $\tau_{100}=50\,$yr, $T_{\rm S}=200\,$yr,
$l_c\simeq1\,$Mpc, $d=50\,$Mpc, corresponding to $B
\simeq3\times10^{-11}\,$G. Contours are in steps of a factor
$10^{0.4}=2.51$; (b) Example of a cluster in the
arrival time-energy plane resulting from the cut indicated in
(a) by the dashed line at $\tau\simeq100\,$yr; (c)
The likelihood function, marginalized over $N_0$
and $\gamma$, for $d=50\,$Mpc, $l_c\simeq\,$Mpc, for the cluster
shown in (b), in the $T_{\rm S}-\tau_{100}$
plane. The contours shown go from the maximum down to about 0.01
of the maximum in steps of a factor $10^{0.2}=1.58$. Note that
the likelihood clearly favors $T_{\rm S}\simeq4\tau_{100}$. For
$\tau_{100}$ large enough to be estimated from the angular
image size, $T_{\rm S}\gg T_{\rm obs}$ can, therefore, be
estimated as well.}
\label{F4.11}
\end{figure}

An emission timescale much larger than the experimental lifetime
may be estimated if a lower cut-off in the spectrum is
observable at an energy $E_{\rm C}$, indicating that $T_{\rm
S}\simeq\tau_{E_{\rm C}}$. The latter may, in turn, be estimated
from the angular image size via Eq.~(\ref{delay}), where the
distance can be estimated from the spectrum visible above the
GZK cut-off, as discussed above. An example
of this scenario is shown in Fig.~\ref{F4.11}. For angular
resolutions $\Delta\theta$, timescales in the range
\begin{equation}
  3\times10^3\,\left(\frac{\Delta\theta}{1^\circ}\right)^2
  \left(\frac{d}{10\,{\rm Mpc}}\right)\,{\rm yr}
  \la T_{\rm S}\simeq\tau_E\la10^4\cdots10^7\,
  \left(\frac{E}{100\,{\rm EeV}}\right)^{-2}\,{\rm yr}
  \label{tsscale}
\end{equation}
could be probed. The lower limit follows from the requirement
that it should be possible to estimate $\tau_E$ from $\theta_E$,
using Eq.~(\ref{delay}), otherwise only an upper limit on $T_{\rm
S}$, corresponding to this same number, would apply.
The upper bound in Eq.~(\ref{tsscale}) comes from constraints on
maximal time delays in cosmic magnetic fields, such as the Faraday
rotation limit in the case of cosmological large-scale field
(smaller number) and knowledge on stronger fields associated
with the large-scale galaxy structure (larger
number). Eq.~(\ref{tsscale}) constitutes an interesting range of
emission timescales for many conceivable scenarios of ultra-high
energy cosmic rays. For example, the hot spots in certain
powerful radio galaxies that have been suggested as ultra-high
energy cosmic ray sources~\cite{rb}, have a size of only several
kpc and could have an episodic activity on timescales of
$\sim10^6\,$yr.

A detailed comparison of analytical estimates for the distributions
of time delays, energies, and deflection angles of nucleons in
weak random magnetic fields with the results of Monte Carlo
simulations has been presented in Ref.~\cite{agnm}. In this
work, deflection was simulated by solving a stochastic differential
equation and observational consequences for the two major
classes of source scenarios, namely continuous and impuslive
UHECR production, were discussed. In agreement with earlier
work~\cite{mw} it was pointed out that at least in the impulsive
production scenario and for an EGMF in the range
$0.1-1\times10^{-9}\,$G, as required for cosmological GRB sources
(see Sect.~5.3 below), there is a typical energy scale
$E_b\sim10^{20.5}-10^{21.5}\,$eV below which the flux is
quasi-steady due to the spread in arrival times, whereas above
which the flux is intermittent with only a few sources
contributing.

{\bf General Case}

\noindent
Unfortunately, neither the diffusive limit nor the limit of
nearly rectilinear propagation is likely to be applicable
to the propagation of UHECR around $10^{20}\,$eV in general.
This is because in magnetic fields in the range of a few
$10^{-8}\,$G, values that are realistic for the Supergalactic
Plane~\cite{bkrproc,rachen1}, the gyro radii of charged particles
is of the order of a few Mpc which is comparable to the distance to
the sources. An accurate, reliable treatment in this regime
can only be achieved by numerical simulation.

To this end,
the Monte Carlo simulation approach of individual trajectories
developed in Refs.~\cite{slo,sl} has recently been generalized
to arbitrary deflections~\cite{slb}. The Supergalactic Plane
was modeled as a sheet with a thickness of a few Mpc and a
Gaussian density profile. The same statistical description
for the magnetic field was adopted as in Refs.~\cite{slo,sl},
but with a field power law index $n_H=-11/3$, representing a
turbulent Kolmogorov type spectrum, and weighted with the
sheet density profile. It should be mentioned, however, that
other spectra, such as the Kraichnan spectrum~\cite{kraichnan},
corresponding to $n_H=-7/2$, are also possible. The largest mode
with non-zero power was taken to be the largest turbulent eddy
whose size is roughly the sheet thickness. In addition, a coherent
field component $B_c$ is allowed that is parallel to the sheet
and varies proportional to the density profile.

\begin{figure}[ht]
\centering\leavevmode
\epsfxsize=6.0in
\epsfbox{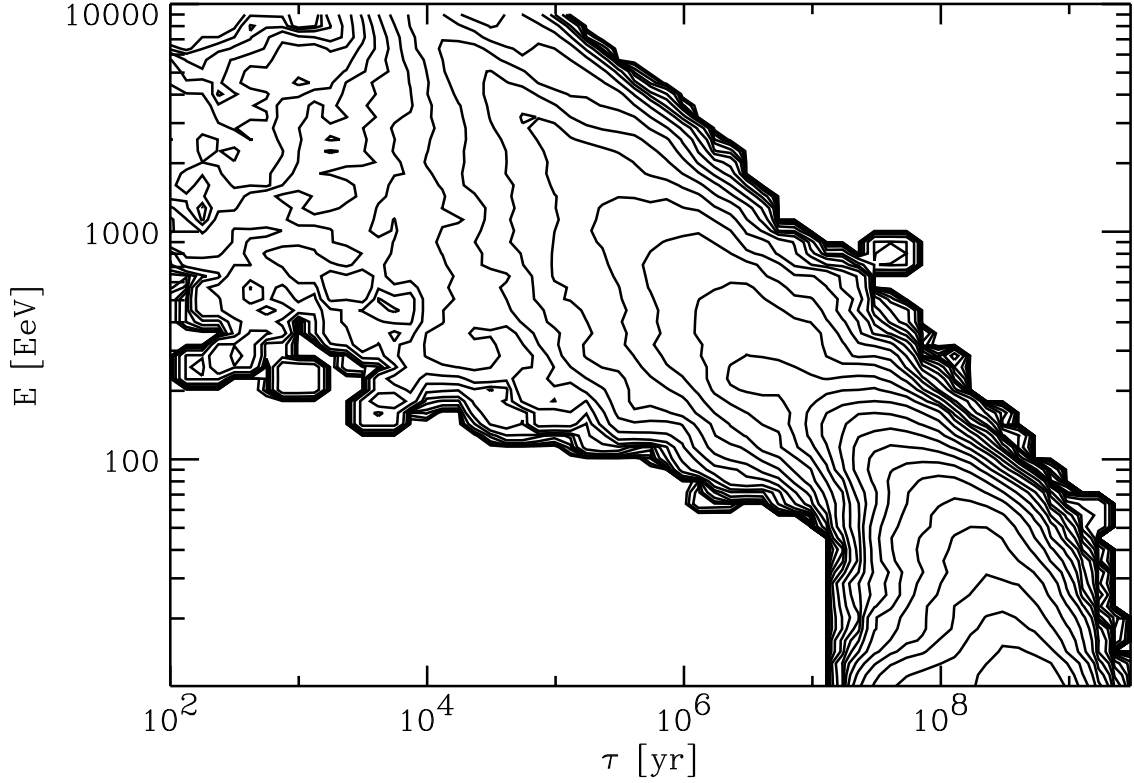}
\bigskip
\caption[...]{The distribution of time delays $\tau_E$ and energies
$E$ for a burst with spectral index $\gamma=2.4$ at a distance
$d=10\,$Mpc, similar to Fig.~\ref{F4.8}, but for the Supergalactic
Plane scenario discussed in the text. The turbulent magnetic field component
in the sheet center is $B=3\times10^{-7}\,$G. Furthermore,
a vanishing coherent field component is assumed.
The inter-contour interval is 0.25 in the logarithm to base 10 of the
distribution per logarithmic energy and time interval.
The three regimes discussed in the text,
$\tau_E\propto E^{-2}$ in the rectilinear regime $E\ga200\,$EeV,
$\tau_E\propto E^{-1}$ in the Bohm diffusion regime $60\,{\rm EeV}
\la E\la200\,$EeV, and $\tau_E\propto E^{-1/3}$ for
$E\la60\,$EeV are clearly visible.}
\label{F4.12}
\end{figure}

When CR backreaction on the weakly turbulent magnetic field
is neglected, the diffusion coefficient of CR of energy $E$
is determined by the magnetic field power on wavelenghts
comparable to the particle Larmor radius, and can be approximated
by~\cite{wentzel}
\begin{equation}
  D(E)\simeq\frac{1}{3}\,r_g(E)\,
  \frac{B}{\int_{1/r_g(E)}^\infty
  \,dk\,k^2\left\langle B^2(k)\right\rangle}\,.\label{Danaly}
\end{equation}
As a consequence, for the Kolmogorov spectrum, in the diffusive
regime, where $\tau_E\ga d$, the diffusion coefficient
should scale with energy as $D(E)\propto E^{1/3}$ for
$r_g\la L/(2\pi)$, and as $D(E)\propto E$ in the so called
Bohm diffusion regime,$r_g\ga L/(2\pi)$. This should be
reflected in the dependence of the time delay $\tau_E$ on
energy $E$: From the rectilinear regime, $\tau_E\la d$,
hence at the largest energies, where $\tau_E\propto E^{-2}$,
this should switch to $\tau_E\propto E^{-1}$ in the regime of
Bohm diffusion, and eventually to $\tau_E\propto E^{-1/3}$
at the smallest energies, or largest time delays. Indeed,
all three regimes can be seen in  Fig.~\ref{F4.12} which shows
an example of the distribution of arrival times and energies
of UHECR from a bursting source.

In a steady state situation, diffusion leads to a modification
of the injection spectrum by roughly a factor $\tau_E$, at
least in the absence of significant energy loss and for a
homogeneous, infinitely extended medium that can be described
by a spatially constant diffusion coefficient. Since in the
non-diffusive regime the observed spectrum repeats the shape of
the injection spectrum,
a change to a flatter observed spectrum at high energies is
expected in the transition region~\cite{diff_refs}. From
the spectral point of view this
suggests the possibility of explaining the observed UHECR
flux above $\simeq10\,$EeV including the highest energy
events with only one discrete source~\cite{bo}.

\begin{figure}[ht]
\centering\leavevmode
\epsfxsize=6.0in
\epsfbox{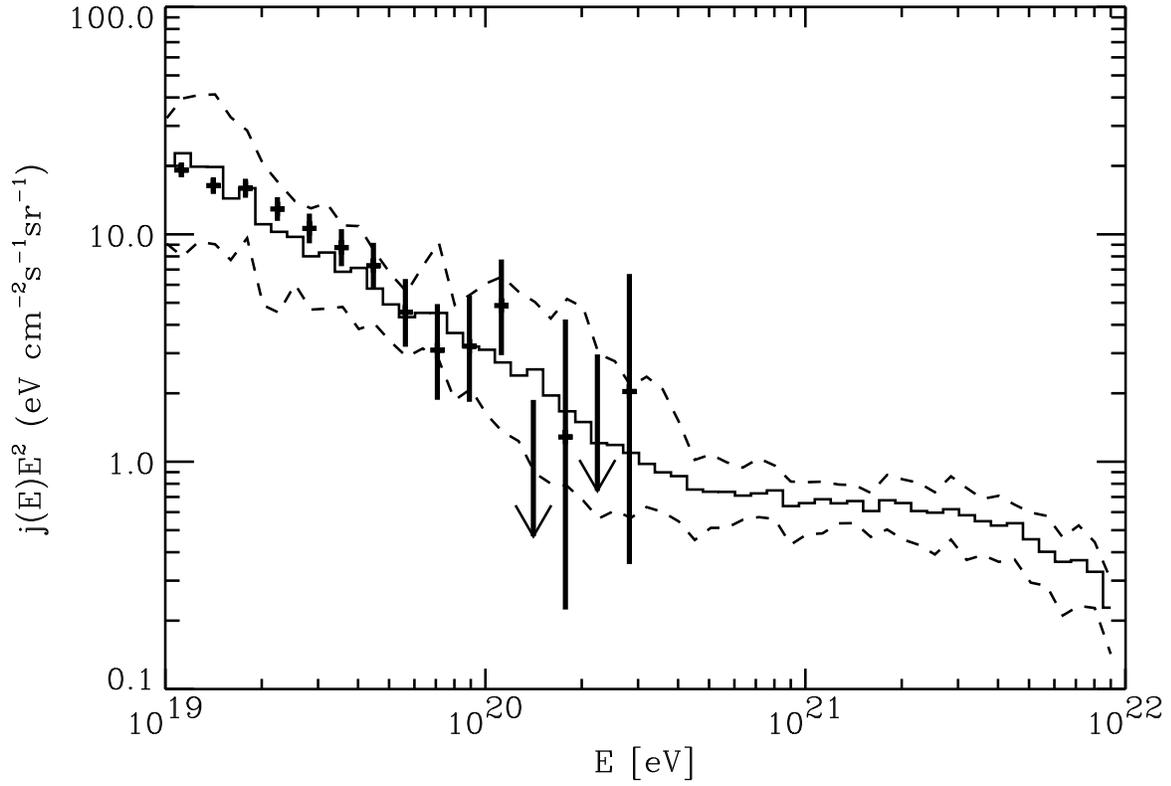}
\bigskip
\caption[...]{The average (solid histogram) and standard
deviation (dashed lines) with respect to 15 simulated
magnetic field realizations of the best fit spectrum to the
data above $10^{19}\,$eV for the scenario of a single source 
in a magnetized Supergalactic Plane. This best fit corresponds to
a maximal magnetic field in the plane center, 
$B_{\rm max}=10^{-7}\,$G, with all other parameters as in
Fig.~\ref{F4.12}. 1 sigma error bars are the combined data
from the Haverah Park~\cite{haverah}, the Fly's Eye~\cite{fe},
and the AGASA~\cite{agasa} experiments above $10^{19}\,$eV.}
\label{F4.13}
\end{figure}

The more detailed Monte Carlo simulations reveal the
following refinements of this qualitative picture: The presence
of a non-trivial geometry where the magnetic field falls
off at large distances, such as with a sheet, tends to deplete
the flux in the diffusive regime as compared to the case
of a homogeneous medium. This is the dominant effect as
long as particles above the GZK cutoff do not diffuse,
this being the case, for example, for an r.m.s. field
strength of $B\la5\times10^{-8}\,$G, $d\simeq10\,$Mpc.
The simple explanation is that the fixed total amount of
particles injected over a certain time scale is distributed
over a larger volume in case of a non-trivial geometry due
to faster diffusion near the boundary of the strong field
region. With increasing field strengths the diffusive regime
will extend to energies beyond the GZK cutoff and the
increased pion production losses start to compensate for the
low energy suppression from the non-trivial geometry.
For very strong fields, for example, for
$B\ga10^{-7}\,$G, $d\simeq10\,$Mpc, the
pion production effect will overcompensate the geometry
effect and reverse the situation: In this case, the
flux above the GZK cutoff is strongly suppressed due
to the diffusively enhanced pion production losses and
the flux at lower energies is enhanced. Therefore, there
turns out to be an optimal field strength that depends on
the source distance and provides an optimal fit to the data
above $10\,$EeV. The optimal case for $d=10\,$Mpc, with
a maximal r.m.s. field strength of $B_{\rm max}=10^{-7}\,$G
in the plane  center is shown in Fig.~\ref{F4.13}.

Furthermore, the numerical results indicate an effective
gyroradius that is roughly a factor 10 higher than the
analytical estimate, with a correspondingly larger diffusion
coefficient compared to Eq.~(\ref{Danaly}). In addition,
the fluctuations of the resulting spectra 
between different magnetic field realizations
can be substantial, as can be
seen in Fig.~\ref{F4.13}. This is a result of the fact that
most of the magnetic field power is on the largest scales
where there are the fewest modes. These considerations mean that 
the applicability of analytical flux estimates of discrete sources in
specific magnetic field configurations is rather limited. 

Angular images of discrete sources in a magnetized Supercluster
in principle contain information on the magnetic field structure.
For the recently suggested field strengths between
$\sim10^{-8}\,$G and $\simeq1\mu\,$G the angular images are
large enough to exploit that information with instruments
of angular resolution in the degree range. An example where
a transition from several images at low energies to one image
at high energies allows one to estimate the magnetic field
coherence scale is shown in Fig.~\ref{F4.14}.

The newest AGASA data~\cite{agasa2}, however, indicate an
isotropic distribution of EHECR. To explain this with only
one discrete source would require the magnetic fields to be
so strong that the flux beyond $10^{20}\,$eV would most likely
be too strongly suppressed by pion production, as discussed above.
This suggests a more continuous source distribution which
may also still reproduce the observed UHECR
flux above $\simeq10\,$EeV with only one spectral
component~\cite{medinatanco3}.
A more systematic parameter study of sky maps and spectra
in UHECR in different scenarios is therefore now being
pursued~\cite{medinatanco4,lsb}.

Intriguingly, scenarios in which a diffuse
source distribution follows the density in the Supergalactic
Plane within a certain radius, can accomodate both the large scale
isotropy and the small scale clustering revealed by AGASA if
a magnetic field of strength $B\ga0.05\mu\,$G
permeates the Supercluster~\cite{lsb}.

Fig.~\ref{F4.15} shows the angular distribution in Galactic
coordinates in such a scenario for different field strengths
and source distribution radii. The integral angular
distributions with respect to the Supergalactic Plane
for two such cases are shown in Fig.~\ref{F4.16}.

Detailed Monte Carlo simulations performed on these distributions
reveal that the anisotropy decreases with increasing magnetic field
strength due to diffusion and that small scale clustering increases
with coherence and strength of the magnetic field due to magnetic
lensing. Both anisotropy and clustering also increase with
the (unknown) source distribution radius.
Furthermore, the discriminatory power
between models with respect to anisotropy and clustering strongly
increases with exposure~\cite{lsb}.

Finally, the corresponding solid angle integrated spectra show
negligible cosmic variance for diffuse sources and fit the data
well both for $B_{\rm max}=0.05\,\mu\,$G and for
$B_{\rm max}=0.5\,\mu\,$G, as shown in Fig.~\ref{F4.17}.

As a result, a diffuse source distribution associated with the
Supergalactic Plane can explain most of the currently observed 
features of ultra-high
energy cosmic rays at least for field strengths close to $0.5\,\mu\,$G.
The large-scale anisotropy and the clustering predicted
by this scenario will allow strong discrimination against other models
with next generation experiments such as the Pierre Auger Project.

\subsection{Anomalous Kinematics, Quantum Gravity Effects,
Lorentz symmetry violations}
The existence of UHECR beyond the GZK cutoff has prompted several
suggestions of possible new physics beyond the Standard Model. We have
already discussed some of these suggestions in Sect.~4.3 in the
context of propagation of UHECR in the extragalactic space. Further,   
in Sect.~6 we will discuss suggestions regarding possible
new sources of EHECR that also involve postulating new physics beyond the
Standard Model. In the present section, to end our discussions on the
propagation and interactions
of UHE radiation, we briefly discuss some examples
of possible small violations or modifications of certain fundamental
tenets of physics (and constraints on the magnitude of those
violations/modifications) that have also been discussed in the literature
in the context of propagation of UHECR. 

For example, as an interesting consequence of the very existence of UHECR,
constraints on possible violations of Lorentz invariance (VLI)  
have been pointed out~\cite{colgla}. These constraints rival
precision measurements in
the laboratory: If events observed around $10^{20}\,$eV are indeed
protons, then the difference between the maximum attainable proton
velocity and the speed of light has to be less than about 
$1\times10^{-23}$, otherwise the proton would lose its energy by
Cherenkov radiation within a few hundred centimeters. Possible
tests of other modes of VLI with UHECR
have been discussed in Ref.~\cite{g-m}, and in Ref.~\cite{cowsik-sreek} in
the context of horizontal air-showers generated by cosmic rays in general.  
Gonzalez-Mestres~\cite{g-m}, Coleman and Glashow~\cite{colgla2}, and
earlier, Sato and Tati~\cite{sato-tati} and Kirzhnits and
Chechin~\cite{kirzh-che} have also suggested that due to modified
kinematical constraints the GZK cutoff could even be evaded by allowing
a tiny VLI too small to have been
detected otherwise. Similar consequences apply to other
energy loss processes such as pair production by photons above
a TeV with the low energy photon background~\cite{kifune}.
It seems to be possible to accomodate such effects within
theories involving generalized Lorentz transformations~\cite{bogo-goenner}
or deformed relativistic kinematics~\cite{g-m1}.
Furthermore, it has been pointed out~\cite{halprin-kim} that violations of
the principle of equivalence (VPE), while not dynamically equivalent, also
produce the same kinematical effects as VLI for particle processes 
in a constant gravitational potential, and so the constraints on VLI  
from UHECR physics can be translated into constraints
on VPE such that the difference between the couplings of protons and
photons to gravity must be less than about $1\times10^{-19}$. Again, this 
constraint is more stringent by several orders of magnitude than
the currently available laboratory constraint from E\"{o}tv\"{o}s
experiments. 

As a specific example of VLI, we consider an energy dependent photon group
velocity $\partial E/\partial k=c[1-\chi E/E_0+{\cal O}(E^2/E_0^2)]$
where $c$ is the speed of light in the low energy limit, $\chi=\pm1$,
and $E_0$ denotes the energy scale where this modification
becomes of order unity. This corresponds to a dispersion relation
\begin{equation}
  c^2k^2\simeq E^2+\chi\frac{E^3}{E_0}\,,\label{mod-dispersion}
\end{equation}
which, for example, can occur in quantum gravity and string
theory~\cite{emn}. The kinematics of electron-positron pair production
in a head-on collision of a high energy photon of energy $E$ with a
low energy background photon of energy $\varepsilon$ then leads to the
constraint
\begin{equation}
  \varepsilon\simeq\frac{E}{4}\left(\frac{m_e^2}{E_1E_2}+
  \theta_1\theta_2\right)+\chi\frac{E^2}{4E_0}\,,\label{mod-kinematics}
\end{equation}
where $E_i$ and $\theta_i\sim{\cal O}(m/E_i)$ are respectively the energy
and outgoing momentum angle (with respect to the original photon
momentum) of the electron and positron ($i=1,2$). 
For the case considered by Coleman and Glashow~\cite{colgla} in which the
maximum attainable speed $c_i$ of the matter particle is different
from the photon speed $c$, the kinematics can be obtained by
substituting $c_i^2-c^2$ for $\chi E/E_0$ in Eq.~(\ref{mod-kinematics}).

Let us define a critical energy $E_c=(m_e^2E_0)^{1/3}
\simeq15(E_0/m_{\rm Pl})^{1/3}\,$TeV in the case of the energy dependent
photon group velocity, and  
$E_c=m_e/|c_i^2-c^2|^{1/2}$ in the case considered by Coleman and
Glashow.
If $\chi<0$, or $c_i<c$, then $\varepsilon$ becomes negative for
$E\ga E_c$. This signals that the photon can spontaneously
decay into an electron-positron pair and propagation of photons across
extragalactic distances 
will in general be inhibited. The observation of extragalactic
photons up to $\simeq20\,$TeV~\cite{mrk421,mrk501} therefore puts the limits
$E_0\ga M_{\rm Pl}$ or $c_i^2-c^2\ga-2\times10^{-17}$.
In contrast, if $\chi>0$, or $c_i>c$,
$\varepsilon$ will grow with energy for $E\ga E_c$ until
there is no significant number of target photon density available
and the Universe becomes transparent to UHE photons.
A clear test of this possibility would be the observation of
$\ga100\,$TeV photons from distances
$\ga100\,$Mps~\cite{kluzniak}.

In addition, the dispersion relation Eq.~(\ref{mod-dispersion})
implies that a photon signal at energy $E$ will be spread out
by $\Delta t\simeq (d/c)(E/E_0)\simeq1(d/100\,{\rm Mpc})(E/\,{\rm TeV})
(E_0/M_{\rm Pl})^{-1}\,$s. The observation of $\gamma-$rays
at energies $E\ga2\,$TeV within $\simeq300\,$s from the AGN Markarian
421 therefore puts a limit (independent of $\chi$) of
$E_0\ga4\times10^{16}\,$GeV, whereas the possible observation
of $\gamma-$rays at $E\ga200\,$TeV within $\simeq200\,$s from a
GRB by HEGRA might be sensitive to $E_0\simeq M_{\rm Pl}$~\cite{acemns}.
For a recent detailed discussion of these limits see Ref.~\cite{efmmn}.

A related proposal originally due to Kosteleck\'{y}
in the context of CR suggests the electron neutrino to be a
tachyon~\cite{kostelecky}.
This would allow the proton in a nucleus of mass $m(A,Z)$ for mass
number $A$ and charge $Z$ to decay via $p\to n+e^++\nu_e$ above
the energy threshold $E_{th}=m(A,Z)[m(A,Z\pm1)+m_e-m(A,Z)]/|m_{\nu_e}|$
which, for a free proton, is $E_{th}\simeq1.7\times10^{15}/
(|m_{\nu_e}|/{\rm eV})\,{\rm eV}$. Ehrlich~\cite{ehrlich} claims
that by choosing $m^2_{\nu_e}\simeq-(0.5\,{\rm eV})^2$ it is
possible to explain the knee and several other features
of the observed CR spectrum, including the high energy end,
if certain assumptions about the source distribution are made.
The experimental best fit values of $m^2_{\nu_e}$ from tritium beta
decay experiments are indeed negative~\cite{p-data}, although this is most
likely due to unresolved experimental issues. In addition, the values of 
$|m^2_{\nu_e}|$ from tritium beta decay experiments are typically larger
than the value required to fit the knee of the CR spectrum. This scenario
also predicts a neutron line around the knee energy~\cite{ehrlich1}.

\section{Origin of UHECR: Acceleration Mechanisms and Sources}
As mentioned in Sect.~3.3, the first-order Fermi acceleration in the form
of DSAM when applied to shocks in supernova remnants can accelerate
particles to energies perhaps up to $\sim10^{17}\ev$ (see, e.g.,
Ref.~\cite{bier-rev1}), but probably not much beyond. Thus, SNRs are
unlikely to be the
sources of the UHECR above $\sim10^{17}\ev$. Instead, for
UHECR, one has to invoke shocks on larger scales, namely extragalactic
shocks. Several papers have, therefore, focussed on extragalactic
objects such as AGNs and radio-galaxies as possible sites of
UHECR acceleration. Reviews on this topic can be found, e.g., in
Refs.~\cite{ssb,norman,blandford,takahara-rev,bier-rev1,bier-rev2,bier-rev3,
proth-rev,kirk-duffy}. 
Below, we briefly discuss the issue of maximum achievable energy within
DSAM and then discuss the viability or otherwise of the extragalactic
sources that have been proposed as acceleration sites for CR up to the
highest energies of $\ga10^{20}\ev$. We also briefly mention
acceleration of UHECR in pulsars.   

\subsection{Maximum Achievable Energy within Diffusive Shock Acceleration
Mechanism} 
There is a large body of literature on the subject of DSAM. We urge the
reader to consult the reviews in
Refs.~\cite{drury,bland-eich,jones-ellison,takahara-rev,proth-rev,kirk-duffy}
for details and original references. In the simplest version of DSAM,
one adopts a so-called test-particle approximation in which the
shock structure is given {\it a-priori} and is not affected by
the particles being accelerated. One also assumes a
non-relativistically moving plane-parallel shock front.
The magnetic field is assumed parallel to the shock normal.
The inhomogeneities of the magnetic field are
assumed to scatter particles efficiently so as to result in a
nearly isotropic distribution of the particles. Under these
circumstances, one gets (see, for example, Ref.~\cite{gaisser},
for a text-book derivation) a universal power-law energy spectrum
of the accelerated particles, $N(E)\propto E^{-q}$, with index
$q=(r+2)/(r-1)$, where $r=u_1/u_2$ is the shock compression
ratio, $u_1$ and $u_2$ being the upstream- and downstream
velocities of the fluid in the rest-frame of the shock. 
The shock-compression ratio $r$ is related to the adiabatic
index of the fluid. For typical astrophysical situations,
one gets $r<4$ and hence $q>2$. 

There are several issues that complicate this simple picture. Among
these are issues associated with (a) the effects of a more realistic
shock geometry, (b) back-reaction of the accelerated particles
on the shock structure and its effect on the resulting
particle spectrum, (c) ultra-relativistic shocks (which may be
relevant for acceleration of particles in GRBs,
for example), (d) the question of generation of the magnetic
fluctuations which are necessary for scattering of
particles and which determine the mean-free path of particles
and hence the relevant diffusion coefficients which, in turn,
effect the spectral shape of the accelerated
particles, and so on. For example, it has been recently
claimed~\cite{malkov} that for strong shocks inclusion of
back-reaction effects can result in a significantly harder
spectrum with $q=1.5$ compared to the canonical spectrum
with $q>2$. On the other hand, for ultrarelativistic shocks
in the limit $\Gamma\to\infty$ ($\Gamma$ being the Lorentz
factor of the shock), the spectrum becomes softer ($q\simeq 2.2$)
than the canonical $q=2$ spectrum~\cite{bednarz-ostrow}.
Furthermore, it has been claimed recently that Fermi-type shock
acceleration by relativistic blast waves leads to an energy
gain of a factor $\simeq\Gamma^2$ in the first shock crossing
cycle, but only by a factor $\simeq2$ in following cycles
because particles do not have time to re-isotropize upstream
before the next cycle~\cite{gallant-achterberg}. For a recent
review of particle acceleration at relativistic shocks
see Ref.~\cite{kirk-duffy}.

The above issues are, however, subjects of considerable on-going
debates and discussions. Here we will not go into the subtleties
associated with these issues. Instead, we focus directly on the
question of the maximum achievable energy, following the analysis of
Ref.~\cite{ssb}. 

In relativistic shocks the cutoff energy $E_c$ for the source spectrum
of accelerated cosmic rays is, in the test particle approximation, 
generally given by $ZeBR$, the product of the charge $Ze$ of the
cosmic ray particle, the magnetic field $B$ and the size $R$ of
the shock, multiplied by some factor of order
unity~\cite{hillas-araa,gaisser,bier-rev2,takahara-rev,cesarsky,quenby-naidu}. 
However, for the highest energies the mean free path of the particle
becomes comparable to the shock size $R$, which has to be properly
taken into account in calculating $E_c$.

The acceleration of a particle of energy $E$ in an astrophysical shock is
governed by the equation 
\begin{equation}
  \frac{dE}{dt}=\frac{E}{T_{\rm acc}}\,,\label{Eacc}
\end{equation}
where $T_{\rm acc}$ is the energy dependent acceleration time. For DSAM, 
the slope $q$ of the resulting power-law energy
spectrum, $dN/dE\propto E^{-q}$, of the particles is related
to $T_{\rm acc}$ and $T_{\rm esc}$, the mean (in general also energy
dependent) escape time by~\cite{gaisser,cesarsky}
\begin{equation}
  q=1+\frac{T_{\rm acc}}{T_{\rm esc}}\,.\label{q}
\end{equation}
For first-order Fermi acceleration at nonrelativistic shocks,
$T_{\rm acc}$ is usually given by
\begin{equation}
  T_{\rm acc}=\frac{3}{u_1-u_2}\left(\frac{D_1}{u_1}+
          \frac{D_2}{u_2}\right)\,,\label{Tacc}
\end{equation}
where $u_1$, $u_2$ are the up- and downstream velocities of material
flowing through the shock in its rest frame, and $D_1$ and $D_2$ are the
corresponding diffusion coefficients, respectively. Diffusion is dominated
by magnetic pitch angle scattering caused by inhomogeneities in the
magnetic field~\cite{bland-eich}. Therefore, the mean free path
$\lambda$ is bounded from below by the gyroradius
$r_g=E/(ZeB)$ multiplied by some factor $g$, and so $D_1$ and $D_2$
(for ultra-relativistic particles) can be estimated as 
\begin{equation}
  D_1,D_2\sim\lambda/3\ga\frac{gE}{3ZeB}\,.\label{Diff}
\end{equation}
For nonrelativistic shocks, $g$ is usually set equal to
1~\cite{gaisser,cesarsky}. However, as we deal with the UHECR, we have to
consider relativistic shocks because they provide the most powerful
accelerators. Monte-Carlo simulations of such
relativistic shocks show that $g$ can be as large as
$\simeq40$~\cite{quenby-lieu},
leading to an effective slowing-down of acceleration compared to 
a naive extrapolation of Eq.~(\ref{Tacc}) (with $g\simeq1$) to the
relativistic shock case. This is, however, partially compensated for
by an additional factor which reaches a value of about 10 in highly 
inclined, and by about a factor of 13.5 in
parallel~\cite{quenby-lieu}, relativistic shocks, respectively, in the
limit $u_1\to 1$. There is, however, some disagreement on this
issue~\cite{ejr}. 

Putting everything together and minimizing $T_{acc}$ from
eq.~(\ref{Tacc}) as a function of $u_1$ and $u_2$ in the interval [0,1] we
arrive at
\begin{equation}
  T_{\rm acc}\ga\frac{g}{2.25}\,\frac{E}{ZeB}\,.\label{Taccmin}
\end{equation}
On the other hand, as long as the diffusion approximation is
valid, {\it i.e.,} as long as $\lambda<R$ corresponding to
$E<E_{\rm diff}\equiv ZeBR/g$, the escape time is given by
$T_{\rm esc}=R^2/\lambda$, whereas for $E\geq E_{\rm diff}$ the particles
are, to a good approximation, freely streaming out of the shock region 
and $T_{\rm esc}=R$. Using eqs.~(\ref{q}) and (\ref{Taccmin}), we
thus get
\begin{equation}
  q(E> E_{\rm diff})\sim1+\frac{E}{2.25E_{\rm diff}}\,.\label{gam}
\end{equation}
Defining the cutoff energy $E_c$ as the energy where the source
spectral index becomes 3 (recall that the slope of the CR 
spectrum observed at the earth is around 2.7 in the UHE region) yields
\begin{equation}
  E_c\equiv E_{q=3}\sim10^{17}\ev\,Z
     \left(\frac{R}{{\rm kpc}}\right)
     \left(\frac{B}{\mu{\rm G}}\right)\,.\label{Ec}
\end{equation}
We have assumed here that the magnetic field is parallel to the shock
normal. If that is not the case there will be an electric field
${\bf E}={\bf u_1}\times{\bf B}$ in the shock rest frame. Together
with diffusion effects this causes a drift
acceleration of charged particles along the shock
front~\cite{jokipii-drift} to a maximal energy $E_{\rm max}$ which, for
magnetic field $B$ substantially inclined to the shock normal, is given by
\begin{equation}
  E_{\rm max}=Zeu_1BR\sim10^{18}\ev\,Zu_1
  \left(\frac{R}{{\rm kpc}}\right)
  \left(\frac{B}{\mu{\rm G}}\right)\,.\label{Emax}
\end{equation}
This is about one order of magnitude larger than
eq.~(\ref{Ec}) if $u_1$ approaches the speed of light.
However, the electric field ${\bf E}$ is expected
to be much smaller in general due to plasma effects so that
rather special conditions have to be fulfilled in order that
such high energies can be approached. We shall, therefore, take the
estimate in Eq.~(\ref{Ec}) as a ``benchmark'' estimate for the maximum
achievable energy within DSAM. 

\subsection{Source Candidates for UHECR} 
Irrespective of the precise acceleration mechanism,
there is a simple dimensional argument, given by Hillas~\cite{hillas-araa},
which allows one to restrict attention to only a few classes of
astrophysical objects as possible sources capable of accelerating  
particles to a given energy. In any statistical acceleration mechanism, 
there must be a magnetic field ($B$) to keep the particles confined
within the acceleration site. Thus, the size $R$ of the acceleration
region must be larger than the diameter of the orbit of the
particle $\sim2r_g$. Including the effect of the characteristic
velocity $\beta c$ of the magnetic scattering centers 
one gets the general condition~\cite{hillas-araa}  
\beq
\left(\frac{B}{\mu\G}\right)\left(\frac{R}{\kpc}\right)>2
\left(\frac{E}{10^{18}\ev}\right)\frac{1}{Z\beta}\,.\label{hillas-eqn}
\eeq

\begin{figure}[ht]
\centering\leavevmode
\epsfxsize=6.0in
\epsfbox{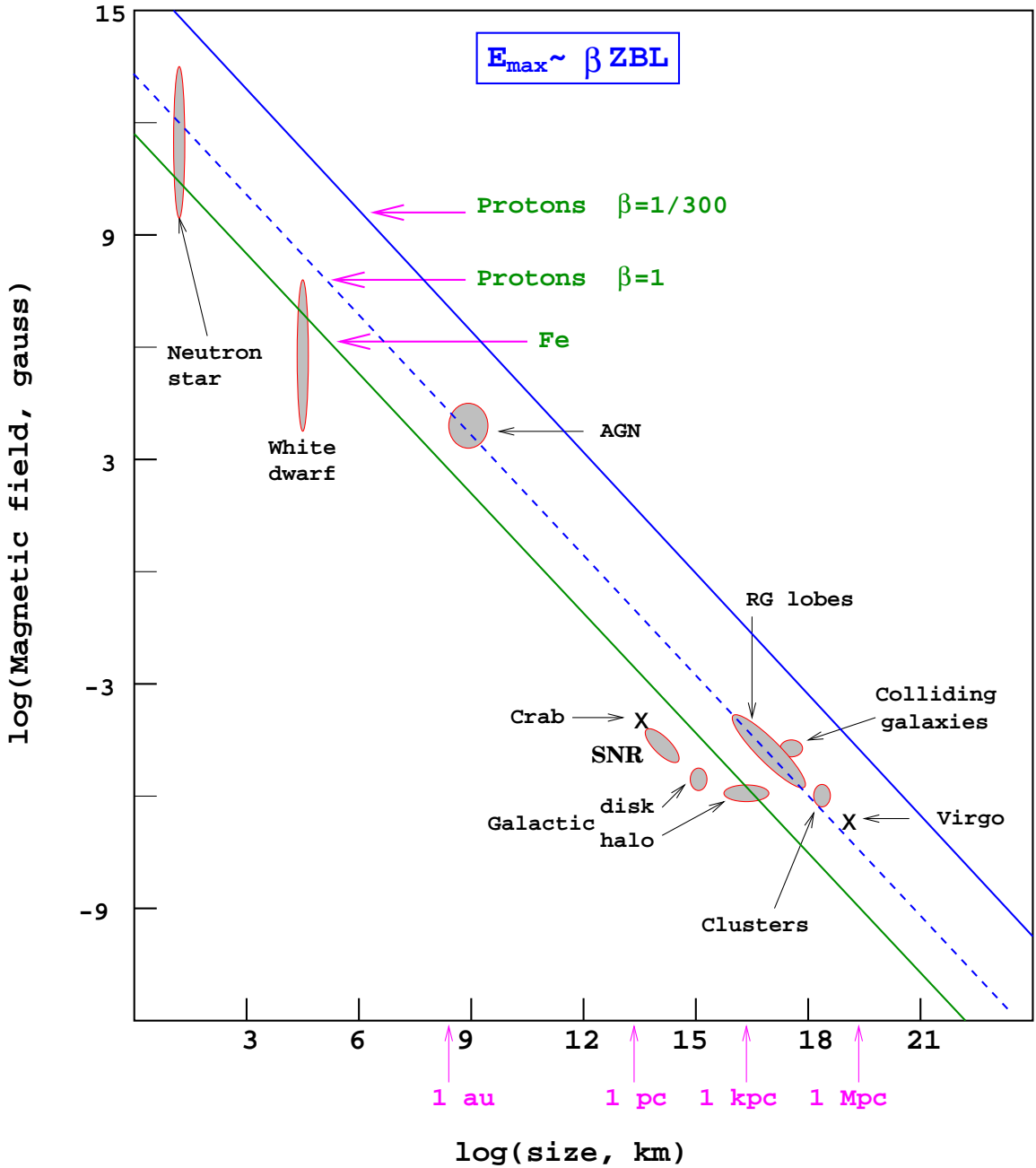}
\bigskip
\caption[...]{The Hillas diagram
showing size and magnetic field strengths of possible sites of particle
acceleration. Objects below the corresponding diagonal lines cannot
accelerate protons (iron nuclei) to $10^{20}\,$eV. $\beta c$ is the
charateristic velocity of the magnetic scattering centers. 
(This version courtesy Murat Boratav.)
\label{F5.1}}
\end{figure}

As argued by Hillas~\cite{hillas-araa}, the above condition also applies to 
direct acceleration scenarios (as may operate for example in
the pulsar magnetosphere), in which the electric field arises due
to a moving magnetic field. The dimensional argument
expressed by Eq.~(\ref{hillas-eqn}) is often presented in the form
of the famous ``Hillas diagram'' shown in Fig.~\ref{F5.1},
which shows that to achieve a given
maximum energy, one must have acceleration sites that have either a large
magnetic field or a large size of the acceleration region. Thus,
for example, only a few  astrophysical sources --- among them,
AGNs, radio-galaxies, and pulsars --- satisfy the
conditions necessary (but may or may not be sufficient) for
acceleration up to $\sim10^{20}\ev$. Currently, therefore, most
discussions of astrophysical acceleration
mechanisms for EHECR have focussed on these objects. Below we
briefly summarize the status of these objects as possible
acceleration sites for CR up to EHECR energies. 

\subsubsection{AGNs and Radio-Galaxies}

AGNs and radio-galaxies could be the main contributors to extragalactic
CR. Several arguments support this possibility: First, at least
two BL Lacertae objects, a certain class of AGNs, have been observed
in $\gamma-$rays above $\simeq10\,$TeV, namely Markarian
421~\cite{mrk421} and Markarian 501~\cite{mrk501}. Photons of such
high energies may be produced by the decay of pions produced in  
interactions of the accelerated protons with the ambient matter in these
sources (see, e.g., Ref.~\cite{proton-blazar}) rather than by inverse
Compton scattering of low-energy photons by accelerated
electrons (e.g.,~\cite{ssc}) in these sources.
Second, it has been pointed out, that the energy content in the
diffuse $\gamma-$ray background measured by EGRET is comparable
to the one required for an extragalactic proton injection
spectrum proportional to $E^{-2}$ up to $10^{20}\,$eV if it
is to explain the observed UHECR spectrum above the ankle at
$10^{18.5}\,$eV~\cite{mannheim2}. This is expected if the
diffuse photons again result from the decay of pions produced by the
accelerated protons and the subsequent propagation (cascading) of those
photons. 

On the other hand, the fast variability of flares recently
observed may favor the acceleration of electrons as an explanation
of the highest energy photons observed by ground Cherenkov
telescopes~\cite{ssc_vs_hadronic}, which has triggered reconsideration
of theoretical flux predictions in these models~\cite{ps3,ddem}.
More generally, both proton
and electron acceleration could provide energy dependent
contributions to the $\gamma-$ray flux. To settle this question
will require more data from the optical up to the TeV range,
whose current status is reviewed in Ref.~\cite{krennrich}. Recently, a
flare with hour-scale variability was observed simultaneously in X-rays
and very high energy $\gamma-$rays from Markarian 421~\cite{mrk421-flare};
the implications of this observation for the emission mechanism(s) of the
radiation in the different wavebands are, however, not clear at this
stage. 

The physics of AGNs and radio-galaxies in the context of the possibility
of these objects being the sources of UHECR
have been reviewed extensively in recent literature; see, e.g.,
Refs.~\cite{norman,blandford,takahara-rev,bier-rev1,bier-rev2} for
original references. 
Estimates of the typical values of $R$ and $B$ for the central regions of AGNs
give~\cite{hz} $R\sim 0.02\,$pc and $B\sim5\,$G. These values
when substituted into Eq.~(\ref{Ec}) above yield
$E_c\sim10^{19}\ev$ for protons. This number can be
uncertain perhaps by a factor of few. So, although AGNs are unlikely to 
be the sources of the EHECR above $10^{20}\ev$, a good part of
the UHECR below $\sim {\rm few}\times10^{19}\ev$ could in
principle originate from AGNs. However, the major problem here
is that the accelerated protons are severely degraded due to
photo-pion production on the intense radiation field in and around
the central engine of the AGN. In addition, there are energy losses
due to synchrotron and Compton processes. Taking into account the
energy losses simultaneously with the energy gain due to
acceleration, Norman et al~\cite{norman} conclude that neither
protons nor heavy nuclei are likely to escape from the central
regions of AGNs with energies much above $\sim10^{16}\ev$.
There is also a suggestion (see, e.g., Szabo and Protheroe in
Ref.~\cite{agn-nu}) that the
photo-pion production process could convert protons into neutrons,
which could then escape from the central region of the AGN, and
the neutron could later decay to protons through $\beta$-decay.
However, neutrons themselves are also subject to photo-pion
production in the dense radiation field, and so neutrons above
$\sim10^{16}\ev$ also cannot escape from the central regions
of AGNs~\cite{norman}. One can thus conclude that
the central regions of AGNs are unlikely to be the sources of the
observed UHECR.\footnote{Although UHE nucleons cannot escape from AGN
cores, the associated UHE neutrinos from the decay of the photo-produced
pions can. The integrated
contribution from all AGNs may then produce a diffuse high energy
neutrino background that may be detectable~\cite{agn-nu}. Clearly,
from the discussion above, this potential contribution to the UHE neutrino
background would, however, be unrelated to the sources of the observed
UHECR, and would also not be subject to the Waxman Bahcall
bound which does not apply to "hidden" sources~\cite{wb2,bw}(see below).}

Currently, perhaps the most promising acceleration sites for UHECR
are the so-called hot-spots of Fanaroff-Riley type II 
radio-galaxies; for reviews and references see
Refs.~\cite{takahara-rev,bier-rev2,norman}. The issue of maximum
energy achievable in this case has also been reviewed by Norman
et al~\cite{norman}. The energy loss due to photo-pion production
at the source is not significant at hot spots because the density of
the ambient soft photons at the hot-spots is thought to be not
high enough. Depending on the magnetic field strength at the
hot-spots (which is crucial but happens to be the most uncertain
parameter in this consideration), a maximum energy of even up
to $\sim10^{21}\ev$ seems to be possible. So these radio-galaxies
could, in principle, be sources of UHECR
including the EHECR above $\sim10^{20}\ev$. However, the main
problem with radio-galaxies as sources of the EHECR is their
locations: the radio-galaxies that lie along the arrival
directions of individual EHECR events are situated at large
cosmological distances ($\ga100\mpc$) from Earth~\cite{elb-som}
(see also Sect.~4.6),
in which case, because of the GZK effect discussed earlier,
the particles do not survive with EHECR energies even if they
are produced with such energies at the source.
Thus, at the present time, although it seems hot-spots of
radio-galaxies may well be the sources of UHECR above
$\sim10^{17}\ev$, it seems difficult to invoke them as sources of
the observed EHECR events above $10^{20}\ev$\footnote{The
distance problem with radio-galaxies may, however, be avoidable
if the EHECR are a possible new kind of supersymmetric
particle ($S^0$)~\cite{farrar,cfk,fb} which could 
be produced by accelerated protons through the photo-production
process in the dense regions of some compact radio-galaxies.
These $S^0$ particles being electrically neutral would be
able to escape from the source, and their specific particle
physics properties may allow them to avoid the drastic loss
process associated with the GZK effect; 
see Sect.~4.3.2 for a discussion.}. 

The ultimate test for the case of AGNs and radio-galaxies as
proton accelerators and for the origin of CR at least up to
the GZK cutoff is expected to come from neutrino astronomy:
Practically no neutrinos are produced in the non-hadronic
AGN models with electron acceleration. In contrast, if jets
(as opposed to cores) of AGNs and
radio-galaxies are the main sources of extragalactic CR, the
secondary $\gamma-$rays and neutrinos are created by pion
production and the energy content in the diffuse neutrino
flux can be normalized to the $\gamma-$ray flux
produced by these AGNs~\cite{mannheim,hz}. It has recently been pointed
out by Waxman and Bahcall~\cite{wb2} that a comparison with the UHECR
flux around $10^{19}\,$eV leads to another bound on the
diffuse neutrino flux that is more stringent by about two
orders of magnitude, as long as accelerated
protons are not absorbed in the source. This upper bound has
become known as the Waxman-Bahcall upper bound. A subsequent
more detailed numerical study by Mannheim, Protheroe, and
Rachen pointed out possible loopholes to this bound and
claim that it only applies to neutrino energies between
$\sim10^{16}\,$eV and $\sim10^{18}\,$eV~\cite{mpr}.
However, according to Bahcall and Waxman~\cite{bw}, the attempts
presented in Ref.~\cite{mpr} to evade the bound on diffuse neutrino
fluxes from optically thin sources are in conflict with observational
evidence and the Waxman-Bahcall bound is robust. We recall,
however, that the Waxmann Bahcall bound does not apply to
sources that are optically thick for protons, such as could
be the case for AGN cores (see above). Also, this bound does 
not directly apply to top-down scenarios because there neutrinos are
produced as primaries, not secondaries, with fluxes that are considerably
higher than the nucleon fluxes. As will be discussed in
Sect.~7.4, in top-down scenarios the diffuse neutrino fluxes
are still bounded by the observed diffuse GeV $\gamma-$ray background.

Recently, Boldt and Ghosh~\cite{boldt-ghosh} have advanced the interesting
suggestion that EHECR particles may be accelerated near the 
event horizons of spinning
supermassive black holes associated with presently {\it inactive} quasar
remnants. The required emf is generated by the black hole induced rotation
of externally supplied magnetic field lines threading the horizon. This
suggestion avoids the problem of the 
dearth of currently {\it active} galactic nuclei, quasars and/or radio
galaxies within acceptable distances ($\la50\mpc$) to serve as possible
sources of EHECR events. Boldt and Ghosh estimate the number of
supermassive black holes of required mass $\ga10^9M_\odot$ associated with
``dead'' (no jet) quasars within a volume of radius $\sim50\mpc$ to be
sufficient to explain the observed EHECR flux. The exact process by which
the rotational energy of the spinning supermassive black hole goes into
accelerating protons to EHECR energies and the process by which the
required magnetic field is generated and sustained remain to be spelled
out, however. If the expectations of Ref.~\cite{boldt-ghosh} are borne out
by more detailed modeling, acceleration of protons to a maximum energy 
of $\sim10^{21}\ev$ would be possible, and in that case, dead quasars
in our local cosmological neighborhood would indeed be one of the most
promising sources of EHECR. 

\subsubsection{Pulsars} 
As seen from the Hillas diagram, Fig.~\ref{F5.1}, pulsars are
potential acceleration sites for UHECR. Most of the acceleration
scenarios involving pulsars rely upon direct acceleration of
particles in the strong electrostatic potential drop induced at
the surface of the neutron star by unipolar induction due to the
axially symmetric rotating magnetic field configuration of
the rotating neutron star. The maximum potential drop for
typical pulsars can in principle be as high as
$\sim10^{21}\ev$ (see, e.g., Ref.~\cite{vmo}). The component
of a particle's momentum perpendicular to the local magnetic
field line is damped out due to synchrotron radiation,
and so the particles are forced to move along and are accelerated
by the electric field component along the magnetic field lines.
However, in any realistic model, the large
potential drop along the magnetic field lines is significantly
short-circuited by electrons and positrons moving in the
opposite directions along the field lines --- the source of the
electron-positron pairs being the pair-cascade initiated by
strong curvature radiation from seed electrons accelerated
along the curved magnetic field lines. 
Acceleration models with pulsars have been reviewed, for
example, in Refs.~\cite{bbdgp,takahara-rev} and more recently in
Ref.~\cite{vmo}. The general conclusion
seems to be that, for isolated neutron stars
(without companion), acceleration of particles to energies beyond
$\sim10^{15}\ev$ is difficult. 

Another class of acceleration models~\cite{lovelace}
utilize large electric fields produced by unipolar induction
in accretion disks around rotating neutron stars or
black holes. An accretion disk threaded by a large-scale poloidal
magnetic field produces a radial component of an electric field
in the disk, which can be utilized for particle
acceleration. However, energy loss through interaction of the
accelerated particles in the ambient photon field around the
central compact object prevents the maximum achievable
energy from exceeding beyond about $\sim10^{15}\ev$ (see, e.g.,
the review by Takahara~\cite{takahara-rev}). Several other
accretion disk-based models are reviewed, for example, in
Refs.~\cite{bbdgp,takahara-rev}. None of these models is, however, 
capable of accelerating particles to UHECR energy regions.

While the pulsar acceleration models mentioned above all deal
with direct acceleration, there exists another
class of models which attempt to utilize the statistical
shock acceleration mechanism in accretion shocks around compact
objects such as neutron stars or black holes; for a review,
see, e.g., Ref.~\cite{takahara-rev}. Again, it is difficult
to go past $\sim10^{15}\ev$ when energy loss processes
are taken into account.

Finally, we mention that recent discovery~\cite{kouveliotou}
of a ``magnetar'' --- a pulsar with a very high magnetic
field --- associated with the Soft Gamma Repeater SGR 1900+14
indicates the existence of a class of pulsars with dipole
magnetic fields approaching $\sim10^{15}\,$G. Obviously,
for pulsars with such high magnetic fields, the energy budget
available for particle acceleration can be 2 to 3 orders of
magnitude larger than the canonical estimates based on pulsar
magnetic field of $\sim10^{12}\,$G, although it is not
clear if the energy loss processes --- a generic problem
with acceleration around compact objects --- can be
gotten around.

Recently, it has been suggested~\cite{oeb} that iron ions from the
surfaces of newly formed strongly magnetized pulsars may be accelerated
through relativistic
MHD winds. It is claimed that pulsars whose initial spin periods are
shorter than $\sim4(B_s/10^{13}\,{\rm G})^{1/2}\,$ms, where $B_s$ is the
surface magnetic field, can accelerate iron ions to greater than
$\sim10^{20}\,$eV. These ions can pass through the remnant of the
supernova explosion that produced the pulsar without suffering significant
spallation reactions. Clearly, in this scenario, the composition of
EHECR is predicted to be dominantly iron nuclei (the relativistic wind
may also accelerate some lighter nuclei though). These predictions will be
testable in the up-coming experiments. 

\subsubsection{Other Candidate Sources}
A variety of other candidate UHECR acceleration sites have
been studied in literature. Among these are acceleration in
Galactic wind termination shocks~\cite{jokipii-morfil}, in
shocks created by colliding galaxies~\cite{colliding-galaxy},
in large-scale shocks resulting from structure formation in
the Universe~\cite{norman}, such as 
shocks associated with acceretion flow onto galaxy clusters and cluster
mergers~\cite{krb,miniati-etal}, and so on. While for some
of these sites $E_{\rm max}$ can reach the UHE region (depending on the
magnetic field strength, which is highly uncertain), 
it is generally difficult to stretch $E_{\rm max}$
beyond $10^{20}\ev$. 

\subsection{A Possible Link Between Gamma-Ray Bursts and
Sources of $E>10^{20}\ev$ Events?}

Cosmological GRBs most likely contribute a negligible
fraction to the low energy CR flux around $100\,$GeV~\cite{sp},
as compared to SNRs, the favorite CR source below the knee.
In contrast, a possible common origin of UHECR and cosmological
GRBs was pointed out in Refs.~\cite{wax2,vietri1}, mainly based
on the observation that the average rate of energy emission required to
explain the observed UHECR flux is comparable to the average rate of
energy emitted by GRBs in $\gamma-$ rays. 
In addition, the predicted spectrum seems to be consistent
with the observed spectrum above $\simeq10^{19}\,$eV for proton
injection spectra $\propto E^{-2.3\pm0.5}$~\cite{wax1}, typical
for the Fermi acceleration mechanism which is supposed to operate in
dissipative wind models of GRBs. Because the rate of cosmological
GRBs within the cone observed by the experiments out to the
maximal range of EHECR beyond the GZK cutoff ($\simeq50\,$Mpc,
see Sect.~4.1) is only about one per century, the likelihood
of observing such UHECR from GRBs within the few years over which these
UHECR experiments have been active is very small, unless cosmic magnetic
fields lead to time delays of at least hundred years. The
cosmological GRB scenario for UHECR therefore necessarily
implies a lower limit on magnetic fields which in case of
a large scale field characterized by an r.m.s. strength $B$
and a coherence scale $l_c$ [see Sect.~4.4, Eq.~(\ref{delay})]
reads
\begin{equation}
  B\ga10^{-10}\left(\frac{E}{10^{20}\,{\rm eV}}\right)
  \left(\frac{d}{30\,{\rm Mpc}}\right)^{-1}
  \left(\frac{l_c}{1\,{\rm Mpc}}\right)^{-1/2}\,{\rm G}
  \,.\label{BGRBlim}
\end{equation}
This is an important observational test of this scenario.
In particular, the observation of $N$ arrival
directions that are different within the typical
deflection angle given by Eq.~(\ref{deflec}), strengthens the bound
in Eq.~(\ref{BGRBlim}) by a factor $N^{1/2}$. The recently
observed isotropy of arrival directions up to the
highest energies~\cite{agasa} may in that respect already
pose a problem to this scenario if one takes into
account observational limits on the large scale EGMF~\cite{dar-grb}.
In addition, the energetic requirements have also become
more severe due to recent GRB observations that indicate
a larger distance scale than assumed at the time when
the UHECR$-$GRB connection was proposed~\cite{dar-grb}.

In the dissipative wind model of GRBs a plasma of photons,
electrons, positrons,
with a small load of baryons, is accelerated to ultrarelativistic
Lorentz factors $\gamma\gg1$, and at some dissipation radius $r_d$,
a substantial part of the kinetic energy is converted to internal
energy by internal shocks~\cite{piran}. At this point the plasma
is optically thin and part of the internal energy is radiated in
the form of the $\gamma-$rays that give rise to the GRB. In addition,
the highly relativistic random motion in the wind
rest frame resulting from dissipation is expected to build up magnetic
fields close to equipartition with the plasma, which in turn
give rise to second order Fermi acceleration of charged particles.
In the following we briefly review the conditions derived in
Ref.~\cite{wax2} on the wind parameters that are required to
accelerate protons to energies beyond $100\,$EeV.

The most crucial constraint arising from acceleration itself
comes from the requirement that the acceleration time in the
wind rest frame, $t_a\simeq r_g$, should be smaller than the
comoving expansion time $t_d\simeq r_d/\gamma$. In terms of
the comoving magnetic field strength $B$, this condition
reads
\begin{equation}
  B\ga\frac{E}{er_g}\simeq3\times10^4
  \left(\frac{E}{10^{20}\,{\rm eV}}\right)
  \left(\frac{r_d}{10^{13}\,{\rm cm}}\right)^{-1}
  \,{\rm G}\,,\label{cond1}
\end{equation}
where $E$ is the proton energy in the observer frame, and we have used 
$r_g\simeq E/(\gamma eB)$.

The time scale for pion production losses on the $\gamma-$rays
in the plasma is given by $l_{E,\gamma}\simeq10/(n_\gamma\sigma_\pi)$,
where $\sigma_\pi\simeq10^{-28}\,{\rm cm}^2$ is the asymptotic
high-energy pion production cross section (see Sect.~4.1),
the factor 10 takes into account the inelasticity of $\simeq0.1$,
and the comoving $\gamma-$ray density $n_\gamma$ can be expressed
in terms of the $\gamma-$ray luminosity $L_\gamma$ and typical
energy $\epsilon_\gamma$ in the observer frame, $n_\gamma\simeq
L_\gamma/(4\pi r_d^2\gamma\epsilon_\gamma)$. The condition
$l_{E,\gamma}>t_a$ then leads to an additional lower limit on $B$,
\begin{equation}
  B\ga20\,\left(\frac{L_{\gamma}}{10^{51}\,{\rm erg}\,{\rm s}^{-1}}\right)
  \left(\frac{r_d}{10^{13}\,{\rm cm}}\right)^{-2}\,
  \left(\frac{\gamma}{300}\right)^{-2}\,{\rm G}\,.\label{cond2}
\end{equation}

Finally, the condition that the synchrotron loss length
$l_{E,{\rm syn}}\equiv E/(dE/dt)_{\rm syn}$ be larger than
$t_a$, where $(dE/dt)_{\rm syn}$ is given by Eq.~(\ref{synch}),
leads to an upper limit on $B$,
\begin{equation}
  B\la3\times10^5\,\left(\frac{\gamma}{300}\right)^2
  \left(\frac{E}{10^{20}\,{\rm eV}}\right)^{-2}\,.\label{cond3}
\end{equation}

The three conditions Eqs.~(\ref{cond1})$-$(\ref{cond3}) can be
satisfied simultaneously, provided
\begin{equation}
  r_d\ga10^{12}\left(\frac{\gamma}{300}\right)^{-2}
  \left(\frac{E}{10^{20}\,{\rm eV}}\right)^3\,{\rm cm}
  \,.\label{cond4}
\end{equation}
These values have to be set in relation to the time scale
$t_{\rm GRB}$ of a GRB, via $r_d\la\gamma^2
t_{\rm GRB}$. Therefore, eventually all conditions can be
satisfied, provided $\gamma\ga40(E/10^{20}\,{\rm eV})^{3/4}
(t_{\rm GRB}/{\rm s})^{-1/4}$,
which are reasonable values within most of the dissipative
wind models. By rewriting Eq.~(\ref{cond1}) in terms of the
equipartition field $B_{\rm ep}$ in the comoving frame,
$(B/B_{\rm ep})^2\ga0.15(\gamma/300)^2(E/10^{20}\,{\rm eV})^2
(L/10^{51}\,{\rm erg}\,{\rm s}^{-1})^{-1}$, where $L$ is
the total luminosity, it is obvious that for reasonable
wind luminosities and Lorentz factors, the magnetic field
is not far from equipartition.

The main proton energy losses in the GRB scenario summarized
above are synchrotron radiation and pion production. Both
processes give rise to secondaries, photons in the first,
and photons and neutrinos from pion decay in the second case,
and the resulting fluxes of these secondaries have
subsequently been estimated in the
literature~\cite{wb,vietri2,vietri3,bd,totani}.

Refs.~\cite{wb,rct}
computed the neutrino flux around $10^{14}\,$eV correlated
with GRBs in the dissipative wind model and showed that
several tens of events should be observed with a km scale
neutrino observatory, with a large fluctuation of the
number of detected events from burst to burst~\cite{hh}.
This neutrino flux has been proposed to be used as an
extremely long neutrino baseline to test neutrino
properties such as flavor mixing (see also discussion in Sect~4.3.1)
and the limits of validity of the relativity principles (see also
discussion in Sect.~4.8).
Ref.~\cite{vietri2} investigated the neutrino flux from the same
process at energies above $10^{19}\,$eV and found it to be detectable
by AIRWATCH-type experiments~\cite{linsley} such as the
MASS~\cite{mass}. A correlation of a fraction of all UHE
neutrinos with GRBs would thus be a strong test of the GRB scenario of
UHECR origin. An experimental upper limit of $0.87\times10^{-9}
\,{\rm cm}^{-2}$ upward going neutrino induced muons per average
GRB has been set by the MACRO detector~\cite{macro-grb}.
The question of the maximal possible neutrino
energies from GRBs and also blazars was reconsidered recently,
resulting in typical numbers of $\sim 10^{19}\,$eV:
Ref.~\cite{rm} gives a detailed account of loss processes
and Ref.~\cite{vietri4} focuses on neutrino production
associated with external shocks in GRB fireball models in
the afterglow phase.

In addition, there would be a background
of UHE neutrinos from the interaction of UHECR with the
CMB outside of the GRB which was found to be detectable
as well~\cite{lee,vietri2}. The distribution of this background
was argued to be an indicator of the distribution of the
source population of UHECR which could be used to
distinguish between the major theoretical scenarios.

Similarly to the case of AGN and
radio-galaxy sources, the neutrino flux level from GRBs is
limited by the diffuse GeV $\gamma-$ray background
(see Sect.~5.2.1).

The synchrotron emission associated with proton acceleration
to UHE in the cosmological GRB scenario has been found
to carry away a fraction of about a percent of the total
burst energy.
At energies around $10\,$MeV this signal should be detectable
with the proposed GLAST satellite
experiment, while above a few hundred GeV, ground-based air
Cherenkov telescopes should be sensitive enough to detect
this flux~\cite{vietri3,bd}. Ref.~\cite{totani} even claims
that synchrotron emission should give rise to afterglows
that extend into the TeV range. These
$\gamma-$ray fluxes above $\sim10\,$MeV would thus provide another
strong signature of proton acceleration up to UHE in GRBs
that should be testable in the near future. This signature
may already have been observed in the 10--20 TeV range~\cite{totani2}
(see also Ref.~\cite{vernetto}),
and the resulting cascade $\gamma-$ray flux in the GeV range has
been pointed out as a possible explanation of the diffuse $\gamma-$ray 
flux observed at these energies~\cite{totani3,vazquez}. This would
imply the phenomenal total liberated energy of $\sim10^{56}\,$erg
per GRB (assuming isotropic radiation) and the lower limit
$\gamma\ga500$ on the Lorentz factor, which is consistent
with the fireball model outlined above.

Recently it was claimed that an explanation of the observed
UHECR spectrum in the context of acceleration in GRBs requires
specific GRB progenitors such as binary pulsars~\cite{gallant-achterberg}.
Emission of high energy $\gamma-$rays and neutrinos in GRBs
associated with supernova explosions in massive binary
systems, whose existence was recently suggested by observations,
has been discused in Ref.~\cite{pb1}.

\section{Non-acceleration Origin of Cosmic Rays above $10^{20}\ev$}
\subsection{The Basic Idea}
As discussed in the Sect.~5, the shock acceleration mechanism
is a self-limiting process: For any given set of values of
dimension of the acceleration region (fixed by, say, the
radius $R$ of the shock) and the magnetic field strength ($B$), simple
criterion of Larmor containment of a particle of charge $Ze$ within the
acceleration region implies that there is a maximum energy
$E_{\rm max}\sim ZeBR$ up to which the particle can be accelerated before
it escapes from the acceleration region, thus preventing further
acceleration. The observed EHECR events above $10^{11}\gev$,
therefore, pose a serious challenge for any acceleration mechanism
because a value of $E_{\rm max}\ga10^{11}\gev$ can barely be achieved
in even the most powerful astrophysical
objects for reasonable values of $R$ and 
$B$ associated with these objects~\cite{hillas-araa,ssb,norman}. 
The problem becomes more acute when
one recognizes that the actual energy at the source has to be
significantly larger than the observed energy of the particles because
of energy loss during propagation as well as in the immediate
vicinity of the source. In addition, there is the 
problem of absence of any obviously identifiable sources for the
observed EHECR events, as discussed in Sect.~4.6. 

Because of these difficulties, there is currently much interest in the
possibility that the EHECR events may represent a 
fundamentally different component of cosmic rays in the sense that
these particles may not be produced
by any acceleration mechanisms at all; instead, these particles may simply
be the result of {\it decay} of certain massive particles
(generically, ``X'' particles) with mass
$m_X > 10^{11}\gev$ originating from high energy processes
in the early Universe. As we shall discuss below, such {\it
non-acceleration} or ``top-down'' decay 
mechanism (as opposed to conventional ``bottom-up''
acceleration mechanism) of production of extremely energetic particles
in the Universe today are possible and may indeed be naturally realized
within the context of unified theories of elementary particle interactions 
in the early Universe. 

The basic idea of a top-down origin of cosmic rays can be 
traced back to Georges Lema\^{\i}tre~\cite{lemaitre} and his theory of 
``Primeval Atom'', the precursor to the Big Bang model of the
expanding Universe. The entire material content of the Universe and 
its expansion, according to Lema\^{\i}tre, originated from the
``super-radioactive disintegration'' of a single Atom of extremely
large atomic weight, the Primeval Atom, which 
successively decayed to ``atoms'' of smaller and smaller atomic
weights. The cosmic rays were envisaged as the energetic particles 
produced in intermediate stages of decay of the Primeval Atom 
--- they were thus ``glimpses of the primeval fireworks''~\cite{lemaitre}. 
Indeed, Lema\^{\i}tre regarded cosmic rays as the main evidential 
relics of the Primeval Atom in the present Universe. 

Of course, we now know that the bulk of the observed cosmic rays are of recent 
(post-galactic) origin, and not cosmological. 
In particular, as far as the EHECR are concerned, we now know that 
the existence of CMBR, which was unknown in Lema\^{\i}tre's time,  
precludes the origin of the observed EHECR particles 
in very early cosmological epoch (or equivalently at very large
cosmological distances) because of the GZK effect discussed in section
4. Nevertheless, it is interesting that one of the earliest scenarios
considered for the origin of cosmic rays was a cosmological top-down,
non-acceleration scenario, rather than a bottom-up acceleration scenario. 

In the modern version of the top-down scenario of cosmic ray origin, 
the X particles (the possible sources of
which we shall discuss below) typically decay to quarks and
leptons. The quarks hadronize, i.e., produce jets of hadrons
containing mainly light mesons (pions) with a small percentage of
baryons (mainly nucleons). The pions decay to photons, neutrinos (and
antineutrinos) and electrons (and positrons). Thus, energetic photons, 
neutrinos and charged leptons, together with a small fraction of nucleons, 
are produced directly with energies up to $\sim m_X$ 
{\it without any acceleration mechanism}. 

In order for the decay products of the X particles to be observed as
EHECR particles today, three basic conditions must be satisfied: (a) The X
particles must decay in recent cosmological epoch, or equivalently at
non-cosmological distances ($\la 100$ Mpc) from Earth --- otherwise the
decay products of the X particles lose all energy by interacting with the
background radiation, and hence do not survive as EHECR
particles. A possible exception is the case where neutrinos of
sufficiently high energy originating from X particle decay at
large cosmological distances $\gg100$ Mpc give rise to EHE nucleons
and/or photons within 100 Mpc from Earth through decay of Z bosons 
resonantly produced through interaction of the original EHE neutrino with
the thermal relic background (anti)neutrinos, if neutrinos have a small
mass of order $\sim$eV; see Sect.~4.3.1., (b) the X
particles must be sufficiently massive with mass $m_X\gg10^{11}\gev$,
and (c) the number density and rate of decay of the X particles must
be large enough to produce detectable flux of EHECR particles. 

In the present section, we first discuss the nature of the expected
production spectra of
observable particles (nucleons, photons, neutrinos) resulting from
the decay of X particles in general. We then discuss in detail    
a particular realization of the top-down scenario in which
the X particles are the supermassive gauge bosons, Higgs bosons and/or 
superheavy fermions produced from cosmic topological
defects (TDs) like cosmic strings, magnetic monopoles, superconducting
cosmic strings etc., which
could be formed in symmetry-breaking phase transitions associated with
Grand Unified Theories (GUTs) in the early
Universe. We then discuss the possibility that the X particles
could be some long-lived metastable supermassive relic particles
of non-thermal origin produced, for example, through vacuum
fluctuations during a possible inflationary phase in the early
Universe. It has been suggested that such metastable supermassive
long-lived particles could constitute (a part of) the dark matter
in the Universe, and  a fraction of these particles decaying in the recent
epochs may give rise to the EHECR. 

\subsection{From X Particles to Observable Particles: Hadron spectra
in Quark $\to$ Hadron Fragmentation}
The precise decay modes of the X particles would depend on the specific
particle physics model under consideration. In the discussions below we
shall assume that X particles decay typically into quarks and leptons,
irrespective of the sources of the X particles\footnote{In some
supersymmetric models, the decay products of the X particles may also
include squarks and/or sleptons. However, for $m_X$ much above the typical
supersymmetry breaking scale of order TeV, the supersymmetric and
non-supersymmetric particles would behave essentially similarly.}. By far
the largest number of ``observable'' particles (nucleons,
photons, neutrinos) resulting from the decay of the X particles are
expected to come through the hadronic channel, i.e., through 
production of jets of hadrons by the quarks. The process of 
``fragmentation'' of the quarks into hadrons is described by QCD. The
spectra of various particles
at production are, therefore, essentially determined by QCD, and not
by any astrophysical processes. The actual decay mechanism of
the X particles into quarks and leptons, and the multiplicities and
the spectra of these quarks and leptons may depend upon the origin and
nature of the X particles themselves. Nevertheless, the
spectra of hadrons in the jets created by individual quarks should be
reasonably independent of the origin of the quarks themselves. We,
therefore, first discuss the expected spectra of hadrons (and the
resulting spectra of nucleons, photons and neutrino --- the latter
two from decay of pions) in individual jets created by individual quarks. 

The hadron spectra under discussion should be very similar to those
measured for $e^+e^-\to \bar{q}q\to{\rm hadrons}$ process in
colliders. 
The actual process by which a single high energy quark gives rise to
a jet of hadrons is not understood fully yet --- it involves the
well-known (and unsolved) ``confinement'' problem of QCD. However,
various semi-phenomenological approaches have 
been developed which describe the fragmentation spectra 
that are in good agreement with the currently available experimental data on
inclusive hadron spectra in quark/gluon jets in a variety of high
energy processes. 

In these approaches, the process of production of a jet containing a large
number of hadrons by a single high energy quark (or gluon)   
is ``factorized'' into three stages.  
The first stage involves ``hard'' processes involving large momentum 
transfers, whereby the initial high energy
quark emits ``bremsstrahlung'' gluons which in turn create more 
quarks and gluons through various QCD processes ($\bar{q}q$ pair
production by gluons, gluon bremsstrahlung by the
produced quarks, gluon emission by gluons, and so on). 
These hard processes are
well described by perturbative QCD. Thus a single high energy quark gives
rise to a ``parton cascade'' --- a shower of quarks and gluons --- which,
due to QCD coherence effects, is confined in a narrow cone
or jet whose axis lies along the direction of propagation of the
original quark. 

In the semi-phenomenological approaches to the jet fragmentation
process, the first stage of 
the process, i.e., the parton cascade development, described by
perturbative QCD, is terminated at a cut-off value, $\langle
k_\bot^2\rangle^{1/2}_{\rm cut-off}\sim 1\gev$, of the typical transverse
momentum. Thereafter, the second stage involving 
the non-perturbative confinement process is allowed to take over, binding
the quarks and gluons into ``primary'' color neutral objects. 
In the third stage, the unstable primary ``hadrons'' decay into known
hadrons. The second and the third stages are usually
described by the available phenomenological hadronization models 
such as the LUND string fragmentation
model~\cite{lund} or the cluster fragmentation model~\cite{cluster}. 
Detailed Monte Carlo numerical codes now
exist~\cite{cluster,jetset,ariadne} which incorporate the three-stage
process outlined above. These codes provide a reasonably good
description of a variety of relevant experimental 
data. 

\subsubsection{Local Parton-Hadron Duality}  
There is an alternative approach that is essentially analytical 
and has proved very fruitful in terms of its ability to
describe the gross features of hadronic jet systems, such
as the inclusive spectra of particles, the particle multiplicities and
their correlations, etc., reasonably well. This approach is based on the
concept of ``Local Parton Hadron Duality'' (LPHD)~\cite{lphd}. Basically,
in this approach, the second stage involving the non-perturbative
hadronization process mentioned above is ignored, and the primary 
hadron spectrum
is taken to be the same, up to an overall normalization constant, as
the spectrum of partons (i.e., quarks and gluons) in the parton
cascade after evolving the latter all the way
down to a cut-off transverse momentum $\langle
k_\bot^2\rangle^{1/2}_{\rm cut-off}\sim R^{-1}\sim$ few hundred MeV,
where $R$ is a typical hadronic size. A rigorous ``proof'' of LPHD at 
a fundamental theoretical level is not yet available. However, the
fact that LPHD gives a remarkably good description of the experimental data
including recent experimental results from LEP, HERA and TEVATRON
~\cite{lphdreview} gives strong indications of 
the general correctness of the LPHD hypothesis in some average sense. 

The fundamental basis of LPHD is that the actual hadronization process,
i.e., the conversion of the quarks and gluons in the parton cascade into
color neutral hadrons, occurs at a low virtuality scale of order of a
typical hadron mass independently of the energy of the cascade initiating 
primary quark, and involves only low momentum transfers and 
local color re-arrangement which do not drastically alter the
form of the momentum spectrum of the particles in the parton cascade
already determined by the ``hard'' (i.e., large momentum transfer)
perturbative QCD processes. Thus, the non-perturbative hadronization
effects are lumped together in an ``unimportant'' overall normalization
constant which can be determined phenomenologically.  

A good quantitative description of the perturbative QCD stage of the
parton cascade evolution is provided by 
the so-called Modified Leading Logarithmic
Approximation (MLLA)~\cite{mlla} of QCD, which allows the parton
energy spectrum (which is a solution of the so-called
DGLAP evolution equations) to be expressed analytically 
in terms of 
functions depending on two free parameters, namely, the
effective QCD scale $\Lambda_{\rm eff}$ (which fixes the effective
running QCD coupling strength $\alpha_s^{\rm eff}$) and
the transverse momentum cut-off $\widetilde{Q}_0$. 
For the case $\widetilde{Q}_0=\Lambda_{\rm eff}$, the analytical result
simplifies
considerably, and one gets what is referred to as the ``limiting
spectrum''~\cite{lphd,lphdreview} for the energy distribution of the
partons in the cascade, which has the following form: 
\beqarray
x\,\frac{dN_{\rm part}}{dx}&=&\frac{4C_F}{b}\,\Gamma(B)
\int_{-\pi/2}^{\pi/2}\frac{d\ell}{\pi}\,
e^{-B\alpha}\left[\frac{\cosh\alpha + \left(2\xi/Y-1\right)
\sinh\alpha}{(4N_c/b)Y(\alpha/\sinh\alpha)}\right]^{B/2}\nonumber \\ 
&&\times
I_B\left(\left[\frac{16N_c}{b}\,Y\,\frac{\alpha}{\sinh\alpha}
\left[\cosh\alpha + \left(2(\xi/Y)-1\right)\sinh\alpha\right]
\right]^{1/2}\right)\,.\label{mlla-spectrum}
\eeqarray
Here $dN_{\rm part}$ is the number of partons carrying a fraction between  
$x$ and $x+dx$ of the energy $E_{\rm jet}=E_q$ of the 
original jet-initiating quark $q,\, $ $\xi=\ln(1/x)$, $Y=\ln(E_{\rm
jet}/\Lambda_{\rm eff})$, $\alpha=\left[\tanh^{-1}\left(1-2\xi/
Y\right)+ i\ell\right]\, $, $I_B$ is the modified Bessel function of order
$B$,
where $B=a/b\, $ with $a=\left[11N_c/3 + 2n_f/(3N_c^2)\right]\, $ and
$b=\left[(11N_c-2n_f)/3\right]\,
$, $n_f$ being the number of flavors of quarks and $N_c=3$ the number of
colors, and $C_F=(N_c^2-1)/2N_c=4/3$. 

Eq.~(\ref{mlla-spectrum}) gives us the spectrum of the partons in
the jet. By LPHD hypothesis, the hadronic fragmentation function (FF),
i.e., the hadron spectrum, 
$dN_h/dx$ (with $x=E_h/E_{\rm jet}\,,\,\,$ $E_h$ being the energy of a
hadron in the jet), due to hadronization of a quark $q$, is given by
the same form as in Eq.~(\ref{mlla-spectrum}), except for an overall
normalization constant $K(Y)$  
that takes account of the effect of conversion of partons into
hadrons: 
\beq
x\,\frac{dN_h}{dx} = K(Y)x\frac{dN_{\rm part}}{dx}\,,\label{q-h}
\eeq
with now $x=E_h/E_{\rm jet}$ on {\it both} sides of the equation. 

Phenomenologically, for given values of
$\Lambda_{\rm eff}$ and $E_{\rm jet}$, the normalization constant
can be determined simply from overall energy conservation,
i.e., from the condition $\int_0^1 x[dN_h(Y,x)/dx]\, dx = 1\,$.
The value of 
$\Lambda_{\rm eff}$ is not known {\it a priori}, but a fit to the 
inclusive charged particle spectrum
in $e^+e^-$ collisions at center-of-mass energy $E_{\rm cm}=2 E_{\rm
jet}\sim 90\gev$ (Z-resonance) gives $\Lambda_{\rm eff}^{\rm ch}\sim$ 250
MeV, while the value of $K$ is found to be typically $\sim 1.3$ at LEP
energies. 

An important characteristic of the MLLA spectrum (\ref{mlla-spectrum})
treated as a function of the variable $\xi$ is 
the existence of a maximum at $\xi_{\rm max}$ given by 
\beq
\xi_{\rm max}=Y \left[\frac{1}{2}+\sqrt{\frac{C}{Y}} - \frac{C}{Y}
\right]\,,\label{ximax} 
\eeq
with $C=a^2/(16bN_c)$. The existence of this maximum is directly
related to the suppression of soft gluon multiplication in the cascade
due to QCD color coherence effect. Recent analysis~\cite{khoze-lupia-ochs} 
of relevant LEP data have provided experimental 
confirmation of the energy evolution of $\xi_{\rm max}$ as
predicted by Eq.~(\ref{ximax}). 

For asymptotically high energies of interest, i.e., for 
$E_{\rm jet}\gg \Lambda_{\rm eff}$, and near the maximum $\xi_{\rm
max}$, the
limiting spectrum can be approximated by a Gaussian in $\xi$: 
\beq 
x\,\frac{dN_h}{dx}\propto\frac{1}{\sigma\sqrt{2\pi}}
\exp\left[-\frac{(\xi-\xi_{\rm max})^2}{2\sigma^2}\right]\,,\label{gauss}
\eeq
where $2\sigma^2=\left[bY^3/(36 N_c)\right]^{1/2}$. The full MLLA
spectrum (\ref{mlla-spectrum}) can be approximated well by a ``distorted
gaussian''~\cite{lphdreview} in terms of calculable higher moments
of $\xi$. 

Note that, within the LPHD picture, there is no way of distinguishing
between various different species of hadrons. Phenomenologically, the
experimental data can be fitted by using different values of $\Lambda_{\rm
eff}$ for different species of particles depending on their masses. For
our consideration of
particles at EHECR energies, all particles under consideration will be
extremely relativistic, and since, in our case, $E_{\rm jet}\sim
m_X/{\rm few}\gg
\Lambda_{\rm eff}$, the hadron spectrum will be relatively insensitive to
the exact value of $\Lambda_{\rm eff}$. 

At the energies of our interest, all six quark flavors ($n_f=6$) should be
involved in the QCD cascade. 
The top quark usually decays before it fragments. On the
other hand, the $s$, $c$, and $b$ quarks fragment before decaying, but the
resulting heavy hadrons will eventually decay to the lighter hadrons
(pions and nucleons). Within the LPHD picture, 
at the asymptotically high energies of interest, all hadrons ---
mesons as well as baryons --- have roughly the same spectrum,
although the
dominant species of particles in terms of their overall number will be the
light mesons (mostly pions);  
the baryons (mostly nucleons) typically
constitute a fraction of $\sim$ (3 -- 10)\% 
as indicated by existing collider data. The
distortion of the MLLA + LPHD spectrum of primary hadrons due to
decays of baryonic and mesonic resonances 
can in principle be taken into account by using the
phenomenological jet fragmentation codes mentioned earlier. However, since
pions and nucleons dominate the hadron spectrum, those 
effects can be neglected as a first approximation. Also, to this
approximation, the difference between quark- and gluon jets can be
ignored. For more details on
various phenomenological aspects of the MLLA + LPHD hypothesis, see the
reviews~\cite{lphdreview}. 

Within the analytical QCD (+LPHD) framework, 
most of the recent calculations of the expected spectra of observable
particles in top-down scenarios have generally employed either the full
MLLA spectrum given by Eq.~(\ref{mlla-spectrum}) \cite{slsc,slby} or
the Gaussian approximation, Eq.~(\ref{gauss}) \cite{bkv,necklace}. 
More recently, spectra obtained by using Monte Carlo event generators as
incorporated in the HERWIG~\cite{cluster} and JETSET~\cite{jetset}
programs have been used~\cite{birkel-sarkar}. 
Earlier calculations~\cite{hill,hsw,cusp,pbcusp,br,pbkofu,bhs,bhatsig}
used
a quark $\to$ hadron fragmentation spectrum originally suggested by
Hill~\cite{hill}, which was based on leading-log QCD formula for
average charged hadron multiplicity. The Hill spectrum
for the total hadron spectrum can be written as~\cite{hill,hsw} 
\beq
\frac{dN_h}{dx}
\simeq\frac{3}{2}\times0.08\exp\left[2.6\sqrt{\ln(1/x)}\right](1-x)^2
\left(x\sqrt{\ln(1/x)}\right)^{-1}\,.\label{hill-spectrum}
\eeq
In this formula, the number of quark flavors involved in the QCD
cascade is taken to be $n_f=6$. The factor of 3/2 is to include 
the neutral particles (dominantly
neutral pions), whereas the original Hill formula was given for the
charged hadrons only.  
Sometimes, a simpler formula, based on a simple $E^{1/2}$ behavior
of the average
multiplicity is also used~\cite{hill,hsw}: 
\beq
\frac{dN_h}{dx}\simeq\frac{15}{16}\,x^{-3/2}
(1-x)^2\,.\label{hill-15-16}
\eeq
At EHECR energies, for which $x\sim E/m_X \ll 1$ (since $m_X$ can be as
large as $\sim10^{16}\gev$ and we are interested in EHECR with energies
$E\sim {\rm few}\times10^{11}\gev$), the spectrum
(\ref{hill-spectrum}) can be well approximated~\cite{pbkofu} by a
power law, $dN_h/dE \propto E^{-\alpha}$ with $\alpha\sim1.3$. The 
spectrum (\ref{hill-15-16}) is also approximated by a power-law with
$\alpha\sim 1.5$. The MLLA spectrum (\ref{mlla-spectrum}) is less well
approximated by a single power-law, but can be approximated by two
or more segmented power-laws. 

It should be emphasized here that, in using the MLLA+LPHD   
hadron spectrum in the calculation of particle spectra in the top-down
scenario, one should keep it in mind that 
there is a great deal of uncertainty involved in extrapolating the QCD
(MLLA + LPHD) spectra --- which have been tested so far only at relatively  
``low'' energies of $\sim$ 100 GeV --- to the extremely high energies  
of our interest, namely, $\ga10^{14}\gev$. For example, there could be
thresholds associated with new physics beyond the standard model which
may alter the spectra as well as content of the particles in the jets. 

One example of possible new physics is supersymmetry (SUSY). 
If SUSY ``turns on'' at an energy scale of say 
$M_{\rm SUSY}\sim 1$ TeV, then the QCD cascade development process
is expected to involve not only the usual quarks and gluons, but also
their supersymmetric partners (squarks, gluinos) with equal probability as 
long as $\tilde{Q}^2 > M^2_{\rm SUSY}$, where $\tilde{Q}^2$ is the 
``virtuality'' (i.e., the 4-momentum transfer squared) involved in
various sub-processes contributing to the cascade. The virtuality
of the cascade particles steadily decreases as the cascade process
progresses. Once
$\tilde{Q}^2$ falls below $M^2_{\rm SUSY}$, the SUSY particles in the
cascade would decouple from the cascade process and eventually
decay into the stable lightest supersymmetric particles (LSPs). Thus, as
pointed out by Berezinsky and Kachelrie\ss\ ~\cite{berez-kachel1}, some
fraction of the particles in the QCD cascade resulting from the decay of X
particles may be in the form of high or even ultrahigh energy
LSPs. Indeed, Berezinsky and Kachelrie\ss\ ~\cite{berez-kachel1} have
claimed that LSPs may take away a significant fraction ($\sim
40\%$) of the total energy of the jet, which must be properly taken into
account in normalizing the spectra and calculating the yield of the
``observable'' hadrons (pions and nucleons). 
Inclusion of SUSY in the parton cascading process also changes the shape
of the fragmentation spectrum~\cite{berez-kachel2}. The limiting QCD MLLA
spectrum is still given by Eq.~(\ref{mlla-spectrum}) with constant $a$
replaced by $a_{\rm SUSY}=11 N_c/3$ and $b$ replaced by 
$b_{\rm SUSY}=9-n_f$, and the maximum
of the spectrum is given by Eq.~(\ref{ximax}) with $C$ replaced by 
$C_{\rm SUSY}=a_{\rm SUSY}^2/(16b_{\rm SUSY}N_c)$. Thus the maximum of the
SUSY-QCD spectrum is shifted to higher $\xi$ (lower energy) 
relative to the non-SUSY MLLA QCD spectrum. 

The properties of LSPs are model dependent, but due to a variety of
phenomenological reasons~\cite{berez-kachel1,mohap-nussi} the
UHE LSPs themselves are unlikely to be the candidates for the observed
EHECR events except possibly for the case of a neutral bound state of
light gluino and uds hadron~\cite{farrar,cfk}. 

Supersymmetry is by no means the only possible kind of new physics beyond
the standard model. Nevertheless, from the discussions above, we see 
that EHECR may in fact provide an interesting probing ground
for search for new physics beyond the standard model. 
For some recent discussions of the effects of the SUSY
versus non-SUSY QCD spectra on the final evolved particle spectra, see 
Ref.~\cite{slby}. 

\begin{figure}[ht]
\centering\leavevmode
\epsfxsize=6.0in
\epsfbox{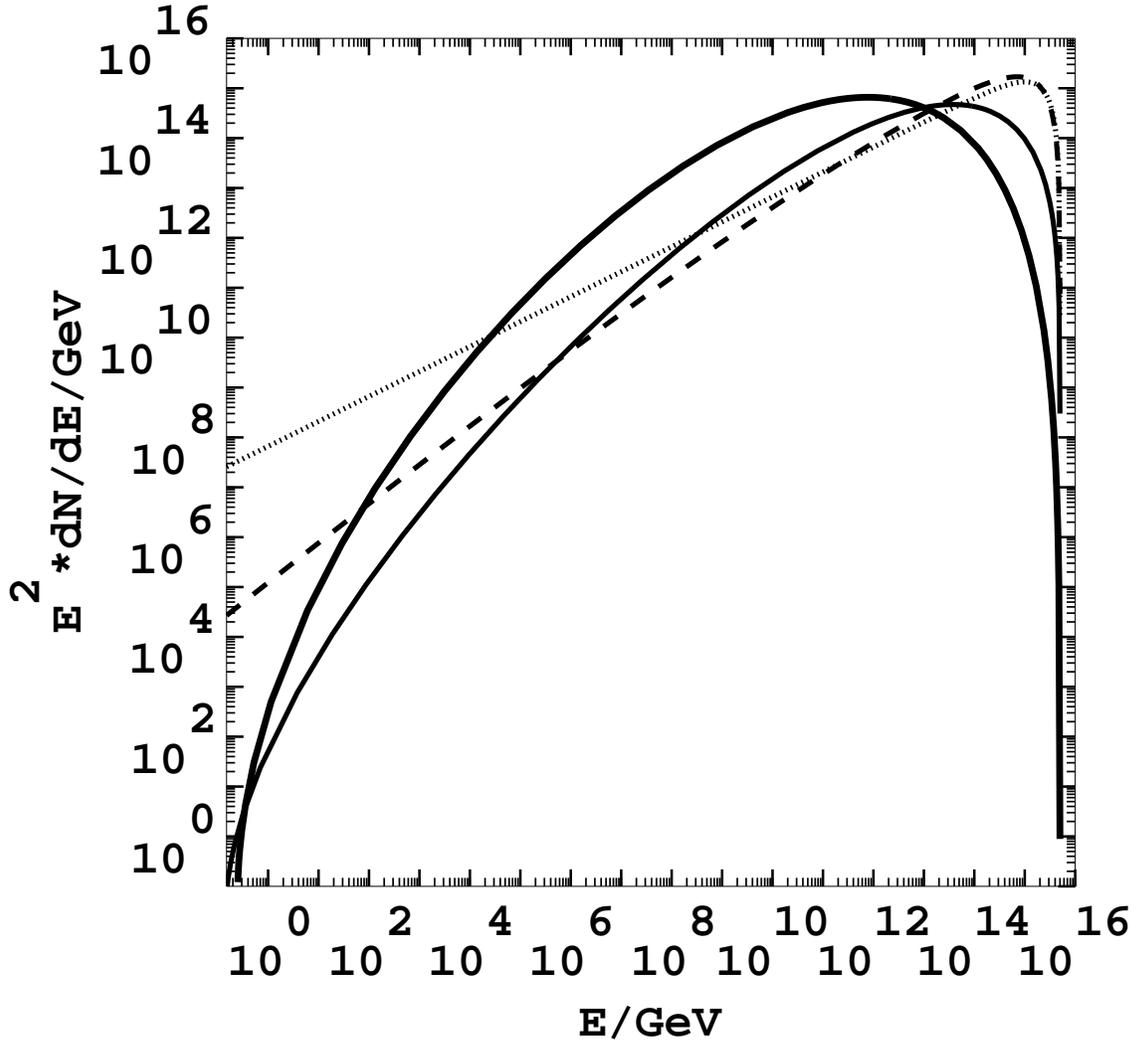}
\bigskip
\caption[...]{The fragmentation function for
$E_{\rm jet}=5\times10^{15}\,$GeV in MLLA approximation
with SUSY (thick solid line peaking at $10^{12}\,$GeV) and
without SUSY (thin solid line)
[see Eq.~(\ref{mlla-spectrum})] in comparison to the older
expressions Eq.~(\ref{hill-spectrum}) (dashed line) and
Eq.~(\ref{hill-15-16}) (dotted line).}
\label{F6.1}
\end{figure}

The various hadronic fragmentation spectra discussed above
are displayed in Fig.~\ref{F6.1} for comparison. 

More recently, attempts have been made~\cite{birkel-sarkar} to
calculate the injection spectra of nucleons, photons and neutrinos
resulting from X particle decay by directly using
numerical Monte Carlo event generators as incorporated in the
HERWIG~\cite{cluster} and JETSET~\cite{jetset} programs. 
Results of Ref.~\cite{birkel-sarkar}
indicate that although for $m_X\sim10^3\gev$ the Monte Carlo results agree
with the MLLA+LPHD predictions (which is not surprising since both
MLLA+LPHD as well as Monte Carlo event generators are suitably
parametrized to fit the existing collider data)\footnote{The Monte Carlo
calculations of Ref.~\cite{birkel-sarkar} are done with standard,
non-SUSY QCD.}, significant
differences with the spectra predicted from MLLA+LPHD appear for
$m_X\gg10^3\gev$. In particular, the spectra of photons and neutrinos seem
to differ significantly from the nucleon spectrum at high $x$ values
(whereas in the MLLA+LPHD picture they are assumed to be roughly
similar; see below). More importantly, nucleons
seem to be almost as abundant as photons and neutrinos in certain ranges
of $x$ values (specifically, in the range 
$0.2\la x \la 0.4$), contrary to the general expectation that baryon
production should be suppressed relative to meson (and hence to 
photon and neutrino) production at all x, independently of
$m_X$\footnote{In the string fragmentation scheme of hadronization
as implemented in the LUND Monte Carlo program JETSET~\cite{jetset}, for 
example, the yield of mesons (and hence the photons and neutrinos
resulting from their decay) is always expected to dominate over baryons. 
This is because, meson formation involves breaking of a color flux tube
through nucleation of a quark-antiquark pair whereas baryon formation
involves formation of a diquark-antidiquark pair, the probability for
which is considerably suppressed compared to that for quark-antiquark pair
formation. In the HERWIG program too, in which the hadronization scheme
involves initial formation of clusters of partons which subsequently break
up into color-neutral 3-quark states (baryons) and quark-antiquark states
(mesons), the predicted baryon/meson ratio is generally always 
considerably less than unity, at least so at currently accessible
accelerator energies (see, e.g., Ref.~\cite{deAngelis} for a review of
baryon production in $e^+e^-$ annihilations).}. 
Unfortunately, due to the very nature of these Monte Carlo calculations,
it is difficult to understand the precise physical reason for the
unexpectedly high relative baryon yield for certain values of $x$ in the
case of large $m_X$. Clearly, more work needs to be done on this 
important issue. 

An important point to note about the particle spectra displayed in
Fig.~\ref{F6.1} is that all these spectra are generally harder
than the production spectra predicted in conventional shock acceleration
theories, which, as discussed in the previous section, by and large
predict power-law differential production spectra with $\alpha\geq 2$.
This fact has important consequences; it leads to the prediction 
of a pronounced ``recovery''~\cite{bhs} of the evolved nucleon spectrum
after the GZK ``cut-off'' and the consequent flattening of the spectrum
above $\sim10^{11}\gev$ in the top-down scenario . Under
certain circumstances, a relatively hard production spectrum may also
naturally give rise to a ``gap'' in the measured EHECR 
spectrum~\cite{slsb}.     

The importance of a relatively hard production spectrum of 
EHECR from possible ``fundamental'' processes was first emphasized
by Schramm and Hill~\cite{schramm-hill-icrc}. For a power-law
differential spectrum $\propto E^{-\alpha}$, the index $\alpha=2$ is a
natural dividing line between what can be characterized as ``soft''
and ``hard'' spectra: For soft spectra ($\alpha > 2$), the total
particle multiplicity ($\propto E^{1-\alpha}$) as well as the total energy
($\propto E^{2-\alpha}$) are both dominated by the lower limits of the
relevant integrals, which means that most of the energy is carried by the large
number of low energy particles. Such a spectrum is inefficient
in producing a significant flux of extremely high energy particles.  
For hard spectra ($1 < \alpha <2$), on the other hand, 
although the total particle multiplicity is still dominated by very low
energy particles, the energy is mainly carried off by a few extremely
energetic particles. Thus, hard spectra such as the ones generically
predicted within the top-down scenario involving QCD cascade 
mechanism discussed above, seem to be more ``natural'' from the point
of view of producing EHECR than soft spectra generally predicted in shock
acceleration scenarios. 

\subsubsection{Nucleon, Photon and Neutrino Injection Spectra}
With a given hadronic fragmentation function, $dN_h/dx$, 
we can obtain the nucleon, photon and neutrino injection
spectra due to decay of all X particles at any time $t_i$ as described
below. We shall assume that nucleons and pions are produced
with the same spectrum; however, see Ref.~\cite{birkel-sarkar} and
the discussions above. 

Let $\dot{n}_X(t)$ denote the rate of decay of X particles per unit
volume at any time
$t$. Let us assume that each X particle, on average, undergoes
$\tilde{N}$-body decay to $N_q$ quarks (including antiquarks) and $N_\ell$
leptons (neutrinos and/or charged leptons), so that $\tilde{N}=N_q +
N_\ell$, and that the available energy $m_X$ is shared roughly equally by
the $\tilde{N}$ primary decay products of the $X$. Then, the 
nucleon injection spectrum, $\Phi_N(E_i,t_i)$, from the decay of all
X particles at any time $t_i$ can be written as 
\beq
\Phi_N(E_i,t_i)=\dot{n}_X(t_i)N_q f_N\frac{\tilde{N}}{m_X}
\frac{dN_h}{dx}\,,\label{nucl-inj}
\eeq
where $E_i$ denotes the energy at injection, 
$f_N$ is the nucleon fraction in
the hadronic jet produced by a single quark, and $x=\tilde{N} E_i/m_X$.  

The photon injection spectrum from the decay of the neutral pions ($\pi^0
\to 2\gamma$) in the jets is given by (see, for example, 
Ref.~\cite{stecker-book}) 
\beq
\Phi_\gamma(E_i,t_i)\simeq 2\int_{E_i}^{m_X/\tilde{N}}\frac{dE}{E}
\,\Phi_{\pi^0}(E,t_i)\,,\label{photon_inj}
\eeq
where $\Phi_{\pi^0}(E,t_i)\simeq\frac{1}{3}\frac{1-f_N}{f_N}
\Phi_N(E,t_i)$ is the neutral pion spectrum in the jet. 

Similarly, the neutrino ($\nu_\mu+\bar{\nu}_\mu$) injection spectrum
resulting from the charged pion decay 
[$\pi^\pm \to \mu^\pm\nu_\mu(\bar{\nu}_\mu)$] 
can be written, for $E_i\gg m_\pi$, as ~\cite{stecker-neut,bhs} 
\beq
\Phi_{\left(\nu_\mu+\bar{\nu}_\mu\right)}^{\pi\to\mu\nu_\mu}(E_i,t_i)\simeq
2.34\int_{2.34E_i}^{m_X/\tilde{N}}\frac{dE}{E}
\,\Phi_{(\pi^++\pi^-)}(E,t_i)\,,\label{nu_inj}
\eeq
where $\Phi_{(\pi^++\pi^-)}\simeq2\Phi_{\pi^0}$. 

The decay of each muon (from the decay of a charged pion) produces two
more neutrinos and an electron (or positron): $\mu^\pm\to e^\pm
\nu_e(\bar{\nu}_e)\, \bar{\nu}_\mu (\nu_\mu)$. Thus each charged pion
eventually gives rise to three neutrinos: one $\nu_\mu$, one
$\bar{\nu}_\mu$ and one $\nu_e$ (or $\bar{\nu}_e$), all of roughly the
same energy. So the total $\nu_\mu + \bar{\nu}_\mu$ injection spectrum 
will be roughly twice the spectrum given in Eq.~(\ref{nu_inj}), while the
total $\nu_e + \bar{\nu}_e$ spectrum will be roughly same as that in
Eq.~(\ref{nu_inj}). The full $\nu_\mu$ and $\nu_e$ spectra resulting from
decay of pions and the subsequent decay of muons can be calculated in
details following the procedure described in the book by
Gaisser~\cite{gaisser}.   

Note that, if the hadron spectrum in the jet is generally approximated by
a power-law in energy, then nucleon, photon and neutrino injection spectra 
will also have the same power-law form all with the same power-law
index~\cite{stecker-book,stecker-neut}. 

As mentioned earlier, it is generally expected that mesons (pions) should
be the most numerous particles in the
hadronic jets created by quarks coming from the decay of X
particles. Thus, $\Phi_{\pi^0}/\Phi_N\simeq\frac{1}{3} 
\frac{1-f_N}{f_N}\simeq 10$ and  
$\Phi_{(\pi^++\pi^-)}/\Phi_N\sim 20$ for $f_N\sim3\%$
This means that, in terms of number of particles at production, the 
decay products of the pions, i.e., photons and neutrinos, dominate
over nucleons at least by factors of order 10. Since neutrinos suffer little
attenuation and can come to us unattenuated from large cosmological
distances (except for absorption due to fermion pair production 
through interaction with the cosmic thermal neutrino background, the path 
length for which is $\gg 100\mpc$; see section 4.3), 
their fluxes are expected to be the largest among
all particles at the highest energies. However, their detection
probability is much lower compared
to those for protons and photons\footnote{The EHE neutrinos of TD origin 
would, however, be potentially detectable by the proposed 
space-based detectors like OWL and AIRWATCH, and ground based
detectors like Auger, Telescope Array, and so on; see discussions in
section 7.4.}. Photons also far outnumber nucleons at
production. However, the propagation of extragalactic 
EHE photons is influenced by a number
of uncertain factors such as the level of the URB and the strength
of the EGMF (see section 4.2). Depending on the level of the
URB and EGMF and the distance of the X particle source, the photon flux
may dominate over the nucleon
flux and thus dominate the ``observable'' diffuse particle flux, at EHECR
energies. Indeed, the prediction~\cite{abs} of a possible large photon/nucleon
ratio ($> 1$) at sufficiently high EHECR energies is a distinguishing
feature of the top-down scenario 
of origin of EHECR, and can be used as a signature for testing the
scenario in forthcoming experiments. This has been discussed recently in
more details in Ref.~\cite{bbv,slby}. 

In this context, note that photons and neutrinos in the top-down 
scenario are {\it primary} particles in the sense that they are 
produced directly
from the decay of the pions in the hadronic jets. In contrast, photons and
neutrinos in conventional acceleration scenarios can be produced only
through {\it secondary} processes --- they are mainly produced by the
decay of photo-produced pions resulting from the GZK interactions of
primary EHECR nucleons with CMBR
photons. Of course, these
secondary neutrinos and photons would also be there in the top-down 
scenario, but their fluxes are sub-dominant to the primary
ones. 

Once the injection spectra for nucleons, photons, and neutrinos are
specified, the evolution of these spectra and the final predicted fluxes
of various particles can be obtained by considering various
propagation effects discussed in detail in Sect.~4. The predicted
particle fluxes in the top-down scenario are discussed later in
Sect.~7, where we also discuss various signatures of the scenario 
in general and constraints from various observations. 

Obviously, while the shapes of the final particle spectra are determined
by the injection spectra (which are fixed by QCD as explained above) and
various propagation effects, the absolute magnitudes of the fluxes will be
fixed by the source function $\dot{n}_X$, the production and/or decay rate
of the X particles, in different realizations of the
non-acceleration scenario\footnote{In the topological defect
models discussed below, the X particles are generally assumed to
have extremely short life-times, so they decay essentially
instantaneously as soon as they are released from the
defects. Therefore, in the topological defect models, $\dot{n}_X$,
for all practical purposes, refers to the production rate of X
particles from the defects. In contrast, in models in which X
particles are metastable, long-lived particles (with lifetime $\ga$
age of the Universe) possibly produced during an inflationary epoch in the
early Universe, $\dot{n}_X$ refers to the {\it decay} rate of the X
particles.}. Unfortunately, $\dot{n}_X$ is highly model dependent and
depends on free parameters of the particular top-down model under
consideration. Because of this reason, it has not been possible to predict
the absolute flux levels in the top-down models with certainty;
only certain plausible models have been identified. 
Below, we shall discuss the expected values of $\dot{n}_X$ in
some specific top-down models and examine their efficacy with regard to
EHECR. But, first, in order to have some idea of the kind of numbers
involved, we perform a simple (albeit crude) benchmark
calculation of $\dot{n}_X$ required to obtain a significant contribution
to the measured EHECR flux. 

\subsubsection{ X Particle Production/Decay Rate Required to Explain the
Observed EHECR Flux: A Benchmark Calculation}
Since in top-down models photons are expected to dominate the
``observable''  EHECR
flux , let us assume for simplicity that the highest energy events are due
to photons. To be specific, let us assume a typical 2-body decay
mode of the X into a quark and a lepton: $X \to q\ell$. The quark will
produce a hadronic jet. The photons from the decay of neutral pions
in the jet carry a total energy
$E_{\gamma,{\rm Total}}\simeq
\left(\frac{1}{3}\times0.9\times\frac{1}{2}\right)m_X (f_\pi/0.9) 
= 0.15 m_X (f_\pi/0.9)$, 
where $f_\pi$ is the fraction of the total energy of the jet carried by
pions\footnote{We assume that the jet consists of only pions and
nucleons, all with the same spectrum, and that all particles are
ultrarelativistic. Thus, $f_\pi$ is also the total pion fraction in terms
of number of particles in the jet.}, and we have assumed that the quark and
the lepton share the energy $m_X$ equally. Assuming a power-law photon
injection spectrum with index $\alpha$, $dN_\gamma/dE_\gamma\propto
E_\gamma^{-\alpha}$, with $0 < \alpha < 2$, and normalizing with the total
photon energy $E_{\gamma,{\rm Total}}$, we get the photon injection
spectrum from the decay of a single X particle as 
\beq
\frac{dN_\gamma}{dE_\gamma} = \frac{0.6}{m_X} (2-\alpha)
\left(\frac{f_\pi}{0.9}\right)\, \left(\frac{2
E_\gamma}{m_X}\right)^{-\alpha}\,.\label{photon_bench}
\eeq
We can neglect cosmological evolution effects and take the
present epoch values of the relevant quantities involved, since photons of
EHECR energies have a cosmologically negligible path length of only 
few tens of Mpc for absorption through pair production on the
universal radio background. 

With these assumptions, and assuming that the (sources of the) X
particles are distributed uniformly in the Universe, the photon flux
$j_\gamma (E_\gamma)$ at the observed energy $E_\gamma$ is simply given by 
\beq
j_\gamma (E_\gamma)\simeq\frac{1}{4\pi}l(E_\gamma)\,
\dot{n}_X\, 
\frac{dN_\gamma}{dE_\gamma}\,,\label{flux_bench}
\eeq
where again $l(E_\gamma)$ is the pair production absorption path length of
a photon of energy $E_\gamma$. 

Normalizing the above flux to the measured EHECR flux, we get  
\beqarray
\left(\dot{n}_{X,0}\right)_{\rm EHECR}&\simeq&1.2\times10^{-46}
\left(\frac{l(E_\gamma)}{10\mpc}\right)^{-1}
\left(\frac{E^2j(E)}{1\ev \cm^{-2} \sec^{-1} 
\sr^{-1}}\right)
\left(\frac{2 E}{10^{16}\gev}\right)^{\alpha-1.5}\nonumber\\
\nopagebreak[3] 
&& \,\,\times\left(\frac{m_X}{10^{16}\gev}\right)^{1-\alpha}
\left(\frac{0.5}{2-\alpha}\right)\left(\frac{0.9}{f_\pi}\right) 
\cm^{-3} \sec^{-1} \,.\label{Xrate_required}
\eeqarray
The subscript 0 stands for the present
epoch. In terms of energy injection ($Q$), defined by 
$\dot{n}_X= Q/m_X$, the above requirement reads 
\beqarray
\left(Q_0\right)_{\rm EHECR}\simeq&&1.2\times10^{-21}
\left(\frac{l(E_\gamma)}{10\mpc}\right)^{-1}
\left(\frac{E^2j(E)}{1\ev \cm^{-2} \sec^{-1}
\sr^{-1}}\right)
\left(\frac{2 E}{10^{16}\gev}\right)^{\alpha-1.5}\nonumber \\ 
&& \,\,\times\left(\frac{m_X}{10^{16}\gev}\right)^{2-\alpha}
\left(\frac{0.5}{2-\alpha}\right)\left(\frac{0.9}{f_\pi}\right)
\ev \cm^{-3} \sec^{-1}\,.\label{Q_0_required}
\eeqarray
For a fiducial value of the observed EHECR flux given by $E^2j(E)\simeq
1\ev \cm^{-2} \sec^{-1} \sr^{-1}$ at the fiducial energy $E=10^{11}\gev$,
and for $m_X=10^{16}\gev$, $\alpha=1.5$,
and $f_\pi=0.9$, the above estimates indicate that 
in order for a generic top-down mechanism 
to explain the measured EHECR flux, the X particles must (be produced and
/or) decay in the present epoch at a rate of $\sim
1\times 10^{35}\mpc^{-3}\yr^{-1}$, or in more ``down-to-earth'' units,
about $\sim 13 {\rm AU}^{-3}\yr^{-1}$, i.e., about 10 X particles within
every solar system-size volume per year over a volume of radius 10 Mpc. 

The above numbers are uncertain (most likely
overestimate~\cite{slby}) by
perhaps as much as an order of magnitude or so, depending on the decay
mode of the X particle, the fraction of energy $m_X$ that goes into
nucleons vs. pions, the form of the hadronic fragmentation function that
determines the injection spectra of various particles, the absorption path
length of EHECR photons, electromagnetic cascading
effect (which is discussed in Sect. 4 and which we have neglected
here), and so on. Nevertheless, we think the numbers derived above do 
serve as crude benchmark numbers.  

The above rough estimate assumes that the
(sources of the) X particles are distributed uniformly in the Universe,
so the flux above refers to the diffuse flux. In principle, EHECR
events could also be produced by isolated nearby bursting sources of X
particles, in which case, depending on the distance of the source, the
above energy and number density requirements would be different. 
In addition, in the case of long-lived X particles of primordial
origin, as also in the case of certain kinds of topological defects such
as monopolonium (see below), the (sources of) X particles could be
clustered in our Galactic halo~\cite{bkv,bbv,birkel-sarkar}
as well as in clusters of galaxies, in which cases the required values of
$\dot{n}_{X,0}$ and $Q_0$ will depend on the clustering factor and
clustering length-scale. 

With the above rough estimates in mind, we now proceed to discuss possible
sources of the X particles. 

\subsection{Cosmic Topological Defects as Sources of X Particles:
General Considerations}
Cosmic Topological Defects (TDs)\cite{kibble,vil-shell,tdrev} --- magnetic
monopoles, cosmic strings, domain walls, superconducting cosmic strings,
etc., as well as various hybrid systems consisting of these 
TDs --- are predicted to form in the early Universe as a result of
symmetry-breaking phase transitions envisaged in GUTs. Topological
defects associated with symmetry-breaking phase transitions are
well-known in condensed matter systems; examples there 
include vortex lines in superfluid helium, magnetic flux tubes in
type-II superconductors, disclination lines and ``hedgehogs'' in nematic
liquid crystals, and so on. Recent laboratory
experiments (see \cite{tdrev} for references)
on vortex-filament formation in the superfluid
transition of $^3{\rm He}$ (which occurs at a temperature of a few
millikelvin) have provided striking confirmation of the basic
Kibble-Zurek~\cite{kibble,vil-shell,tdrev} picture of TD formation in
general that was initially developed within the context of TD formation in
the early Universe. 

It is sometimes thought that the existence of TDs in the Universe today is
inconsistent with the idea of an inflationary early Universe --- after
all, one of the motivations behind the development of the inflationary
paradigm (for reviews, see, e.g., Refs.~\cite{linde-book,kolb-turner})
was to get rid of the unwanted TDs like
superheavy magnetic monopoles and domain walls by diluting their abundance
through the exponential expansion of the Universe characteristic of an
inflationary phase. However, it has been recently realized that TDs could
be produced in non-thermal phase transitions occurring during the
preheating stage {\it after} inflation~\cite{kklt,kuz-tkachev-rev}, and
models can be constructed in which interesting abundances of ``harmless''
TDs so formed can exist in the Universe today. This mechanism of TD
production involves explosive particle production due to stimulated decay
of the inflaton oscillations through parametric resonance effect, which
leads to large field variances for certain fields coupled to the inflaton
field. These large field variances in turn lead to symmetry restoration
for those fields even if the actual reheat temperature after inflation is
small. Subsequent reduction of the field variances due to the continuing
expansion of the Universe would again cause a symmetry breaking phase
transition at which TDs could be formed. Thus, TDs can exist even if there
was an early inflationary phase of the Universe.  

In general, a TD has a ``core'' of size $\sim\eta^{-1}$ , $\eta$ being the
vacuum expectation value of the Higgs field in the broken symmetry phase. 
The Higgs field is zero and the symmetry is unbroken at the center of this
core --- the center being a line for a cosmic string, a point for a
monopole and a 2-dimensional surface for a domain wall --- while far
outside the core the symmetry is broken and the gauge and the Higgs fields
are in their true ground states. It is in this sense that the object is
referred to as a ``defect'' --- it is a region of unbroken symmetry
(``false vacuum'') surrounded by broken symmetry regions (``true
vacuum''). 

The energy densities associated with the
gauge and the Higgs fields are higher within the defect core than
outside. TDs are topologically stable due to
non-trivial ``winding'' of the Higgs field around the
defect cores. This topological stability ensures the ``trapping''
of the excess energy density associated with the gauge and the Higgs
fields inside the defect cores, which is what makes TDs massive
objects.    

The mass scale of a defect is fixed by the energy (or temperature) scale of
the symmetry breaking phase transition at which the defect is formed.
Thus, if we denote by $T_c$ the critical temperature of the defect-forming
phase transition in the early Universe, then the mass of a monopole formed
at that phase transition is roughly of order $T_c$, 
the mass per unit length of a cosmic string is of order $T_c^2$, and the
mass per unit area of a domain wall is of order $T_c^3$. 
For generic symmetry breaking potentials of the Higgs field,
$T_c\sim\eta$.

The TDs can be thought of as ``constituted''
of the ``trapped'' quanta of massive gauge- and Higgs fields of the
underlying spontaneously broken gauge 
theory\footnote{We shall restrict ourselves to
considerations of ``local'' TDs, that is, the TDs arising from
breaking of local symmetries only. ``Global'' topological defects 
are possible in theories with spontaneously broken global 
symmetries. There are no massive gauge bosons for global defects;  
a large portion of the energy density of a global defect resides in
the form of massless Goldstone bosons, and massive scalar particles play
only a subdominant role.}. Under certain circumstances, 
quanta of some massive fermion fields could also
be trapped inside the defects due to their coupling with the
defect-forming gauge and Higgs fields. We shall generically denote the
massive particles trapped inside TDs as ``X'' particles which can be
``supermassive'' with mass $m_X$ that can be as large as
$\sim10^{16}\gev$ if the TDs under consideration are associated with
breaking of a GUT symmetry. 

Due to their topological stability, 
once formed in the early Universe, the TDs can survive forever with 
X particles trapped inside them. However, from time to
time, some TDs, through collapse, annihilation or other processes, can
release the trapped X particles~
\cite{bkt,nussinov,hill,hsw,cusp,pbcusp, 
br,pbkofu,gk,bhatsig,necklace,vincent,susytd}. 
Decays of these X particles can give rise to extremely energetic nucleons,
neutrinos and photons with energies up to $\sim m_X$~\cite{bhs,abs}. 
Depending on the parameters involved, some of these processes can give
a significant contribution to, and possibly explain, the measured EHECR
flux above $\sim 10^{20}\ev$~\cite{ssb,slsb,ps1,slsc,bbv,slby}.  

There is a large body of literature on the subjects of nature and
classification of topological defects of various kinds and their 
formation and evolution in the early Universe. We will not
attempt to review these topics here; instead we refer the reader to 
Refs.~\cite{vil-shell,tdrev} for a comprehensive review and list of
references. 

Historically, much of the early considerations of cosmic topological
defects had to do with their gravitational effects, namely, the possible role
of cosmic strings in providing the seeds for formation of galaxies and
large scale structure in the Universe, the possibility of magnetic
monopoles being a candidate for the dark matter in the Universe, the
recognition of the disastrous role of massive domain walls in
cosmology, and so on. The particle aspect of TDs --- that TDs harbor
massive quanta of gauge, Higgs, and possibly other fields inside them,
and that these massive particles could, under certain circumstances, be
released from TDs with possibly important consequences --- did not
receive much attention early on. In 1982, two independent 
works~\cite{bkt,nussinov} pointed out that collapsing closed loops of
cosmic strings~\cite{bkt}, and monopole-antimonopole
annihilations~\cite{nussinov} could be sources of massive X particles
of GUT scale mass $\sim10^{16}\gev$ 
in the Universe, and that baryon number violating decays of these X
particles could be responsible for the observed baryon asymmetry of the 
Universe; however, the possible connection of these massive X
particles with EHECR was not explored then\footnote{Recently, however, 
it has been pointed out~\cite{pbbau} that {\it both} EHECR and   
the baryon asymmetry of the Universe may arise from the 
decay of X particles released from TDs. For a brief discussion 
of this possibility, see Sect.~6.10.}. 
In 1983 Hill~\cite{hill} pointed out that decay of
supermassive X particles released from monopole-antimonopole annihilation
through formation and eventual collapse of metastable 
monopolonium (monopole-antimonopole bound
states) could give rise to very energetic particles, and soon it was
also pointed out~\cite{schramm-hill-icrc} that these energetic
particles could be observable as extremely high energy cosmic
rays. After the discovery of superconducting cosmic string solutions
by Witten~\cite{witten-scs} in 1985, Ostriker, Thompson and
Witten~\cite{otw} pointed out that massive charge carriers whose mass
can be as large as the GUT scale, $\sim10^{16}\gev$, could be
spontaneously emitted from superconducting cosmic string loops which 
attained a certain maximum current (beyond which the string
would lose its superconductivity). 
Following this, Hill, Schramm and Walker~\cite{hsw}
calculated the expected ultrahigh energy nucleon and neutrino 
spectra resulting
from the decay of the massive charge carriers (X particles) released
from superconducting cosmic string loops. Soon thereafter, ultrahigh
energy cosmic ray spectrum resulting from decay of X particles
released from ordinary (i.e., non current-carrying) 
cosmic strings due to cusp evaporation
process~\cite{cusp,pbcusp}, as well as due to collapse or multiple
self-intersections of closed cosmic string loops~\cite{pbkofu,br} were
considered. 

A general formulation of calculating the flux of ultrahigh energy
particles from decay of X particles released from TDs in general was
given in Ref.~\cite{bhs} in which a convenient parametrization of the
production rate of X particles from any kind of TDs was proposed. This
facilitated calculation of expected UHE particle spectra due to decay of X
particles released from TDs in general without considering specific TD
models. The importance of the expected high flux of photons relative
to nucleons from the decay of X particles at the highest energies 
as a signature of the TD scenario in general was pointed out in 
Ref.~\cite{abs}. The recent detections of the cosmic ray
events above $10^{20}\ev$ by the Fly's Eye as well as the AGASA
experiments and the realization of the difficulties faced by conventional
acceleration scenarios in explaining these
events~\cite{ssb,elb-som,slsb,norman} have led to a renewed interest 
in the TD scenario of origin of EHECR.  
 
In what follows, we discuss various X particle production
processes involving various different kinds of TDs that have been
studied so far, and discuss their efficacies
with regard to EHECR keeping in mind the rough estimates of the
X particle production rate required to explain EHECR discussed 
above. Among the various kinds of TDs, cosmic strings are perhaps the
most well-studied, analytically as well as numerically, in terms of
their properties pertaining to their formation and evolution in the
Universe. It is, therefore, possible to make some quantitative estimates
of X particle production rates due to various cosmic string processes
with relatively less number of free parameters as compared to processes
involving other TDs. Because of this, and for illustrative reasons, we
first discuss the cosmic string processes in some details, and then
discuss other defects briefly. 

\subsection{X particle production from Cosmic Strings}
The release of X particles from cosmic strings requires removal of
local topological stability of (segments of) cosmic strings. 
The topological stability of cosmic strings is due
to non-trivial winding of the phase of the relevant (complex) Higgs field
around the string. This local stability is removed whenever
the string is in such a configuration that conflicting demand is
placed on the phase of the Higgs field, leading to ``unwinding'' of the
phase. This happens, for example, (a) at the point of intersection 
of two string segments, (b) if a closed loop of string
shrinks (due to energy loss through, e.g., gravitational radiation) 
down to a
radius of the order of the width of the string, and (c) in the ``cusp
evaporation'' process. Brief descriptions of these processes are given
below. 

{\it Intersection and Intercommuting of String Segments}: 

The cosmic string has a small but finite width $w\sim\eta^{-1}$,
where $\eta$ is the vacuum expectation value (VEV) of the relevant 
Higgs field. The energy per unit length of the string is
$\mu\sim\eta^2$. Because of the finite width of the string, 
two intersecting string segments have to overlap on top of each other
over a length scale of order $w\sim\eta^{-1}$ near 
the point of intersection. The Higgs field phase becomes undefined in
the overlapping region, and a topology removal event takes place,
whereby the energy contained in the overlapping region is released in
the form of massive X particles. The remaining string segments then
``exchange partners'' and reconnect so as to maintain the continuity of
the phase of the Higgs field. This process of ``intercommuting'' of
strings has been verified by numerical simulations of cosmic string
interactions. This is also the fundamental process that leads to
formation of closed loops of string from self-intersections of long
strands of string and to 
splitting of any closed loop into smaller daughter loops when the parent
loop self-intersects. The
energy released in the form of X particles at each intercommuting
event is $\sim\mu w\sim\eta$, and since $m_X\sim\eta$, we see that, 
modulo factors of order unity, roughly of order one X particle is
released at each intercommuting event. The contribution of this
process to the cosmic ray flux will depend on the rate of occurrence
of these intercommuting events which, as we shall see below, is negligibly
small. 

{\it Final Stage of Loop Shrinkage}: 

Similarly, each closed loop of string after
shrinking down to a radius of order the width of the string will
produce roughly of order one X particles, most of the energy of the
original loop having been radiated away in the form of gravitational
radiation.  

{\it Cusp Evaporation}: 

The cusp evaporation is another process by which X particles
can be released from cosmic strings. Cusps are points on a string at
which the string at an instant of time moves with the speed of
light. The radius of curvature of the string at a cusp point 
becomes very small. Existence of cusps is a generic
feature~\cite{turok-cusp} of equations of motion of closed loops of string
described by the Nambu-Goto action~\cite{vil-shell}. Strictly speaking,
the Nambu-Goto action is valid only for a mathematically
infinitesimally thin string with no width, and in fact, in this case,
the radius of curvature of the string at the cusp is mathematically
undefined. For a realistic string with a small but finite width,
the Nambu-Goto action provides a good description of
the string motion as long as radius of curvature of the string is
much larger than the width of the string. Clearly, the Nambu-Goto
action breaks down at the cusp points; nevertheless, the Nambu-Goto
description of a string with initially no cusp shows that ``almost'' cusp
points tend to develop, at which, due to finite width of the string,
overlapping of two string segments takes place, leading to
``evaporation''~\cite{branden-cusp} of the overlapped regions
of the string in the form of X particles. 

The existence of cusps was initially shown for closed
loops~\cite{turok-cusp}, but cusps can
also occur on long strands of strings due to the presence of
small-scale structures such as ``kinks'' (where the tangent vector to
the string changes discontinuously). Cusps can be formed when kinks  
propagating on the string in opposite directions 
collide with each other~\cite{mohazzab-branden}. The length of
string involved in a cusp is~\cite{branden-cusp,mohazzab-branden}   
$\ell_c\sim\zeta^{2/3}w^{1/3}$,
where $\zeta$ is a characteristic length of the small-scale structure
on strings (basically the interkink distance); for a loop of length 
$L$ with no or few kinks, the cusp length $\ell_c$ is given by the
same expression as above with $\zeta$ replaced by $L$. The energy
released in the form of X particles due to a single cusp event is
$\sim\mu\ell_c$, and hence the number of X particles released in a
single cusp evaporation event on a long string is 
$\sim (\eta\zeta)^{2/3}$, while that for a closed loop of length $L$ is
$\sim(\eta L)^{2/3}$. 

Very recently, it has been pointed out in Ref.~\cite{blanco-cusp} that
the value of cusp length~\cite{branden-cusp} $\ell_c\sim
L^{2/3}w^{1/3}$ mentioned above is actually valid for only a special
class of cusps. For a more generic cusp, including the effects of the
Lorentz contraction of the string core, Ref.~\cite{blanco-cusp} finds 
$\ell_c\sim(Lw)^{1/2}$, which is smaller than the previous value by a
factor of $(w/L)^{1/6}$. Since $L$ will typically be a cosmological
length scale whereas $w$ is a microscopic length scale, the factor 
$(w/L)^{1/6}$ will be an extremely small factor which will drastically 
reduce the effectiveness of the cusp evaporation process in producing
observable cosmic ray flux. We shall continue to use the ``old''
estimate~\cite{branden-cusp} of $\ell_c$ in the calculations below as
a sort of ``upper limit'' on $\ell_c$. We shall see that even this
upper limit will be too low to yield observable cosmic ray flux.

For a given $\ell_c$, the contribution of the cusp evaporation
process to the cosmic ray flux will depend upon the number of cusp
evaporation events on loops and long strings occurring per unit volume per
unit time, which can be determined only if we know the number
densities of closed loops and long strings and the rate of occurrence
of cusps on each loop. 
In order to estimate the number densities of closed loops and long
strings, let us briefly recall the salient features of evolution of cosmic
strings in the Universe~\cite{vil-shell,tdrev}. 

\subsubsection{Evolution of Cosmic Strings}
Immediately after
their formation, the strings would be in a random tangled configuration.
One can define a characteristic length scale, $\xi_s$, of the 
string configuration in terms of the overall mass-energy density,
$\rho_s$, of strings through the relation 
\beq
\rho_s = \mu/\xi_s^2\,,\label{string-density1}
\eeq
where $\mu$ denotes the string mass (energy) per unit length. 
Initially, the strings find themselves in a dense medium, so 
they move under a strong frictional damping force. The damping remains
significant~\cite{vil-shell} until the temperature falls to $T\la
(G\mu)^{1/2}\eta$, where $G\equiv 1/M_{\rm Pl}^2$ is Newton's constant
and $\eta$ is the symmetry-breaking scale at
which strings were formed. [Recall, for GUT scale cosmic strings, for
example, $\eta\sim10^{16}\gev$, 
$\mu\sim \eta^2\sim(10^{16}\gev)^2$, and so $G\mu\sim10^{-6}$.] In the
friction dominated epoch, a curved string segment of radius of curvature
$r$ quickly achieves a terminal velocity $\propto 1/r$. The small scale
irregularities on the strings are, therefore, quickly smoothed out. As a
result, the strings are straightened out and their total length shortened.   
 This means that the characteristic length scale $\xi_s$ describing
the string configuration increases and consequently the energy density
in strings decreases with time as the Universe expands.
Eventually $\xi_s$ becomes comparable to the causal horizon distance
$\sim t$.  At about this time, the ambient density of the Universe
also becomes dilute enough that damping becomes unimportant so that
the strings start moving relativistically. 

Beyond this point, there are two possibilities.    
Causality prevents the length scale $\xi_s$ from growing faster than the
horizon length. So, either (a) $\xi_s$ keeps up with the horizon length,
i.e., $\xi_s/t$ becomes a constant, or (b) $\xi_s$ increases less rapidly
than $t$. In the latter case, the 
string density falls less rapidly than $t^{-2}$. On the other
hand, we know that the radiation density in the radiation-dominated
epoch as well as matter density in the matter-dominated epoch both
scale as $t^{-2}$. Clearly, therefore, in case (b) the strings would 
come to dominate the density of the Universe at some point of time. 
It can be shown that this
would happen quite early in the history of the Universe unless the
strings are very light, much lighter  
than the GUT scale strings. A string
dominated early Universe would be unacceptably inhomogeneous conflicting
with the observed Universe\footnote{However, a string dominated {\it 
recent} Universe --- dominated by ``light'' strings formed at a phase
transition at about the electroweak symmetry breaking scale --- is
possible. Such a string dominated recent Universe may even have some
desirable cosmological properties~\cite{string-dom}. Such light strings
are, however, not of interest to us in this discussion.}.       

The other possibility, the case (a) above, 
which goes by the name of ``scaling'' hypothesis,
seems to be more probable, as suggested by detailed numerical
as well as analytical studies~\cite{vil-shell,tdrev}. The numerical simulations
generally find that the string density does reach the scaling regime given
by $\rho_{s,{\rm scaling}}\propto 1/t^2$, and then continues to be in this
regime. It is, however, clear that in order for this to happen, strings
must lose energy at a certain rate. This is because, in
absence of any energy loss, the string configuration would only be
conformally stretched by the expansion of the Universe on scales larger
than the horizon so that $\xi_s$ would only scale as the scale factor
$\propto t^{1/2}$ in the radiation dominated Universe, and $\propto
t^{2/3}$ in the matter dominated Universe. In both cases, this would fail
to keep the string density in the scaling regime, leading back to
string domination. In order for the string density
to be maintained in the scaling regime, energy must be lost by the string 
configuration per unit proper volume at a rate
$\dot{\rho}_{s,{\rm loss}}$ satisfying the equation 
\beq
\dot{\rho}_{s,{\rm total}} = -2\,\frac{\dot{R}}{R}\,\rho_s + 
\dot{\rho}_{s,{\rm loss}}\,,\label{string-energy-loss-rate}
\eeq
where the first term on the right hand side is due to expansion of the
Universe, $R$ being the scale factor of the expanding Universe. In
the scaling regime $\dot{\rho}_{s,{\rm total}} = -2 \rho_s/t$, which gives 
$\dot{\rho}_{s,{\rm loss}} = - \rho_s/t$ in the radiation dominated
Universe, and $\dot{\rho}_{s,{\rm loss}} = -(2/3) \rho_s/t$ in the matter
dominated Universe. 

The important question is, in what form does the string configuration lose
its energy so as to maintain itself in the scaling regime? One possible
mechanism of energy loss from strings is 
formation of closed loops. Occasionally, a segment of 
string may self-intersect by curling up on itself. The
intersecting segments may intercommute, 
leading to formation of a closed loop which pinches off the string. 
The closed loop would then oscillate and lose energy by emitting
gravitational radiation and eventually disappear. It can be shown that
this is indeed an efficient mechanism of extracting energy from 
strings and transferring it to other forms. The string energy loss rate
estimated above indicates that scaling could be maintained by roughly of
order one closed loop of horizon size ($\sim t$) formed in a horizon size
volume ($\sim t^3$) in one hubble expansion time ($\sim t$) at any time
$t$. In principle, as far as energetics is concerned, one can have the
same effect if, instead of one or few large loops, a large number of
smaller loops are formed. Which one actually happens depends on the
detailed dynamics of string evolution, and can only be decided by means of
numerical simulations.         

Early numerical simulations seemed to support the large (i.e., $\sim$
horizon size) loop formation
picture. Subsequent simulations with improved resolution, however, found a
lot of small-scale
structure on strings, the latter presumably being due to kinks left on the
strings after each crossing and intercommuting of string segments.
Consequently, loops formed were found to be much smaller in size than
horizon size and correspondingly larger in number. Further simulations
showed that the loops tended to be formed predominantly on the scale of
the cut-off length imposed for reasonable resolution of the smallest
size loops allowed by the given resolution scale of the simulation. 
It is, however, generally thought
that the small-scale structure cannot continue to build up indefinitely,
because the back-reaction of the gravitational radiation from the 
kinky string itself would
eventually stabilize the small-scale structure at a scale  
$\zeta\sim\Gamma G\mu t$, where $\Gamma\sim100$ is a geometrical factor. 
The loops would,
therefore, be expected to be formed predominantly of size $\sim\zeta$,
at any time $t$. 
Although much smaller than the horizon size, these loops
would still be of ``macroscopic'' size, much larger than the microscopic
string width scale ($\sim\eta^{-1}\sim\mu^{-1/2}$). These loops would,
therefore, also oscillate and eventually disappear by emitting
gravitational radiation. Thus, according to above picture, the dominant
mechanism of energy loss from strings responsible for maintaining the
string density in the scaling regime would be formation of 
macroscopic-size ($\gg \eta^{-1}$) loops. 

With the above picture of string evolution in mind, we can estimate
the number density of closed loops born at any time $t_b$ as follows:
The scaling solution for the long
string energy density\footnote{Long strings are defined as string segments with
radius of curvature $\ga$ the horizon length $\sim t$.} 
gives $\rho_s=\mu/(xt)^2$ with $x$ determined by
numerical simulations; typically, in the matter-dominated epoch
relevant for our considerations, $x$ lies in the range~\cite{vil-shell,tdrev}
from about 0.4 to about 0.7. A recent simulation~\cite{vincent} indicates
$x\sim 0.3$. 
This scaling solution is maintained by closed loops being chopped off
from the long strings. Since loops are born on the scale of the
small-scale structure $\zeta$, it is reasonable to 
assume that the loops born at any time
$t_b$ all have the (average) length 
\beq
L_b=K\zeta(t_b)=K\Gamma G\mu t_b\,,\label{loop-birth-length}
\eeq 
where $K$ is a numerical factor of
order unity. Then from Eq.~(\ref{string-energy-loss-rate}) we see that
these loops must be born at a rate (i.e., per unit volume per unit
time) given by 
\beq
\frac{dn_b}{dt}=\frac{2}{3x^2}\,(\Gamma
G\mu)^{-1}K^{-1}t^{-4}\,\label{loop-birth-rate} 
\eeq 
in the matter-dominated epoch. (For radiation-dominated epoch, the
prefactor 2/3 should be replaced by unity.) 

After their birth these loops oscillate freely and lose energy, and
hence shrink, due to emission of gravitational radiation. A closed loop of
length $L$ oscillates with a period~\cite{kibble-turok} $L/2$ and
loses energy through gravitational radiation at a rate 
\beq
\dot{E}_{\rm grav}=\Gamma G\mu^2\,.\label{grav-loss-rate}
\eeq
Thus, a loop born with a length $L_b$ at a time
$t_b$ has a length 
\beq
L(t)=L_b-\Gamma G\mu (t-t_b)\label{loop-length-after-birth}
\eeq 
at any later time $t$, and so a loop of length $L$ has a lifetime 
$\tau_{\rm grav}=(\Gamma G\mu)^{-1} L$.  

From Eqs.~(\ref{loop-birth-length}), (\ref{loop-birth-rate}) and 
(\ref{loop-length-after-birth}), we can write the loop length
distribution (= number density of loops per unit length) at any time
$t$ in the matter dominated era as 
\beq
\frac{dn}{dL}(L,t)=\cases{\frac{2}{3 x^2}\frac{K+1}{K}
\frac{1}{t^2 (L+\Gamma G\mu t)^2}, & $L \le K\Gamma G\mu t$\,,\nonumber\cr
0, & $L > K\Gamma G\mu t$\,.\cr}\label{loop-ldf}
\eeq
It is often convenient to define $n_L(t)=L(dn/dL)(L,t)$ 
as the number density of
loops in a length interval $\Delta L\sim L$ around $L$. Using
Eq.~(\ref{loop-ldf}) we see that $n_L(t)\propto t^{-4}L$ for
$L\ll\Gamma G\mu t$ and $n_L(t)\propto t^{-2}L^{-1}$ for $L\gg\Gamma
G\mu t$, while $n_L(t)$ has a peak at $L=\Gamma G\mu
t$. Thus, for $K=1$, the most abundant loops today have a typical length  
$\sim 200 (G\mu/10^{-6}) (\Omega_0 h^2)^{-1/2}\kpc$ and number
density $\sim 4.6\times 10^{-6} (G\mu/10^{-6})^{-1}(\Omega_0
h^2)^{3/2} \mpc^{-3}$. The typical separation between these loops
is $\sim 60 (G\mu/10^{-6})^{1/3}(\Omega_0 h^2)^{-1/2}\mpc$. Note the
dependence of the above quantities on the parameter $G\mu$. 

We are now ready to estimate the X particle production rate due to
various cosmic string processes. 
\subsubsection{Intercommuting of Long Strings}
The rate of intercommuting events may be estimated as
follows~\cite{gk}: Self-intersections of the long string segments
must occur on the length scale $\xi_s$ of the long-string
configuration as defined by Eq.~(\ref{string-density1}). The time
scale of intersection is, therefore, also $\sim\xi_s$. The total length of
string in a given volume $V$ is $L\sim V/\xi^2$. Therefore, the number
of self-intersections and intercommuting events in the long-string
network per unit time per unit volume is 
\beq
n_{\rm ic}(t)=\chi/\xi_s^4\,,\label{ic-rate}
\eeq
where $\chi\la1$ is the probability of intercommuting in each
intersection. Using $\xi_s(t)\simeq x t$ with $x\simeq$ 0.3 -- 0.7,
and recalling that each intercommuting event releases of the order of
one X particle, we see by comparing with Eq.~(\ref{Xrate_required})
that the X particle production rate through the intercommuting process
is utterly negligible compared to that required to explain the
EHECR flux. 
\subsubsection{Final Stage of Loop Shrinkage}
From Eqs.~(\ref{loop-birth-length}) and
(\ref{loop-length-after-birth}) we see that loops born at time $t_b$
disappear (due to gravitational radiation) at the time
$t_d=(K+1)t_b$. Taking into account the expansion of the Universe
between the times $t_b$ and $t_d$, Eq.~(\ref{loop-birth-rate}) gives
the number density of loops disappearing per unit time per unit volume
at any time $t$ as 
\beq
\frac{dn_d}{dt}(t)=\frac{2}{3x^2}\,(\Gamma G\mu)^{-1}\frac{K+1}{K}
\,t^{-4}\,.\label{loop-expiry-rate} 
\eeq
Each expiring loop produces $\sim$ one X particle. Again, comparing
with Eq.~(\ref{Xrate_required}) we see that the production rate of X
particles through this process is also negligible in the context of
cosmic rays~\cite{pb90-unpub,gk}. 
\subsubsection{Cusp Evaporation}
Let us consider cusp evaporation from closed 
loops~\cite{branden-cusp,cusp,pbcusp,gk} first and then consider 
long strings. The number of X
particles emitted in a single cusp evaporation event on a loop of
length $L$ is $\sim\mu\ell_c m_X^{-1} \sim\mu
L^{2/3}w^{1/3}m_X^{-1}$. The time scale of the cusp evaporation
process itself is $\Delta t_{\rm cusp}\sim\ell_c$, but the cusps
recur~\cite{turok-cusp} 
on the loop once every oscillation period ($=L/2$) of the loop. Since
$\Delta t_{\rm cusp}\ll L$, the rate of X particles emitted through
cusp evaporation from a loop of length $L$ at any time $t$ can be
written as 
\beq
\frac{dN_X^{\rm cusp}}{dt}=4 f_c m_X^{-1}\mu L^{-1/3}w^{1/3}\,,
\eeq
where $f_c$ is a numerical factor accounting for the efficiency of the
cusp evaporation process, and we have assumed that two cusps appear
every oscillation period of the loop~\cite{turok-cusp}. Folding this
with the loop length distribution function of Eq.~(\ref{loop-ldf}), we
get the number density of X particles produced per unit time per unit
volume due to all loops at any time $t$: 
\beq
\dot{n}_X^{\rm cusp}=4 f_c m_X^{-1}\mu w^{1/3}\int dL\,L^{-1/3}
\frac{dn}{dL}(L,t)\,.\label{cusp-Xrate1}
\eeq
Using Eq.~(\ref{loop-ldf}) we see that the dominant contribution to
the integral in the above equation comes from loops of length
$L\sim\Gamma G\mu t$, giving 
\beqarray
\dot{n}_X^{\rm cusp} & \simeq 4 f_c m_X^{-1}\mu w^{1/3}\,\frac{2}{3
x^2}\,\frac{K+1}{K}\,(\Gamma G\mu)^{-4/3}t^{-10/3}\nonumber\\ 
& = 4 f_c\,\frac{2}{3x^2}\,\frac{K+1}{K}\,\Gamma^{-4/3}(G\mu)^{-1}M_{\rm
Pl}^{2/3} t^{-10/3}\,.\label{cusp-Xrate2}
\eeqarray
Taking $f_c=1$, $x=0.3$, $K=1$, $\Gamma=100$, and with $t_0\simeq
2.06\times10^{17}(\Omega_0 h^2)^{-1/2}\sec$ for the age of the Universe,
we get the rate of X particle production in the present epoch due to cusp
evaporation from all cosmic string loops as 
\beq
\dot{n}_{X,0}^{\rm cusp}\simeq
6.4\times10^{-56}\left(\frac{G\mu}{10^{-6}}\right)^{-1}
\cm^{-3}\sec^{-1}\,.\label{cusp-Xrate-today} 
\eeq
Comparing this rate with that in Eq.~(\ref{Xrate_required}), we see
that the X particle production rate due to cusp evaporation from
cosmic string loops is too small (by about ten orders of magnitude) to
give any significant contribution to cosmic ray flux. 

Note that $\dot{n}_X^{\rm cusp}$ generally increases
with decreasing value of $\mu$. This is due to the fact that for lower
values of $\mu$ (lighter strings) the gravitational radiation rate is
reduced, so loops survive longer and consequently the number density
of loops present at any time is larger, giving a larger contribution
to X particle production rate. This at first suggests~\cite{cusp} 
that for
sufficiently light strings (i.e., for sufficiently small values of
$G\mu$) the X particle production rate due to cusp evaporation may
even exceed the rate required for sufficient cosmic ray
production, thereby giving a lower limit on $G\mu$, i.e., a lower
limit on the energy scale of any string-forming phase transition. 
However, this turns out not to be the case~\cite{pbcusp}: 
For too small values of $G\mu$, the energy loss of the string loops 
through gravitational radiation becomes so small that the cusp
evaporation process itself becomes the dominant energy loss mechanism
which then determines the number density of loops. Detailed
calculations~\cite{pbcusp} show that when the loop length distribution
function, Eq.~(\ref{loop-ldf}), is modified (for small values of $G\mu$)
to include the energy loss of the loop due to cusp evaporation itself,
then there is no lower limit on $G\mu$. Indeed, $\dot{n}_X^{\rm cusp}$
increases with decreasing values of $G\mu$ reaching a peak at 
around $G\mu\sim10^{-15}$, and $\dot{n}_X^{\rm cusp}$ then decreases
with further decrease in the value of $G\mu$. The peak value of
$\dot{n}_X^{\rm cusp}$, however, still remains about four orders of
magnitude below the value required for producing sufficient cosmic ray
flux. 

Let us now consider cusp evaporation from long strings. Cusps can be
formed on long strings due to collisions of kinks traveling on long
strings~\cite{mohazzab-branden}. The total length of string in a given
volume $V$ is $L\sim V/\xi_s^2$. The inter-kink separation is
$\sim\zeta\sim\Gamma G\mu t$, so the number of kinks on the string is
$\sim L/\zeta$. The time-scale of kink collisions is
$\sim\zeta$. Therefore, the number of cusps formed per unit time per
unit volume is $\sim\chi(\zeta\xi_s)^{-2}$, where $\chi\la1$ is the
probability of cusp formation in each kink
collision. (Ref.\cite{mohazzab-branden} finds that $\chi$ can be
$\sim0.5$). As explained earlier, the number of X particles released
in a single cusp evaporation event on a long string is 
$\sim(\eta\zeta)^{2/3}$. Thus, the production rate of X particles due to 
cusp evaporation from the long string network in the scaling solution 
($\xi_s=xt$) can be written as 
\beqarray
\dot{n}_X^{{\rm cusp},LS}&&\simeq \chi
(\eta\zeta)^{2/3}(\zeta\xi_s)^{-2} \nonumber\\
&&=\chi x^{-2}\Gamma^{-4/3}(G\mu)^{-1}M_{\rm
Pl}^{2/3}t^{-10/3}\,.\label{cusp-Xrate-long-string} 
\eeqarray
Except for an overall factor of order unity, this is same as the
X particle production rate in the case of cusp evaporation from
loops. Thus, X particle production from cusps on cosmic strings gives
negligible contribution to cosmic ray flux. 

As already mentioned earlier, the cusps assumed above are special
ones. Taking into account the 
fact that for a generic cusp the energy released is actually much
smaller~\cite{blanco-cusp}, one can conclude that the cusp evaporation
process leads to utterly negligible cosmic ray flux. 

\subsubsection{Collapse or Repeated Self-intersections of Closed
Loops} 
It is clear from the above discussions that in order to produce 
X particles with a large enough rate so as to be
relevant for cosmic rays, {\it macroscopically large} lengths of
strings are required to be involved in the X particle production
process. One such
process is complete collapse or repeated self-intersections of closed
loops~\cite{br}. It is known~\cite{kibble-turok} that any initially 
static non-circular loop, or any loop
configuration that can be described by single-frequency Fourier modes,
collapses into a double-line configuration at a time $L/4$ after its
birth, $L$ being the length of the loop. (Recall, period of
oscillation of a loop of length $L$ is $L/2$.) In such overlapped
configurations, the entire string would annihilate into X
particles~\cite{bkt}. Such completely collapsing
configurations are, however, likely to be very rare. 
Nevertheless, this kind of
collapsing loops serve as an example of a general class of situations
in which 
macroscopically large fraction of the energy of cosmic string loops is
dissipated in the form of X particles on a time-scale much shorter
than the time-scale $\tau_g\sim(\Gamma G\mu)^{-1}L$ of decay of the loops  
due to energy loss through gravitational radiation. For example, one
can think of a situation in which a large loop self-intersects and
splits into two smaller loops, and each daughter loop self-intersects
and splits into two further smaller loops, and so on. Under such a
circumstance, it can be seen~\cite{vil-physrep} that a single
initially large loop of length $L$ can break up into a debris of tiny
loops (of size $\sim\eta^{-1}$, thereby turning into X particles)
on a time-scale $\tau_{\rm debris}\sim L$.  
Since, as discussed earlier, loops
are expected to be born at any time $t$ with typical size 
$L \sim K\Gamma G\mu t \ll t$, we see that the above time-scale of
break-up of a large loop into X particles is much less than the Hubble
time, and very much less than the gravitational decay time-scale
$\tau_g$. 

Following 
Ref.~\cite{br} let us suppose that a fraction $f_X$ of the total energy
in all newly born loops at any time $t$ goes into non-relativistic X
particles of mass $m_X$  
on a time-scale much shorter than the gravitational decay time-scale
$\tau_g$. Using Eqs.~(\ref{loop-birth-length}) and
(\ref{loop-birth-rate}) we then get 
\beq
\dot{n}_X(t)=f_X\frac{\mu}{m_X}\,\frac{2}{3x^2}\,t^{-3}
\,,\label{Xrate-macro-string} 
\eeq
where $x\simeq 0.3$. Eq.~(\ref{Xrate_required}) then implies that in
order to explain EHECR we require 
\beq
f_X \eta_{16}^{3/2}\simeq2.8\times10^{-5}\,,\label{f_X-constraint-ehecr}
\eeq
where we have taken $l(E_\gamma=300\,{\rm EeV})=50\mpc$, 
$m_X\sim\eta\sim\mu^{1/2}$, and defined $\eta_{16}=(\eta/10^{16}\gev)$. 
One should keep in mind that Eq.~(\ref{f_X-constraint-ehecr}) is valid
provided the cosmic string loops producing the X particles are distributed
in a spatially homogeneous manner. 

For a given value of $f_X$, there is an independent {\it constraint} 
on $\eta_{16}$ which comes from the fact that any
electromagnetic radiation injected at the EHECR energies would initiate an
electromagnetic cascade (see Sect.~4.2) whereby a part of the injected
energy would show
up at lower energies (in the 10 MeV -- 100 GeV region) where existing
measurements of the diffuse gamma ray background constrain any such energy
injection. The energy density that would go into the cascade radiation is
approximately given by~\cite{bbv} (see Sect.~7.1)
\beq
\omega_{\rm cas}\simeq\frac{1}{2}\,m_X\dot{n}_X t_0\,,\label{w-cascade}
\eeq
and the measured gamma ray background in the 10 MeV -- 100 GeV
region~\cite{cdkf} imposes the constraint~\cite{bbv} 
\beq
\omega_{\rm cas}\leq
2\times10^{-6}\ev\cm^{-3}\,.\label{w-cascade-constraint}
\eeq
Eqs. (\ref{Xrate-macro-string}), (\ref{w-cascade}), and
(\ref{w-cascade-constraint}) together imply the condition 
\beq
f_X \eta_{16}^2\leq 9.6\times10^{-6}\,.\label{f_X-constraint-cascade}
\eeq
The requirement of Eq.(\ref{f_X-constraint-ehecr}) can be satisfied (so
that we are able to explain the EHECR) without violating the cascade 
constraint (\ref{f_X-constraint-cascade}) only for $\eta$ lying in the
range $9.2\times 10^{12}\gev \la \eta \la 1.2\times10^{15}\gev$ with $f_X$
in the corresponding range $6.7\times10^{-4}\la f_X \la 1$ satisfying
$f_X\simeq2.8\times10^{-5}\eta_{16}^{-3/2}$. 

The above discussions indicate that that a cosmic string scenario of
EHECR with $m_X$ much above $\sim 10^{15}\gev$ may be difficult to
reconcile with the low energy diffuse gamma ray constraint. On the other
hand, for $m_X\sim 10^{15}\gev$, cosmic strings can be
responsible for EHECR without violating the gamma ray background
constraint provided that a fraction ${\rm few}\times 10^{-4}$ of the newly
born loops at any time $t$ goes into X particles on a time-scale much
smaller than the Hubble time $\sim t$. 

The above crude analytical estimates indicate that the measured ``low''
energy diffuse gamma ray background provides an important constraint on
the mass (or energy) of the decaying particle if the number densities of
these particles are normalized so as to explain the EHECR. This important
fact was first pointed out in Refs.~\cite{chi} and has subsequently been
emphasized in many studies~\cite{ps1,lee,slsc,susytd,bbv,slby}. 

As discussed earlier below Eq.~(\ref{Q_0_required}), the energy 
injection rate needed to explain the EHECR is uncertain, and the estimate 
of Eq.~(\ref{Q_0_required}) is probably an overestimate in which case the
upper limit on $m_X$ derived above (for the viability of cosmic string
scenario of EHECR) may be pushed up to $\sim10^{16}\gev$, a typical GUT
scale. Indeed, recent detailed numerical calculations~\cite{slby} show
that, for a
large range of other parameters, TD scenarios of EHECR in general
are consistent with all observational data for $m_X$ up to
$\sim10^{16}\gev$, but not much above this value.  

Coming now to the question of $f_X$, it is not known what fraction of
loops may be born in collapsing and/or repeatedly
self-intersecting configurations such that essentially all their
energy eventually turns into X particles. 
In principle, numerical simulations
of loop self-intersections should be able to answer this question, but
in practice the simulations have so far lacked the necessary
resolution. On the theoretical side, Siemens and
Kibble~\cite{siemens} have shown that self-intersection
probability of a loop increases exponentially with the number of harmonics
needed to describe the loop configuration. In particular, since kinks
are high harmonic configurations, loops having kinks have high probability
of self-intersection. Since the loops formed from
self-intersection of long strings (or from splitting of existing 
loops) invariably have kinks on them, it is not inconceivable that an
interesting 
fraction ($\sim {\rm few} \times 10^{-4}$) of loops at any time may indeed
undergo repeated self-intersections and rapidly
deteriorate into tiny loops which decay into X particles. Note that if
loops always self-intersect and thus quickly turn into X particles, i.e.,
if $f_X\sim 1$,
then the the constraint (\ref{f_X-constraint-ehecr}) gives $\eta\leq
9.2\times10^{12}\gev$, which would rule out the existence of GUT-scale
cosmic
strings~\cite{br} because of excessive X particle production and the
resulting overproduction of EHECR. 
At the same time, for $f_X\simeq 1$, cosmic strings
with $\eta\sim10^{13}\gev$ can explain the EHECR flux without violating
the low energy cascade $\gamma-$ray constraint.   

Note that in the process of repeated self-intersection and splitting off 
of loops, a fraction of the total energy of a parent loop is likely to
go into kinetic energy of the daughter loops at each
splitting~\cite{vil-physrep}. Depending on this fraction 
the daughter loops may get substantial kicks at their birth. 
The smallest loops which eventually turn into X particles may, therefore,
be relativistic~\cite{vil-physrep} and hence well dispersed in
space. Thus, although at any time there might be relatively few initially
large loops within our Hubble volume so that the distribution of those
initially large loops might be highly inhomogeneous and anisotropic,  
the X particles themselves resulting from repeated 
self-intersection and splitting of those loops and the resulting 
cosmic rays may be more isotropically and uniformly distributed in the
sky. 

However, there may be a problem if the X particles are too
relativistic, and the authors of Ref.~\cite{bbv}, in
particular, have argued within the context of a specific loop
fragmentation model that in the relativistic X particle case it is hard
to obtain sufficient number of X particles to explain the EHECR flux
without violating the cascade constraint (\ref{f_X-constraint-cascade}) 
for any reasonable values of $\eta$. However, this conclusion seems to be
specific to the particular loop fragmentation model considered in
Ref.~\cite{bbv}, and can be evaded in other loop fragmentation scenarios.
For example, in the Siemens-Kibble scenario~\cite{siemens} 
mentioned above in which all cosmic string loops quickly break up into
X particles, thus giving $f_X\sim 1$, it can be
shown~\cite{pb-relativistic-X} that the EHECR
flux can be explained without violating the $\gamma-$ray cascade
constraint, provided the strings are sufficiently light, 
$\eta\la3.1\times10^{13}\gev$,  
and $f_{\rm KE}$, the fraction of energy of any parent loop (in its rest
frame) that goes into the kinetic energy of daughter loops, is not too
large, $f_{\rm KE}\la\,$ few percent. 

It may be mentioned here that an EHECR scenario involving lighter (i.e.,
lighter than GUT scale) cosmic strings 
with, e.g., $\eta\sim {\rm few}\times10^{13}\gev$, has  
an advantage over one with heavier strings because the  
number density of loops of such light strings would be larger than
that for heavier strings. Recall (from the discussions following
Eq.~(\ref{loop-ldf})) that the number density of loops at any time
is proportional to $(G\mu)^{-1}$ and the average separation between
the loops is proportional to $(G\mu)^{1/3}$, while the typical length
of a loop is proportional to $G\mu$. Thus, while for GUT-scale
strings with $\eta\sim10^{16}\gev$ (i.e., $G\mu\sim10^{-6}$), there are
only about $2.4 (\Omega_0h^2)^{3/2}$ loops within a typical ``GZK
volume'' of radius $\sim 50\mpc$, the number would be
larger by a factor of $10^{6}$ for strings with $\eta\sim10^{13}\gev$
($G\mu\sim10^{-12}$). Thus the problem of lack of enough cosmic string  
loops~\cite{sato-kofu,pb90-unpub,gk} encountered in the GUT-scale cosmic
string scenario of origin of EHECR may be solved with sufficiently light
string loops undergoing repeated splittings and thereby producing
sufficiently energetic X particles relevant for EHECR. 

\subsubsection{Direct Emission of X Particles from Cosmic Strings}
Finally, we consider the recent suggestion~\cite{vincent} that X
particles may be directly radiated from cosmic strings. 
From the results of their new numerical simulations of
evolution of cosmic strings, authors of Ref.~\cite{vincent}
have claimed that if loop production is not artificially restricted by
imposing a cutoff length for loop size in the simulation, then loops 
tend to be produced predominantly on the smallest allowed length scale in
the problem, namely, on
the scale of the width of the string. Such small loops promptly collapse 
into X particles. In other words, according to Ref.~\cite{vincent}, there
is essentially
no loop production at all --- the string energy density is maintained in
the scaling regime by energy loss from strings predominantly in the form
of direct X particle emission, rather than by formation of large loops
and their subsequent gravitational radiation. It should be mentioned here
that this result, which implies a radical departure from the results of
earlier numerical simulations of evolution of cosmic 
strings~\cite{vil-shell,tdrev}, has been questioned 
recently~\cite{moore-shellard}.
However, the basic issues involved here are quite complex and
currently rather ill-understood, and as such 
the results of Ref.~\cite{vincent} cannot be ruled out 
at this time. 
{\it If} the results of Ref.~\cite{vincent} are correct, then X
particles are directly produced by cosmic strings at a rate given by
Eq.~(\ref{Xrate-macro-string}) with $f_X=1$. Comparing this rate with that
in Eq.~(\ref{Xrate_required}), we see that in order for cosmic rays from
cosmic strings not to exceed the observed cosmic ray flux, the
string-forming symmetry breaking scale $\eta$ is
constrained~\cite{br,vincent,susytd,wmgb} as $\eta\la 10^{13}\gev$.  
Thus, in this case, GUT scale cosmic strings with
$\eta\sim10^{16}\gev$ will be ruled out~\cite{br,vincent,susytd,wmgb}
--- because they would necessarily overproduce EHECR --- while at the
same time cosmic strings formed at a phase transition with $\eta\sim
10^{13}$ -- $10^{14}\gev$ would be a ``natural'' source of EHECR. 

Note, however, that the typical radius of curvature of long strings
today, and hence the typical distance between neighboring long
strings is of the order of the Hubble distance $\sim t_0$. Therefore,
the observed EHECR can be due to direct emission of X particles from
long strings only in the case of accidental proximity of a long string
segment lying within say $\sim$ 50 Mpc from
us~\cite{bbv}. In this case, however, the observed
arrival directions of the EHECR events is predicted to be highly
anisotropic. In particular, one should expect the sources of
individual EHECR events to trace out a linear or filamentary region of
sky~\cite{br} corresponding to the long string configuration. This
prediction should be testable with the upcoming large-area EHECR
detectors.  

Before closing this discussion on cosmic strings as possible sources
of EHECR, it is worthwhile to mention that 
cosmic string formation at a phase transition with desired 
$\eta\sim10^{13}$ -- $10^{14}\gev$ rather than at the GUT
scale transition with $\eta\sim 10^{16}\gev$ is not hard to envisage. 
For example, the
symmetry breaking SO(10) $\to$ SU(3) $\times$ SU(2) $\times$
U$(1)_{\rm Y} \times$ U(1) can take place at the GUT unification scale
$M_{\rm GUT}\sim10^{16}\gev$; with no U(1) subgroup broken, this phase
transition produces no strings. However, the second U(1) can be
subsequently broken with a phase transition at a scale $\sim
10^{14}\gev$ or lower to yield the cosmic strings relevant for EHECR. Note
that these strings would be too light to be relevant for structure
formation in the Universe and their signature on the CMBR sky would
also be too weak to be detectable.
Instead, the extremely high energy end of the cosmic ray
spectrum may offer a probing ground for signatures of these
``light'' cosmic strings. We add that cosmic strings of mass scales
somewhat below the GUT scale could also be produced in non-thermal phase
transitions associated with the preheating stage after inflation, as
mentioned in Sect.~6.3. 

\subsection{X Particles from Superconducting Cosmic Strings}
Superconducting cosmic strings (SCSs)~\cite{witten-scs} are cosmic strings
carrying persistent electric currents. The current can be carried either
by a charged Higgs field having a non-zero vacuum expectation value inside
the string (thus breaking the electromagnetic gauge invariance inside
the string and thereby making the string superconducting), or by a
charged fermion field living as a ``zero mode'' inside 
the string due to coupling of the fermion to the string-forming Higgs
field.  (Here zero mode refers to the fact that the relevant fermions are
massless inside the string whereas they have a finite mass outside the
string.) In the case of non-abelian cosmic strings the superconductivity
of the string can also be due to a charged vector field condensate inside
the string. Superconducting strings would also in general carry electric
charges due to charges trapped at the formation of the string by
the Kibble mechanism and/or due to inter-commuting of string segments with
different currents. A review of basic properties of SCSs can be found in
Ref.~\cite{vil-shell}.

Superconducting strings cannot sustain currents beyond a certain critical 
current $J_c$. In the case of fermionic superconductivity, this
happens because the density of the charge carriers at the critical current
becomes degenerate enough that the fermi momentum inside the string 
exceeds the mass of the
fermion in the vacuum outside the string, at which point the fermions
above the fermi sea cease to be trapped on the string and
begin to be ejected into the vacuum outside the string. Similarly, in
the bosonic case, the energy density in
the charged scalar field condensate inside the string at the
maximum current becomes high enough to cause restoration of the broken
electromagnetic symmetry inside the string, as a result of which the
string loses its superconductivity. 

The magnitude of $J_c$ is model dependent, but the string forming
symmetry-breaking scale $\eta$ provides an upper bound on $J_c$,
namely, $J_c\leq J_{\rm max}\simeq e\eta$, both for bosonic as well as
fermionic superconductivity. (Here $e$ is the elementary electronic
charge.) If an  SCS achieves the critical current, the charge
carriers will be expelled from the string. Outside the string, the
charge carriers are massive with a mass that --- depending on the particle
physics model ---  can be as large as the GUT-scale $\sim10^{16}\gev$.
These massive charge carriers would then be the X particles whose decay
may give rise to extremely energetic cosmic ray particles. 

There can be a variety of mechanisms of setting up the initial current on
the string. Apart from small-scale current and charge fluctuations
induced on the string at the
time of the superconducting phase transition, large scale coherent (dc)
currents can be induced on macroscopically large string segments (of
horizon scale $\sim t$) as the string moved through a possible primordial
magnetic field in the Universe. Or a string can pick up a current due 
to its motion through the Galactic magnetic field, for example. In
addition, any closed loop of oscillating SCS in an external magnetic field
would have an ac current, and there would also be short-wavelength ac
contribution to the current on the scale of the small-scale wiggles on
long strings. 

The evolution of current-carrying SCSs is considerably more complicated
than that of ``ordinary'' non current-carrying cosmic strings, and is
rather poorly understood at the present time. It is therefore difficult to
make concrete predictions about contributions of SCSs to EHECR. One
possible model of X particle production from SCSs with fermionic
superconductivity and the resulting EHECR
flux was first studied by Hill, Schramm and Walker (HSW)~\cite{hsw}. Their
model was based on the scenario of evolution of current-carrying SCS
closed loops suggested by Ostriker, Thompson and Witten (OTW)~\cite{otw}.    

In the OTW scenario, initial currents on closed SCS loops are induced due
to the changing magnetic flux of a primordial magnetic field linked by
the loops --- the flux change being due to the expansion of the Universe.    
Once an initial current is set up, a current-carrying,
oscillating closed loop of SCS loses energy
through electromagnetic as well as gravitational radiation, and as a
result the loop shrinks in size. This, in turn, leads to an increase
of the dc component of the current on the loop ($J\propto L^{-1}$, $L$
being the instantaneous length of the loop) due to conservation of the
initial magnetic flux linked by the loop. 
Eventually, the current on the loop would reach the saturation
value $J_s=J_c$, at which point charge carriers --- the massive X
particles --- would be emitted from the loop. 

The X particle emission rate from a current saturated SCS loop in the case
of fermionic superconductivity can be estimated as follows~\cite{hsw}: 
The number of fermions of a given chirality plus antifermions of
the opposite chirality per unit length of the string is $n_F=p_F/\pi$,
where $p_F$ is the Fermi momentum. The current on the string is related to
the Fermi momentum through the relation $J=en_F=ep_F/\pi$. The string is
saturated when $p_F=m_F$, where $m_F$ is the mass of the fermion (to 
be identified with the X particle in this case). Thus $J_s=em_F/\pi$. 
The mass of the fermion arises from its Yukawa coupling with the
symmetry-breaking Higgs field responsible for
the formation of the string. Thus $m_F=g\eta$, where $g$ (assumed $\la
1$) is the Yukawa coupling constant and $\eta$ is the VEV of the
string-forming Higgs field. 
The total number of fermions plus antifermions inside a SCS loop of length
$L$ in the saturated regime is simply $N_F(t)=(m_F/\pi)L(t)$. After it is
saturated, the loop continues to radiate energy and shrink. As
the saturated loop shrinks, the fermi momentum remains constant at
$p_F=m_F$, the current remains constant at $J_s$, and so the shrinkage of
the saturated SCS loop is accompanied by fermion emission at a rate given
by $\dot{N}_F=(m_F/\pi)\dot{L}$. The loop shrinkage rate $\dot{L}$ in
the saturated regime is in general determined by the combined rate of
electromagnetic (e.m) plus gravitational energy radiation from the loop.
However, depending on the
values of the Yukawa coupling $g$ and the symmetry-breaking scale $\eta$, 
either the e.m. or the gravitational radiation may dominate. 

The e.m. radiation power from
a SCS loop (without cusp\footnote{Loops with cusps may have significantly
higher radiated electromagnetic power~\cite{vil-vach-scs}}) with current
$J$ is given by~\cite{otw} $P_{\rm em}=\gamma_{\rm em}J^2$, where
$\gamma_{\rm em}\simeq100$. The gravitational energy loss rate, $P_{\rm
grav}$, is given by Eq.~(\ref{grav-loss-rate}). In order to have EHECR
particles, we shall require that $m_F=g\eta\geq10^{12}\gev$, and since one
generally expects $g\la1$, we shall require $\eta\ga10^{12}\gev$, i.e.,
$G\mu\ga10^{-14}$ and $10^{-7}(G\mu)^{-1/2}\la g\leq 1$. From these
conditions one can see that, for a saturated SCS loop with $10^{-14}\la
G\mu\la10^{-8}$, e.m. radiation dominates over gravitational
radiation for all allowed values of $g$ that satisfy the requirement
mentioned above. On the other hand, for a saturated loop with
$G\mu > 10^{-8}$, e.m. radiation dominates if $10 (G\mu)^{1/2}\la g\leq
1$, and gravitational radiation dominates if $10^{-7}(G\mu)^{-1/2}\la g <
10(G\mu)^{1/2}$. If $\dot{L}$ is determined by the e.m. radiation, then
the fermion emission rate from a saturated SCS loop is given by 
\beq
\dot{N}_F=\frac{4}{\pi^2}\,\alpha_{\rm em}\gamma_{\rm
em}g^3\eta\,,\label{N_F-dot-em}
\eeq
where $\alpha_{\rm em}=1/137$ is the e.m. fine-structure
constant, and we have used $\mu\simeq\eta^2$ and $m_F=g\eta$. 
If gravitational radiation dominates (which requires $g$ to be
sufficiently small), then 
\beq
\dot{N}_F=\frac{g}{\pi}\,(\Gamma G\mu)\eta\,.\label{N_F-dot-grav}
\eeq
 
Eq.~(\ref{N_F-dot-em}) or (\ref{N_F-dot-grav}) gives the fermion emission
rate from a single saturated loop. To find the total number density
of fermions (X particles) produced by all saturated SCS loops per unit
time at any time $t$, we need to know the number density of saturated SCS
loops in the Universe as a function of cosmic time $t$. It is here that
things become rather complicated and model-dependent. The evolution of the
length distribution function for current-carrying SCS loops is not known.
HSW~\cite{hsw} assumed that the loop formation rate for SCSs is same as
that for ordinary cosmic strings and their evolution in the
pre-saturation regime is governed by gravitational radiation loss as
discussed in Sect. 6.4.1\footnote{The evolution of the current in 
SCS loops under combined e.m. plus gravitational radiation 
is discussed in a different context in Ref.~\cite{pbscs}.}. One expects
that at any given cosmic time $t$, all existing loops of length below a
certain ``saturation length'' $L_s$ will have achieved the saturation
current, and each of these loops would be emitting fermions at a rate
given by Eq.~(\ref{N_F-dot-em}) or (\ref{N_F-dot-grav}), as the case may
be. The saturation length $L_s$ depends on the details of the manner in
which the initial current is induced on loops, the magnitude of the
initial current on a loop (which depends on the strength of the ambient
magnetic field), the subsequent magnetic field history experienced by the
loop, and so on. 

In the OTW scenario, in which the
initial currents on loops are induced by the removal (due to expansion of
the Universe) of a primordial magnetic field whose energy density scales
as that of the universal radiation background, $L_s$ is roughly constant
in time~\cite{hsw}. In this case, with the loop length distribution
function given by that for ordinary cosmic strings 
[Eq.~(\ref{loop-ldf})], one sees that the X particle production rate,
$\dot{n}_X(t)\propto t^{-4}$. HSW~\cite{hsw} showed that for certain
ranges of parameter values, this scenario can produce EHECR flux
comparable with observed flux. 

In a more general situation, depending on the magnetic field history,
the saturation length $L_s$ can increase or decrease with time, and
consequently the time dependence of $\dot{n}_X$ would be different. 
According to Ref.~\cite{hsw}, in scenarios where $L_s$ grows with time ---
such a scenario may obtain, for example, if the intergalactic magnetic
fields are increased by dynamo effects --- the value of saturation length
in the present epoch $L_s(t_0)$ would be smaller than its value in the
OTW scenario, and the resulting absolute value of $\dot{n}_X$ in the
present epoch would be insufficient to explain the observed EHECR flux. On
the other hand, as we shall discuss in Sect. 7, there is a general
problem in situations where $L_s$ decreases with time because then the
energy injection in the early epochs would turn out to be unacceptably
large from considerations of distortion of the CMBR and primordial
nucleosynthesis~\cite{sjsb} if the absolute value of $\dot{n}_X$ in the
present epoch were such as to explain the EHECR flux. Thus, in general
it seems difficult to invoke SCS loops (at least in the case of fermionic
superconductivity) as possible sources of EHECR. 

We should stress that the above conclusion hinges upon the assumption that
evolution of SCSs in the unsaturated regime is similar to that of 
ordinary cosmic strings. This is highly uncertain, and as of now, no
detailed numerical simulations comparable to those available for ordinary
cosmic strings have been done for the study of evolution of a network of
superconducting strings. Even the dynamics of a single current-carrying
SCS loop is uncertain. The assumption of homologous shrinkage of
saturated SCS loops assumed above is probably too simplistic. Indeed, the
loop can fold onto itself in complicated shapes with one or more
self-intersections, leading to enhanced emission of the charge carriers. A
variety of other instabilities can appear (see, e.g.,
Ref.~\cite{martin-peter} and
references therein for a recent discussion of these issues). In addition,
the effects of the ac current modes as well as the effects of the plasma
in which the strings move are highly uncertain. It is also possible
that SCS loops may in fact never achieve the saturation current at all; 
instead they may form stable ``vortons'' --- charge- and current-carrying
SCS loops stabilized against shrinkage by angular
momentum~\cite{davis-shell-89} --- which themselves may be relevant
for EHECR (see below). 

Apart from the uncertainties inherent to the physics of SCSs in general,
there are several ``astrophysical'' uncertainties associated with the
proposal of SCSs as possible sources of EHECR. First, as already mentioned
above, the mechanism of induction of initial current on SCSs is uncertain
due to uncertainties in our knowledge of the magnetic field history of the
Universe. Second, it has been pointed out~\cite{berez-rubin} that even if
SCS loops achieve saturation current and emit the massive charge carriers
(X particles), the energetic particles resulting from the decay of the X
particles are likely to be quickly degraded in energy due to synchrotron
losses and other processes occurring in the high magnetic field region
around the string. Within the framework of the Standard Model it has
been claimed that most of the energy is radiated as thermal neutrinos
with a temperature of roughly 10 MeV that may be observable by
underground detectors and be comparable to the atmospheric neutrino
flux at these energies, whereas the emitted $\gamma-$rays
may give rise to coincident GRBs~\cite{plaga3}. The problem of
degradation of energetic particles may, however, be avoided~\cite{hsw} if the
charge carriers have a lifetime sufficiently long that they may be able to
drift into the weak-field region far away from the string before decaying.
Another possibility arises if the string has mainly ac current: In this
case, there can be sections of the string with large electric charge but
small current, and high energy particles can escape through those 
regions~\cite{bbv}. 

To summarize this discussion on SCSs, then, the range of possibilities
here are so large and our current state of knowledge of evolution of SCSs
is so uncertain that a definite conclusion regarding the viability or
otherwise of SCSs as sources of EHECR cannot be made at this stage. The
simplest models that have been studied so far generally fail to produce
sufficient EHECR flux. However, more work will be needed in this regard.  

\subsection{X Particles from Decaying Vortons} 
A SCS loop possessing both a net charge as well as a current can, under
certain circumstances, be stabilized against collapse by the angular
momentum of the charge carriers. Such stable SCS loops of microscopic
dimension, called ``vortons''~\cite{davis-shell-89}, do not radiate
classically and essentially behave like particles with quantized charge and
angular momentum. 

A vorton can be characterized by essentially two integer ``quantum''  
numbers: (a) $N$, the total winding number of the phase of the charge
carrier scalar field condensate along the length of the loop, which is
responsible for the conserved current on the loop\footnote{We consider  
here the case of bosonic superconductivity of the string. However, similar
arguments apply for fermionic case also because of formal equivalence of
bosons and fermions in (1+1) dimensional field theory on the string
world-sheet; for more details, see Ref.~\cite{vil-shell}.}, and (b) $Z$,
which is related to the total charge
$Q=Ze$ on the loop. A vorton generally tends to evolve towards a
``chiral'' state\footnote{The name chiral refers to the fact that in this
case the rotation velocity of the vorton approaches the speed of light.} 
in which $|Z|\simeq |N|$, and angular momentum ${\mathcal L}\simeq
ZN\simeq N^2$. The characteristic vorton radius $R_v$ obtained by
minimizing the total energy of a SCS loop is given by
$R_v\simeq(2\pi)^{-1/2}|NZ|^{1/2}\eta_s^{-1}$, where $\eta_s$ is the
string-forming symmetry breaking scale. Note that $\eta_s$ may in general
be different from (larger than) the symmetry breaking scale $\eta_\sigma$
associated with the appearance of superconductivity in the string. 
However, the most favorable conditions for vorton formation
occur when $\eta_s$ and $\eta_\sigma$ are not too widely
different. For GUT-scale vortons with
$\eta_s\sim\eta_\sigma\sim10^{16}\gev$, a rough
estimate~\cite{bcdt} gives $N\simeq Z\sim100$, and so
$R_v\sim10^{-28}\cm$, but these estimates can be off by several orders of
magnitude depending on the detailed dynamics of the vorton formation
process. 

(Meta)stable vortons with lifetime greater than the age of the Universe
can be a dark matter candidate. However, in some cases, their predicted 
abundance in the early Universe is so large as would overclose
the Universe at early times, in which case the particle physics models
under consideration (which predict vorton formation) have to be ruled
out. These considerations strongly constrain the vorton formation energy
scale in the early Universe. For more details on these issues and for
detailed discussions and references on properties, formation and evolution
of vortons , see, e.g.,
Refs.~\cite{bcdt,martins-shellard1,martins-shellard2}. 

Vortons can be relevant for EHECR in two ways: Although classically
stable, a vorton can decay by quantum mechanical tunnelling process. Such
metastable vortons decaying in the present epoch can release the massive
charge carrier particles with mass $m_\sigma\la\eta_\sigma$ which can act
as the X particles of the top-down scenario of EHECR if
$m_\sigma\ga10^{12}\gev$ and if vortons exist today with a sufficient
abundance. This possibility has been studied in
Ref.~\cite{masperi-silva}. Alternatively, vortons, being highly charged
particles, could be accelerated to extremely high energies
in some astrophysical sites, and thus vortons themselves could act as the
EHECR particles~\cite{bonazzola-peter}. Here we briefly discuss the first
possibility (decaying vortons); the second possibility will be discussed
briefly in Sect.~6.12.2.  

Authors of Ref.~\cite{masperi-silva} have studied an approximate
semiclassical model of vorton decay through quantum tunnelling originally
suggested by Davis~\cite{davis-vorton-decay}. This involves calculating 
the tunnelling probability of a chiral vorton configuration with $N\simeq
Z$ units of topological winding number to change to a configuration 
with $N-1$ units (with the accompanying emission of one quantum of the
charge carrier field of mass $m_\sigma\sim\eta_\sigma$), through a barrier
of height $\Delta E$ in energy and spatial width $\Delta R$, where $R$ is
the radius of the vorton. The tunnelling rate, or the inverse of the
lifetime of the vorton, is generally given by $\tau_v^{-1}\sim
m_v\exp{-(\Delta E\Delta R)}$. From simple consideration of energy
conservation, one can show that~\cite{masperi-silva} $\Delta E\Delta 
R\simeq N$. Thus vortons
with larger initial $N$ have longer lifetime. In order to be present and
decaying in the present epoch, the vortons must have $N$ larger than
a certain minimum value $N_{\rm min}\sim \ln(t_0\eta_\sigma)$ (we have
assumed $\eta_s\sim\eta_\sigma$). On the other hand, from the point of
view of obtaining sufficient EHECR flux, the vorton lifetime (and hence
$N$) should not be too large, for a given vorton abundance in the
Universe. The vorton abundance and the typical values of $N$ (which as
explained above determines the vorton lifetime) depend on the detailed
dynamics of the vorton formation process and are rather poorly understood
at present. But, in general, it turns out that
the joint requirements on the vorton lifetime and abundance (in order to
obtain sufficient EHECR flux) place conflicting demands on the vorton
formation energy scale. Thus at the present time it seems rather difficult
to explain EHECR with decaying vortons. 

According to a recent study~\cite{martins-shellard2}, the vorton density
is most sensitive to the order of the string forming phase transition and
relatively insensitive to the details of the subsequent superconducting
phase transition. For a second-order string forming phase transition,
vorton production is cosmologically disastrous (because they overclose
the Universe) and hence unacceptable for
$\eta_s (\sim\eta_\sigma)$ in the range $10^5\gev\la\eta_s\la10^{14}\gev$.
For a first-order string forming phase transition, the exclusion range is
somewhat narrower: $10^9\gev\la\eta_s\la10^{12}\gev$. For
$\eta_s\gg10^{14}\gev$, no vortex are expected to form. On the other hand,
vortons formed at $\eta_s\la10^5\gev$ ($10^9\gev$ for a first-order phase
transition) can provide a (part) of the dark matter and are, therefore,
cosmologically interesting. However these low mass-scale vortons are
unlikely to be relevant for EHECR because the typical mass of the emitted
charge carriers (X particles) are then too low to produce EHECR particles.
Note, however, that analysis of Ref.~\cite{martins-shellard2} still
leaves open a window of potentially interesting vorton density for
$\eta_s\simeq\eta_\sigma\sim10^{12}$--$10^{14}\gev$, which may then be
relevant for EHECR if the appropriate requirement on the vorton lifetime
can be met. 

We mention here that, as far as EHECR are concerned, vortons as
possible EHECR sources would behave very much like the possible 
long-lived, superheavy metastable relic X particles which are discussed in
a general way in Sect.~6.13.  
Thus, like standard cold dark matter, vortons would be expected to cluster
in the Galactic halo and so their density in the Galactic halo would be
significantly enhanced over their average cosmological density. The
dominant contribution of vortons to the EHECR flux would then come from
this clustered component with the concomitant advantages and disadvantages
that are discussed later in Sect.~6.13. We add, however, that vortons
are highly charged particles and as such
they should be subject to a variety of cosmological constraints applicable
to massive highly charged particles. The vortons also have a circulating
current and hence behave essentially as point magnetic dipoles. These
attributes may or may not have dramatic effects on their clustering
properties, but remain to be studied in detail. 

\subsection{X particles from Monopoles}
Compared to vortons and superconducting strings, magnetic monopoles as
topological defects are perhaps 
somewhat more well-studied in terms of their formation and evolution in
the Universe. Formation of magnetic monopoles is essentially inevitable in 
most realistic GUT models. They lead to the well-known monopole
overabundance problem, which historically played a major role in the
development of the idea of inflationary cosmology. For a review of
monopoles and their cosmological implications, see. e.g.,
Ref.~\cite{kolb-turner}.

The relevance of monopoles as possible sources of X particles in a
top-down scenario of EHECR arose from the works of Hill~\cite{hill} and
Schramm and Hill~\cite{schramm-hill-icrc}. If monopoles were formed at a
phase transition in the early Universe,
then, as Hill~\cite{hill} suggested in 1983, formation of metastable 
monopole-antimonopole bound states --- ``monopolonium'' --- is possible. 
At any temperature $T$, monopolonia would be formed with binding 
energy $E_b\ga T$. The initial radius $r_i$ of a monopolonium would be 
$r_i\sim g_m^2/(2E_b)$, where $g_m$ is the magnetic charge
(which is related to the electronic charge $e$ through the Dirac
quantization condition $eg_m=N/2$, $N$ being the Higgs field winding
number characterizing the monopole as a TD). 
Classically, of course, the monopolonium is unstable. Quantum
mechanically, the monopolonium can exist only in certain ``stationary''
states characterized by the principal quantum number $n$ given by $r=n^2
a_m^{\rm B}$, where $n$ is a positive integer, $r$ is the instantaneous
radius, and $a_m^{\rm B}=8\alpha_e/m_M$ is the ``magnetic'' Bohr radius of
the monopolonium. Here $\alpha_e=1/137$ is the ``electric'' fine-structure
constant, and $m_M$ is the mass of a monopole. 

Since the Bohr
radius of a monopolonium is much less than the Compton wavelength (size) 
of a monopole, i.e., $a_m^{\rm B}\ll m_M^{-1}$, the
monopolonium does not exist in the ground ($n=1$) state, because then the
monopole and the antimonopole would be overlapping, and so would
annihilate each other. However, a monopolonium would initially be formed
with $n\gg1$. It would then undergo a series of transitions through a
series of tighter and tighter bound states by emitting initially photons
and subsequently gluons, Z bosons, and finally the GUT X bosons.
Eventually, the cores of the monopole and the antimonopole would overlap, 
at which point the monopolonium would annihilate into X particles. Hill
showed that the life time of a monopolonium is proportional to the cube of
its initial radius. Depending on the epoch of formation, some of the
monopolonia formed in the early Universe could be surviving in the
Universe today, and some would have collapsed in recent epochs including
the present epoch. The resulting X particles could then be a source of 
EHECR. 

The monopolonia collapsing in the
present epoch would have been formed in the early Universe at around the
epoch of primordial nucleosynthesis~\cite{hill,bhatsig}. At that epoch, the
monopole-plasma energy exchange time scale would still be smaller than the
expansion time scale of the Universe~\cite{bhatsig}, so the relevant
monopolonium abundance at formation can be reasonably well described in
terms of the classical Saha ionization formalism. On the other hand, the
$e^+e^-$ annihilations at a temperature of $\sim$ 0.3 MeV (i.e., shortly
after the nucleosynthesis epoch) significantly reduces the effectiveness
of monopole-plasma scatterings in maintaining thermal equilibrium of the
monopoles. Thus although the relevant monopolonia are formed when the
monopoles are still in thermal equilibrium, their subsequent ``spiraling
in'' and collapse mostly occurs in a situation when the monopoles are
effectively decoupled from the background plasma. Thus the lifetime of the
relevant monopolonia can be calculated to a good approximation by using
the standard ``vacuum'' dipole radiation formula given by Hill~\cite{hill}

The X particle production from collapsing monopolonia and the resulting
EHECR flux was studied in details in Ref.~\cite{bhatsig}. 
As in the case of collapse and/or successive self-intersections of
cosmic string loops, the X particle production rate $\dot{n}_X(t)$ due to
monopole-antimonopole annihilations through monopolonia formation
turns out to be proportional to $t^{-3}$. 
The efficacy of the process with regard to EHECR, however, depends on two
parameters, namely, (a) the monopolonium-to-monopole fraction at formation 
($\xi_f$) and (b) the monopole abundance. The latter is unknown. However,
for a given monopole abundance, $\xi_f$ is in principle calculable by
using the classical Saha ionization formalism. 

Phenomenologically, since a monopole mass is typically
$m_M\sim 40 m_X$ (so that each monopolonium collapse can release $\sim$ 
80 X particles), we see from Eq.~(\ref{Xrate_required}) that one
requires roughly (only!) a few monopolonium collapse per decade
within roughly every Solar system-size volume over a volume of radius
$\sim$ few tens of Mpc centered at Earth. Whether or not this can
happen depends, as already mentioned, on $\xi_f$ as well as on the
monopole abundance, the condition~\cite{bhatsig} being $(\Omega_M
h^2)h\xi_f \simeq 1.7\times10^{-8} (m_X/10^{16}\gev)^{1/2}
[10\mpc/l(E_\gamma=300\,{\rm EeV})]$, where $\Omega_M$ is the mass density
contributed by monopoles in units of closure density of the Universe. 
Thus, as expected, larger the monopole abundance, smaller is
the monopolonium fraction $\xi_f$ required to explain the EHECR flux.

Note that, since $\xi_f$ must be less than
unity, the above requirements can be satisfied as long as 
$(\Omega_M h^2) h > 1.7\times10^{-8} (m_X/10^{16}\gev)^{1/2}$.  
Recall, in this context, that 
the most stringent bound on the monopole abundance is given by the Parker
bound (see Ref.~\cite{kolb-turner}), 
$(\Omega_M h^2)_{\rm Parker}\la 4\times10^{-3}(m_M/10^{16}\gev)^2$. 
The estimate of $\xi_f$ obtained by using the Saha
ionization formalism~\cite{hill,bhatsig} shows that the resulting requirement
on the monopole abundance (in order to explain the EHECR flux) is
well within the Parker bound mentioned above. The monopolonium collapse, 
therefore, is an attractive scenario in this regard. 
A detailed study of the kinetics of monopolonium formation is needed to
determine the monopolonium fraction at formation. 

The above scenario assumes, of course, that the well-known monopole
overabundance problem (see, e.g., Ref.~\cite{kolb-turner}) is
``solved'' by some mechanism, e.g., inflation, but at the same time
the scenario also assumes that a small but 
interesting relic abundance of monopoles was somehow left behind. 
Such a relic abundance could have been produced, for
example, thermally during the reheating stage after the inflationary
phase. Also, various out-of-equilibrium processes, such as a phase
transition to a transient superconducting phase~\cite{gkt,hkr} giving rise
to transient magnetic flux tubes connecting monopole-antimonopole pairs,
could well enhance monopolonium abundances beyond the equilibrium
estimates mentioned above and be more easily compatible with the required
numbers derived above for explaining the EHECR flux. These possibilities
remain to be studied in detail. 

Recently, the kinetics of monopolonium formation process has been
studied in Ref.~\cite{new-mono} by solving the relevant Boltzmann
equation. The authors of Ref.~\cite{new-mono} claim that the resulting
monopolonium abundance is too low to be able to explain the EHECR flux.
This is based on the observation that the typical energy loss time
scale of monopolonium with the plasma due to friction
is smaller than the Hubble time by a factor $\simeq10m_M/M_{\rm Pl}\gg1$
before recombination such that bound states can be formed only
after recombination.
Instead, Ref.~\cite{new-mono} suggests a different (non-thermal) mechanism
of monopolonium formation in which essentially {\it all} monopoles and
antimonopoles are connected by strings formed at a relatively low energy
phase transition (at $\sim$ 100 GeV). The monopole ``magnetic'' flux is
assumed to be completely confined inside the strings --- the monopoles are
also assumed to have no other unconfined charges ---so that the
monopolonia decay mainly through emission of gravitational radiation
(rather than electromagnetic radiation), with lifetimes comparable to the
age of the Universe. It is claimed that in this scenario the relic
abundance of monopolonia can be sufficient to 
explain the EHECR flux. This mechanism, however, remains to be studied in
detail.  

An interesting possibility is that monopolonia, unlike monopoles, may be
clustered in the Galactic halo. Monopoles may be
accelerated by the Galactic magnetic field causing them to escape 
(``evaporate''~\cite{kolb-turner}) from the halo even if they were
initially clustered there. However, monopolonia, being magnetically
neutral, should be immune to the Galactic magnetic field: it is easy to
check that typical Galactic magnetic field strength of $\sim$ few $\mu\G$
is too weak to ``ionize'' a monopolonium. In this respect, the clustering 
properties of monopolonia in the Galactic halo should be very similar to
the standard Cold Dark Matter. Thus, as mentioned in the case of
vortons above, the density of monopolonia in
the Galactic halo may be significantly enhanced over their average
cosmological density in the Universe~\cite{bbv}. 
This means that the actual universal monopolonium abundance required for
explaining the EHECR could be even lower than the estimates obtained
above assuming unclustered distribution of monopolonia in the
Universe. The signatures of clustered monopolonia as sources of EHECR
will in many respects be similar to those of metastable massive relic
particles discussed in Sect.~6.13. 

\subsection{X Particles from Cosmic Necklaces}
A cosmic necklace is a possible hybrid topological defect consisting of a
closed loop of cosmic string with monopole ``beads'' on it. Such a hybrid
defect was first considered by Hindmarsh and 
Kibble~\cite{beads}. Such hybrid defects could be formed in a
two stage symmetry-breaking scheme such as $G\to H \times U(1) \to H
\times Z_2$. In such a symmetry breaking, monopoles are formed at the
first step of the symmetry breaking if the group is semisimple. In the
second step, ``$Z_2$'' strings are formed, and then each monopole gets
attached to two strings, with monopole magnetic flux channeled along the
string. Possible production of massive X particles from necklaces has 
been pointed out in Ref.~\cite{necklace}.   

The evolution of the necklace system is not well understood. The crucial
quantity is the dimensionless ratio $r\equiv m_M/(\mu d)$, where $m_M$
denotes the monopole mass, $\mu$ is the string energy per unit length, and
$d$ is the average separation between a monopole and its neighboring
antimonopole along the string. For $r\ll 1$, the monopoles play a
subdominant role, and the evolution of the system is similar to that of
ordinary strings. For $r\gg 1$, the monopoles determine the behavior of
the system. Authors of Ref.~\cite{necklace} assume that the system evolves
to a configuration with $r\gg 1$. This is a crucial assumption, which
remains to be verified by numerical simulations. If this assumption holds,
then one may expect that the monopoles sitting on the strings would tend
to make the motion of the closed necklaces aperiodic, leading to frequent
self-intersections of these necklaces and to eventual rapid annihilation 
of the monopoles and antimonopoles trapped on necklaces. This would
lead to X particle production. 

The X particle production rate from necklaces is given by~\cite{necklace} 
\beq
\dot{n}_X \sim\frac{r^2\mu}{m_X t^3}\,.\label{Xrate-necklace}
\eeq
Except for numerical factors, this equation has the same form as
Eq.~(\ref{Xrate-macro-string}) for cosmic string loops with $\mu$
replaced by $r^2\mu$. For
suitable choices of values of $r^2\mu$, necklaces can explain the observed
EHECR. One advantage of the necklace scenario is that, for sufficiently
large values of $r$, the distance between necklaces can be small enough
that sufficient number of necklaces may be expected within a typical
``GZK'' radius of few tens of Mpc. For sufficiently
large $r$, necklaces may also cluster within the Local Supercluster, and
may even cluster on galactic scales~\cite{bbv}. Again, necklaces clustered
in the Galactic halo could be an attractive source of EHECR.

A somewhat related monopole-string system, namely, a network of monopoles
connected by strings~\cite{vach-vil2} as possible sources
of EHECR was studied in Ref.~\cite{bmv}. This system is obtained by
replacing the factor $Z_2$ in the symmetry breaking scheme mentioned above
by $Z_N$ with $N>2$. In this case, after the second stage of symmetry
breaking, each monopole gets attached to $N>2$ strings. Each monopole is
pulled in different directions because of the tension in the strings
attached to it. The net acceleration suffered by a monopole due to
these pulls causes it to radiate gauge boson quanta, mostly photons and
gluons (monopoles carry both ordinary magnetic charge as well as
color-magnetic charge). However, the predicted flux at EHECR energies 
turns out to be too low to explain the observed flux for all reasonable
values of the parameters of the system~\cite{bmv,bbv}. 

\subsection{A General Parametrization of Production Rate of X Particles
from Topological Defects}
From the above discussions on various different kinds of topological
defects as possible sources of EHECR, it is clear that different kinds of
defects in general produce X particles at different rates. It was
suggested in Ref.~\cite{bhs} that X-particle production rate
for any general TD process may, on dimensional grounds, be expressible in
terms of the two fundamental parameters entering in the problem, namely,
the mass-scale $m_X$ (which, in turn, is related to the symmetry-breaking
scale at which the relevant TDs were formed) and the Hubble
time $t$ in the form (in natural units with $\hbar=c=1$)  
\beq
\dot{n}_X (t)=\kappa m_X^p t^{-4+p}\,,\label{Xrate-general-1}
\eeq
where $\kappa$ and $p$ are dimensionless constants whose values depend on
the specific process involving specific TDs under consideration, or
alternatively, in the form 
\beq
\dot{n}_X (t)=\frac{Q_0}{m_X} \left(\frac{t}{t_0}\right)^{-4+p}
\,,\label{Xrate-general-2}
\eeq
where $Q_0$ is the rate of energy injected in the form of X
particles of mass (energy) $m_X$ per unit volume in the present epoch, and
$t_0$ denotes the present age of the Universe. The quantity $Q_0$ depends
on the specific TD process under consideration. 

The above forms for $\dot{n}_X$ are expected to be valid for any TD
systems for which there is no intrinsic time and energy scales involved
other than the Hubble time $t$ and mass scale $m_X$. This is the case in
situations in which the TDs under consideration evolve in a
scale-independent way. As discussed above, this ``scaling'' is indeed a
property of evolution of cosmic strings. The same is true for X 
particle production from monopolonia and necklaces\footnote{In the case 
of hybrid defects such as necklace, there are more than one mass
scales involved. However, the time dependence of $\dot{n}_X$ is still
expressible in the form of Eq.~(\ref{Xrate-general-1}) or
(\ref{Xrate-general-2}); cf. Eq.~(\ref{Xrate-necklace}).}. 

For a given TD process, the quantity $Q_0$ is in principle calculable.
However, in practice, for essentially all kinds of TDs as discussed above,
the evolutionary properties of the TD systems are not known well enough to
allow us to calculate the values of $Q_0$ {\it a priori} in a
parameter-free manner. Nevertheless, for a given value of the parameter
$p$, the above parametrization of the X particle production rate allows us
to study the TD scenario of EHECR in a general way (i.e., without
referring to any specific TD process) by suitably normalizing the value of
$Q_0$ so as to explain the observed EHECR data and then checking to see if
the value of $Q_0$ so obtained is consistent or not with other relevant
data (such as the diffuse gamma ray background in the 10 MeV -- 100 GeV
region; see section 7). 

Except for the cusp evaporation process, other relevant X particle
production processes involving cosmic strings studied so far are  
characterized by Eq.~(\ref{Xrate-general-2}) with $p=1$, as are the
processes involving monopolonia and necklaces. The decaying vorton
scenario is characterized by $p=2$. On the other hand, superconducting
cosmic string scenarios studied so far correspond to $p<1$. 
As we shall discuss in section 7, TD processes with $p < 1$ generally lead
to unacceptably high rate of energy injection in the early cosmological
epochs, which would cause excessive ${}^4{\rm He}$ photo-disintegration
and CMBR distortion~\cite{sjsb}, and are, therefore, currently unfavored
in the context of EHECR. 

\subsection{TDs, EHECR, and the Baryon Asymmetry of the Universe}
In the TD scenario of EHECR origin, the X particles typically belong
to some Grand Unified Theory. The decay of the X particles may,
therefore, involve baryon number violation. Based on this observation,
it has been suggested~\cite{pbbau} that there may be a 
close connection between 
the EHECR and the observed baryon asymmetry of the Universe (BAU). Indeed
it may be the case that both arise from the decay of X particles
released from TDs. Production of X particles from TDs is an
irreversible process, so the standard out-of-thermal-equilibrium
condition necessary for the creation of BAU is (and hence the famous
Sakharov conditions are) automatically satisfied.   

In this scenario, the X particles released from TDs in the early
epochs produce the BAU, whereas the EHECR are due to X
particle decay in the recent epoch. As indicated by 
Eq.~(\ref{Xrate-general-2}), the rate of X particle 
production by TDs in the early epochs was higher than it is
now. Normalizing the present-day rate of X particle production from
the requirement to explain the EHECR flux, the total integrated baryon
asymmetry produced by decays of all X particles released from TDs at
all epochs in the past can be calculated. As pointed out in
Ref.~\cite{pbbau}, depending on
the amount of baryon number violation in each X particle decay, which
unfortunately is unknown and is model dependent, the net baryon
asymmetry produced can account for or at least be a significant
fraction of the observed BAU. Thus, if this scenario is correct, then not
only the extremely {\it high} energy cosmic rays, but the entire {\it low}
energy baryonic content of the Universe today may at some stage or another
have arisen from decay of massive particles from TDs, and the EHECR
observed today would then represent the baryon creation process ``in
action'' in the Universe today. The baryon asymmetry should in principle
be reflected in the observed EHECR, but it will be extremely difficult, if
not impossible, to detect this in EHECR. Realistic calculations including 
all relevant baryon number violating processes within specific 
GUT models will be needed to explore this idea further. 

\subsection{TeV-Scale Higgs X Particles from Topological Defects in
Supersymmetric Theories} 
We have so far dealt with X particles of mass $\gg10^{11}\gev$
produced by topological defects. Recently, however, it has been
realized (see Ref.~\cite{susytd} for details and other references)
that in a wide class of supersymmetric gauge theories, the relevant 
Higgs boson can be ``light'', of mass $m_H\sim\tev$ (the ``soft''
supersymmetry breaking scale), whereas the gauge boson can be much
heavier with mass $m_V\la10^{16}\gev$, the GUT scale. In these
theories, therefore, topological defects can simultaneously be sources
of the Tev mass-scale Higgs bosons {\it as well as} the GUT mass-scale
gauge bosons. It has been suggested~\cite{susytd} that while the
superheavy gauge bosons may act as the X particles generating the
EHECR, there is now an additional direct source of energy injection at
the TeV scale due to decays of the Higgs bosons, which may contribute
significantly to the extragalactic diffuse $\gamma-$ray background
above $\sim$ 10 GeV which also seems to be difficult to explain in terms of
conventional sources. Some implications of this are discussed further
in Sect.~7. 

\subsection{TDs Themselves as EHECR Particles}
Strictly speaking, the subject of this section belongs to Sect.~5 because
the basic ideas discussed below involve acceleration mechanisms rather
than any top-down decay mechanism. Nevertheless, since the objects which
are accelerated are topological defects themselves, we discuss them here.
Two situations have been discussed in literature, involving monopoles and 
vortons: we discuss them in turn. 

\subsubsection{Monopoles as EHECR Particles} 
It has been suggested by Kephart and
Weiler~\cite{kep-wei,wei-kep,weiler-owl},
following an earlier suggestion by Porter~\cite{porter}, that magnetic
monopoles of mass $m_M\sim10^{9}$ -- $10^{10}\gev$ may themselves act as
the EHECR particles. This is an attractive suggestion because, from the
point of view of energetics, monopoles
can indeed be easily accelerated to the requisite EHECR energies by the
Galactic magnetic field. A monopole of minimum Dirac magnetic charge
$q_M=e/2\alpha$ (where $\alpha=e^2/4\pi\simeq1/137$ is the
fine-structure constant) will typically acquire a kinetic energy 
\beqarray
E_K&\sim&q_MBL_c\sqrt{N}\nonumber\\ 
&&\simeq 5.7\times10^{20}\left(\frac{B}{3\times10^{-6}\G}\right)
\left(\frac{L_c}{300\pc}\right)^{1/2}\left(\frac{R}{30\kpc}\right)^{1/2} 
\ev\,\label{monop-ke}
\eeqarray
in traversing through the Galactic magnetic field region of size
$R\sim30\kpc$ containing a coherent magnetic field $B\sim3\times10^{-6}\G$
with a coherence length $L_c\sim300\pc$, where $N\sim R/L_c$ is the
average number of coherent magnetic domains encountered and the $\sqrt{N}$
factor takes account of the random difference in the magnetic field
orientations within different coherent domains. 

In order to ensure that
air-showers induced by monopoles contain relativistic particles, the
monopoles themselves must be sufficiently relativistic which requires that
the monopole mass be $m_M\la10^{10}\gev$. Such relatively low mass
monopoles must be formed at a symmetry-breaking scale $\la10^{9}\gev$.
These monopoles would also be interesting because they would be free of
the usual monopole over-abundance problem associated with GUT-scale
monopoles of mass $\sim10^{17}\gev$ formed at the GUT symmetry-breaking
phase transition at a scale of $\sim10^{16}\gev$. In fact, it is a curious 
coincidence~\cite{kep-wei,wei-kep} that the observed EHECR flux lies just 
three to four orders of magnitude below the ``Parker limit'' (see,
e.g., \cite{kolb-turner}) on the Galactic monopole flux for monopoles of
the required mass $\sim10^{10}\gev$. This interesting coincidence 
has prompted Kephart and Weiler to speculate that
this possible connection between EHECR and monopoles may be a hint towards
some dynamical reason that forces the monopole flux to saturate the Parker
bound\footnote{A scenario in which monopoles ``naturally'' occur with an 
abundance at the level of the Parker saturation limit was discussed
earlier in connection with certain phase transitions in some superstring
theories~\cite{lps}.}.  

There are, however, several uncertainties in this monopole scenario of
EHECR. The precise mechanism behind, and the nature of,  monopole-induced
air showers are largely unknown. A monopole is expected to have an
intrinsic strongly interacting hadronic ``cloud'' around it, of typical
dimension $\sim\Lambda_{\rm QCD}^{-1}\sim$ few fm. Thus monopoles, like
protons, are expected to have a typical strong interaction cross section
for interaction with air nuclei. In addition, a variety of other
monopole-nucleus interactions are possible, such as enhanced
monopole-catalyzed baryon number violating processes with a strong cross
section of $\sim10^{-27}\cm^2$~\cite{rubakov-callan}, possible binding of
nuclei to monopoles~\cite{giacomelli} (in which case the monopole-air
interaction would resemble a relativistic nucleus-nucleus collision),
and so on. Bound states of charged particles and monopoles as EHECR
primaries have also been considered recently in Ref.~\cite{hp}
where it has been suggested that the EAS spectrum created by such
primaries should exhibit a line spectrum component. This specific
prediction should be easy to test with next generation UHECR
experiments. Several other possible strong as well as
electromagnetic interactions of monopoles with nuclei
are mentioned by Weiler~\cite{weiler-owl}. One of the major problems,
however, is that although in many cases the relevant cross sections can be
large, the required large inelasticities (i.e., large energy
transfers) are generally difficult to realize for a massive particle like
the monopole~\cite{mohap-nussi}. 

Another problem arises from considerations of distributions of arrival
directions of individual EHECR events~\cite{escobar}. The arrival
directions of monopole primaries are expected to show preference for the
local Galactic magnetic field directions. However, the arrival
directions of the observed EHECR events seem to show no such
preference~\cite{escobar}. Moreover, Monte Carlo   
calculations~\cite{escobar} indicate that the expected spectrum of
monopoles accelerated in the Galactic magnetic field is very different
from that of the observed UHECR. However, the spectrum beyond $10^{20}\ev$
is not yet well-measured, and only future measurements of the EHECR
spectrum by the up-coming large-area detectors will hopefully be able to
settle this issue. Thus, it is not yet possible to completely rule out
the monopole scenario of EHECR. However, it seems rather 
unlikely~\cite{mohap-nussi} at the present time. 
Clearly, more theoretical work especially on the air-showering
aspects of monopoles are needed. 

\subsubsection{Vortons as EHECR Particles}
Vortons were already discussed in Sect.~6.6. Vortons are highly charged
particles and can, therefore, be accelerated in powerful astrophysical 
objects such as AGNs, radio galaxy hot-spots, and so on, if vortons are
present in those objects with sufficient abundance --- a possibility if
vortons are (at least a part of) the ubiquitous dark matter. Authors of
Ref.~\cite{bonazzola-peter} have proposed vortons of mass $m_V\sim Zm$
with $Z\sim100$ and $m\sim\eta_s\sim\eta_\sigma\sim10^{9}\gev$ as the
EHECR particles. Like the case of monopoles discussed above, 
accelerating vortons to the requisite energies seems to be no problem. The
main problem, however, is that the interaction properties of vortons with
ordinary matter and the kind of atmospheric air-showers they are likely to 
generate are highly uncertain, and so nothing much definite can be
said about this possibility. For details on one particular model studied
so far, see Ref.~\cite{bonazzola-peter}, which suggests that vortons
should produce a line spectrum component of EAS, similar to the
case of charged particle-monopole bound state primaries~\cite{hp}.
 
\subsection{EHECR from Decays of Metastable Superheavy Relic Particles } 
\subsubsection{General Considerations}
It has been suggested recently by Kuzmin and Rubakov~\cite{kuz-rub} and by
Berezinsky, Kachelrie\ss, and Vilenkin~\cite{bkv} (see also
Ref.~\cite{frampton}) that
EHECR may be produced from decay of some metastable superheavy relic
particles (MSRPs) of mass $m_X\ga10^{12}\gev$ and lifetime larger than or
comparable to the age of the Universe\footnote{This possibility, with a
specific MSRP candidate from superstring theory called
``crypton''~\cite{crypton} in mind, was noted several years ago in a {\it
footnote} in Ref.~\cite{pbkofu}. However, it was not explored there further.}. 
The long but finite lifetime of MSRPs could be due to
slow decay of MSRPs through non-perturbative instanton 
effects~\cite{kuz-rub} or through quantum gravity (wormhole)
effects~\cite{bkv} which induce small violation of some otherwise 
conserved quantum number associated with the MSRPs. The (almost) conserved
quantum number could be due to some global discrete symmetry, for example.
In other words, but for the instanton and/or quantum gravity effects, the
MSRPs would have been absolutely stable, and their long but finite
lifetime is then due to small violation of some protector symmetry. 

Possible candidates for MSRPs and their possible decay
mechanisms giving them long lifetime have been discussed in the
context of specific particle physics/superstring theory models in 
Refs.~\cite{frampton,benakli,birkel-sarkar,hamaguchi,nano-talk}.
Several non-thermal mechanisms of production of MSRPs in the
post-inflationary epoch in the early Universe have been studied. They
include gravitational production through the effect of the expansion of
the background metric on the vacuum quantum fluctuations of the
MSRP field~\cite{ckr1}, or creation during reheating at
the end of inflation if the MSRP field couples to the inflaton
field~\cite{kuz-rub,ckr2}. The latter case can be divided into
three subcases, namely ``incoherent'' production with an
abundance proportional to the MSRP annihilation cross section,
non-adiabatic production in broad parametric resonances with
the inflaton field during preheating, and creation in bubble wall 
collisons if inflation is completed by a first
order phase transition~\cite{wimpzillas}. A recent review of MSRP
production mechanisms associated with inflation is given in
Ref.~\cite{kuz-tkachev-rev}. 

Under certain circumstances MSRPs can even exist in the present-day
Universe with sufficient abundance so as to act as non-thermal
superheavy dark matter. The MSRPs of mass $\ga10^{12}\gev$, 
if they are to act as superheavy
dark matter, would have to have been created in non-thermal processes,
because the unitarity limit on the self-annihilation cross section
imposes~\cite{greist-kamion} an upper bound of $\sim$500 TeV on the 
mass of any relic dark matter particle species that 
was once in thermal equilibrium in the early Universe. Another
possibility is the thermal production of such MSRPs with a
subsequent substantial entropy production, for example,
by thermal inflation~\cite{aky}.

Various phenomenological aspects of the MSRP decay 
scenario of EHECR origin has recently been studied further in 
Refs.~\cite{birkel-sarkar,dub-tin,bbv,bsw,berez-mikhail,ck}.
In order to explain the observed EHECR flux, a
specific relationship between the abundance of MSRPs and their lifetime
must be satisfied. We can see this from the
discussions already presented in Sect.~6.2.3. Let us assume for simplicity
that each MSRP decay produces roughly a similar number of quarks and/or
leptons as in the case of X particles from topological defects (however,
this need not~\cite{kuz-rub} be the case ---see
below). Also let us first assume that the MSRPs (we shall call them 
X particles throughout this section) are
distributed uniformly across the Universe, 
and as in Sect.~6.2.3 we consider decay of the X particles in
the present epoch only. The decay rate is simply $\dot{n}_X=n_X/\tau_X$,
where $\tau_X$ is the lifetime as defined through the usual exponential
decay law. The quarks from the decay of these X particles will hadronize
in the same way as discussed in Sect.~6.2, producing a photon and neutrino
dominated injection spectrum. 
Then using Eq.~(\ref{Xrate_required}) with $\alpha=1.5$ we see
that in order to produce the observed flux of EHECR (assuming
they are photons), we require 
\beq
\tau_X\simeq2.8\times10^{21}(\Omega_X\,h^2)\left(\frac{l_E}{100\mpc}
\right)\left(\frac{10^{12}\gev}{m_X}\right)^{1/2} 
\yr\,,\label{tau-omega-rel} 
\eeq
where $\Omega_X$ is the cosmic average mass density contributed by the
MSRPs in units of the critical density
$\rho_c\simeq1.05\times10^{-4}h^2\gev\cm^{-3}$, $h$ is the present
value of Hubble constant in units of $100\km\sec^{-1}\mpc^{-1}$, and
$l_E$ is the effective penetration depth of UHE radiation.

Thus, for $\Omega_Xh^2\sim1$ (maximal value allowed), one requires
$\tau_X\sim3\times10^{21}\yr$ for $m_X=10^{12}\gev$ and
$l_E=100\mpc$. Recall that in order to
produce the EHECR particles (assuming they are nucleons and/or photons),
the relic X particles must decay in the present epoch. Thus $\tau_X$
cannot be much smaller than $\sim10^{10}\yr$ because then most of the
MSRPs would have already decayed in earlier epochs\footnote{On the other
hand, $\tau_X\gg t_0$ does not of course mean that no MSRPs are decaying
currently. Indeed, as mentioned already, the current decay rate of the X
particles is simply $n_X/\tau_X$.}. With $\tau_X\sim t_0\sim10^{10}\yr$,
one needs only a tiny MSRP abundance $\Omega_Xh^2\sim3\times10^{-12}$
(for $m_X=10^{12}\gev$) to explain the EHECR flux.  

Currently, neither the abundance $\Omega_X$ nor the lifetime $\tau_X$ of 
the proposed MSRPs is known with any degree of confidence. Clearly, in
order to produce (but not overproduce) the EHECR flux, some fine-tunning
between
$\Omega_X$ and $\tau_X$, which should approximately satisfy 
the condition given by Eq.~(\ref{tau-omega-rel}), is needed. Note that
since $m_X$ has to be $\ga10^{12}\gev$ (to explain the EHECR), the
MSRP lifetime cannot exceed $\sim10^{22}\yr$. 

It is expected that the MSRPs would behave like the standard cold dark
matter (CDM) particles in the Universe. If $\Omega_Xh^2\sim1$, then MSRPs
would {\it be} the CDM or at least a significant part of it. The important
point noticed in Ref.~\cite{bkv} is
that irrespective of the value of the universal MSRP density $\Omega_X$,
the ratio $\xi_X=\rho_X/\rho_{\rm CDM}$ of mass density contribution from 
the X particles to that from CDM particles 
should be roughly same everywhere in the Universe, since both X
particles and CDM particles would respond to gravity in the same way.
Thus, the value of $\xi_X$ within individual galactic halos should be same
as that in the extragalactic space. Thus, since CDM particles (by
definition) are clustered on galactic halo scales, so will be the X
particles, and that by the same factor. In particular, MSRPs are expected
to be clustered in our own Galactic Halo (GH). 
Using the solar neighborhood value of the DM density in the GH, 
$\rho_{{\rm CDM},\odot}^{\rm H}\simeq0.3\gev\cm^{-3}$ (see, e.g.,
Ref.~\cite{kolb-turner}) as a reference value for the average CDM density
$\rho_{\rm CDM}^{\rm H}$ in the GH, one can write the average number
density of X particles in the GH as 
\beq
n_X^{\rm H}\simeq3\times10^{-13}\left(\frac{\Omega_X}{\Omega_{\rm
CDM}}\right)\left(\frac{10^{12}\gev}{m_X}\right)
\left(\frac{\rho_{\rm CDM}^{\rm H}}{0.3\gev\cm^{-3}}\right)
\cm^{-3}\,,\label{n_X-halo} 
\eeq
where $\Omega_{\rm CDM}=\rho_{\rm CDM}/\rho_c$ gives the
cosmic average mass density of CDM. Noting that the cosmic average
number density of X particles is $n_X^{\rm cos}\simeq10^{-16}
(\Omega_Xh^2)(10^{12}\gev/m_X)\cm^{-3}$, we see that the GH 
contribution to the predicted EHECR flux from MSRP decay exceeds the
extragalactic (EG) MSRP decay contribution by a factor $f$ given by 
\beq
f\equiv\frac{j_{\rm H}}{j_{\rm ex}}=\zeta\frac{n_X^{\rm H}}{n_X^{\rm
cos}}\frac{R_{\rm H}}{l_E}\simeq 15\zeta\left(\frac{0.2}
{\Omega_{\rm CDM}h^2}\right)\left(\frac{R_{\rm H}}
{100\kpc}\right)\left(\frac{100\mpc}{l_E}\right)
\,,\label{halo-flux-enhance}
\eeq
where $R_{\rm H}$ is the radius of the GH, and $\zeta$ is a geometric
factor of order unity determined by the spatial distribution of the 
X particles in the Halo. If the X particles 
constitute a dominant fraction of the CDM in the GH, then 
their density must fall off as $1/r^2$ at large Galactocentric
distances in order to explain the flat rotation curve of the Galaxy. 
In the situation in which the X particles do not constitute the dominant
component of the CDM in the Galaxy, the spatial distribution of X
particles in the GH may, depending on the nature of their interaction
with other matter, be different from that of CDM. However,
if the X particles are only very weakly interacting with other
particles --- interacting mainly through gravity --- then they will be
expected to be distributed like the CDM irrespective of their density
contribution. Thus, for an isothermal density profile~\cite{caldwell}, for
example~\cite{dub-tin,bbv,bsw,berez-mikhail}, $\zeta\simeq6.7$. For
another model~\cite{navarro}, $\zeta\simeq 2$. 

From the above discussion, it is clear that in this scenario the
contribution of X particles to the EHECR flux will be dominated by the GH 
component over the EG component by a factor of order 10 or more {\it
except} probably for the case of neutrino flux for which one should replace
$l_E$ in the above equation by the neutrino
absorption length, $\gg100\mpc$. For neutrinos 
the EG contribution would be comparable to or greater than the GH
contribution. For UHECR protons with energy $E\ll E_{\rm
GZK}\sim5\times10^{19}\ev$, the effective energy attenuation length scale
becomes large ($\gg100\mpc$), so the EG proton component can also be
comparable to the GH proton component at these energies. 

The immediate conclusion~\cite{bkv,bbv} that follows from the above
discussion is that in this GH dominated X particle decay (GHXPD) scenario,
the spectrum of UHECR, which is predicted to be photon dominated, should
show (almost) complete {\it absence} of the GZK suppression. The required
energy injection rate in the form of X particles
is also not constrained by the EGRET measured $\gamma-$ray
background because there is practically no EHE $\gamma-$ray cascading 
--- $R_{\rm H}$ is too small compared to EHE photon interaction length of
$\ga10\mpc$ (see Fig.~\ref{F4.4}). The apparent absence of the GZK
suppression in the recent UHECR data from the AGASA
experiment~\cite{agasa} may be taken as indicative~\cite{hillas-nature}
of support to the GHXPD scenario, although the AGASA EHECR data can also
be consistent with a general EG X particle decay scenario (involving
topological defects, for example) with $p=1$ of
Eq.~(\ref{Xrate-general-2}) (see Ref.~\cite{slby} for the most
recent calculations). Note in this context that a general top-down
scenario in which X particles are MSRPs can be effectively described 
by Eq.~(\ref{Xrate-general-2}) with $p=2$. Also, as mentioned earlier,
some kinds of topological defect sources of (unstable) X particles 
such as monopolonium, vortons, and possibly necklaces can also be
clustered in the GH, and will produce EHECR spectrum having roughly
the same characteristics as that in the MSRP related GHXPD scenario
being discussed here. An example of the fluxes predicted in the
GHXPD scenario is shown in Fig.~\ref{F7.4}.

Note that with the GH contribution dominating over the
EG component, the condition expressed by
Eq.~(\ref{tau-omega-rel}) should be modified --- the r.h.s of
Eq.~(\ref{tau-omega-rel}) will now get multiplied by the factor $f$ of
Eq.~(\ref{halo-flux-enhance}). Thus, for a given $\tau_X$ the required
universal abundance of MSRPs, $\Omega_X$, will be a factor $f$ lower than
what is indicated by Eq.~(\ref{tau-omega-rel}). 

Finally, we note that the number of quarks and/or leptons resulting
from the decay of the MSRP X particle has been assumed to be around 2
or 3 in practically all 
calculations done so far. This is reasonable in a perturbative
decay picture. On the other hand, as pointed out by Kuzmin and
Rubakov~\cite{kuz-rub}, instanton mediated decay processes 
(which may be needed for making the MSRPs sufficiently
long-lived) typically lead to multiparticle final states. Thus
instanton mediated X particle decays will typically give a relatively
large number ($\sim$10) of quarks with a fairly flat distribution in
energy, rather than typically 1 or 2 quarks expected in the
perturbative X decay scenario implicitly assumed, for example, 
in the topological defect scenario. In other words, the numbers
$\tilde{N}$ and $N_q$ used in the discussion of Sect.~6.2.2 may be
quite different from what are usually assumed in the top-down
scenario. This may leave a distinguishing 
signature in the predicted spectrum which may help in distinguishing
the topological defect scenario from the MSRP scenario~\cite{kuz-rub}. 

\subsubsection{Anisotropy}
An important signal of the GHXPD scenario of EHECR, pointed out 
by Dubovsky and Tinyakov~\cite{dub-tin}, is the expected anisotropy of
the EHECR flux. Because of the off-center location of the
Solar system with respect to the (assumed spherical) GH, some amount
of anisotropy of the EHECR flux measured at Earth is expected. This is
important because
an experimental confirmation of the {\it absence} of the predicted
anisotropy will be a definite evidence {\it against} the GHXPD model. The
expected anisotropy has been calculated in
Refs.~\cite{dub-tin,bbv,bsw,berez-mikhail} in several different GH models.
The predicted anisotropy varies from about 10\% to 40\% depending on the
parameters of the GH model. The strongest measure of the anisotropy in
this model is associated with the large ratio of the predicted UHECR flux from
the direction of the Galactic Center (GC) to that from the
direction of the Galactic Anticenter (GA)~\cite{dub-tin,bbv}.
Unfortunately, except for one southern-hemisphere EAS array, namely, 
the now-defunct SUGAR array in Sydney, Australia, none of the
major currently operating as well as closed EAS arrays (AGASA, Fly's Eye,
Haverah Park, Yakutsk --- all located in the northern hemisphere) 
can ``see'' the GC. So this strongest signal of the
predicted anisotropy will have to wait to be tested until the southern
hemisphere detector of the Pierre Auger project is built and made
operational. In the meantime, large northern hemisphere EAS arrays 
should see
a dip in the measured EHECR flux from the direction of the GA relative to
the direction perpendicular to the Galactic plane~\cite{bbv}. 

Another strong signal of the GHXPD model are the expected ``dips'' in the
GH flux enhancement factor $f$ of Eq.~(\ref{halo-flux-enhance}), due
to direct flux from relatively close-by EG dark matter clumps
represented by objects such as the the Andromeda galaxy or the Virgo
cluster~\cite{bkv,bbv,dub-tin,bsw}. 

According to Ref.~\cite{berez-mikhail}, in the
case in which the GH component becomes dominant over the EG component at
$E\ga10^{19}\ev$, the latest AGASA data already rule out the GHXPD model. 
On the other hand, in the case in which the GH component makes the
dominant contribution to the flux measured by AGASA only at
$E\ga10^{20}\ev$, the model cannot yet be ruled out even for the largest
predicted anisotropy in the model: the number of events ($\sim 6$) is too
low to allow any statistically significant conclusion to be drawn. 
On the other hand, according to the analysis of Ref.~\cite{bsw} (BSW) which
includes the data from the now defunct SUGAR array in the analysis,
the GHXPD model is ruled out, because these authors claim that the
predicted anisotropies are much higher than those observed and that
the predicted flux from the Andromeda galaxy is not seen. On the basis
of their analysis, BSW claim that the GHXPD scenario cannot contribute
more than $\sim10\%$ to the observed UHECR flux. However, Berezinsky and
Mikhailov~\cite{berez-mikhail} have pointed out that in the light of
the latest AGASA data~\cite{agasa} the GHXPD contributes dominantly
to the observed flux only at energies $E>10^{20}\ev$ where the SUGAR
array most probably observed no events, so the inclusion of the SUGAR
array data could bias the analysis significantly. Clearly, this
important question of anisotropy remains unresolved. In addition,
the recent evidence for small scale clustering found by the AGASA
experiment~\cite{agasa2} may require a significant clustering
of the dark matter component consisting of the decaying X particles
on scales of the Galactic halo.
As already mentioned, the up-coming large arrays, especially the
southern hemisphere array of the Pierre Auger project will have the
capability to unambiguously resolve the issue. 
For a recent more detailed study of the question of anisotropy in the
GHXPD model, see, Ref.~\cite{tanco-watson}. 

In scenarios where neutrinos have a mass in the eV range and
UHE neutrinos from MSRP decays at large
cosmological distances $\gg100\,$Mpc give rise to EHECR
within 100 Mpc from Earth through decay of Z bosons resonantly
produced by interactions of the primary neutrino with the
relic neutrinos (see Sect.~4.3.1.), the expected EHECR
angular distribution is not expected to correlate with the
GH. Rather, the EHECR anisotropy is then expected to correlate
with the large scale structure at a certain redshift range
which is determined by the resonance condition for the
annihilating primary neutrinos~\cite{gk2}.

\subsection{Cosmic Rays from Evaporation of Primordial Black Holes}
A recent proposal that does not involve the production of
X particles and therefore, in the strict sense does not belong
to the top-down scenarios, concerns the production of UHECR during
evaporation of primordial black holes~\cite{barrau}. It was
claimed that the required black hole abundance is consistent
with observational constraints on the local rate of
black hole evaporation. This mechanism may, however, not
contribute to the UHECR flux at all due to the formation
of a photosphere around the black hole whose typical
temperature is of the order of the QCD scale~\cite{heckler}.
The photosphere strongly reduces the possibility of observing individual
black holes with temperatures greater than the QCD scale, since the
high energy particles emitted from the black hole are recycled
to lower energies in the photosphere. However, primordial
black holes may provide a significant contribution to the
Galactic flux of protons, electrons, their antiparticles,
and $\gamma-$rays in the 100 MeV range, a fact that
has extensively been used to in turn constrain the number
of black holes formed in the early Universe~\cite{cm-review}.

\section{Constraints on the Topological Defect Scenario}

Scenarios of UHECR production that are related to new physics
near the Grand Unification scale exhibit a striking difference
to conventional acceleration models: Whereas acceleration is
tied, in one form or another, to astrophysical objects and
magnetized shocks associated with them and took place at
redshifts not greater than a few, energy release associated with
Grand Unification scale physics takes place not only today, but
already in the early Universe up to temperatures corresponding
to the Grand Unification scale. Consequently, this energy
release --- that can roughly be estimated by normalizing its present
day rate to the value required to explain the observed UHECR flux --- 
can imply other effects that are constrained by
observations. Such effects are a diffuse $\gamma-$ray background
below $\sim100\,$GeV (see Sect.~7.1) and its influence on light
element abundances through photo-disintegration of light nuclei 
(Sect.~7.2), and on
distortions of the CMB (Sect.~7.3), as well as comparatively
high diffuse neutrino fluxes above $\sim10^{18}\,$eV (Sect.~7.4).

\subsection{Low-Energy Diffuse $\gamma-$ray Background: Role
of Extragalactic Magnetic Field and Cosmic Infrared Background}

In Sect.~4.2 the development of EM cascades and the resulting
diffuse $\gamma-$ray background were discussed qualitatively,
and Sect.~4.4.1 addressed how magnetic fields can modify the
picture. Here, we demonstrate concrete implications of these
general aspects for the low-energy diffuse $\gamma-$ray
background predicted by TD scenarios.

A simple analytical estimate of the saturated cascade $\gamma-$ray 
spectrum can be given in the case when the CMB is the only relevant low
energy photon background in which the cascading takes place ($E$
is now the photon energy)~\cite{bbdgp,cas1}:
\begin{equation}
  j_\gamma^{\rm cas}(E)\simeq
  \frac{\omega_{\rm cas}}{E_x^2(2+\ln(E_c/E_x))}\times
  \cases{(E/E_x)^{-1.5} & for $E<E_x$ \cr
         (E/E_x)^{-2} & for $E_x<E<E_c$ \cr
         0 & for $E_c>E$ \cr}\,.\label{casspec}
\end{equation}
Here, $\omega_{\rm cas}$ is the total release of EM energy
per unit volume above $E_c$, and $E_c$ and $E_x$ are characteristic
redshift dependent energies given
by~\cite{cas2}
\begin{eqnarray}
  E_c&\simeq&\frac{m_e^2}{22T_{\rm
  CMB}}\simeq\frac{4.9\times10^4\,{\rm GeV}}{1+z}\,,\nonumber\\
  E_x&\simeq&0.04E_c\simeq\frac{2\times10^3\,{\rm GeV}}{1+z}
  \,,\label{ecrit}
\end{eqnarray}
where $T_{\rm CMB}\simeq2.735(1+z)\,$K is the CMB temperature at the epoch
characterized by redshift $z$. 
Ref.~\cite{zdziarski} comes to a very similar result for the
cascade spectrum for all photon background spectra that
fall off steeper than $\varepsilon^{-2}$. It was furthermore
pointed out that near $E_c$ the spectrum steepens to $E^{-5}$
when photon-photon scattering becomes important~\cite{sz}.
This effect, however, becomes important only at redshifts
larger than about 100 and therefore does not play a role in viable
TD scenarios where most of the cascade flux is created at
lower redshifts. The influence of photon backgrounds at energies
above the CMB energies such as the IR/O backgrounds can
approximately been taken into account
by multiplying the right hand side of Eq.~(\ref{casspec}) with 
$\exp[-\tau(E)]$,
where $\tau(E)$ is the optical depth for pair creation
on these backgrounds at energy $E$ over the distance to
the source.

\begin{figure}[ht]
\centering\leavevmode
\epsfxsize=6.0in
\epsfbox{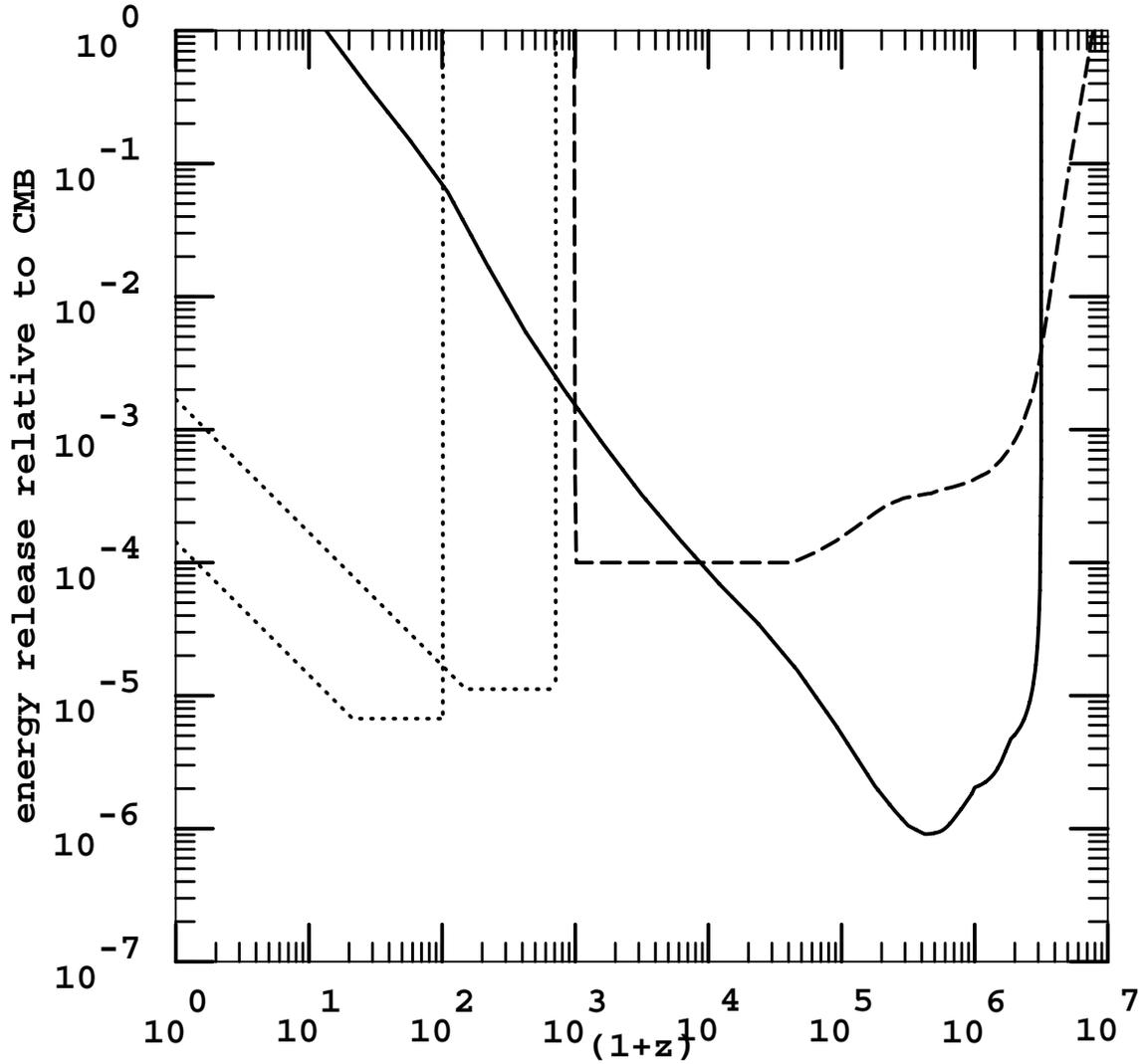}
\bigskip
\caption[...]{Maximal energy release in units of the
CMB energy density allowed by the constraints from the observed
$\gamma-$ray background~\cite{cdkf} at $100\,$MeV (dotted curve
limiting left most range) and $5\,$GeV (dotted curve limiting
next to left most range), CMB distortions
(dashed curve, from Ref.~\cite{wright}), and $^4$He photo-disintegration
as a function of redshift $z$.
These bounds apply for instantaneous energy release at the
specified redshift epoch. 
\label{F7.1}}
\end{figure}

Due to the short interaction lengths for PP and ICS above $E_c$,
the cascade spectrum forms essentially instantaneously on
cosmological time scales. For
$E\la E_c$, PP is strongly suppressed and ordinary
pair production, i.e. the Bethe-Heitler (BH) process, and
Compton scattering (CS) become the dominant processes. The
photon energy attenuation length is then given by
\begin{equation}
  l_E^{-1}(E)=n_B\left[(1-Y)\sigma_{\rm H}^{\rm BH}(E)
  +(Y/4)\sigma_{\rm He}^{\rm BH}(E)+(1-Y/2)
  \eta_{\rm CS}(E)\sigma_{\rm CS}(E)\right]
  \,,\label{lEgamma}
\end{equation}
where $n_B$ is the baryon number density, $Y$ is the mass
fraction of $^4{\rm He}$, $\sigma_{\rm H}^{\rm
BH}$ and $\sigma_{\rm He}^{\rm BH}$ are
the BH cross sections on hydrogen and $^4{\rm He}$, and $\sigma_{\rm
CS}$ and $\eta_{\rm CS}$ are cross section and inelasticity for
CS. For $z\la10^3$, $l_E(E)\ga H_0^{-1}(1+z)^{-3/2}$, and
the Universe is therefore transparent for the cascade photons
which can be directly observed as diffuse $\gamma-$ray
background. In Fig.~\ref{F7.1} the maximal instantaneous
release of energy  $\omega_{\rm cas}$ that results from comparing
Eq.~(\ref{casspec}) with the observed $\gamma-$ray background
at $200\,$MeV is shown as a function of redshift.

The measurement of the diffuse $\gamma-$ray
background between 30 MeV and 100 GeV~\cite{cdkf} imposes
constraints on the total amount of EM energy density 
$\omega_{\rm cas}\simeq4.5\times10^{-6}\,{\rm eV}\,{\rm cm}^{-3}$
injected and recycled by cascading to lower energies. These constraints
and their role for TD models have
first been pointed out in Ref.~\cite{chi}. There, cascading in
the CMB and the IR/O backgrounds was considered for EM injection
at comparatively low redshifts. An analytical estimate based on
Eq.~(\ref{casspec}) with $\omega_{\rm cas}$ determined from the EHE
$\gamma-$ray fluxes in CEL approximation, Eq.~(\ref{cel_diff}),
neglecting any EGMF, was given in Ref.~\cite{sjsb}. Since the
diffuse $\gamma-$ray spectrum observed between $\simeq100\,$MeV
and $\simeq100\,$GeV is roughly $\propto
E^{-2.1}$~\cite{cdkf}, the most stringent limit is
obtained at the highest energy. As a result, scenarios with
$p=0$ such as the simplest TD models involving superconducting
cosmic strings are ruled out altogether~\cite{sjsb}. Further, in
case of a power law fragmentation function, scenarios
with $p=1$ such as models involving ordinary cosmic strings or
annihilation of magnetic monopoles and antimonopoles require the
power law index to satisfy
\begin{equation}
  q\ga2-\frac{3/2}{3+\log_{10}(m_X/10^{23}\,{\rm eV})}
  \,,\label{qconstr}
\end{equation}
e.g., $q\ga1.7$ for a GUT scale mass $m_X=10^{16}\,$GeV.

\begin{figure}[ht]
\centering\leavevmode
\epsfxsize=6.0in
\epsfbox{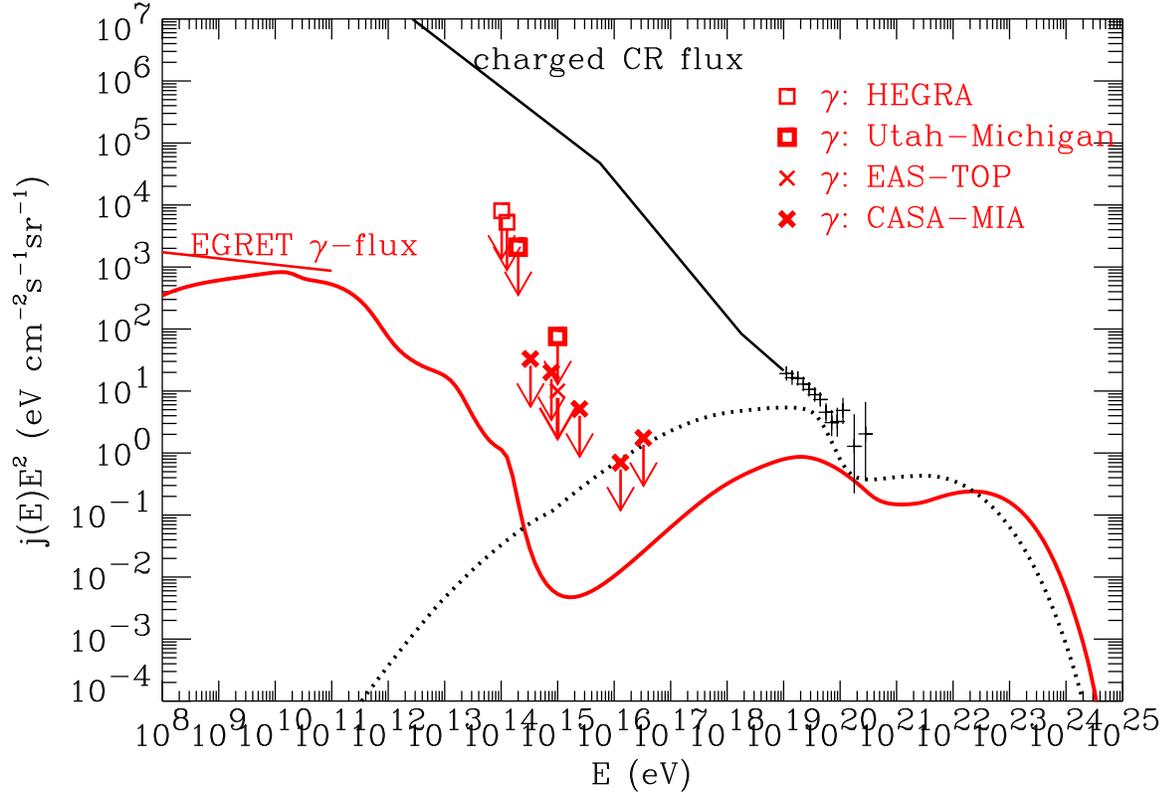}
\bigskip
\caption[...]{Predictions for the differential fluxes of
$\gamma-$rays (solid line) and protons and neutrons (dotted
line) in a TD model characterized by $p=1$, $m_X = 10^{16}\,$GeV,
and the decay mode $X\to q+q$, assuming the supersymmetric modification of
the 
fragmentation function, Eq.~(\ref{mlla-spectrum}), with a fraction of about
10\% nucleons. The calculation
used the code described in Ref.~\cite{slby} and assumed the strongest
URB version shown in Fig.~\ref{F4.3} and an EGMF $\ll10^{-11}\,$G.
1 sigma error bars are the combined data from the Haverah Park~\cite{haverah},
the Fly's Eye~\cite{fe}, and the AGASA~\cite{agasa} experiments
above $10^{19}\,$eV. Also shown are piecewise power law fits to the observed
charged CR flux (thick solid line) and the EGRET measurement
of the diffuse $\gamma-$ray flux between 30 MeV and 100 GeV~\cite{cdkf}
(solid line on left margin). Points with arrows
represent upper limits on the
$\gamma-$ray flux from the HEGRA~\cite{hegra2}, the
Utah-Michigan~\cite{utahmich}, the EAS-TOP~\cite{eastop1}, and
the CASA-MIA~\cite{casa2} experiments, as indicated.
\label{F7.2}}
\end{figure}

\begin{figure}[ht]
\centering\leavevmode
\epsfxsize=6.0in
\epsfbox{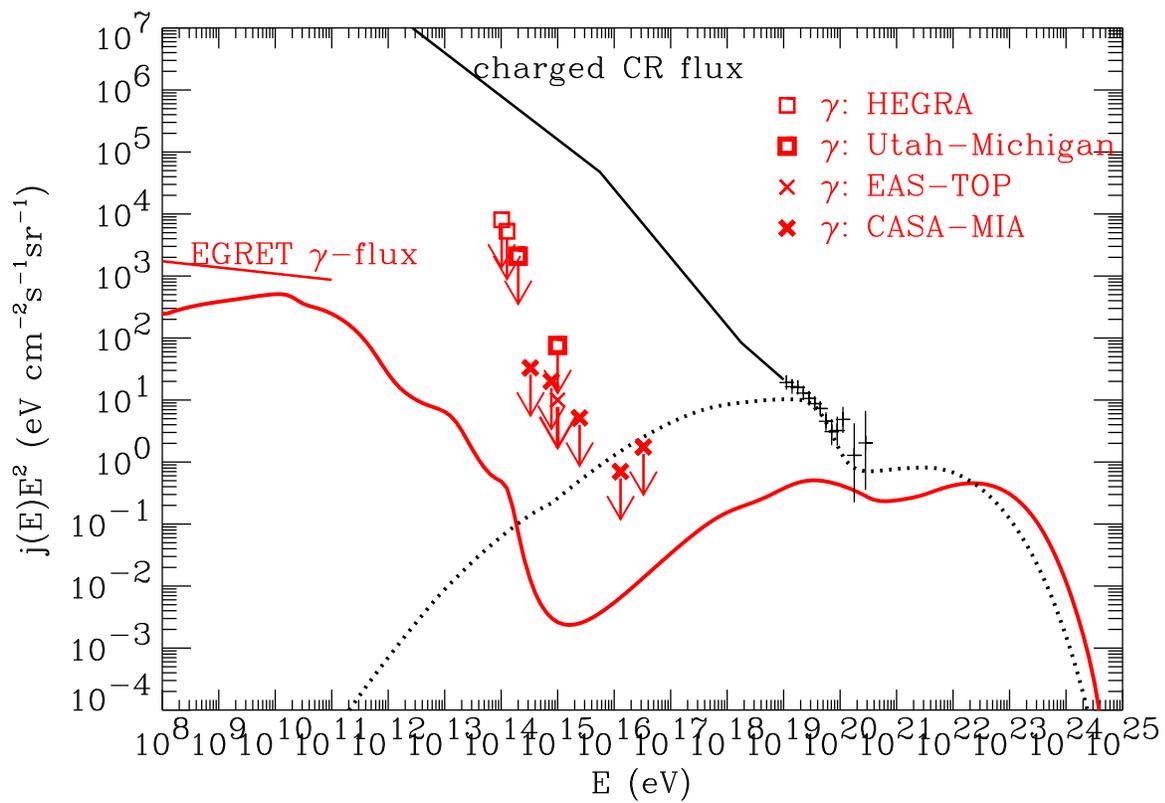}
\bigskip
\caption[...]{Same as Fig.~\ref{F7.2}, but for an EGMF of
$10^{-9}\,$G.
\label{F7.3}}
\end{figure}

More accurate numerical calculations take into account
the IR/O background as well as any EGMF and the development of
unsaturated cascades at UHE and its impact on the normalization
of the energy release. For the following the new estimates
of the IR background~\cite{irb} are assumed. In addition, 
to be conservative in terms of scenarios obeying all
constraints, the strongest URB version from Ref.~\cite{pb} 
( see Fig.~\ref{F4.3}) is assumed.
Fig.~\ref{F7.2} shows results from Ref.~\cite{slby} for the
time averaged $\gamma-$ray and nucleon
fluxes in a typical TD scenario, assuming no
EGMF, along with current observational constraints on
the $\gamma-$ray flux. The spectrum was optimally normalized 
to allow for an explanation of the observed EHECR
events, assuming their consistency with a nucleon or
$\gamma-$ray primary. The flux below $\la2\times10^{19}\,$eV
is presumably due to conventional
acceleration in astrophysical sources and was not fit. Similar spectral
shapes have been obtained in Ref.~\cite{ps1}, where the normalization
was chosen to match the observed differential flux at
$3\times10^{20}\,$eV. This normalization, however, 
leads to an overproduction of the
integral flux at higher energies, whereas above $10^{20}\,$eV,
the fits shown in Figs.~\ref{F7.2} and~\ref{F7.3} have
likelihood significances above 50\% (see Ref.~\cite{slsb} for
details) and are consistent with the integral flux above
$3\times10^{20}\,$eV estimated in Refs.~\cite{fe,agasa}.
The PP process on the CMB depletes the photon flux above 100 TeV, and the
same process on the IR/O background causes depletion of the photon flux in
the range 100 GeV--100 TeV, recycling the absorbed energies to
energies below 100 GeV through EM cascading (see Fig.~\ref{F7.2}).
The predicted background is {\it not} very sensitive to
the specific IR/O background model, however~\cite{ahacoppi}.
The scenario in Fig.~\ref{F7.2} obviously
obeys all current constraints within the normalization
ambiguities and is therefore quite viable.

We mention at this point, however, that TD scenarios are
constrained by the true extragalactic
contribution to the diffuse $\gamma-$ray background which might
be significantly smaller than limits on or measurements of the
diffuse $\gamma-$ray flux. This was indeed suggested
recently~\cite{swz} for the EGRET measurements near $1\,$GeV which
could be dominated by contributions from the Galactic halo. In
addition, the bulk of the extragalactic $\gamma-$ray background
may be caused by unresolved blazars~\cite{ss}, although there is some
disagreement on this issue~\cite{mp}; for example, a recent analysis
indicates that only $\simeq25\%$ of the diffuse extragalactic
emission can be attributed to unresolved $\gamma-$ray
blazars~\cite{muk-chiang}. Other sources such as
secondary $\gamma-$rays from the interactions of CR confined in
galaxy clusters also contribute to the extragalactic $\gamma-$ray
background~\cite{bbp}. In any of these cases the
constraints on energy injection in TD models discussed here
would consequently become more stringent, typically by a factor
2--3. However, above $\simeq10\,$GeV, another component might
be necessary and a
cascade spectrum induced by decay of heavy particles of energy
beyond $\simeq100\,$GeV seems to fit the observed spectrum between
10 and $100\,$GeV~\cite{susytd}, as can also be seen in Fig.~\ref{F7.2}.
Furthermore, the
$\gamma-$ray background constraint can be circumvented by
assuming that TDs or the decaying long lived X particles
do not have a uniform density throughout the
Universe but cluster within galaxies~\cite{bkv}. An example for this
case will be discussed further below.

Fig.~\ref{F7.3} shows results for the same TD scenario as in
Fig.~\ref{F7.2}, but for a high EGMF $\sim 10^{-9}\,$G
(somewhat below the current upper limit~\cite{kronberg}). 
In this case, rapid synchrotron cooling of the initial cascade pairs quickly
transfers energy out of the UHE range. The UHE $\gamma-$ray flux then depends
mainly on the absorption length due to pair production and is typically
much lower~\cite{abs,los}. (Note, though, that for $m_X\ga
10^{25}$ eV, the synchrotron radiation from these pairs
can be above $10^{20}\,$eV, and the UHE flux
is then not as low as one might expect.) We note, however,
that the constraints from the EGRET measurements do not
change significantly with the EGMF strength as long as the
nucleon flux is comparable
to the $\gamma-$ray flux at the highest energies, as is the
case in Figs.~\ref{F7.2} and~\ref{F7.3}. A more detailed
discussion of viable TD models depending on uncertainties
in the fragmentation function, the mass and dominant decay channel
of the X particle, the URB, and the EGMF, has been given
in Ref.~\cite{slby}. The results of Ref.~\cite{slby} differ from
those of Ref.~\cite{ps1} which obtained more stringent
constraints on TD models because of the use of the older
fragmentation function Eq.~(\ref{hill-15-16}), and a stronger dependence
on the EGMF because of the use of a weaker EGMF which lead
to a dominance of $\gamma-$rays above $\simeq10^{20}\,$eV.

Dubovsky and Tinyakov~\cite{dub-tin2} have recently pointed out that there
could be an extra component of $\gamma-$ rays that would dominate the flux
shown in Figs.~\ref{F7.2} and \ref{F7.3} around the trough at
$\simeq10^{15}\,$eV due to synchrotron emission from the electron
component of the extragalactic cascade hitting the Galactic magnetic
field. For TD models, such a component is predicted to be close to the
observational upper limit from the CASA-MIA experiment~\cite{casa2} (see
Figs.~\ref{F7.2} and \ref{F7.3}). The detection of such a component would,
however, not necessarily be a signature of the top-down origin of EHECR
because a similar component would for the same reason be expected as an
extension of the diffuse $\gamma-$ray background around $\sim100\,$GeV if
this background is produced by sources whose spectrum extends to
$\ga100\,$TeV. Observations~\cite{mrk421,mrk501} suggest that to be the
case for blazars which are also expected to significantly contribute to
the diffuse $\gamma-$ray flux observed by EGRET~\cite{ss,mp}.  

The energy loss and absorption lengths for UHE nucleons and photons
are short ($\la100$ Mpc). Thus, their predicted UHE fluxes are
independent of cosmological evolution. The $\gamma-$ray flux
below $\simeq10^{11}\,$eV, however, scales as the
total X particle energy release integrated over all redshifts
and increases with decreasing $p$~\cite{sjsb}. For
$m_X=2\times10^{16}\,$GeV,
scenarios with $p<1$ are therefore ruled
out (as can be inferred from Figs.~\ref{F7.2} and ~\ref{F7.3}), whereas
constant comoving injection models ($p=2$) are well within the
limits. Since the EM flux above $\simeq10^{22}$ eV is
efficiently recycled to lower energies, the constraint on $p$
is in general less sensitive to $m_X$ then expected from
earlier CEL-based analytical estimates~\cite{chi,sjsb}.

\begin{figure}[ht]
\centering\leavevmode
\epsfxsize=6.0in
\epsfbox{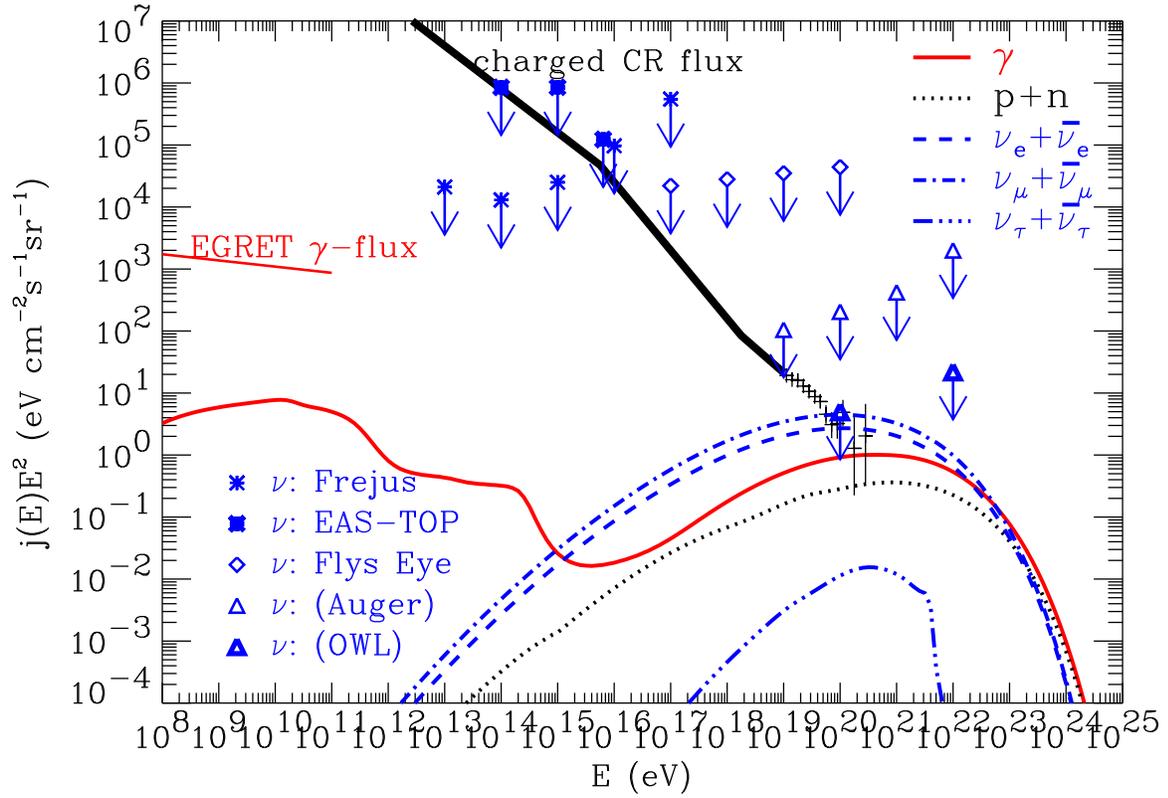}
\bigskip
\caption[...]{Same as Fig.~\ref{F7.2}, but for the case $p=2$,
where the decaying X particles are long lived and contribute
to cold dark matter, assuming an overdensity of $10^4$ on the
scale of the Galactic halo, $\simeq100\,$kpc. Points with arrows represent
approximate upper limits on the
diffuse neutrino flux from the Frejus~\cite{frejus}, the
EAS-TOP~\cite{eastop2}, and the Fly's
Eye~\cite{baltrusaitis} experiments, as indicated. The projected
sensitivity for the Pierre Auger project is using the acceptance
estimated in Ref.~\cite{pz}, and the one for the OWL concept
study is based on Ref.~\cite{owl}, both assuming observations
over a few years period.
\label{F7.4}}
\end{figure}

A specific $p=2$ scenario is realized in the case where the
supermassive X particles have a lifetime longer than the age of
the Universe and contribute to the cold dark matter and cluster
with the large scale structure, as discussed in Sect.~6.13. An
example of the expected fluxes in this scenario is shown in
Fig.~\ref{F7.4}. In this context, we mention that a possible
additional contribution to the $\gamma-$ray flux from the
synchrotron emission (in the Galactic magnetic field) of the
electrons/positrons produced in the decays of the MSRPs clustered in
the Galactic halo has recently been calculated in Ref.~\cite{blasi}. This
contribution, however, seems to be smaller than the contribution from
cascading of the extragalactic $\gamma-$ray component shown in
Fig.~\ref{F7.4}. Moreover, at least below $\simeq10^{15}\,$eV, both
components are unlikely to be detectable even with next generation
experiments because of the dominant well established fluxes
that are higher by at least a factor of 1000 at the relevant energies
and are presumably due to conventional sources such as blazars.

We now turn to signatures of TD models at UHE.
The full cascade calculations predict
$\gamma-$ray fluxes below 100 EeV that are a factor $\simeq3$
and $\simeq10$ higher than those obtained
using the CEL or absorption approximation often used in the
literature~\cite{bbv}, in the case of strong and weak URB,
respectively. Again, this shows the importance
of non-leading particles in the development of unsaturated EM
cascades at energies below $\sim10^{22}\,$eV.
Our numerical simulations give a $\gamma$/CR flux ratio at
$10^{19}\,$eV of $\simeq0.1$. The experimental exposure
required to detect a $\gamma-$ray flux at that level is
$\simeq4\times10^{19}\,{\rm cm^2}\,{\rm sec}\,{\rm sr}$, about a
factor 10 smaller than the current total experimental exposure.
These exposures are well within reach of the
Pierre Auger Cosmic Ray Observatories~\cite{auger}, which may be able to
detect a neutral CR component down to a level of 1\% of the total
flux. In contrast, if the EGMF exceeds $\sim 10^{-11}\,$G, then UHE
cascading is inhibited, resulting in a lower UHE
$\gamma-$ray spectrum. In the $10^{-9}$ G scenario of Fig.~\ref{F7.3},
the $\gamma$/CR flux ratio at $10^{19}\,$eV is $0.02$,
significantly lower than for no EGMF. 

\begin{table}
\caption[...]{\label{tab7.1}
Some viable $p=1$ TD scenarios explaining EHECR at least above 100 EeV.}
\smallskip
\begin{tabular}[9]{cccccccccc}
$m_X$$^a$ & Fig. & URB & EGMF$^b$ & FF & $f_N$ & mode & $Q^0$$^c$ &
 $\la$ GZK$^d$ & $\ga$ GZK$^d$ \\
\noalign{\vskip3pt\hrule\vskip3pt}
$10^{13}$ &
 & \multicolumn{4}{c}{$f_\nu l_\nu\ga400$ Mpc for high URB, no EGMF$^e$}
 & $\nu\nu$ & $\la$31 & $\gamma$ & $\gamma$ \\
 & & high & any & no-SUSY & 10\% &
 $qq$ & 1.4 & $N$ & $N$ \\
 & & $\la$ med & $\la10^{-11}$ & no-SUSY & $\la10\%$
 & $qq$ & 1.4 & $N$ & $\gamma$ \\
 & & high & $\la10^{-11}$ & no-SUSY & 10\% & $ql$
 & 0.88 & $N$ & $\gamma$ \\
 & & $\la$ med & $\la10^{-11}$ & any & $\la10\%$ & $ql$
 & 0.93 & $\gamma$ & $\gamma$ \\
 & & any & $\la10^{-11}$ & -- & -- & $ll$, $l\nu$
 & 1.3 & $\gamma$ & $\gamma$ \\
\noalign{\vskip3pt\hrule\vskip3pt}
$10^{14}$ &
 & \multicolumn{4}{c}{$f_\nu l_\nu\ga150$ Mpc for high URB, no EGMF$^e$}
 & $\nu\nu$ & $\la$19 & $\gamma$ & $\gamma$ \\
 & & high & any & no-SUSY & 10\% & $qq$ & 1.3 & $N$
 & $\gamma+N$, $N^f$ \\
 & & $\la$ med & $\la10^{-10}$ & no-SUSY & $\la10\%$ &
 $qq$, $q\nu$ & 1.3 & $\gamma+N$ & $\gamma $ \\
 & & any & $\la10^{-11}$ & any & $\la10\%$ & $ql$
 & 0.97 & $N$ & $\gamma$ \\
 & & any & $\la10^{-11}$ & -- & -- & $ll$, $l\nu$
 & 1.4 & $\gamma$ & $\gamma$ \\
\noalign{\vskip3pt\hrule\vskip3pt}
$10^{15}$ &
 & \multicolumn{4}{c}{$f_\nu l_\nu\ga500$ Mpc for high URB, no EGMF$^e$}
 & $\nu\nu$ & $\la$25 & $\gamma$ & $\gamma$ \\
 & & any & any & any & $10\%$ & $qq$, $ql$, $q\nu$ & 1.3 & $N$ & \\
 & & $\la$ med & $\la10^{-11}$ & any & $\la10\%$
 & $qq$, $ql$, $q\nu$ & 1.3 & & \\
 & & any & $\la10^{-11}$ & -- & -- & $ll$, $l\nu$
 & 1.3 & $\gamma$ & $\gamma$ \\
\noalign{\vskip3pt\hrule\vskip3pt}
$10^{16}$ &
 & \multicolumn{4}{c}{$f_\nu l_\nu\ga3000$ Mpc for high URB, no EGMF$^e$}
 & $\nu\nu$ & $\la$2.0 & $\gamma$ & $\gamma$ \\
 & \ref{F7.2}, \ref{F7.3} & high & any & SUSY & 10\% & $qq$
 & 1.6 & $N$ & $\gamma+N$, $N^f$ \\
 & & high & $\la10^{-9}$ & no-SUSY
 & 10\% & $qq$ & 1.3 & $\gamma$, $N^f$
 & $\gamma$, $\gamma+N^g$ \\
 & & any & $\la10^{-11}$ & any & $\la10\%$ &
 $qq$, $ql$, $q\nu$ & 1.9 & & \\
 & & $\la$ med & $\la10^{-11}$ & -- & -- & $ll$, $l\nu$
 & 1.6 & $\gamma$ & $\gamma$ \\
\end{tabular}
\smallskip
$^a$ in GeV.\\
$^b$ in Gauss.\\
$^c$ maximal total energy injection rate at zero redshift in units of 
$10^{-23}\,h\,{\rm eV}\,{\rm cm}^{-3}\,{\rm sec}^{-1}$.\\
$^d$ dominant component of ``visible'' TD flux below and above GZK cutoff
at $\simeq70$ EeV; no entry means different composition is possible,
depending on parameters.\\
$^e$ viable for eV mass neutrinos if their overdensity $f_\nu$ over
a scale $l_\nu$ obeys the specified condition, for the case of high URB
and vanishing EGMF; for weaker URB/stronger EGMF, the condition 
on $f_\nu l_\nu$ relaxes/becomes more stringent, respectively. \\
$^f$ for EGMF $\ga10^{-10}$ G.\\
$^g$ for EGMF $\ga10^{-9}$ G.
\end{table}

It is clear from the above discussions that the predicted particle fluxes
in the TD scenario are currently uncertain to a large extent due to 
particle physics uncertainties (e.g., mass and decay modes of the X
particles, the quark fragmentation function, the nucleon fraction $f_N$,
and so on) as well as astrophysical uncertainties (e.g., strengths of the
radio and infrared backgrounds, extragalactic magnetic fields, etc.). 
More details on the dependence of the predicted UHE particle spectra and
composition on these particle physics and astrophysical
uncertainties are contained in Ref.~\cite{slby}. We summarize in
Table~\ref{tab7.1}, (adapted from Ref.~\cite{slby}), some of the
scenarios that are capable of explaining the observed EHECR flux at least
above 100 EeV, without violating any observational constraints. The
predicted compositions of the particles below and above the GZK cutoff are
also indicated. We stress here that there are viable TD scenarios which
predict nucleon fluxes that are comparable to or even higher than
the $\gamma-$ray flux at all energies, even though $\gamma-$rays
dominate at production.
This occurs, e.g., in the case of high URB
and/or for a strong EGMF, and a nucleon fragmentation fraction of
$\simeq10\%$; see, for example, Fig.~\ref{F7.3}. Some of these TD 
scenarios would therefore remain viable even if EHECR induced EASs
should be proven inconsistent with photon primaries (see,
e.g., Ref.~\cite{hvsv}).

The normalization procedure to the EHECR flux described above
imposes the constraint $Q^0_{\rm EHECR}\la10^{-22}\,{\rm eV}\,{\rm
cm}^{-3}\,{\rm sec}^{-1}$ within a factor of a
few~\cite{ps1,slby,slsc} for the total energy release rate $Q_0$
from TDs at the current epoch (see also the benchmark calculation
in Sect.~6.2.3). In most TD models, because of the unknown values of the
parameters involved, it is currently not
possible to calculate the exact value of $Q_0$ from first principle,    
although it has been shown that the required values of $Q_0$ (in order to
explain the EHECR flux) mentioned above are quite possible for
certain kinds of TDs (see Sect.~6 for details). Some cosmic
string simulations suggest that strings may lose most of
their energy in the form of X particles and estimates of this
rate have been given~\cite{vincent}. If that is the case, the
constraint on $Q^0_{\rm EHECR}$ translates into a limit on the symmetry
breaking scale $\eta$ and hence on the mass $m_X$ of the X particle: 
$\eta\sim m_X\la10^{13}\,$GeV~\cite{susytd,wmgb}. Independently 
of whether or not this scenario explains EHECR, the EGRET measurement
of $\omega_{\rm cas}$ leads to a similar bound,
$Q^0_{\rm EM}\la2.2\times10^{-23}\,h
(3p-1)\,{\rm eV}\,{\rm cm}^{-3}\,{\rm sec}^{-1}$, which leaves
the bound on $\eta$ and $m_X$ practically unchanged.
We may mention here that in most supersymmetric (SUSY) GUT models, X
particle (gauge and Higgs boson) masses smaller than about $10^{16}\gev$
are disfavored from proton decay and other considerations; see, e.g.,
Refs.~\cite{arno-nath,lukas-raby}. However, if the relevant topological
defects are formed not at the GUT phase transition, but at a subsequent
phase transition after the GUT transition (see the final paragraph of
Sect.~6.4.6), then depending on the model, 
the associated X particles may not mediate baryon number violating
processes, and so the proton decay constraints may not apply to them, and
then X particles of mass $m_X < 10^{16}\gev$ are not ruled out. 

\subsection{Constraints from Primordial Nucleosynthesis}

For $z\ga10^3$, the Universe becomes opaque to the cascade
photons discussed in the previous section~\cite{cas1}. These
photons typically survive for a time $\simeq l_E(E)$
during which they have a certain chance to photo-disintegrate a
$^4{\rm He}$ nucleus, with corresponding effective cross sections
$\sigma_{\rm D/^3He}^{\rm eff}(E)$ for production of D
or $^3$He. This is possible as long as cascade cut off energy
$E_c(z)$ is larger than the $^4$He photo-disintegration
threshold, $E_{\rm th}^{^4{\rm He}}=19.8\,$MeV, or, from
Eq.~(\ref{ecrit}), $z\la3\times10^6$. Instantaneous generation of a
cascade spectrum $j_\gamma^{\rm cas}(E)$ then leads to
creation of D and $^3$He
with number densities $n_{\rm D/^3He}\simeq(Y/4)n_B\int
dE l_E(E)4\pi j_\gamma^{\rm
cas}(E)\sigma_{\rm D/^3He}^{\rm eff}(E)$, or,
using Eq.~(\ref{lEgamma}),
\begin{equation}
  n_{\rm D/^3He}\simeq\int dE
  \frac{4\pi j_\gamma(E)Y\sigma_{\rm D/^3He}(E)}
  {4(1-Y)\sigma_{\rm H}^{\rm BH}(E)+
  Y\sigma_{rm He}^{\rm BH}(E)+(4-2Y)
  \eta_{\rm CS}(E)\sigma_{\rm CS}(E)}
  \,.\label{nDHe}
\end{equation}
The measured photo-disintegration cross sections immediately
imply
\begin{equation}
  \left(\frac{^3{\rm He}}{{\rm D}}\right)_{\rm photo}\ga8
  \,,\label{photoratio}
\end{equation}
i.e., ``cascading nucleosynthesis'' predicts much more $^3$He
than D. On the other hand, data imply $(^3{\rm He}/{\rm
D})_\odot\la1.13$~\cite{geiss} for the abundances at the time of
solar system formation and chemical evolution suggests $(^3{\rm
He}/{\rm D})_p\la(^3{\rm He}/{\rm D})_\odot$ for the primordial
abundances. Therefore, $^4$He photo-disintegration
can not be the predominant production mechanism of primordial D
and $^3$He, and altogether one has~\cite{sjsb}
\begin{equation}
  \left(\frac{^3{\rm He}+{\rm D}}{{\rm H}}\right)_{\rm photo}
  \la5\times10^{-5}\,,\label{photoconstr}
\end{equation}
which translates into an upper bound of EM energy release at
redshifts $3\times10^6\ga z\ga10^3$ by either applying Eq.~(\ref{nDHe})
or more detailed Monte Carlo simulations~\cite{cas1}. Fig.~\ref{F7.1}
shows the resulting maximal allowed instantaneous energy release as
a function of redshift. We see that the resulting constraints 
from primordial nucleosynthesis on the TD models discussed in Sect.~6 are
comparable to but
independent of the constraints from the diffuse $\gamma-$ray
background discussed in Sect.~7.1~\cite{sjsb}.

\subsection{Constraints from Distortions of the Cosmic
Microwave Background}

Early non-thermal electromagnetic energy injection can also lead
to a distortion of the cosmic microwave background. We
focus here on energy injection during the epoch prior to
recombination. A comprehensive discussion of this subject is 
given in Ref.~\cite{silk}. Regarding the character of
the resulting spectral CMB distortions
there are basically two periods to distinguish: First, in the
range $3\times10^6\simeq z_{\rm th}>z>z_{\rm
y}\simeq10^5$ between the thermalization redshift $z_{\rm th}$
and the Comptonization redshift $z_{\rm y}$, a fractional energy
release $\Delta u/u$ leads to a pseudo-equilibrium Bose-Einstein
spectrum with a chemical potential given by
$\mu\simeq0.71\Delta u/u$. This relation is valid for negligible
changes in photon number which is a good approximation for the
Klein-Nishina cascades produced by the GUT particle decays we
are interested in~\cite{silk}. Second, in the range $z_{\rm
y}>z>z_{\rm rec}\simeq10^3$ between $z_{\rm y}$ and the
recombination redshift $z_{\rm rec}$ the resulting spectral distortion is
of the Sunyaev-Zel'dovich type~\cite{szy} with a Compton $y$
parameter given by $4y=\Delta u/u$. The most recent limits on
both $\mu$ and $y$ were given in Ref.~\cite{mather}. The
resulting bounds on $\Delta u/u$ for instantaneous energy release as
a function of injection redshift~\cite{wright} are shown as the
dashed curve in Fig.~\ref{F7.1}.

The resulting bounds for the TD models are less stringent than
the bounds from cascading nucleosynthesis and the observable
$\gamma-$ray background, but they still allow one to rule out the simplest
model for superconducting cosmic strings corresponding to $p=0$.

\subsection{Constraints on Neutrino Fluxes}

As discussed in Sect.~6, in TD scenarios most of the energy is
released in the form of EM particles and neutrinos. If the X
particles decay into a quark and a lepton, the quark hadronizes
mostly into pions and the ratio of energy release into the
neutrino versus EM channel is $r\simeq0.3$. The resulting
diffuse UHE neutrino fluxes have been calculated in the
literature for various situations: Ref.~\cite{yoshida} contains
a discussion of the (unnormalized) predicted spectral shape and
Ref.~\cite{cusp} computes the absolute flux predicted by
specific processes such as cusp evaporation on ordinary cosmic
strings. Ref.~\cite{bhs} calculated absolute fluxes for the
scenarios discussed in Sect.~6 by using a simple
estimate of the neutrino interaction redshift as mentioned in
Sect.~4.3.1, and Ref.~\cite{ydjs} improved on this by taking
into account neutrino cascading in the RNB. None of these works,
however, took into account cosmological constraints on TD
models such as the ones discussed in Sect.~7.1$-$7.3.

\begin{figure}[ht]
\centering\leavevmode
\epsfxsize=6.0in
\epsfbox{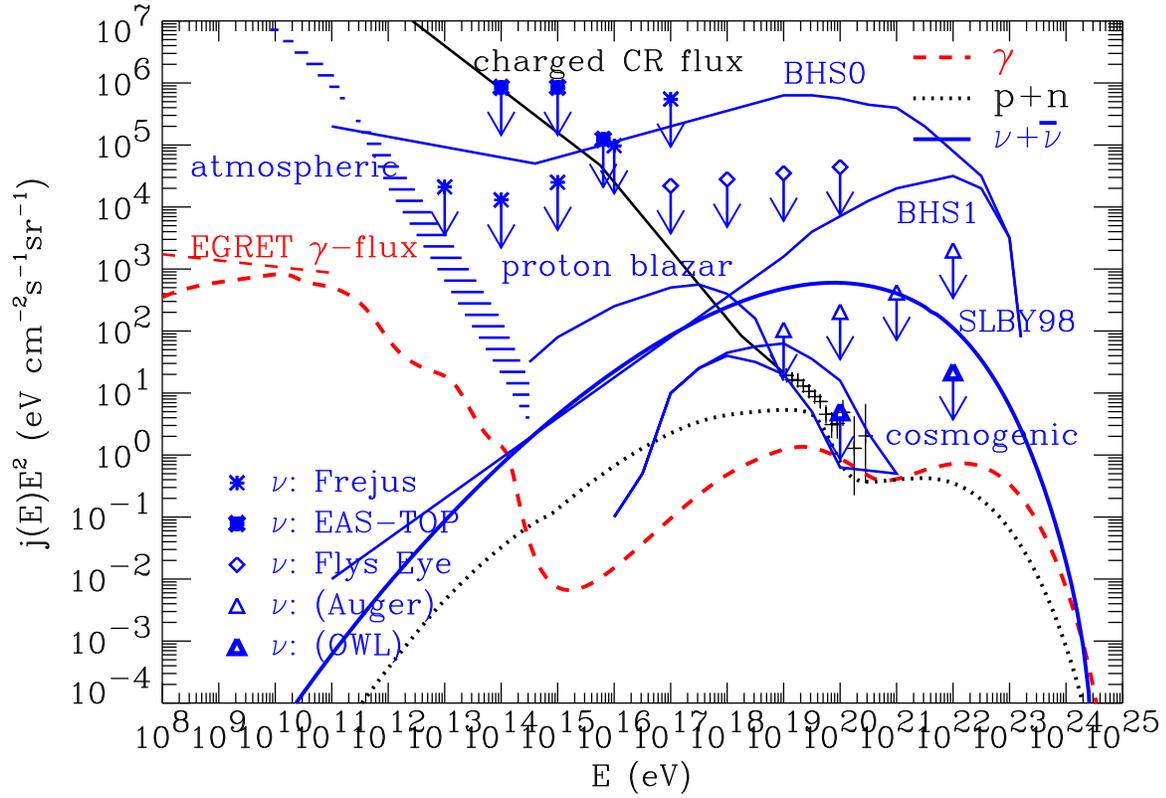}
\bigskip
\caption[...]{Predictions for the summed differential fluxes of all
neutrino flavors (solid lines) from the atmospheric
background for different zenith angles~\cite{lipari} (hatched
region marked ``atmospheric''), from proton blazars that are
photon optically thick to nucleons but contribute to the diffuse
$\gamma-$ray flux~\cite{protheroe2}
(``proton blazar''), from UHECR interactions with the CMB~\cite{pj}
(``cosmogenic''), for the TD model from Ref.~\cite{bhs} with
$p=0$ (``BHS0'') and $p=1$ (``BHS1''), and for the TD model from
Fig.~\ref{F7.2}, assuming an EGMF of
$\la10^{-12}\,$G (``SLBY98'', from Ref.~\cite{slby}).
Also shown are the fluxes of $\gamma-$rays (dotted line), and
nucleons (dotted lines) for this latter TD model.
The data shown for the CR flux and the diffuse $\gamma-$ray flux
from EGRET are as in Figs.~\ref{F7.2} and~\ref{F7.3}.
Points with arrows represent approximate upper limits on the
diffuse neutrino flux from the Frejus~\cite{frejus}, the
EAS-TOP~\cite{eastop2}, and the Fly's
Eye~\cite{baltrusaitis} experiments, as indicated. The projected
sensitivity for the Pierre Auger project is using the acceptance
estimated in Ref.~\cite{pz}, and the one for the OWL concept
study is based on Ref.~\cite{owl}, both assuming observations
over a few years period.
\label{F7.5}}
\end{figure}

Fig.~\ref{F7.5} shows predictions of the total neutrino
flux for the same TD model on which
Fig.~\ref{F7.2} is based, as well as some of the older
estimates from Ref.~\cite{bhs}. In the absence of neutrino
oscillations the electron neutrino and
anti-neutrino fluxes that are not shown are about a factor of 2
smaller than the muon neutrino and anti-neutrino fluxes,
whereas the $\tau-$neutrino flux is in general negligible.
In contrast, if the
interpretation of the atmospheric neutrino deficit in terms
of nearly maximal mixing of muon and $\tau-$neutrinos proves
correct, the muon neutrino fluxes shown in Fig.~\ref{F7.5} would
be maximally mixed with the $\tau-$neutrino fluxes. To put the TD
component of the neutrino flux in perspective with contributions
from other sources, Fig.~\ref{F7.5} also shows the atmospheric
neutrino flux, a typical prediction for the diffuse flux from
photon optically thick proton blazars~\cite{protheroe2} that are not
subject to the Waxman Bahcall bound and were normalized to recent
estimates of the blazar contribution to the diffuse $\gamma-$ray
background~\cite{muk-chiang}, and the flux range expected for
``cosmogenic'' neutrinos created as secondaries from the decay
of charged pions produced by UHE nucleons~\cite{pj}. The TD flux
component clearly dominates above $\sim10^{19}\,$eV.

In order to translate neutrino fluxes into event rates,
one has to fold in the interaction cross sections with
matter that were discussed in Sect.~4.3.1. At UHE these
cross sections are not directly accessible to laboratory
measurements. Resulting uncertainties therefore
translate directly to bounds on neutrino fluxes derived from,
for example, the non-detection of UHE muons produced in charged-current
interactions. In the following, we will assume the estimate
Eq.~(\ref{cccross2}) based on the Standard Model
for the charged-current muon-neutrino-nucleon
cross section $\sigma_{\nu N}$ if not indicated otherwise.

For an (energy dependent) ice or water equivalent acceptance
$A(E)$ (in units of volume times solid angle), one can obtain an
approximate expected rate of UHE muons produced by neutrinos
with energy $>E$, $R(E)$, by
multiplying $A(E)\sigma_{\nu N}(E)n_{\rm H_2O}$ (where 
$n_{\rm H_2O}$ is the nucleon density in water) with the
integral muon neutrino flux $\simeq Ej_{\nu_\mu}$. This can be used to
derive upper limits on diffuse neutrino fluxes from a
non-detection of muon induced events. Fig.~\ref{F7.5} shows bounds
obtained from several experiments: The Frejus experiment
derived upper bounds for $E\ga10^{12}\,$eV from
their non-detection of almost horizontal muons with an energy
loss inside the detector of more than $140\,$MeV per radiation
length~\cite{frejus}. The EAS-TOP collaboration published
two limits from horizontal showers, one in the regime
$10^{14}-10^{15}\,$eV, where non-resonant neutrino-nucleon
processes dominate, and one at the Glashow resonance which
actually only applies to $\bar\nu_e$~\cite{eastop2}. The Fly's
Eye experiment derived
upper bounds for the energy range between $\sim10^{17}\,$eV and
$\sim10^{20}\,$eV~\cite{baltrusaitis} from the non-observation
of deeply penetrating particles.
The AKENO group has published an upper
bound on the rate of near-horizontal, muon-poor air
showers~\cite{nagano}. Horizontal
air showers created by electrons or muons that are in turn
produced by charged-current reactions of electron and muon
neutrinos within the atmosphere have recently been pointed out
as an important method to constrain or measure UHE neutrino
fluxes~\cite{bpvz,pz} with next generation detectors.
Furthermore, the search for pulsed radio emission
from cascades induced by neutrinos or cosmic rays above
$\sim10^{19}\,$eV in the lunar regolith with the NASA
Goldstone antenna has lead to an upper limit
comparable to the constraint from the Fly's Eye experiment~\cite{gln}.

The $p=0$ TD model BHS0 from the early work of Ref.~\cite{bhs}
is not only ruled out by the constraints from Sect.~7.1$-$7.3,
but also by some of the
experimental limits on the UHE neutrino flux, as can be seen in
Fig.~\ref{F7.5}. Further, although both the BHS1 and the SLBY98 models 
correspond to $p=1$, the UHE neutrino flux above $\simeq10^{20}\,$eV in
the latter is almost two orders of magnitude smaller
than in the former. The main reason for this are the different
flux normalizations adopted in the two papers: First, the BHS1 model 
was obtained by normalizing the predicted {\it proton} flux to the
observed UHECR flux at $\simeq 4\times10^{19}\ev$, whereas 
in the SLBY98 model the actually ``visible'' sum of the nucleon and
$\gamma-$ray fluxes was normalized in a optimal way. Second, the BHS1
assumed a nucleon fraction about a factor
3 smaller~\cite{bhs}. Third, the BHS1 scenario used the older fragmentation
function Eq.~(\ref{hill-spectrum}) which has more power at larger energies
(see Fig.~\ref{F6.1}). Clearly, the 
SLBY98 model is not only consistent with the constraints
from Sect.~7.1$-$7.3, but also with all existing neutrino
flux limits within 2-3 orders of magnitude.

What, then, are the prospects of detecting UHE neutrino fluxes
predicted by TD models? In a $1\,{\rm km}^3\,2\pi\,$sr size
detector, the SLBY98 scenario from Fig.~\ref{F7.5},
for example, predicts a muon-neutrino event rate
of $\simeq0.15\,{\rm yr}^{-1}$, and an electron neutrino event rate
of $\simeq0.089\,{\rm yr}^{-1}$ above $10^{19}\,$eV, where
``backgrounds'' from conventional sources should be negligible.
Further, the muon-neutrino event rate above 1 PeV should be
$\simeq1.2\,{\rm yr}^{-1}$, which could be interesting if
conventional sources produce neutrinos at a much smaller
flux level. Of course, above $\simeq100\,$TeV, instruments
using ice or water as detector medium, have to look at downward
going muon and electron events due to neutrino absorption in
the Earth. However, $\tau-$neutrinos obliterate this Earth
shadowing effect due to their regeneration from $\tau$
decays~\cite{halzen-saltzberg}. The presence of $\tau-$neutrinos,
for example, due to mixing with muon neutrinos, as suggested
by recent experimental results from Super-Kamiokande,
can therefore lead to an increased upward going event
rate~\cite{mannheim3}; see discussion in Sect.~4.3.1.
For recent compilations of UHE neutrino flux
predictions from astrophysical and TD sources see
Refs.~\cite{protheroe,mc} and references therein.

For detectors based on the fluorescence technique such as the
HiRes~\cite{hires} and the Telescope Array~\cite{tel_array}
(see Sect.~2.6), the sensitivity to UHE neutrinos is often
expressed in terms of an effective aperture $a(E)$ which is
related to $A(E)$ by $a(E)=A(E)\sigma_{\nu N}(E)n_{\rm
H_2O}$. For the cross section of Eq.~(\ref{cccross2}), the
apertures given in Ref.~\cite{hires} for the HiRes correspond to
$A(E)\simeq3\,{\rm km}^3\times2\pi\,{\rm sr}$ for
$E\ga10^{19}\,$eV for muon neutrinos. The expected acceptance
of the ground array component of the Pierre Auger
project for horizontal
UHE neutrino induced events is $A(10^{19}\,{\rm eV})\simeq
20\,{\rm km}^3\,{\rm sr}$ and $A(10^{23}\,{\rm eV})\simeq
200\,{\rm km}^3\,{\rm sr}$~\cite{pz}, with a duty cycle close to
100\%. We conclude that detection of neutrino
fluxes predicted by scenarios such as the SLBY98 scenario shown
in Fig.~\ref{F7.5} requires running a detector of acceptance
$\ga10\,{\rm km}^3\times2\pi\,{\rm sr}$ over a period of a few
years. Apart from optical detection in air, water, or ice, other
methods such as acoustical and radio detection~\cite{ghs}
(see, e.g., Ref.~\cite{radio-technique} and the RICE
project~\cite{rice} for the latter) or even
detection from space~\cite{owl} appear to be interesting
possibilities for detection concepts operating at such scales
(see Sect.~2.6). For example, the OWL satellite concept, which
aims to detect EAS from space, would have an
aperture of $\simeq3\times10^6\,{\rm km}^2\,{\rm sr}$
in the atmosphere, corresponding to $A(E)\simeq6\times10^4
\,{\rm km}^3\,{\rm sr}$ for $E\ga10^{20}\,$eV,
with a duty cycle of $\simeq0.08$~\cite{owl}.
The backgrounds seem to be
in general negligible~\cite{ydjs,price}. As indicated by the
numbers above and by the projected sensitivities shown in
Fig.~\ref{F7.5}, the Pierre Auger Project and especially the
OWL project should be capable of detecting typical TD neutrino
fluxes. This applies to any detector of acceptance
$\ga100\,{\rm km}^3\,{\rm sr}$. Furthermore, a 100 day
search with a radio telescope of the NASA Goldstone type
for pulsed radio emission from cascades induced by neutrinos
or cosmic rays in the lunar regolith could reach
a sensitivity comparable or better to the Pierre Auger
sensitivity above $\sim10^{19}\,$eV~\cite{gln}.

A more model independent estimate~\cite{slsc} for the average
event rate $R(E)$ can be made if the underlying
scenario is consistent with observational nucleon and
$\gamma-$ray fluxes and the bulk of the energy is released above
the PP threshold on the CMB. Let us assume that the
ratio of energy injected into the neutrino versus EM channel is a
constant $r$. As discussed in Sect.~7.1, cascading effectively reprocesses
most of the injected EM energy into low
energy photons whose spectrum peaks at $\simeq10\,$GeV~\cite{ahacoppi}.
Since the ratio $r$ remains roughly unchanged during
propagation, the height of the
corresponding peak in the neutrino spectrum should
roughly be $r$ times the height of the low-energy
$\gamma-$ray peak, i.e., we have the condition
$\max_E\left[E^2j_{\nu_\mu}(E)\right]\simeq
r\max_E\left[E^2j_\gamma(E)\right].$ Imposing the observational
upper limit on the diffuse $\gamma-$ray flux around $10\,$GeV
shown in Fig.~\ref{F7.5}, $\max_E\left[E^2j_{\nu_\mu}(E)\right]\la
2\times10^3 r \,{\rm eV}{\rm cm}^{-2}{\rm sec}^{-1}{\rm
sr}^{-1}$, then bounds the average diffuse neutrino rate above
PP threshold on the CMB, giving 
\begin{equation}
  R(E)\la0.34\,r\left[{A(E)\over1\,{\rm
  km}^3\times2\pi\,{\rm sr}}\right]
  \,\left({E\over10^{19}\,{\rm eV}}\right)^{-0.6}\,{\rm
  yr}^{-1}\quad(E\ga10^{15}\,{\rm eV})\,.\label{r2}
\end{equation}
For $r\la20(E/10^{19}\,{\rm eV})^{0.1}$ this bound is consistent
with the flux bounds shown in Fig.~\ref{F7.5} that are dominated
by the Fly's Eye constraint at UHE. We stress again that TD
models are not subject to the Waxman Bahcall bound because
the nucleons produced are considerably less abundant than
and are not the primaries of produced $\gamma-$rays and
neutrinos.

\begin{figure}[ht]
\centering\leavevmode
\epsfxsize=6.0in
\epsfbox{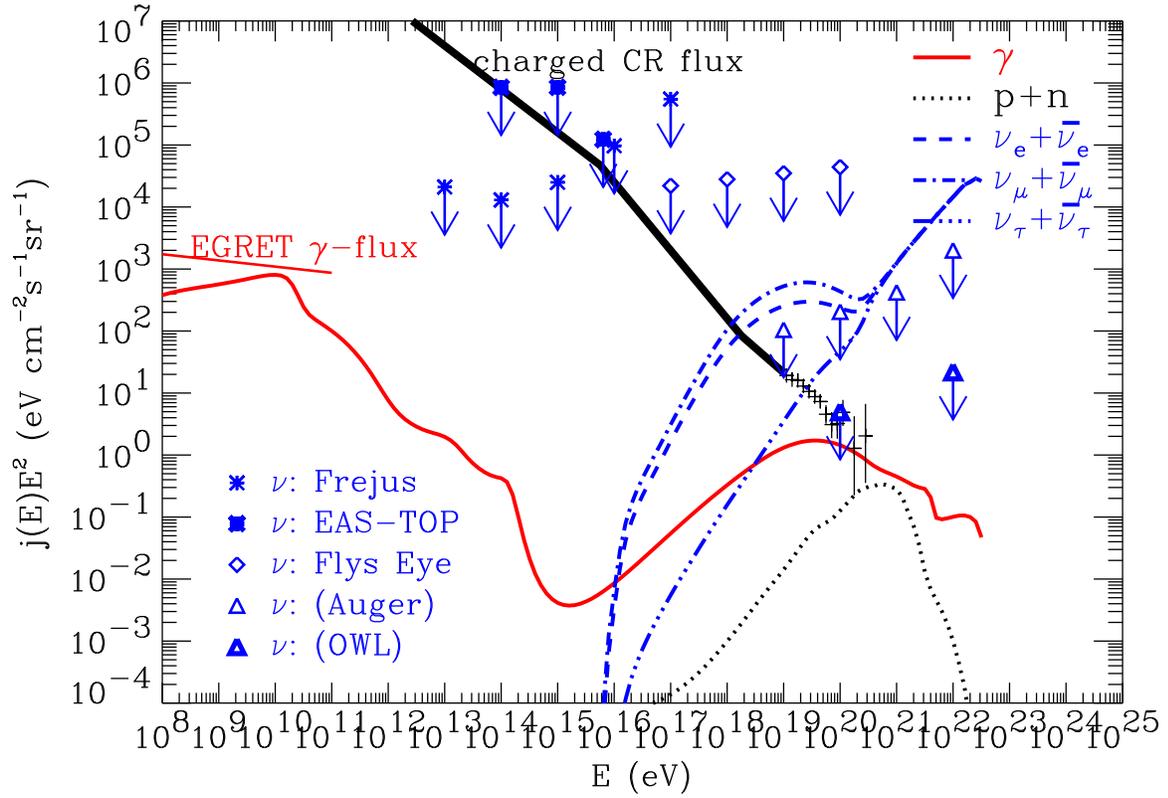}
\bigskip
\caption[...]{Flux predictions for a TD model characterized
by $p=1$, $m_X=10^{14}\,$GeV, with X particles exclusively
decaying into neutrino-antineutrino pairs of all flavors
(with equal branching ratio), assuming a neutrino mass $m_\nu=1\,$eV. For
neutrino clustering, the lower limit from Table~\ref{tab7.1} was
assumed, corresponding to an overdensity of $\simeq30$ over
a scale of $l_\nu\simeq5\,$Mpc. The calculation assumed the strongest
URB version shown in Fig.~\ref{F4.3} and an EGMF $\ll10^{-11}\,$G.
The line key is as in Figs.~\ref{F7.2} and~\ref{F7.4}.
\label{F7.6}}
\end{figure}

In typical TD models such as the one discussed above where
primary neutrinos are produced by pion decay,
$r\simeq0.3$. However, in TD scenarios with $r\gg1$ neutrino
fluxes are only limited by the condition that the {\it secondary}
$\gamma-$ray
flux produced by neutrino interactions with the RNB be below
the experimental limits. An example for such a scenario is given
by X particles exclusively decaying into neutrinos~\cite{slby}
(although this is not very likely in most particle physics
models, but see Ref.~\cite{slby} and Fig.~\ref{F7.6} for a scenario
involving topological defects and Ref.~\cite{gk2} for a scenario
involving decaying superheavy relic particles, both of which explain the
observed EHECR events as secondaries of neutrinos interacting
with the RNB). Another possibility is the existence of hidden sector
(mirror) topological defects radiating hidden (mirror) matter
which interacts only gravitationally or via superheavy particles,
whereas the mirror neutrinos can maximally oscillate into
ordinary neutrinos~\cite{bv1}. Such scenarios could induce
appreciable event rates above $\sim10^{19}\,$eV in a km$^3$
scale detector. A detection would thus open the exciting
possibility to establish an experimental lower limit on $r$.
Being based solely on energy conservation,
Eq.~(\ref{r2}) holds regardless of whether or not the underlying
TD mechanism explains the observed EHECR events.

The transient event rate could be much higher than Eq.~(\ref{r2}) in
the direction to discrete sources which emit particles in
bursts. Corresponding pulses in the EHE
nucleon and $\gamma-$ray fluxes would only occur for sources
nearer than $\simeq100\,$Mpc and, in case of protons, would be
delayed and dispersed by deflection in Galactic and
extragalactic magnetic fields~\cite{wm,lsos}. The recent observation
of a possible correlation of CR above $\simeq4\times10^{19}\,$eV by the
AGASA experiment~\cite{haya2} might suggest
sources which burst on a time scale $t_b\ll1\,$yr.
A burst fluence of $\simeq r\left[A(E)/1\,{\rm km}^3\times2\pi\,{\rm
sr}\right](E/10^{19}\,{\rm eV})^{-0.6}$ neutrinos within a time
$t_b$ could then be expected. Associated pulses
could also be observable in the ${\rm GeV}-{\rm TeV}$
$\gamma-$ray flux if the EGMF is smaller than
$\simeq10^{-15}\,$G in a significant fraction of extragalactic
space~\cite{wc}.

In contrast, the neutrino flux is comparable to (not significantly
larger than) the UHE photon plus nucleon 
fluxes in the models involving metastable superheavy relic
particles discussed in Sect.~6.13; see Fig.~\ref{F7.4}.
This can be understood because the neutrino flux is dominated
by the extragalactic contribution which scales with the
extragalactic nucleon and $\gamma-$ray contribution in exactly
the same way as in the unclustered case, whereas the extragalactic
``visible'' contribution is much smaller in the clustered case.
Such neutrino fluxes would be hardly detectable even with next generation
experiments.

\section{Summary and Conclusions}
It is clear from the discussions in the previous sections that the 
problem of origin of EHECR continues to remain as a major unsolved
problem. 

The EHECR present a unique puzzle: Recall that for
lower energy cosmic rays (below about $10^{16}\ev$) there is a strong
belief that these are produced in supernova remnants (SNRs) in the
Galaxy. However, because of the twists and turns the trajectories of these
particles suffer in propagating through the Galactic magnetic field, it
is not possible to point back to the sources of these particles in the
sky, and one has to take recourse to indirect arguments, such as
those involving $\gamma-$ray production or other secondary particle
abundances, to establish the ``evidence'' that indeed these ``low'' 
energy cosmic rays
are produced in SNRs. The EHECR particles, on the other hand, are hardly
affected by any intergalactic and/or 
Galactic magnetic fields. Also, because of energy losses they suffer
during their propagation through the intergalactic space, the EHECR
particles --- if they are ``standard'' particles such as nucleons, heavy
nuclei and/or photons --- cannot originate at distances much
further away than about 60 Mpc from Earth. So knowing
(observationally) the arrival directions of the individual EHECR events,
we should be able to do ``particle astronomy'', i.e., trace back to the
sources of these EHECR particles in the sky. Yet, all attempts toward this
have generally failed to identify possible  
specific powerful astronomical sources (such as active galactic nuclei,
radio galaxies, quasars, etc., that could in principle produce these
particles). In some cases arrival directions of some EHECR events point
to some quasars or radio galaxies, but those are generally situated at
distances well beyond 60 Mpc.

Apart from the problem of source identification, the other basic question
is that of energetics. What processes are responsible for endowing the
EHECR particles with the enormous energies beyond $10^{11}\gev$?
As we have discussed in this review, conventional
``bottom-up'' acceleration mechanisms are barely able to produce particles
of such energies when energy loss processes at the source as well as
during propagation are taken into account. 

The conventional bottom-up models thus face a
two-fold problem: Energetics as well as source location. It seems to be
becoming increasingly evident --- though this is far from  
established yet --- that to solve both problems together, some
kind of new physics beyond the Standard Model of particle physics may be
necessary. In this report we have discussed several such proposals that
are currently under study. We summarize them here. 

Currently, two types of distinct ``scenarios'' are being studied: 
In one of these scenarios, one still works within the framework of the
bottom-up scenario and assumes that there exist sources where the
requirements of energetics are met somehow; the source distance problem is
then solved by postulating any of the following: 

\begin{itemize}
\item The EHECR are a new kind of particle, for example, a supersymmetric
particle whose interactions with the CMB is such that they suffer little
or no significant energy loss during propagation, so that they can reach
Earth from cosmological distances. However, in this proposal, the new
particles must be produced as secondary particles at the source or during
propagation through interactions of a known accelerated particle such as
proton with the medium. This has other
implications such as associated secondary $\gamma-$ray production which
will have to confront various constraints from observation.
\medskip
\item The EHECR are known particles but they are endowed with new
interaction properties. One possibility along this line is the suggestion
that the neutrino interaction cross section at the relevant energies
receives dominant contributions from physics beyond the Standard Model
which causes neutrinos to interact in the atmosphere. In order not
to violate unitarity bounds, this cross section can not be
a point cross section and/or high partial waves have to
contribute significantly.

\medskip
\item EHECR are ``exotic'' particles such as magnetic monopoles which are
fairly easily accelerated to the requisite energies by the Galactic
magnetic field. This is an attractive proposal, because neither energetics
nor source distance is much of a problem. However, the problem with this
proposal seems to be that magnetic monopoles, because of their heavy mass,
are unlikely to induce relativistic air-showers.  
\medskip
\item EHECR are nucleons and/or $\gamma-$rays that are produced within the
GZK distance limit of $\sim$ 60 Mpc from Earth by interactions of
sufficiently high energy neutrinos with the thermal relic neutrino
background. This mechanism is most effective if some species of neutrinos
have a small mass of $\sim$ eV, in which case the neutrinos play the role
of hot dark matter in the Universe. This mechanism, if it leaves some
characteristic signature in the spectrum and/or composition of EHECR, has
the potential to offer an indirect method of ``detecting'' the
relic massive neutrino dark matter background, but is also
strongly constrained by associated $\gamma-$ray production.
\medskip
\item EHECR are nucleons, but one allows small modifications to the
fundamental laws of physics. One specific proposal along this line is to
postulate a tiny violation of Lorentz invariance, too small to have been
detected so far, which allows avoidance of the GZK cutoff effect for
nucleons so that nucleons can arrive to Earth from cosmological distances. 

\end{itemize}

Clearly, all the above proposals involve new physics beyond the Standard
Model. 

In the other scenario, generally called ``top-down'' scenario, there is no
acceleration mechanism involved: The EHECR particles arise simply from the
decay of some sufficiently massive ``X'' particles presumably originating
from processes in the early Universe. There is thus no problem of
energetics. One clear prediction of the top-down scenario is that EHECR
should consist of ``elementary'' particles such as nucleons, photons, and
even possibly neutrinos, but {\it no heavy nuclei}. 

Two classes of possible sources of these massive X particles have been
discussed in literature. These are: 

\begin{itemize}
\item X particles produced by collapse, annihilation or other processes
involving systems of cosmic topological defects such as cosmic strings,
magnetic monopoles, superconducting cosmic strings and so on, which could
be produced in the early Universe during symmetry-breaking phase
transitions envisaged in Grand Unified Theories. In an inflationary
early Universe, the relevant topological defects could be formed
at a phase transition at the end of inflation. The X particle mass can
be as large as a typical GUT scale $\sim10^{16}\gev$. 
\medskip
\item X particles are some long-living metastable superheavy  relic
particles (MSRP) of mass $\ga10^{12}\gev$ with lifetime comparable to or
much larger than the age of the Universe. 
These X particles could be
produced in the early Universe in particle production processes associated
with inflation, and they could exist in the present day Universe with
sufficient abundance as to be a good candidate for (at least a part of)
the cold dark matter in the Universe. These MSRP X particles are expected
to cluster in our Galactic Halo. Decay of these X particles
clustered in the Galactic Halo (GH) can easily explain the
apparent absence of the GZK cutoff in the recent EHECR data. 
\end{itemize} 

In both of these top-down mechanisms, the decay of the X particles
predominantly produce photons and neutrinos, and their spectra at
production is fixed by particle physics (QCD). 
The topological defect (TD) model, as we have discussed in details in this
review, is severely constrained by existing measurements and limits on the
diffuse gamma ray background at energies below 100 GeV that would 
receive a contribution (through electromagnetic cascading process) from 
decay of TD-produced X particles at large cosmological distances. 
The MSRP X particles clustered in the GH are, however, not constrained by
the diffuse gamma ray measurements. Instead, the
MSRP scenario is (or rather will be) constrained by measurements of the
anisotropy of UHECR, because a rather strong anisotropy is expected in
this model because of the off-center location of the Sun with respect to
the Galactic center. 

In the MSRP decay model, the observed spectrum of EHECR should essentially
be the unmodified injected spectrum resulting from the decay of the MSRPs.
The precise measurements of the EHECR spectrum can then be a probe of QCD 
at energies much beyond that currently available in
accelerators. The same would apply in the case of processes involving
topological defects that can be clustered in the GH (as in the case of
monopolonium). 

In addition to offering a variety of different kinds of probes of new
physics, the measurement of the EHECR spectrum by the next generation 
large detectors will be expected to yield significant new information
about fundamental physical processes in the Universe today. For example, 
observations of images and spectra of EHECR by
the next generation experiments will be able to yield new information
on Galactic and especially the poorly known extragalactic magnetic
fields~\cite{kronberg} and possibly their origin. 
This could lead to the discovery of a large scale primordial
magnetic field which potentially could open still another new window into
processes occurring in the early Universe.

\section*{Acknowledgments}

\addcontentsline{toc}{section}{\protect\numberline{}{Acknowledgments}}

We are most grateful to the late David Schramm whose insights, encouragements
and constant support had been crucial to us in our efforts in
this exciting area of research over the past several years. Indeed it was
he who first urged us to undertake the project of writing this Report. 
We also wish to thank
Felix Aharonian, Peter Biermann, Paolo Coppi, Veniamin Berezinsky,
Chris Hill, Karsten Jedamzik, Sangjin Lee, Martin Lemoine,
Angela Olinto, (the late) Narayan Rana, Qaisar Shafi, Floyd Stecker, and
Shigeru Yoshida for
collaborations at various stages. We wish to thank Shigeru Yoshida for
providing us with the figures containing the results of various UHECR
experiments. In this context we also thank Peter Biermann,
Murat Boratav, Roger Clay, Raymond Protheroe, Andrew Smith,
David Seckel and Masahiro Teshima for allowing us to use various figures
from their papers in this Review. We acknowledge stimulating
discussions with colleagues and collaborators mentioned above, 
as well as with Ivone Albuquerque, John Bahcall, Pasquale Blasi,
Silvano Bonazola, Brandon Carter, Daniel Chung, Jim Cronin,
Arnon Dar, Luke Drury, Torsten Ensslin, Glennys Farrar, Raj Gandhi,
Haim Goldberg, Yuval Grossman, Francis Halzen, Mark Hindmarsh,
Michael Kachelrie\ss, Karl-Heinz Kampert, Edward Kolb, Thomas Kutter,
Vadim Kuzmin, Eugene Loh, Norbert Magnussen, Alfred Mann, Karl Mannheim,
Hinrich Meyer, Motohiko Nagano, Patrick Peter, Rainer Plaga,
Clem Pryke, J\"org Rachen, Georg Raffelt, Esteban Roulet, Wolfgang Rhode,
Ina Sarcevic, Subir Sarkar, Paul Sommers, Parameshwaran Sreekumar,
Leo Stodolsky, Tanmay Vachaspati, Heinz V\"olk, Eli Waxman, Tom Weiler,
Gaurang Yodh, Enrice Zas, Arnulfo Zepeda, and Igor Zheleznyk.
PB wishes to thank
Ramanath Cowsik and Kumar Chitre for their interest and encouragement, and
Arnold Wolfendale for communications. PB was supported at
NASA/Goddard by a NAS/NRC Resident Senior Research Associateship award
during a major part of this work, and he thanks John Ormes and Floyd
Stecker for the hospitality and for providing a friendly and stimulating
work environment at LHEA (Goddard). PB also acknowledges support under the
NSF US-India collaborative research grant (No. INT-9605235) at the
University of Chicago. GS was supported by DOE, NSF and NASA at the
University of Chicago, by the Centre National de la Recherche
Scientifique at the Observatoire de Paris-Meudon, and by the
Max-Planck-Institut f\"ur Physik in Munich, Germany.

\clearpage

\addcontentsline{toc}{section}{\protect\numberline{}{Figures}}

\clearpage

\clearpage

\clearpage

\clearpage

\clearpage

\clearpage

\clearpage

\clearpage

\clearpage

\clearpage

\clearpage

\clearpage

\clearpage

\clearpage

\clearpage

\clearpage

\clearpage

\clearpage

\clearpage

\clearpage

\begin{figure}[ht]
\bigskip
\caption[...]{Angular image of a point-like source in a magnetized
Supergalactic Plane, corresponding to
one particular magnetic field realization with
a maximal magnetic field in the plane center, 
$B_{\rm max}=5\times10^{-8}\,$G,
all other parameters being the same as in Fig.~\ref{F4.13}.
The image is shown in different energy ranges, as indicated,
as seen by a detector of $\simeq1^\circ$ angular resolution.
A transition from several images at lower energies to only one image at the
highest energies occurs where the linear deflection becomes
comparable to the effective field coherence length.
The difference between neighboring shade levels is 0.1 in the
logarithm to base 10 of the integral flux per solid angle.}
\label{F4.14}
\end{figure}

\clearpage

\begin{figure}[ht]
\bigskip
\caption[...]{Angular distribution in Galactic
coordinates in scenarios where the UHECR sources with
spectral index $\gamma=2.4$ are
distributed according to the matter density and r.m.s.
magnetic field strength in the Local
Supercluster, following a pancake profile with scale
height of 5 Mpc and scale length and maximal field
strength $B_{\rm max}$ in the plane center as indicated.
The observer is within 2 Mpc of the Supergalactic Plane whose location is
indicated by the solid line and at a distance $d=20\,$Mpc from the
plane center. The absence of sources within 2 Mpc from the observer
was assumed. The color scale shows the intensity per solid angle,
and the distributions are averaged over 4 magnetic field realizations
with 20000 particles each.}
\label{F4.15}
\end{figure}

\clearpage

\begin{figure}[ht] 
\centering\leavevmode 
\epsfxsize=6.0in
\epsfbox{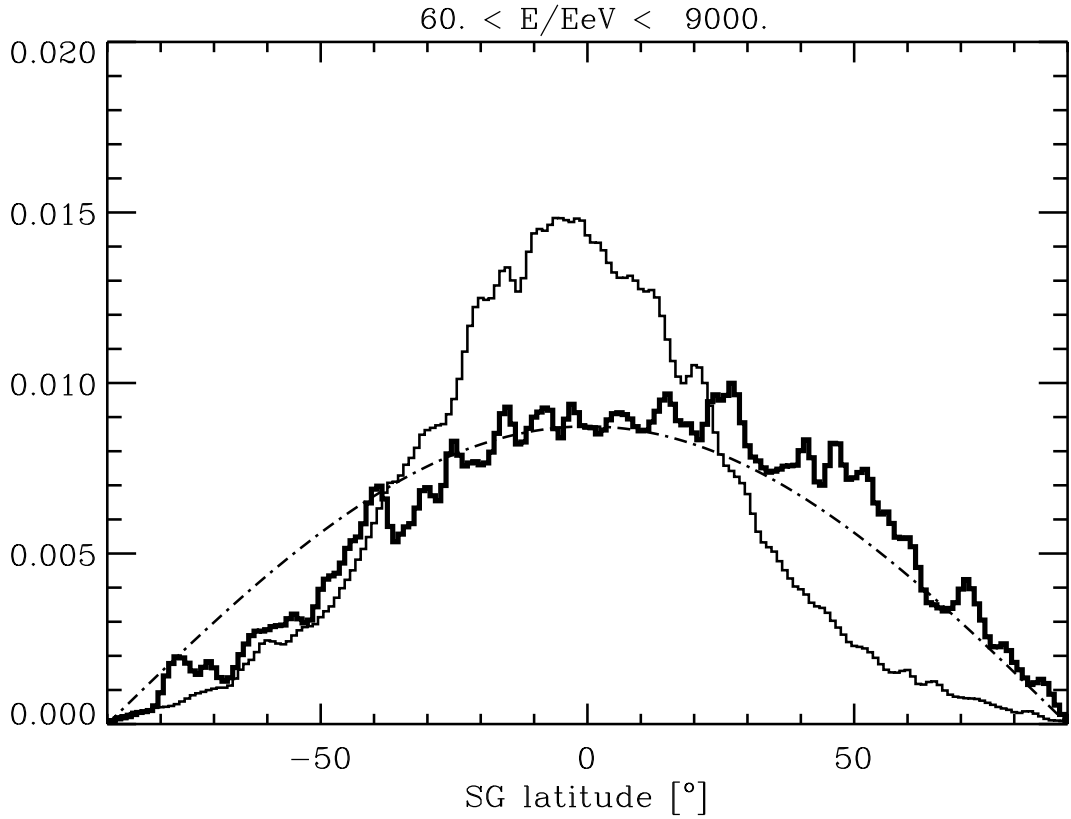}
\bigskip
\caption[...]{The distribution of events above $6\times10^{19}\,$eV
in Supergalactic latitude for two
scenarios with a diffuse source distribution in a magnetized
Supergalactic Plane, for $B_{\rm max}=0.5\,\mu\,$G
(thick histogram), and for $B_{\rm max}=0.05\,\mu\,$G
(thin histogram), assuming
$1.6^\circ$ angular resolution. The full angular distributions
for these cases were shown in the lower and upper panel of
Fig.~\ref{F4.15}, respectively. The dash-dotted curve represents a
completely isotropic distribution.} 
\label{F4.16}
\end{figure}

\clearpage

\begin{figure}[ht]
\centering\leavevmode
\epsfxsize=6.0in
\epsfbox{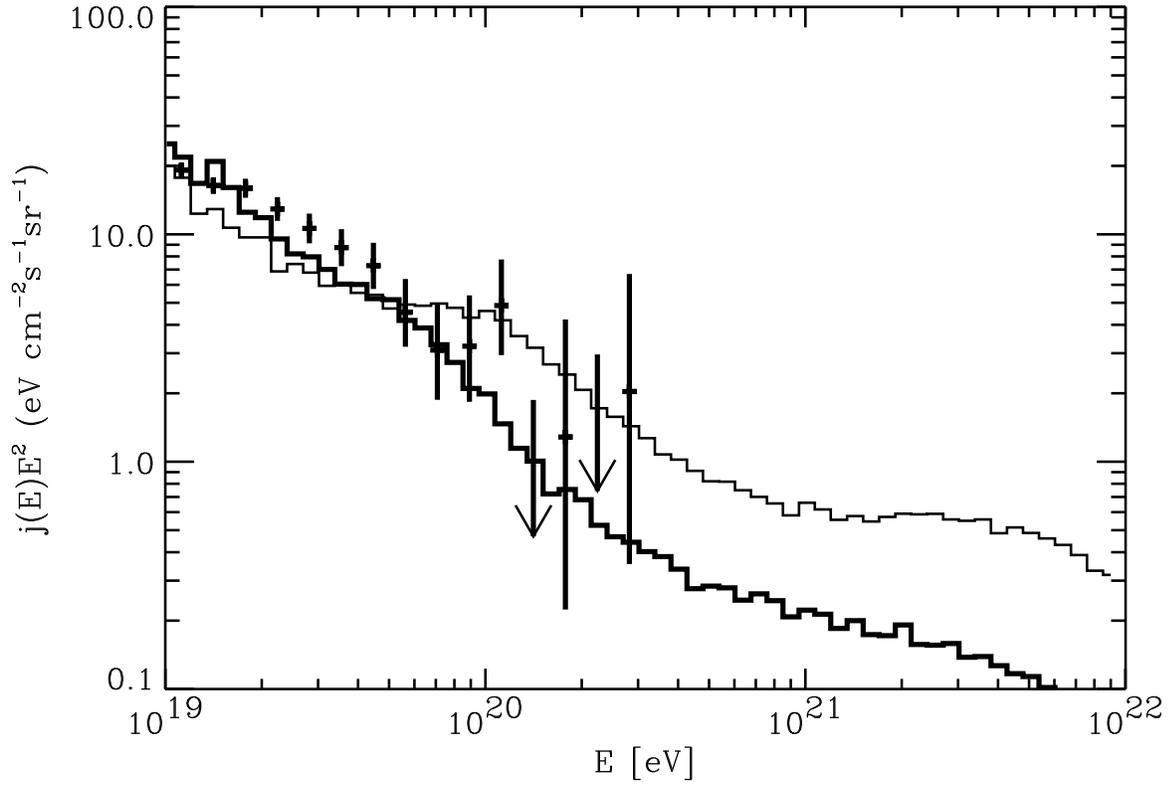}
\bigskip
\caption[...]{Best fit to the data above $10^{19}\,$eV
of the spectra predicted by two scenarios with diffuse sources
in a magnetized Supergalactic Plane whose predicted angular
distributions were shown in the top and bottom panel of
Fig.~\ref{F4.15} and in Fig.~\ref{F4.16}. The thick histogram
is for $B_{\rm max}=0.5\,\mu\,$G,
with the observer $2\,$Mpc above the Supergalactic Plane,
and the thin histogram is for $B_{\rm max}=0.05\,\mu\,$G, with
the observer in the plane center. The cosmic variance between
different realizations is negligible.}
\label{F4.17}
\end{figure}

\clearpage

\clearpage

\clearpage

\clearpage

\clearpage

\clearpage

\clearpage

\clearpage

\end{document}